\documentclass{ws-revol}
\newcommand{\ud}{\mathrm{d}}

\newcommand{\boldarrayrulewidth}{1\p@}
\def\bhline{\noalign{\ifnum0=`}\fi\hrule \@height
\boldarrayrulewidth \futurelet \@tempa\@xhline}

\def\@xhline{\ifx\@tempa\hline\vskip \doublerulesep\fi
      \ifnum0=`{\fi}}

\newcommand{\bs}{\noalign{\vspace{6\p@ plus2\p@ minus2\p@}}}
\newcommand{\es}{\noalign{\vspace{6\p@ plus2\p@ minus2\p@}}\displaystyle}%
\newcommand{\AmS}{{\protect\the\textfont2
  A\kern-.1667em\lower.5ex\hbox{M}\kern-.125emS}}

\newcommand{\ber}{\begin{eqnarray}}
\newcommand{\eer}{\end{eqnarray}}
\newcommand{\be}{\begin{eqnarray}}
\newcommand{\ee}{\end{eqnarray}}

\def\Toprel#1\over#2{\mathrel{\mathop{#2}\limits^{#1}}}

\makeindex

\makeatletter
\def\@dottedtocline#1#2#3#4#5{\ifnum #1>\c@tocdepth \else
  \vskip \z@ \@plus.2\p@
  {\leftskip #2\relax \rightskip \@tocrmarg \parfillskip -\rightskip
    \parindent #2\relax\@afterindenttrue
   \interlinepenalty\@M
   \leavevmode
   \@tempdima #3\relax
   \advance\leftskip \@tempdima \null\hskip -\leftskip
    {#4}\nobreak
    \hfill \nobreak
           \hb@xt@\@pnumwidth{%
             \hfil\normalfont \normalcolor #5}\par}\fi}
\def\numberline#1{\hb@xt@\@tempdima{#1.\hfil}}
\makeatother

\begin{document}

\markboth{Peter Braun-Munzinger, Krzysztof  Redlich, Johanna
Stachel} {Particle Production in Heavy Ion Collisions}

\title{Particle Production in Heavy Ion Collisions }

\author{Peter Braun-Munzinger$^a$, Krzysztof  Redlich$^{b,c}$, Johanna  Stachel$^d$}

\address{$^a$ Gesellschaft f\"ur Schwerionenforschung, GSI, D-64291
Darmstadt, Germany\\
[-2pt] E-mail: p.braun-munzinger@gsi.de
\\
$^b$
Fakult\"at f\"ur Physik, Universit\"at Bielefeld,\\
Postfach 100 131, D-33501 Bielefeld, Germany
 \\
$^{c}$  Institute of Theoretical Physics, University of Wroc\l aw,\\
PL-50204 Wroc\l aw, Poland\\
[-2pt] E-mail: redlich@ift.uni.wroc.pl\\
$^{d}$ Physikalisches Institut der Universit\"at Heidelberg,\\
D 69120 Heidelberg, Germany\\
[-2pt] E-mail: stachel@physi.uni-heidelberg.de}

\maketitle

\begin{abstract}
The status of thermal model    descriptions  of particle
production in heavy ion  collisions is presented.  We discuss the
formulation of statistical models with different implementation of
the conservation laws and indicate their applicability in heavy
ion and elementary particle collisions. We
   analyze experimental data on hadronic abundances obtained in
ultrarelativistic heavy ion collisions, in a very broad energy
range starting from RHIC/BNL ($\sqrt s=200$ A GeV), SPS/CERN
($\sqrt s\simeq 20$ A GeV) up to AGS/BNL ($\sqrt s\simeq 5$ A GeV)
and SIS/GSI ($\sqrt s\simeq 2$ A GeV) to test equilibration of the
 fireball created in the collision. We argue  that
the statistical approach provides a very satisfactory description
of experimental data covering this wide energy range. Any
deviations of the model predictions from the data are  indicated.
 We discuss the unified description of
particle chemical freeze--out and the excitation functions of
different particle species. At SPS and RHIC energy the relation of
freeze--out parameters with the QCD phase boundary is  analyzed.
Furthermore, the application of the extended statistical model to
quantitative understanding of open and hidden charm  hadron yields
is considered.
\end{abstract}
\newpage \tableofcontents

\section{Introduction}

%\section{History and Applications of Statistical Model for Particle Production}
The ultimate goal of the physics program with ultrarelativistic
nucleus--nucleus collisions
 is to study the properties of strongly interacting
matter under extreme conditions of  high energy density. Quantum
Chromodynamics (QCD) predicts\cite{r1}$^-$\cite{gery1} that
strongly interacting matter undergoes a phase transition from a
state of hadronic constituents to a plasma  of deconfined  quarks
and gluons
%\cite{r1,gery,gery1}.
(QGP). By colliding heavy ions at ultrarelativistic energies, one
expects to create  matter under  conditions that  are sufficient
for
%\cite{r1,r3,r4,r5,uh,r28}.
deconfinement\cite{r1}$^-$\cite{r28}.
 Thus, of particular relevance is
finding experimental probes to check whether the produced medium
in its early stage was indeed in the QGP phase. Different probes
have been studied  with the various  SPS/CERN and  RHIC/BNL
experiments. The most promising signals of deconfinement are
related to particular properties of the transverse momentum
spectra of  photons\cite{alam,dinesh},
%\cite{rapp,vesa,vesa1,vesa2,red,hatsuda}
dileptons\cite{rapp}$^-$\cite{hatsuda}, and hadrons
%\cite{uh,r2,wideman,tomasik}
\cite{uh,r2}$^-$\cite{tomasik}. The photon rate is studied to
probe the temperature evolution  from formation to decoupling of
the fireball, implying sensitivity to a high temperature
deconfined phase.
%
%  The
%photon rate is expected to be enhanced if the QGP were  formed in
%the initial state.
 The invariant mass distribution of dileptons is
expected to be modified by in-medium effects related to chiral
symmetry
%\cite{gery,gery1,red,hatsuda,medium,medium1,medium2}.
restoration\cite{gery,gery1,red,hatsuda,medium}$^-$\cite{medium2}.
The modification of charmonium production was argued to be a
consequence of collective effects in the  deconfined
medium\cite{r1,satz}.

Hadron multiplicities  and their correlations are observables
which can provide  information on the nature, composition, and
size of the medium from which they are originating. Of particular
interest is the extent to which the measured particle yields are
showing equilibration. The appearance of the QGP, that is a
partonic medium being at (or close to) local thermal equilibrium
and its subsequent hadronization during the phase transition
should in general drive hadronic constituents towards chemical
equilibrium\cite{r3,r4,uh,knol}. Consequently, a high level of
chemical saturation, particularly for strange
particles\cite{r34,r35}, could be   related to the deconfined
phase created at the early stage of heavy ion collisions.

The level of equilibrium of secondaries in heavy ion collisions
was tested by analyzing the particle
%\cite{r3,r4,uh,r9,r10,r11,r12,r13,r14,anti,rn13,r15,r17,r18,r21,r31,r31n,r32,r33,r42,r53,r54,becalast,r59,r60,r61}
abundances\cite{r3,uh,r9}$^-$\cite{r612} or their momentum spectra
%\cite{uh,r2,r11,wideman,tomasik,r18,r19}.
\cite{uh,r2,wideman}$^-$\cite{xus,r11,r18,r19}. In the {\rm first}
case one establishes the chemical composition of the system, while
in the second  case additional information on dynamical evolution
and collective flow can be extracted.

In this review  we will discuss the formulation of statistical
models and their applications to a phenomenological description of
particle production in nucleus--nucleus collisions. We emphasize
the importance of conservation laws and their different
implementations in the statistical approach. We
   analyze experimental data on hadronic abundances obtained in
ultrarelativistic heavy ion collisions, in a very broad energy
range starting from RHIC/BNL ($\sqrt s=130$ A GeV), SPS/CERN
($\sqrt s\simeq 20$ A GeV) down  to AGS/BNL ($\sqrt s\simeq 5$ A
GeV) and SIS/GSI ($\sqrt s\simeq 2$ A GeV) to test equilibration.
We argue  that the statistical approach  provides a very
satisfactory description of experimental results covering this
wide energy range.
 We further provide arguments for a unified description of
 chemical freeze--out of hadrons and discuss  excitation functions of different
particle species. An extension  of the model for a quantitative
understanding of open and hidden charm particle yields will be
also discussed.

%%%%%%%%%%%%%%%%%%%%%%%%%%%%%%%%%%%%%

\subsection{Initial conditions in A--A  collisions and deconfinement}
\begin{figure}[htb]
%\begin{minipage}[t]{80mm}
%\framebox[79mm]{\rule[-26mm]{0mm}{52mm}}
 {\hskip -0.8cm\vskip -0.5cm
\center\includegraphics[width=28.5pc, height=15.9pc]{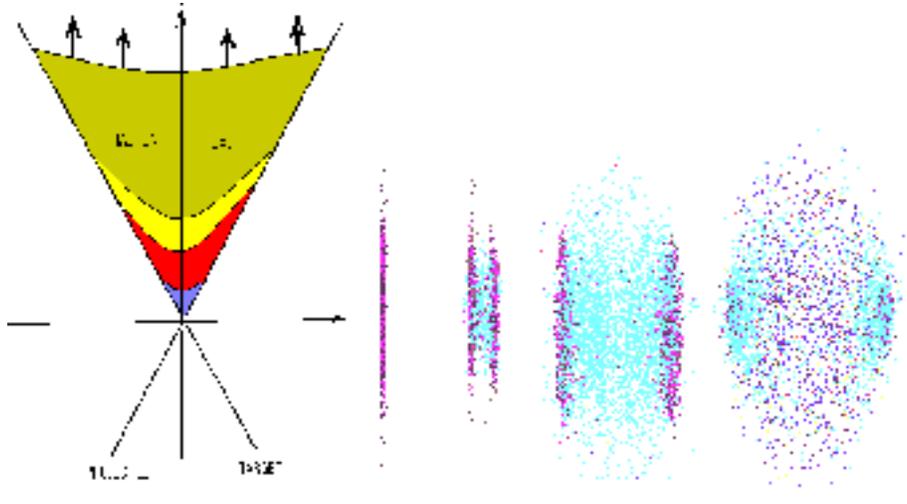}}\\
\vskip -0.5cm { \caption{Schematic space--time view of a  heavy
ion collision that indicates four basic stages in the evolution of
the collision fireball: initial overlap region, pre--equilibrium
partonic system, equilibrated quark-gluon plasma and its
subsequent hadronization to a hadron gas.
 }
  \label{sche}}
%\end{minipage}
\end{figure}

%%%%%%%%%%%%%%%%%%%%%%%%%%%%%%%%%%%%%%%%%%%%%%%%%%%%%%%%%%%%%%%%%%%%%%%%%%%%%%%%%%%%%%
In ultrarelativistic heavy ion collisions, the knowledge of the
critical energy density $\epsilon_c$ required for deconfinement
 as well as the equation of state (EoS)
of strongly interacting matter are of particular importance. The
value of $\epsilon_c$ is needed to establish the necessary initial
conditions in heavy ion collisions to possibly create the QGP,
whereas the EoS is required as an input to describe  the
space-time evolution of the collision fireball\footnote{ In
Fig.~(\ref{sche}) we show a schematic view of the space-time
evolution of heavy ion collisions in the Bjorken
model\cite{r26}.}.

%%%%%%%%%%%%%%%%%%%%%%%%%%%%%%f222222222222222222222222222222222222222222%%%%%%%%%%%%%%%%%%%%
%%%%%%%%%%%%%%%%%%%%%%%%%%%%%%1%%%%%%%%%%%%%%%%%%%%

Both of these  pieces of information can be obtained today from
 first principle  calculations by formulating QCD on the lattice
and performing   Monte-Carlo simulations. In Fig.~({\ref{feqs}})
we show the most recent results\cite{r23} of  lattice gauge theory
(LGT) for the temperature dependence of energy density and
pressure. These results have been obtained in LGT for different
numbers of dynamical fermions. The energy density is seen in
Fig.~({\ref{feqs}) to exhibit the typical behavior of a system
with a phase transition\footnote{ In a strictly statistical
physics sense a phase transition  in two flavour QCD can only
appear in the limit of massless quarks where it is of second
order. In three flavour QCD, with (u,d,s) quarks, the phase
transition and its order depends on the value of the quark masses.
In general it can be a first order, second order or cross--over
transition. For physical quark masses, both the value of the
transition temperature and the order of the deconfinement phase
transition are still not well established. }: an abrupt change in
a very narrow temperature range. The corresponding pressure curve
shows a smooth change with temperature. In the region below $T_c$
the basic constituents of QCD, quarks and gluons, are confined
within hadrons and here the EoS is well parameterized\cite{fkk} by
a hadron resonance gas. Above $T_c$ the system appears in the QGP
phase where quarks and gluons can travel  distances that
substantially exceed    the  typical size of hadrons. The most
recent results of improved perturbative expansion of the
thermodynamical potential in  continuum QCD indicate\cite{r24,r66}
that, at some distance above $T_c$, the EoS of QGP can be well
described by a gas of massive quasi-particles with a temperature
dependent mass. In the vicinity of $T_c$ the relevant degrees of
freedom were argued\cite{r25,r65} to be described by Polyakov
loops.

Lattice Gauge Theory  predicts, in two--flavour QCD, a critical
temperature $T_c=173\pm ~8$MeV and   corresponding critical energy
density $\epsilon_c=0.6\pm 0.3$ GeV/fm$^3$ for the deconfinement
phase transition\cite{r23}. The value of $\epsilon_c$ is
surprisingly low and corresponds quantitatively to the energy
density inside the nucleon.
%%%%%%%%%%%%%%%%%%%%%%%%%%%%%%f11111111%%%%%%%%%%%%%%%%%%%%
%%%%%%%%%%%%%%%%%%%%%%%%%%%%%%f11111111%%%%%%%%%%%%%%%%%%%%
\begin{figure}[htb]
%\vskip 1.0 true cm
\begin{minipage}[t]{60mm}
\includegraphics[width=12.5pc, height=14.3pc]{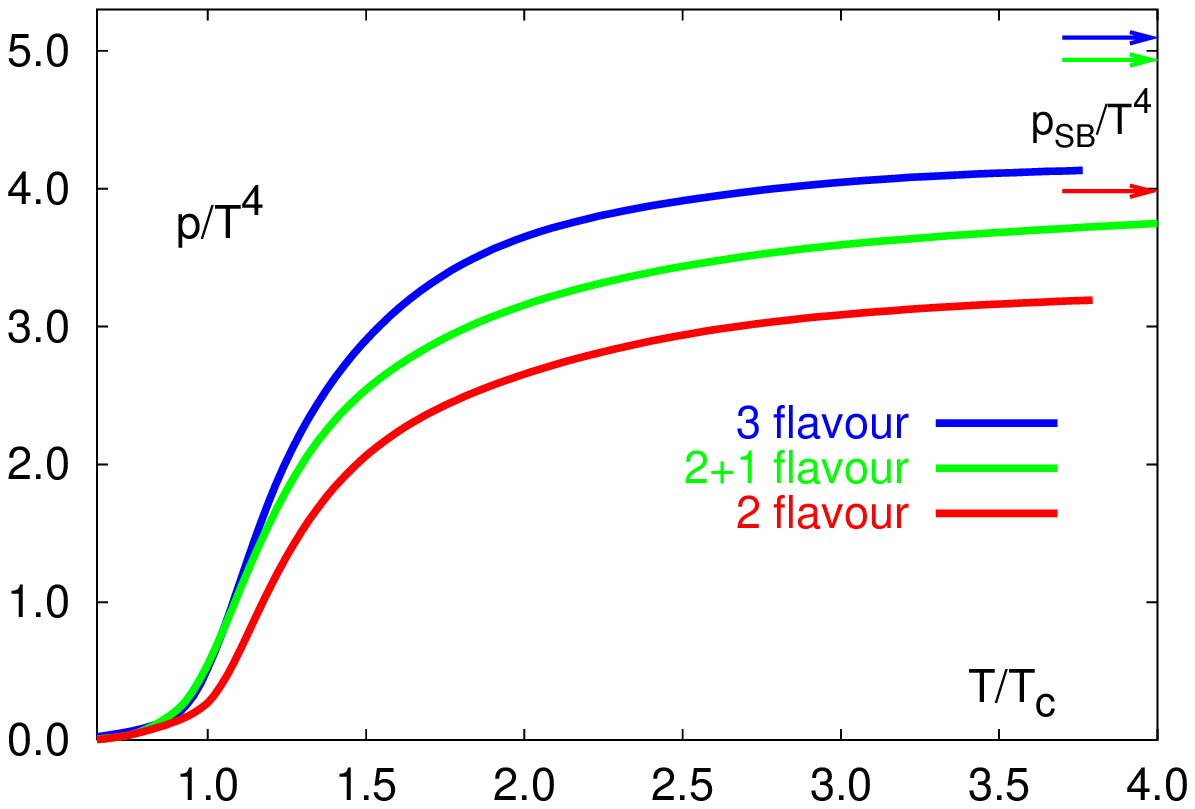}\\
%\caption{ }
\end{minipage}
%
%\hspace{\fill}
%
\begin{minipage}[t]{60mm}
\includegraphics[width=12.5pc, height=14.2pc]{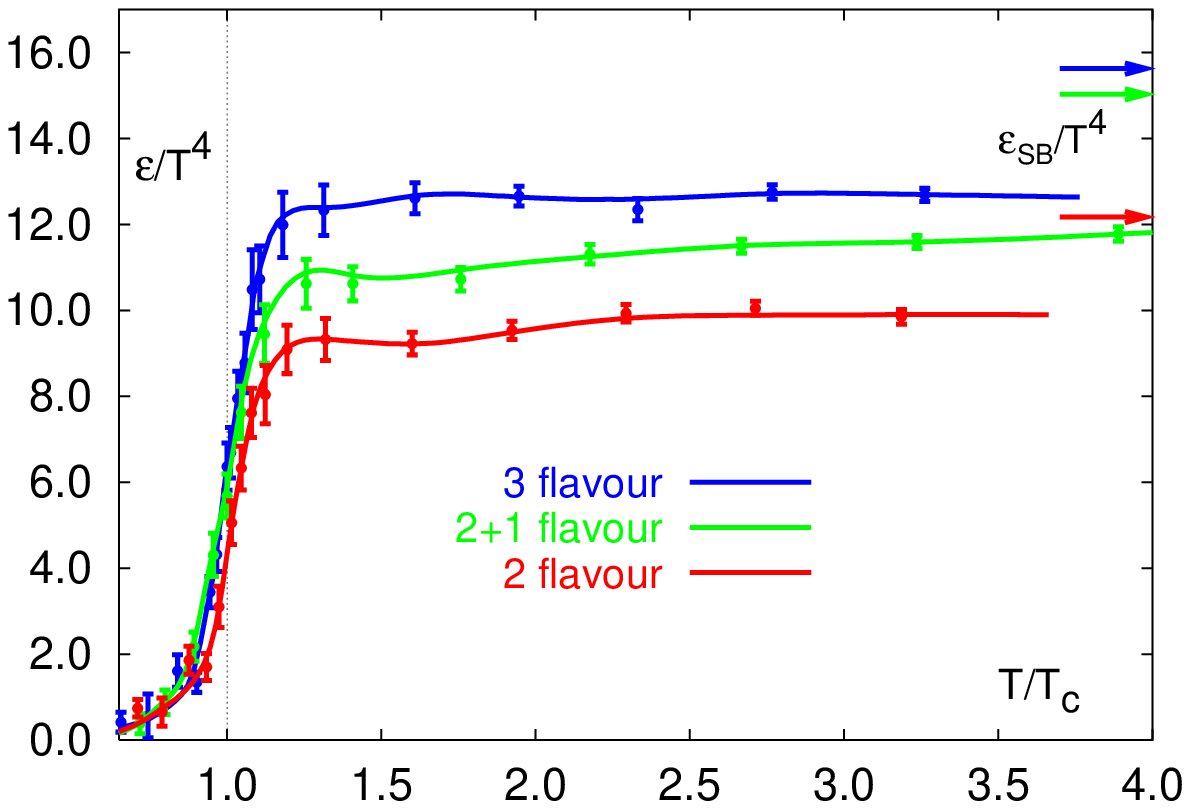}\\
 \end{minipage}
\begin{minipage}[t]{114mm}
\caption{\label{feqs} The pressure $P$ and energy density
$\epsilon$ normalized to the temperature to the fourth power,
versus temperature normalized to its critical value. The
calculations\protect\cite{r23} were performed  within LGT for
different numbers of flavors.
  The values of the corresponding ideal gas results  are indicated by the arrows. }

\end{minipage}
\end{figure}
%%%%%%%%%%%%%%%%%%%%%%%%%%%%%%1%%%%%%%%%%%%%%%%%%%%
 The initial energy density reached in  heavy ion collisions can
be estimated within the Bjorken model\cite{r26}. From
%the
%rapidity distribution of protons and  their
the transverse energy $E_T$ measured in nucleus--nucleus
collisions the initial energy density $\epsilon_0$ is determined
as

%%%%%%%%%%%%%%%%%%%
 \be
\epsilon_0(\tau_0)={1\over {\pi R^2}}{1\over \tau_0}{{dE_T}\over
{dy}}, \label{eqq1}
 \ee
%%%%%%%%%%%%%%%%%%%%%%%%%%%%%%%%%%%%%%%%%%%%%%%%%%%%%%%%%
where the  initially produced collision fireball is considered as
a cylinder of length $dz=\tau_0dy$ and transverse radius  $R\sim
A^{1/3}$. Inserting for $\pi R^2$ the overlap area of colliding Pb
nuclei together with an assumed  initial time of $\tau_0\simeq 1$
fm, and using an average transverse energy at midrapidity
measured\cite{400} at the SPS ($\sqrt s=17.3$ GeV) to be 400 GeV,
one obtains
 \be
\epsilon_0^{SPS}(\tau_0\simeq 1~ {\rm fm})\simeq 3.5\pm 0.5 ~{\rm
GeV/{\rm fm}^3}. \label{eq2}
 \ee

Increasing the collision energy to $\sqrt s=130$ A$\cdot$GeV for
Au--Au at RHIC and keeping the same initial thermalization time as
at the SPS, would  increase  $\epsilon_0$ by only 50--60 $\%$.
However, at RHIC the thermalization time was  argued in terms of
different models\cite{r28,hot}
%within
%saturation models,
to be shorter by a factor of 3--5.

In the context of saturation models\cite{r28,r27}$^-$\cite{r30}
the thermalization time can be possibly  related with the
saturation scale\cite{r5,r28}. The basic concept of the saturation
models is the conjecture  that there is some transverse momentum
scale $p_{\rm sat}$ where the gluon and quark phase space density
saturates\cite{r28,r27}$^-$\cite{r30}.  For an isentropic
expansion of the collision fireball,   the transverse energy at
$p_{\rm sat}$ was  related in  Ref.~(\refcite{r28}) to that
measured in nucleus--nucleus collisions in the final state. The
saturation scale was also used  to fix the thermalization time as
$\tau_{eq}\simeq 1/p_{\rm sat}$. Taking the value of $p_{\rm sat}$
predicted in Ref.~(\refcite{r28}) for RHIC energy, $p_{\rm
sat}\simeq 1.13$ GeV, one gets $\tau_{eq}\simeq 0.2$ fm and a
corresponding energy density $\epsilon_{eq}\simeq 98$ GeV/fm$^3$.
%This value is much larger  than the one obtained  from Eq.2 at
%$\tau_0\sim1$fm.
This is  a larger value than expected for the initial energy
density at RHIC  in
% which is model--dependent.
the McLerran--Venugopalan model\cite{r29} where
$\epsilon_0^{RHIC}\sim 20$ GeV/fm$^3$,  also in agreement  with
the prediction of Ref.~(\refcite{r30}).

 At SPS energy the
saturation model described in Ref.~(\refcite{r28}) leads to
$\epsilon_{eq}^{SPS}\sim 16$ GeV/fm$^3$, a  much higher value than
that obtained from Eq.~(\ref{eq2}). The estimate of
$\epsilon_{eq}$ and initial thermalization time strongly depends
on the value of $p_{\rm sat}$ and the model assumptions. In a $"$
bottom--up $"$ equilibration scenario\cite{r5} the thermalization
time in Au--Au collisions at RHIC energy was estimated to be as
large as 3.2--3.6 fm and the temperature $T\simeq$(210--230) MeV.
Nevertheless, this initial temperature still corresponds to an
energy density by factor of 2--3 larger then that required for
deconfinement. In A--A collisions at the LHC the initial energy
density of the equilibrated partonic medium is
expected\cite{r28,lhci} to be in the range
$400<\epsilon_{eq}^{LHC}< 1300$ GeV/fm$^3$.

%It is thus clear that  there are large uncertainties on the value
%of the initial energy density reached in ultrarelativistic heavy
%ion collisions. In Pb--Pb collisions at the SPS according to the
%models one gets $2.5 ~{\rm GeV/fm^3}< \epsilon_0^{SPS}< 16~ {\rm
%GeV/fm^3}$ whereas in Au--Au collisions at RHIC,~  $20 {\rm
%~GeV/fm^3}< \epsilon_0^{RHIC}< 100~ {\rm GeV/fm^3}$.

The dominant constituents of the partonic medium produced in
ultrarelativistic heavy ion collisions at LHC, RHIC and even at
SPS energy are gluons. The energy density of gluons in thermal
equilibrium scales with the fourth power of the temperature
$\epsilon=gT^4$, where $g$ is related to  the number of degrees of
freedom. For an ideal gluon gas, $g=16\pi^2/30$; in an interacting
system, the effective number of degrees of freedom $g$ is smaller.
The results of LGT shown in Fig.~{\ref{feqs}} indicate deviations
from the Boltzmann limit by 20--25 $\%$. Relating the thermal
energy density with the initial energy density discussed above,
one can make an estimate of the initial temperature reached in
heavy ion collisions. For the SPS, RHIC and LHC energies  this
gives a temperature in the range: $200~  {\rm MeV} < T^{SPS} <
330$ MeV, $210~  {\rm MeV} < T^{RHIC} < 600$ MeV and $1000 ~ {\rm
MeV}< T^{LHC}< 1200 $ MeV, respectively.

Comparing the  initial energy density  expected in heavy ion
collisions with   LGT results, it is clear, that  the initial
energy density at LHC and RHIC  by far exceeds the critical value.
% At  SPS the
%system can appear only slightly above $\epsilon_c$, however,
%still in a quark gluon plasma phase.
%Thus, the necessary conditions to create the partonic medium in a
%deconfined phase are reached  at LHC, RHIC as well as  at the  top
%SPS energy.
A  large energy density is, however,  still not sufficient to
create a QGP. The distribution of initially produced gluons is
very far from being thermal, thus the system needs enough time to
equilibrate. Recently it was shown\cite{r5},  in the framework of
perturbative QCD and kinetic theory, that the equilibration of
partons should definitely happen   at  LHC and most likely at RHIC
energy.

A previous microscopic study\cite{pc} within the Parton Cascade
Model has led to the conclusion that thermalization can also be
reached even at the lower SPS energy. Here, however, due to the
relatively low collision energy, it is not clear whether a model
inspired by perturbative QCD is indeed applicable.

Assuming QGP formation in the initial state in heavy ion
collisions one  expects that the thermal nature of the partonic
medium could be preserved during  hadronization.\footnote{The fact
that the phase transition is a driving force towards equilibration
is found\cite{knol,biro} e.g.  in  different kinetic models for
QGP evolution and hadronization.}  Consequently, the particle
yields measured in the final state should resemble a thermal
equilibrium population.
%In the following, we present the recent results
%related with the question of equilibration of secondaries in heavy
%ion collisions and discuss their  possible relation with
% deconfinement.

%%%%%%%%%%%%%%%%%%%
%\section{Statistical approach - grand canonical  formalism}
\section{Statistical approach - general remarks}

In the approach of Gibbs (see, e.g., Ref.~(\refcite{huang})) the
equilibrium behavior of thermodynamical observables can be
evaluated as an average over statistical ensembles (rather than as
a time average for a particular state). The equilibrium
distribution is thus obtained by an average over all accessible
phase space. Furthermore, the ensemble corresponding to
thermodynamic equilibrium is that for which the phase space
density is uniform over the accessible phase space. In this sense,
filling the accessible phase space uniformly is both a necessary
and sufficient condition for equilibrium. Consequently, the
agreement between observables  and predictions using the
statistical operator  imply equilibrium (to the accuracy with
which agreement is observed). "Filling phase space" is not a
different statement, although it is often and erroneously used in
the literature.

In our further analysis  we use in the statistical operator as
Hamiltonian that leading to the full hadronic mass spectrum. In
some sense this is synonymous with using the full QCD Hamiltonian.
The only parameters in the statistical operator describing the
grand-canonical ensemble are temperature $T$ and baryon chemical
potential $\mu_B$. There is no room here for strangeness
suppression ($\gamma_s$) factors. So the interpretation is that
agreement between data and theoretical predictions implies
statistical equilibrium at temperature T and chemical potential
$\mu_B$. If an additional factor $\gamma_s$ is needed to describe
the data this implies a clear deviation from chemical equilibrium:
a state in which e.g. strangeness is suppressed compared to the
equilibrium value implies additional dynamics not contained in the
statistical operator and not consistent with uniform phase space
density. Similar arguments, of course, apply if one uses canonical
phase space. If, in this regime, canonically calculated particle
ratios agree with those measured, this implies equilibrium at
temperature $T$ and over the canonical volume $V$. To the extent
that this describes data for e+e- or pp collisions, the same
conclusions on thermodynamic equilibrium apply. However, we note
that, in this approach e.g., particles ratios involving particles
with hidden strangeness are generally not well predicted, again
implying non-equilibrium behavior.

\subsection{Statistical approach - grand canonical  formalism}

%\subsection{Partition function and particle yields }
The basic quantity required to compute the thermal composition of
particle yields  measured in heavy ion collisions is   the
partition function $Z(T,V)$. In the Grand Canonical (GC) ensemble,
\be Z^{GC}(T,V,\mu_Q) = {\rm Tr}[e^{-\beta (H -\sum_i\mu_{Q_i}
Q_i)}], \label{eq3} \ee where  $H$ is the Hamiltonian of the
system, $Q_i$ are the conserved charges and  $\mu_{Q_i}$  are the
chemical potentials that guarantee that the charges $Q_i$ are
conserved on the average in the whole system. Finally $\beta =1/T$
is the inverse temperature. The Hamiltonian is usually taken such
as to describe a hadron resonance gas. For practical reasons, the
hadron mass spectrum contains contributions from all mesons with
masses below $\sim$1.5 GeV and baryons with masses below $\sim$2
GeV. In this mass range the hadronic spectrum is well established
and the decay properties of resonances are reasonably well
known\cite{particle}. This mass cut in the contribution of
resonances to the partition function limits, however, the maximal
temperature to $T_{\rm max}\simeq 200$ MeV, up to which the model
predictions may be considered trustworthy\cite{r11,r12,rn13,r60}.
For higher temperatures the contributions of ( in general poorly
known) heavier resonances are not negligible. The interaction of
hadrons and resonances are usually only included by implementing a
hard core repulsions, i.e. a Van der Waals--type interaction.
Details of such a implementation are discussed below. The main
motivation of using the Hamiltonian of a hadron resonance gas in
the partition function is that  it contains all relevant degrees
off  freedom of the confined, strongly interacting medium and
implicitly includes interactions that  result  in resonance
formation. Secondly, this model is consistent with  the equation
of state obtained from the LGT below the critical
temperature\cite{fkk,fkk1}. In a strongly interacting medium, one
includes the conservation of electric charge, baryon number and
strangeness. The GC partition function (\ref{eq3}) of a hadron
resonance gas can then be written as a sum of partition functions
$\ln Z_i$ of all hadrons and resonances \be
 \ln Z(T,V,\vec\mu )=\sum_i \ln Z_i(T,V,\vec\mu),
\label{par} \ee
 where $\epsilon_i =\sqrt {p^2+m_i^2}$ and
$\vec \mu =(\mu_B,\mu_S,\mu_Q)$ with the chemical potentials
$\mu_i$ related to baryon number, strangeness and
 electric charge, respectively.
For particle $i$ of stran\-geness $S_i$, baryon number $B_i$,
electric charge $Q_i$ and spin--isospin degeneracy factor
$g_i$,\cite{r1j} \be \ln Z_i(T,V,\vec\mu) ={{Vg_i}\over
{2\pi^2}}\int_0^\infty \pm p^2dp \ln [1\pm \lambda_i\exp (-\beta
\epsilon_i)], \label{parp} \ee with (+) for fermions,  (-) for
bosons and fugacity \be \lambda_i(T,\vec\mu )= \exp ({{
B_i\mu_B+S_i\mu_S+Q_i\mu_Q}\over T }) \ee Expanding the logarithm
and  performing the momentum integration in Eq.~(\ref{parp}) we
obtain \be \ln Z_i(T,V,\vec\mu)={{VTg_i}\over
{2\pi^2}}\sum_{k=1}^\infty {{(\pm 1)^{k+1}}\over {k^2}}\lambda_i^k
m_i^2 K_2({{km_i}\over T}), \label{z1ii} \ee where $K_2$ is the
modified Bessel function and the upper sign is for bosons and
lower for fermions. The first term in Eq.~(\ref{z1ii}) corresponds
to the Boltzmann approximation. The density of particle $i$ is
obtained from Eq.~(\ref{z1ii}) as \be n_i(T, \vec\mu )={{{\langle
N_{i}\rangle}\over V}}= {{Tg_i}\over {2\pi^2}}\sum_{k=1}^\infty
{{(\pm 1)^{k+1}}\over {k}}\lambda_i^k m_i^2 K_2({{km_i}\over T}),
\label{denj} \ee
%%%%%%%%%%%%%%%%%%%%%%%%%%%%%f66666666666666
%\vskip -0.8cm
\begin{figure}[htb]
 {\vskip -0.5cm
{\hskip .6 cm
\includegraphics[width=22.pc, height=21.pc,angle=-180]{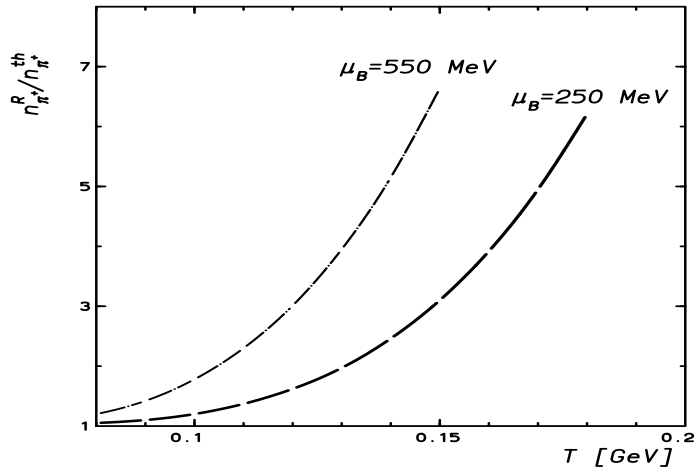}}}\\
{\vskip -1.1cm \caption{\label{res} The ratio of the total density
 of  positively charged
pions that includes all resonance contributions to the density of
thermal pions. The calculations are done in the hadron resonance
gas model for  $\mu_B=$250, 550 MeV   and    for different
temperatures. }}
\end{figure}
The partition function (\ref{par}) is the basic quantity that
allows to describe all thermodynamical properties of a fireball
composed of hadrons and resonances being in thermal and chemical
equilibrium. In view of further application of this statistical
operator to the description of particle production in heavy ion
collisions we write explicitly the results for particle density
obtained from Eq.~(\ref{par}). Of particular importance here  is
to account for resonances and their decay into lighter particles.
The average number $\langle N_i\rangle$  of particles $i$   in
volume $V$ and temperature $T$, that carries strangeness $S_i$,
baryon number $B_i$, and electric charge $Q_i$, is obtained from
Eq.~({\ref{par}}) as \be
 \langle N_i^{}\rangle (T,\vec\mu)~=~\langle N_i\rangle^{th} (T,\vec\mu)
 +{\sum}_j\Gamma_{j\to i}
\langle N_j\rangle^{th,R}(T,\vec\mu)\label{denn} \ee where the
first term describes the thermal average number of particles of
species $i$ and second term describes overall resonance
contributions to particle multiplicity of species $i$. This term
is taken as a sum of all resonances that decay into particle $i$.
The $\Gamma_{j\to i}$ is the corresponding decay branching ratio
of $j\to i$. The corresponding multiplicities in Eq.~(\ref{denn})
are obtained from Eq.~(\ref{denj}). The importance of the
resonance contribution to the total particle yield in
Eq.~(\ref{denn}) is illustrated in Fig.~(\ref{res}) as the ratio
of total to thermal  number of $\pi^+$. From this figure it is
clear that at high temperature (or density) the overall
multiplicity of light hadrons is indeed dominated  by resonance
decays. In the high-density regime, that is for large $T$ and/or
$\mu_B$, the repulsive interactions of hadrons should be included
in the partition function (\ref{par}). To incorporate the
repulsion at short distances one usually uses a hard core
description by implementing excluded volume corrections\cite{r60}.
In a thermodynamically consistent approach\cite{r79} these
corrections lead to a shift of the baryon--chemical potential. .
We discuss below how this is implemented in our calculations. The
repulsive interactions are important when discussing observables
of density type.  Particle density ratios, however, are only
weakly affected\cite{r12} by the repulsive corrections. The
partition function (\ref{par}) depends in general on five
parameters. However, only three are independent, since the isospin
asymmetry in the initial state fixes the charge chemical potential
and the strangeness neutrality condition eliminates the strange
chemical potential. Thus, on the level of particle multiplicity
{\it ratios} we are only left with temperature $T$ and baryon
chemical potential $\mu_B$ as independent parameters. In
Fig.~(\ref{RFS}) we show the relation of $\mu_S=\mu_S (T,\mu_B)$
obtained from the strangeness neutrality condition. For low
temperature this relation is highly non-linear. For larger $T$,
however, $\mu_S$ shows an almost linear dependence on $\mu_B$. One
sees by inspection of Fig.~(\ref{RFS}) that, at $T\sim 200$ MeV
and $\mu_B\sim 300$ MeV, $\mu_S\sim {1\over 3}\mu_B$.  This
relation  is obtained  in a QGP from strangeness neutrality
conditions.  In the present context of a hadron resonance gas this
is a pure accident with no dynamical information.
\begin{figure}[htb]
%\begin{minipage}[t]{80mm}
%\framebox[79mm]{\rule[-26mm]{0mm}{52mm}}
{\hskip -.1cm
\includegraphics[width=17.5pc,height=26.pc,angle=90]{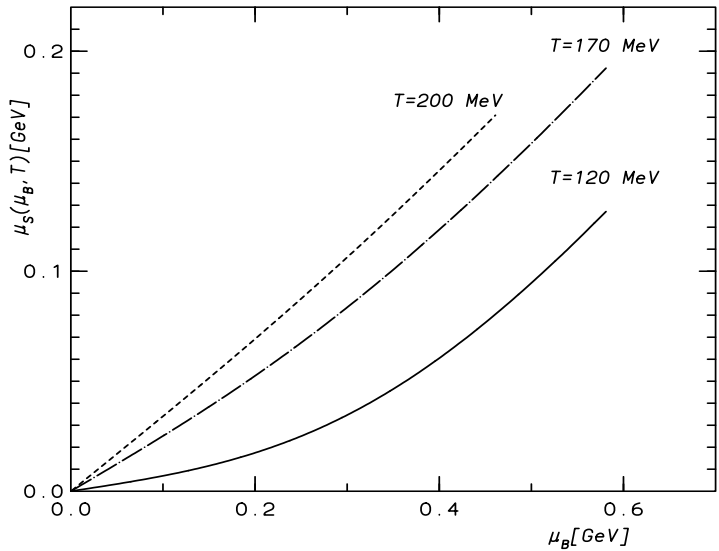}}\\
\vskip -0.5cm { \caption{ \label{RFS} The strange chemical
potential $\mu_S$ as a function of baryon--chemical potential for
T=120,170 and 200 MeV. The results are obtained by imposing the
strangeness neutrality condition in a hadron resonance gas.
 }}
%\end{minipage}
\end{figure}

 At lower energies, in practise for $T< 100$ MeV, the widths of the
resonances have to be included\cite{r31,medium1} in
Eq.~(\ref{denn}). This is because the number of light particles
coming from the decay of resonances is increased by the finite
resonance width. In practice, the width of the $\Delta$ resonance
is most important\cite{medium1,medium2}.
%since  they  can be produced over the mass range
%determined  by  the  width,  this  plays  a  large  role in
%temoperature range relevant for the experiments at SIS.
Thus, the approximation of the resonance width by a $\delta$
function is not justified.  Assuming the validity of Boltzmann
statistics one replaces the  partition function in equation
(\ref{z1ii}) by: \be \ln Z_R &=& N \, {{Vd_R}\over {2\pi^2}} \, T
\, \exp [
(B_R\mu_B+Q_R\mu_Q+S_R\mu_S)/T] \, \nonumber \\
 &&\int_{s_{min}}^{s_{max}} ds \, s \, K_2(\sqrt {s}/T)
\, {1\over \pi} \, {{m_R\Gamma_R}\over {    (s-m_R^2)^2 +
m_R^2\Gamma_R^2 }} \label{eq4} \ee \noindent
%%%%%%%%%%%%%%%%%%%%%%%%%%%%%%%%%%%%%%%%%
where $s_{min}$ is chosen to be the threshold value for the
resonance decay and $\sqrt {s_{max}}\sim m_R +2\Gamma_R$. The
normalization constant $N$ is adjusted such  that the integral
over the Breit-Wigner factor gives 1.

The statistical model, outlined above, was
applied\cite{r9}$^-$\cite{r611,r612} to describe particle yields
in heavy ion collisions. The model was compared with  all
available experimental data obtained in the energy range from AGS
up to RHIC energy. Hadron multiplicities ranging from pions to
omega baryons and their ratios were used to verify that there is a
set of thermal parameters $(T,\mu_B)$ which simultaneously
reproduces all measured yields. In the following Section we
present the most recent analysis of particle production in A--A
collisions at RHIC, SPS and AGS energies.
%%%%%%%%%%%%%%%%%%%%%%%%%%%%%%%%
\subsection{Thermal  analysis of particle yields from AGS to  RHIC energies }

For the analysis of data in the energy range of 40 GeV/nucleon and
upwards\footnote{The results in this section were obtained in
collaboration with D. Magestro  and are published in part in
Refs.~(\refcite{r10,magestro3}).}  we use a grand canonical
ensemble to describe the partition function and hence the density
of the particles following Eqs.~(\ref{par} -\ref{denn}).
% of  species
%\(i\) in an equilibrated fireball:
%\begin{equation}
%n_i= \frac{g_i}{2 \pi^2} \int_0^\infty \frac{p^2 \, {\rm
%d}p}{e^{(E_i(p)-\mu_i)/T} \pm 1} \label{grundgl}
%\end{equation}
%\noindent with particle density \(n_i\), spin degeneracy \(g_i\),
%\(\hbar\) = $c$ = 1, momentum \(p\), total energy \(E\)  and chemical
%potential \(\mu_i = \mu_B B_i-\mu_S S_i-\mu_{I_3} I^3_i\). The
%quantities \( B_i\), \( S_i\) and \( I^3_i\) are the baryon,
%strangeness and three-component of the isospin quantum numbers of the
%particle of species \(i\).
As discussed above the temperature $T$ and the baryochemical
potential \(\mu_B\) are the two independent parameters of the
model, while the volume of the fireball $V$, the strangeness
chemical potential \(\mu_S\), and the charge chemical potential
\(\mu_{Q}\) are fixed by the following additional conditions.
First, overall strangeness conservation fixes $\mu_S$. Note that
this applies strictly for data integrated over 4$\pi$. For slices
near mid-rapidity this condition is, however, also appropriate as
the flow of strangeness in and out of the rapidity slice under
consideration very nearly cancels. Charge conservation implies a
condition on $I^3$ according to: \be \mbox{} &\quad& V \sum_i n_i
I^3_i = \frac{Z-N}{2}. \label{iso} \ee Here, Z and N are the
proton and neutron numbers of the colliding nuclei, $I^3$ and
$I^3_i$ are the third component of the total isospin and that of
particle i. This condition is appropriate (and relevant) at lower
beam energies where there is full stopping and 4$\pi$ yields are
used. For details see Refs.~(\refcite{r12,diplomarbeit}). At
higher energies and for data analyzed in rapidity slices the right
hand side of Eq.~( \ref{iso}) has to be replaced by the neutron
excess of baryons transported into the rapidity slice under
consideration. This number is clearly smaller than the full
neutron excess entering Eq. ~(\ref{iso}) but in general not well
known. However, its precise knowledge is less relevant for higher
beam energies since the isospin balance is dominated by pions. For
practical purposes isospin conservation is important for AGS
energies and below but its effect is small (on the 10 \% level)
already at 40 GeV/nucleon beam energy (where we have used as an
upper limit the full neutron excess of the colliding nuclei,
leading to a slight overestimate of the pion charge asymmetry) and
negligible at top SPS and RHIC energies. Finally, the volume
(which drops out anyway for particle ratios) can be obtained from
total baryon number conservation (for full stopping and quantities
which are evaluated over the complete phase space) or is fixed by
using the measured pion multiplicity in the rapidity slice under
consideration. As discussed above the hadronic mass spectrum used
in the calculations extends over all mesons with masses below 1.5
GeV and baryons with masses below 2 GeV.
% This limits the temperature up to which thermal model
%calculations are trustworthy to $T_{max} < 200$ MeV. We note,
%however, that calculations with higher temperatures should anyway
%be considered with caution as the mass spectrum for heavier
%hadrons is not sufficiently well known.
To take into account a more realistic equation of state we
incorporate the repulsive interaction at short distances between
hadrons by means of the excluded volume correction discussed
above.  A number of different corrections have been discussed in
the literature. Here we choose that proposed in
Refs.~(\refcite{diplomarbeit,exclv2,r79}):
\begin{equation}
 p^{excl.} (T,\mu)= p^{id.gas}(T,\hat{\mu}); \quad \mbox{with }
\hat{\mu} = \mu - v_{eigen} \; p^{excl.}(T,\mu). \label{druck}
\end{equation}
%%%%%%%%%%%%%%%%%%%%%%%%%%%%%%f222222222222222222222222222222222222222222%%%%%%%%%%%%%%%%%%%%
\begin{figure}[htb]
%\begin{minipage}[t]{80mm}
%\framebox[79mm]{\rule[-26mm]{0mm}{52mm}}
 \vskip -0.9cm{\hskip -0.1cm
\includegraphics[width=30.0pc]{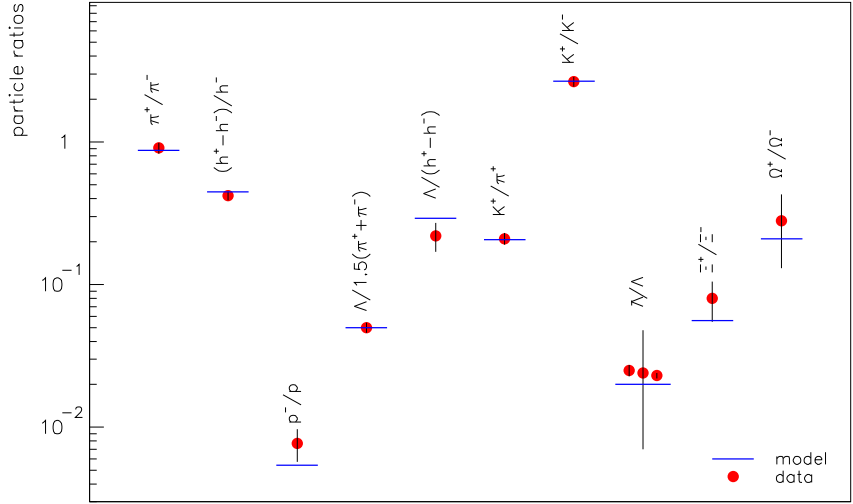}}\\
\vskip -1.5cm { \caption{ \label{ff41}\label{fig:40gev}
 Comparison between thermal model predictions and
experimental particle ratios for Pb--Pb collisions at 40
GeV/nucleon. The   thermal model calculations are obtained with
$T=148$ MeV and $\mu_B=400$ MeV.
 }}
%\end{minipage}
\end{figure}
%%%%%%%%%%%%%%%%%%%%%%%%%%%%%%1%%%%%%%%%%%%%%%%%%%%
%%%%%%%%%%%%%%%%%%%%%%%%%%%%%%f222222222222222222222222222222222222222222%%%%%%%%%%%%%%%%%%%%
\begin{figure}[htb]
%\begin{minipage}[t]{80mm}
%\framebox[79mm]{\rule[-26mm]{0mm}{52mm}}
\vskip -0.7cm {\hskip -0.1cm
 \includegraphics[width=30.0pc]{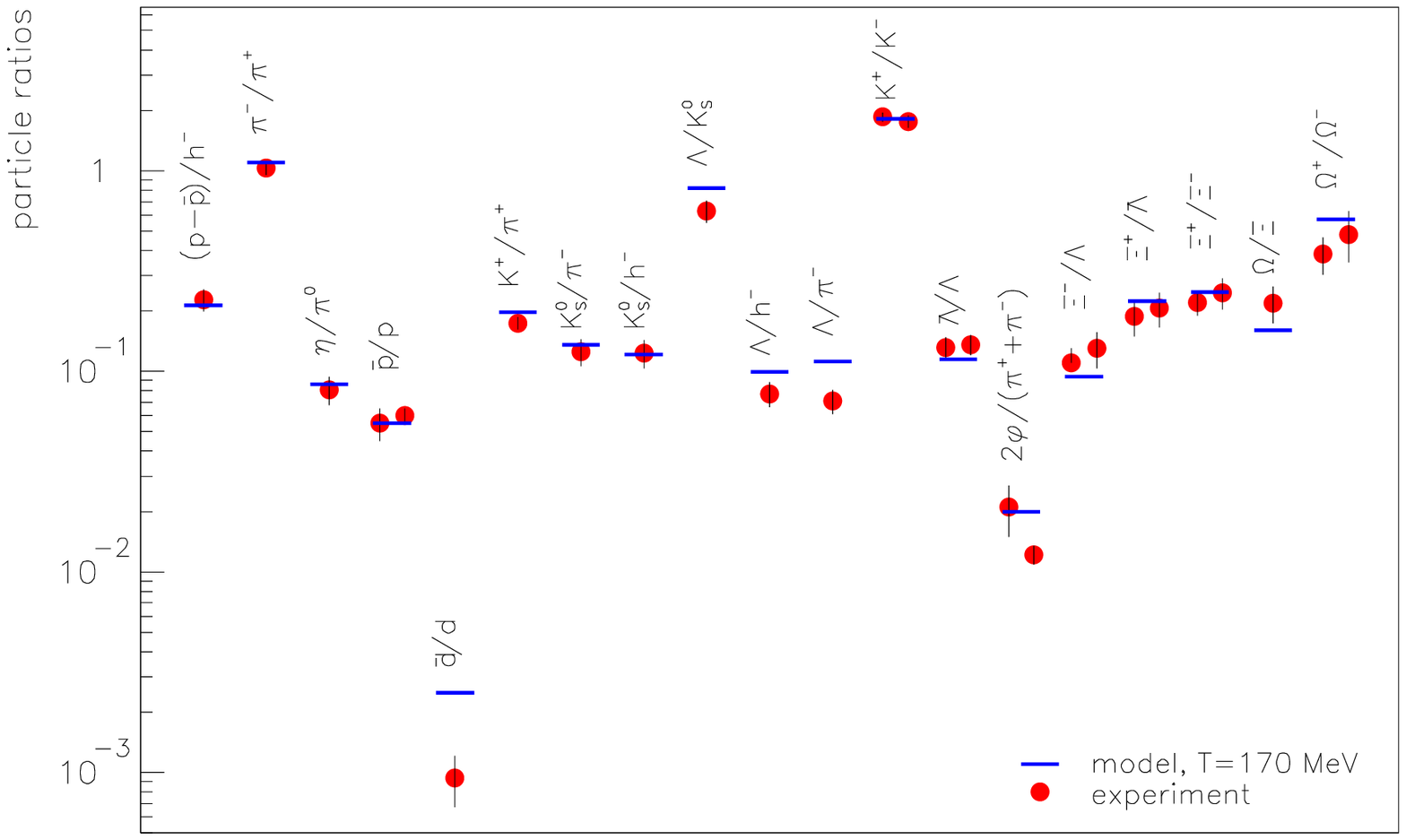}}\\
\vskip -1.5cm { \caption{ \label{ff42}\label{fig:160gev}
Comparison between thermal model predictions and experimental
particle ratios for Pb--Pb collisions at 158 GeV/nucleon. The
thermal model calculations are obtained with $T=170$ MeV and
$\mu_B=255$ MeV.
 }}
%\end{minipage}
\end{figure}
%%%%%%%%%%%%%%%%%%%%%%%%%%%%%%1%%%%%%%%%%%%%%%%%%%%
This thermodynamically consistent approach to simulate
interactions between particles by assigning an eigenvolume
\(v_{eigen}\) to all particles modifies the pressure \(p\) within
the fireball. Equation (\ref{druck}) is recursive, as it uses the
modified chemical potential \(\hat{\mu}\) to calculate the
pressure, while this pressure is also used in the modified
chemical potential, and the final value is found by iteration.
Particle densities are calculated by substituting \(\mu\) in Eq.~
(\ref{denj}) by the modified chemical potential \(\hat{\mu}\). The
eigenvolume has to be chosen appropriately to simulate the
repulsive interactions between hadrons, and we have investigated
the consequences for a wide range of parameters for this
eigenvolume in Ref.~(\refcite{r12,diplomarbeit}). Note that the
eigenvolume is $v_{eigen} =4 \frac{4}{3} \pi R^3$ for a hadron
with radius R. Assigning the same eigenvolume to all particles can
reduce particle densities drastically but hardly influences
particle ratios.  Ratios may differ strongly, however, if
different values for the eigenvolume are used for different
particle species. Our approach here is, to determine, for
nucleons, the eigenvolume according to the hard-core volume known
from nucleon-nucleon scattering\cite{bohrmott}.  Consequently, we
assigned 0.3 fm as radius for all baryons. For mesons we expect
the eigenvolume not to exceed that of baryons.  For lack of better
theoretical guidance we chose also for the mesons a radius of 0.3
fm. For a discussion of the implications of varying these radius
parameters see Ref.~(\refcite{r12,diplomarbeit}). After thermal
``production'', resonances and heavier particles are allowed to
decay, therefore contributing to the final particle yield of
lighter mesons and baryons, as indicated above. Decay cascades,
where particles decay in several steps, are also included.
Systematic parameters regulate the amount of decay products
resulting from weak decays. This allows to simulate the different
reconstruction efficiencies for particles from weak decays in
different experiments.
%%%%%%%%%%%
In the following we compare predictions of the model with results
of measured particle ratios for central Pb-Pb collisions at SPS
energies (40 and 158 GeV/nucleon) and for central Au-Au collisions
at RHIC energies $\sqrt{s_{nn}} =$ 130 and 200 GeV. An important
issue in this context is whether to use data at mid-rapidity or
data integrated over the full phase space. While it is clear that
full 4$\pi$ yields should be used at low beam energies, this is
not appropriate any more as soon as fragmentation and central
regions can be distinguished. In that case the aim is to identify
a boost-invariant region near mid-rapidity and to choose a slice
in rapidity within that region.  For RHIC energies this implies
that an appropriate choice, given the available data, is a
rapidity interval of width $\Delta y =1$ centered at midrapidity.
The anti-proton/proton ratio stays essentially constant within
that interval, but drops rather strongly for larger rapidities and
similar results are observed\cite{brahms} for other ratios.
Furthermore, the rapidity distribution exhibits a boost-invariant
plateau near mid-rapidity\cite{ullrich}. As has been demonstrated
in Ref.~(\refcite{cleymans}), effects of hydrodynamic flow cancel
out in particle ratios under such conditions. At SPS energies a
boost-invariant plateau is not fully developed but stopping is not
complete, either. In addition, the proton and anti-proton rapidity
distributions differ rather drastically, especially near the
fragmentation regions, implying that particle ratios depend on
rapidity (see. e.g., Ref.~(\refcite{appels1})). Under those
circumstances we have decided to use, wherever available, data in
a slice of $\pm 1 $ unit of rapidity centered at mid-rapidity.
This is slightly different from the analysis performed in
Ref.~(\refcite{r12}), where both mid-rapidity and fully integrated
data were used. We note, however, see below, that the fit
parameters T and $\mu_B$ obtained at 158 A GeV are very close to
those determined earlier. The criterion for the best fit of the
model to data was  a minimum in
\begin{equation}
\chi^2=\sum_{i} \frac{( {\cal R}_i^{\rm exp.}-{\cal R}_i^{\rm
model})^2}{\sigma_i^{\rm 2}}.
\end{equation}
In the above equations \( {\cal R}_i^{\rm model}\) and \({\cal
  R}_i^{\rm exp.}\) are the \(i\)th particle ratio as calculated from
  our model or measured in the experiment, and \(\sigma_i\) represent
  the errors (including systematic errors where available) in the
  experimental data points as quoted in the experimental publications.
For the data we used all information available including that
presented at the QM2002 conference in July 2002. Details on the
data selection, corrections for the weak-decay reconstruction
efficiency, as well relevant references are found in
Ref.~(\refcite{magestro3}). Under the conditions discussed above
the data can all be well described, as is detailed below, by a
thermal distribution with T and $\mu_B$ as independent parameters.
There is no need to introduce additional parameters such as as
strangeness suppression factors.
%%%%%%%%%%%%%%%%%%Magestro%%%%%%%%%%%%f222222222222222222222222222222222222222222%%%%%%%%%%%%%%%%%%%%
\begin{figure}[htb]
%\begin{minipage}[t]{80mm}
%\framebox[79mm]{\rule[-26mm]{0mm}{52mm}}
 \vskip -0.2cm{\hskip -.1cm
\includegraphics[width=28.8pc]{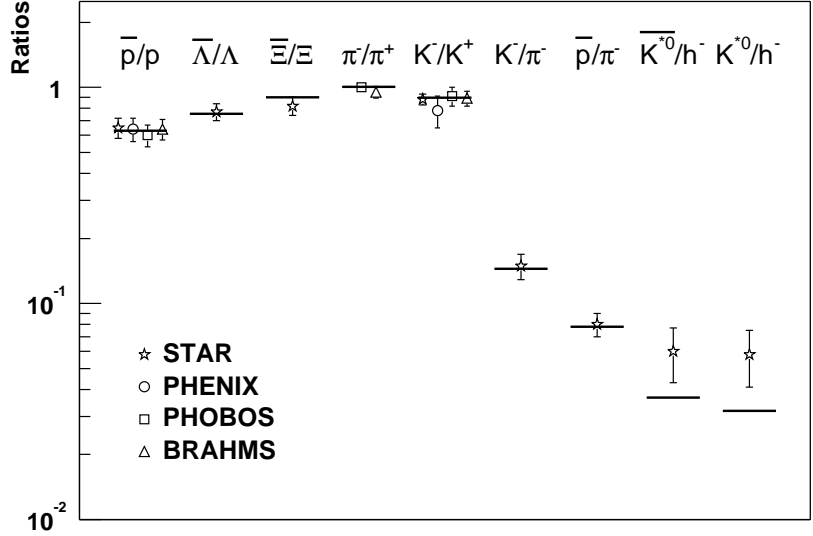}}\\
\vskip -1.cm { \caption{ \label{ff33} Comparison between thermal
model predictions\protect\cite{r10} and experimental particle
ratios for Pb--Pb collisions at $\sqrt{s_{nn}}$ = 130 GeV.
Calculations were performed for $T = 174$ MeV and $\mu_B$ = 46
MeV.}}
%\end{minipage}
\end{figure}
%%%%%%%%%%%%%%%%%%%%%%%%%%%%%%1%%%%%%%%%%%%%%%%%%%%
The results of the fits for central Pb-Pb collisions at 40 and 158
GeV per nucleon are presented in Figs.~(\ref{ff41},\ref{ff42}). At
40 GeV/nucleon 11 particle ratios are included in the fit, while
the number is 24 at 158 GeV/nucleon. We obtain values for
(T,$\mu_B$) of (148$\pm 5$, 400$\pm 10$) and (170$\pm 5$, 255$\pm
10$), respectively, with reduced $\chi^2$ values of 1.1 and 2.0.
%%%%%%%%%%%%%%%%these numbers need to be modified depending on the
%%%%%%%%%%%%%%%%detailed fits!!!!!pbm    done 1/31-pbm
Obviously the fits are quite good. A possible exception is the
$\phi/(\pi^++\pi^-)$ ratio at top SPS energy, where there are
conflicting data from NA49 and NA50. This is already discussed in
detail in Ref.~(\refcite{r12}) and no new information on this
problem has appeared since. Note that this ratio has not been used
in the $\chi^2$ minimization. The somewhat larger values of
$\chi^2$ at full SPS energy and the remaining uncertainty in
$\mu_B$ are due to a systematic problem not yet sufficiently
addressed by the experiments. The contribution of weak decays of
strange baryons to final baryons has been discussed and cuts are
applied to reduce this contribution in the data. However, there is
also a contribution from weak decays to charged pions (or, more
generally) charged hadrons which is up to now poorly quantified by
the experiments. If, e.g., in the ratio $\Lambda/h^-$, feeding of
the $\Lambda$ by decays of $\Xi$'s is suppressed due to cuts, but
the $\pi^-$ measurement has a 50 \% efficiency for detection of
pions from weak decays, the ratio would drop by 15-20 \%, compared
to the case with 0 \% eficiency for weak decays. With this option
the reduced $\chi^2$ value for the 158 GeV fit would drop from 2.0
to 1.5. This discussion indicates that there are sources of
systematic uncertainties not included in the data. The corrections
for weak decays are, consequently, of utmost importance when
discussing the ``precision'' of fits.
%%%%%%%%%%%%%%%%%%Magestro%%%%%%%%%%%%f222222222222222222222222222222222222222222%%%%%%%%%%%%%%%%%%%%
\begin{figure}[htb]
%\begin{minipage}[t]{80mm}
%\framebox[79mm]{\rule[-26mm]{0mm}{52mm}}
 {\hskip -.3cm
\includegraphics[width=28.8pc]{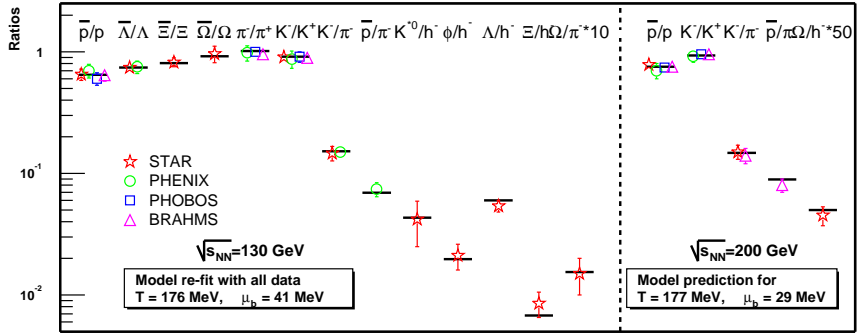}}\\
\vskip -.5cm { \caption{ \label{ff3} Comparison of the
experimental data on different particle multiplicity ratios
obtained  at RHIC at $\sqrt s_{NN} =130$  and 200 GeV  with
thermal model calculations. The thermal model analysis is from
Refs.~(\protect\refcite{r10,r10n}) and recent update by D.
Magestro.
 }}
%\end{minipage}
\end{figure}
%%%%%%%%%%%%%%%%%%%%%%%%%%%%%%1%%%%%%%%%%%%%%%%%%%%

The results for RHIC energies are shown in Figs.~(\ref{ff33},
\ref{ff3}). In Fig.~(\ref{ff33}) we present the results as
published in Ref.~(\refcite{r10}) in the summer of 2001. Since
then, the data at $\sqrt{s_{nn}}=130$ GeV have been consolidated
and extended and first (in some cases still preliminary) results
have been provided for $\sqrt{s_{nn}}=200$ GeV. The current state
of affairs in summarized in Fig.~(\ref{ff33}). The results
demonstrate quantitatively the high degree of equilibration
achieved for hadron production in central Au-Au collisions at RHIC
energies. We obtain values for (T,$\mu_B$) of (174$\pm 7$, 46$\pm
5$) and (177$\pm 7$, 29$\pm 6$), respectively, with reduced
$\chi^2$ values of 0.8 and 1.1.
%%%%%%%%%%%%%%%these numbers need to cross checked!!!pbm
We note that ratios involving multi-strange baryons  are well
reproduced as is  the $\phi/h^-$ ratio. Even relatively wide
resonances such as the K$^*$'s fit well into the picture of
chemical freeze-out. This obviates the need for quark coalescence
models as proposed in Ref.~(\refcite{biro1}) and non-equilibrium
models as proposed in Ref.~(\refcite{rafelski}).

Very recently, the STAR collaboration has provided\cite{fachini}
first data, with about 30 - 50 \% accuracy, on the $\rho^0/\pi$
and $f^0(980)/\pi$ ratios in semi-central Au-Au collisions. These
mesons have been reconstructed in STAR via their decay channel in
2 charged pions. Comparing the preliminary results from STAR with
our thermal model prediction reveals that the measured ratios
exceed the calculated values by about a factor of 2. This is quite
surprising, especially considering that we use a chemical
freeze-out temperature of 177 MeV for the calculation, while one
might expect these wide resonances to be formed near to thermal
freeze-out, i.e. at a temperature of about 120 MeV. At this
temperature, the equilibrium value for the $\rho^0/\pi$ ratio is
about $4 \cdot 10^{-4}$, while it is 0.11 at 177 MeV. Even with a
chemical potential for pions of close to the pion mass and taking
into account the apparent (downwards) mass shift of 60 - 70 MeV
for the $\rho^0$ it seems difficult to explain the experimentally
observed value of about 0.2.

We finally note that the model discussed here was also applied to
the AGS data collected in Ref.~(\refcite{r11}).  The best fit,
obtained for R\(\rm _{baryon}\)=R\(\rm_{meson}\)\-=\-0.3 fm,
yields T = 125 (+3-6) MeV and $\mu_B$ = 540 $\pm 7$ MeV, well in
line with the calculations reported in Ref.~(\refcite{r11}).

In summary, hadron multiplicities produced in central
nucleus-nucleus collisions in the range of AGS to full RHIC energy
can be quantitatively described with a grand-canonical partition
function based on the full hadron resonance spectrum, assuming
complete chemical equilibrium. There is no need to introduce
non-equilibrium parameters or strangeness suppression factors if
data near mid-rapidity are considered. The physical relevance of
the two model parameters T and $\mu_B$ is described in detail in
our discussions below concerning the phase boundary between
hadrons and the quark-gluon plasma.

%%%%%%%%%%%%%%%%%%%%%%%%%%%
\subsection{Comparison of measured particle densities with thermal
  model predictions}
As discussed below, the value for the energy density predicted by
the presently used thermal model, including the excluded volume
correction, agrees well with results from the lattice for
temperatures below the critical temperature. It makes therefore
sense to compare the densities for pions and nucleons predicted by
the model with values determined from experiments. The CERES
collaboration has recently performed an analysis of 2-pion
correlation experiments for the energy range between AGS and RHIC,
from which values for these densities have been
determined\cite{ceres1,ceres2} from data taken at mid-rapidity.
For the nucleon density (at thermal freeze-out) the experimental
numbers are, at 40 and 158 GeV/nucleon\footnote{We take here the
data published in Ref.~(\refcite{appels1}); the data reported in
Ref.~(\refcite{leuwen}) are about 20 \% lower and would not fit
the beam energy systematics.} and at $\sqrt{s_{nn}} = 130 $ GeV,
0.077 $\pm 0.005$/fm$^3$, 0.063 $\pm 0.005$/fm$^3$  and 0.06 $\pm
0.009$/fm$^3$.  From the model we deduce, at chemical freeze-out,
values of 0.10/fm$^3$, 0.10/fm$^3$ and 0.08/fm$^3$. This would
imply a volume increase of about 40 \% from chemical to thermal
freeze-out. For pions the situation could be more complicated
since yield ratios involving pions are (apparently) fixed at
chemical freeze-out, implying the build-up of a pion chemical
potential between chemical and thermal freeze-out. From the data
one deduces\cite{ceres1,ceres2} a  pion density at (thermal)
freeze-out of 0.28 $\pm 0.03$/fm$^3$, 0.43 $\pm 0.03$/fm$^3$, and
0.49 $\pm 0.1$/fm$^3$, at 40 and 158 GeV/nucleon and at
$\sqrt{s_{nn}} = 130 $ GeV. These values should be contrasted with
the calculated (chemical) freeze-out values of 0.35/fm$^3$,
0.59/fm$^3$ and 0.62/fm$^3$. From these numbers one would conclude
a 30 \% volume increase between chemical and thermal freeze-out,
assuming that the pion chemical potential fixes the pion number to
the value obtained at chemical freeze-out. This rather small
volume increase indicates that the time between chemical and
thermal freeze-out cannot be very long at SPS and RHIC energies.
At AGS energy, the corresponding $\pi^+$ and proton densities of
0.051/fm$^3$ and 0.053/fm$^3$ agree well with those
estimated\cite{agsplb,nu} from particle interferometry
(0.058/fm$^3$ and 0.063/fm$^3$, respectively) implying that, at
AGS energy, thermal and chemical freeze-out take place at nearly
identical times and temperatures.

%%%%%%%%%%%%%%%%%%%%%%%%%%%
\subsection{Statistical model and composite particles}

An often overlooked aspect of the thermal model is the possibility
to compute also the yields of composite particles. For example,
the d/p and $\bar{\rm{d}}/\bar{\rm{p}}$ ratios measured at SPS and
AGS energies are well reproduced\cite{col3} with the same
parameters which are used to describe\cite{r11,r12} baryon and
meson ratios. Furthermore, the AGS E864 Collaboration has recently
published\cite{col1} yields for composite particles (light nuclei
up to mass number 7) produced in central Au-Au collisions at AGS
energy near mid-rapidity and at small $p_t$. In this
investigation, an exponential decrease of composite particle yield
with mass is observed over 7--8 order of magnitude, yielding a
penalty factor $P_p$ of about 48 for each additional nucleon.
Extrapolation of the data to large transverse momentum values,
considering the observed mass dependence of the slope constants,
reduces this penalty factor to about 26, principally because of
transverse flow. In the thermal model, this penalty factor can be
related with thermal particle phase--space.
%easily derived.
 In the relevant Boltzmann approximation, we obtain
\be R_p \approx \exp{\frac{m \pm \mu_b}{T}}, \ee  where m is the
nucleon mass and the negative sign applies for matter, the
positive for anti-matter. Small corrections due to the spin
degeneracy and the A$^{3/2}$ term in front of the exponential in
the Boltzmann formula for particle density are neglected.  Using
the freeze--out parameters T=125 MeV and $\mu_b = $ 540 MeV
appropriate\cite{r11} for AGS energy one gets\cite{col3}   R$_p
\approx 23$, in close agreement with the data for the production
of light nuclei. It was  also noted that the anti-matter yields
measured\cite{col2} by the E864 Collaboration yield penalty
factors of about 2$\cdot 10^5$, again close to the
predicted\cite{col3} value of 1.3$\cdot 10^5$ .

This rather satisfactory quantitative agreement between measured
relative yields  for composite particles and thermal model
predictions provides some confidence in the predictions for yields
of exotic objects produced in central nuclear collisions. We
briefly comment here on the results obtained in
Ref.~(\refcite{col3}).

In this investigation, the production probabilities for exotic
strange objects and, in particular, for strangelets were computed
in the thermal model. The results are reproduced in Table
\ref{col44} for temperatures relevant for beam energies between 10
and 40 GeV/nucleon. We first note that predictions of the thermal
model and, where available, the coalescence model of
Ref.~(\refcite{col4}) agree (maybe surprisingly) well particularly
for lighter clusters. Secondly, inspection of Table \ref{col44}
also shows that, in future high statistics experiments which will
be possible at the planned\cite{r84} new GSI facility,
multi-strange objects such as $_{\Xi^0\Lambda \Lambda}^{~~~~~7}$He
should be experimentally accessible with a planned sensitivity of
about 10$^{-13}$ per central collision in a years running, should
they exist and be produced with thermal yields. Investigation of
yields of even the lightest conceivable strangelets will be
difficult, though.

\begin{table}
\tbl{\label{col44} Produced number of nonstrange and strange
clusters and of strange quark matter  per  central Au+Au collision
at AGS energy, calculated in a thermal model for two different
temperatures, baryon chemical potential $\mu_b$= 0.54 GeV and
strangeness chemical potential $\mu_s$ such that overall
strangeness is conserved.}
{\tabcolsep11.1pt \begin{tabular*} %{l l l l }
{\typewidth}{@{}llll@{}} \hline
% \br
 & {~~~~~~~~Thermal Model Parameters} & \\
%\ns
 &  & \\
Particles & $T$=0.120 GeV & $T$=0.140 GeV & Coalescence\\

%\mr
d & 15 & 19 & 11.7  \\
t+$^3$He & 1.5 & 3.0 & 0.8  \\
$\alpha$ & 0.02 & 0.067 & 0.018 \\
$H_0$       & 0.09 & 0.15 & 0.07  \\
$_{\Lambda \Lambda}^{~~~5}$H & 3.5 $\cdot 10^{-5}$ & 2.3 $\cdot
10^{-4}$ &
4$\cdot 10^{-4}$ \\
$_{\Lambda \Lambda}^{~~~6}$He & 7.2 $\cdot 10^{-7}$ & 7.6 $\cdot
10^{-6}$ &
 1.6$\cdot 10^{-5}$ \\
$_{\Xi^0\Lambda \Lambda}^{~~~~~7}$He & 4.0 $\cdot 10^{-10}$ & 9.6
$\cdot
10^{-9}$  & 4 $\cdot 10^{-8}$  \\ %\mr
&  &  & \\
\hline
&  &  & \\
$^{10}_{1}$St$^{-8}$ &  1.6 $\cdot 10^{-14}$ & 7.3 $\cdot 10^{-13}$ & \\
$^{12}_1$St$^{-9}$ &  1.6 $\cdot 10^{-17}$ & 1.7 $\cdot 10^{-15}$  &  \\
$^{14}_1$St$^{-11}$ & 6.2 $\cdot 10^{-21}$ & 1.4 $\cdot 10^{-18}$  & \\
$^{16}_2$St$^{-13}$ & 2.4 $\cdot 10^{-24}$ & 1.2 $\cdot 10^{-21}$   & \\
$^{20}_2$St$^{-16}$ & 9.6 $\cdot 10^{-31}$ & 2.3 $\cdot 10^{-27}$   & \\
&  &  & \\\hline
\end{tabular*}}
%\begin{tabnote}[Note]
%Some data are still preliminary
%\end{tabnote}
\begin{tabnote}[Source] The Coalescence model predictions in the last column are from Table 2 of
Ref.~(\protect\refcite{col4}).
\end{tabnote}
\end{table}

\section{Exact Implementation of the conservation laws  in the statistical models}

The  analysis of particle yields obtained in central heavy ion
collisions from AGS up to LHC energy  has shown that hadron
multiplicities are very well described by assuming a complete
thermalized state at fixed $T$ and $\mu_B$ . In this broad energy
range, particle yields and their ratios are, within experimental
error,  well reproduced by the statistical hadron resonance gas
model that accounts for the conservation laws of baryon number,
strangeness and electric charge in the grand canonical ensemble.
The natural question arising here is whether this statistical
order is a unique feature of high energy central heavy ion
collisions or is it  also there at lower energies as well as in
hadron--hadron and peripheral heavy ion collisions. To address
this question one needs, however, to stress that when going beyond
high energy central heavy ion collisions the grand canonical
statistical operator (\ref{eq3}) has to be modified.

Within  the statistical approach, particle production can only be
described using the grand canonical ensemble with respect to
conservation laws, if the number of produced particles that carry
a conserved charge is sufficiently large. In view of the
experimental data this also means that  the event-averaged
multiplicities are controlled by  the chemical potentials. In this
description the net value of a given  charge (e.g. electric
charge, baryon number, strangeness, charm, etc.) fluctuates from
event to event. These fluctuations can be neglected (relative to
the squared  mean particle multiplicity)  only if the particles
carrying the charges in question are abundant. Here,  the charge
will be conserved on the average and the grand canonical
description developed in the last section is adequate. In the
opposite limit of low production yield  the particle number
fluctuation can be as large as its event averaged value. In this
case  charge conservation has to be implemented exactly in each
event\cite{r20,r22}.

The exact conservation of quantum numbers introduces a constraint
on the thermodynamical system. Consequently, the time dependence
and equilibrium distribution of particle multiplicity can differ
from that expected in the grand canonical limit. To
 see these differences one needs to perform a detailed study
of particle equilibration in a thermal environment. To discuss
equilibration from the theoretical point of view  one needs to
formulate the kinetic equations for  particle production and
evolution. In a partonic medium this requires, in general, the
formulation of a transport equation\cite{r6,r7,r8} involving
colour degrees of freedom and a non-Abelian structure of QCD
dynamics. In the hadronic medium, on the other hand, one
needs\cite{r20,r20a,r22,r37,r38} to account for the charge
conservations related with the U(1) internal symmetry.

\subsection{Kinetics of time evolution and equilibration of charged
particles}

In this section we will discuss  and formulate the  kinetic
equations that include constraints imposed by the conservation
laws of Abelian charges related with U(1) internal symmetry. We
will indicate the importance of the conservation laws  for the
time evolution and chemical equilibration of produced particles
and their probability distributions. In particular, we demonstrate
that the constraints imposed by the  charge conservation  are of
crucial importance for rarely produced  particle species such as
for particles with hidden quantum numbers like e.g. for $J/\psi$.
%This is of physical relevance when discussing strangeness
%production in low energy heavy ion collisions \cite{aa} or

To study chemical equilibration in a hadronic medium we introduce
first a kinetic model that takes into account the production and
annihilation of particle--antiparticle pairs $c\bar c$ carrying
U(1) quantum numbers like strangeness or charm. It is also assumed
that particles $c$ and $\bar c$ are produced according to a binary
process $ab \to c\bar c$ and that all particle momentum
distributions are thermal and described by the Boltzmann
statistics. The  charge neutral particles $a$ and $b$  are
constituents of a thermal fireball  with temperature $T$ and
volume $V$. We will consider the time evolution and equilibration
of particles $c $ and $\bar c$ inside this  fireball, taking into
account the constraints imposed by the U(1) symmetry. First, we
formulate a general master equation for the probability
distribution of particle multiplicity in a medium with vanishing
net charge and consider its properties and solutions. Then we will
discuss two limiting cases of  abundant and rare particle
production. Finally, the rate equation will be extended to a more
interesting situation where there are different particle species
carrying the conserved quantum numbers  inside a thermal fireball
that also has a non vanishing net charge.

\subsubsection{Kinetic master equation for probabilities}
%\subsubsection{Rate equation for charged particle multiplicity
%distributions}

Consider $P_{N_c}(\tau )$ as the probability to find $N_c$
particles $c$, where $0\leq N_c\le \infty$. This probability will
obviously change in time owing to the  production
 $a b\to c\bar c$
and absorption $c\bar c \to ab$ processes. The equation for the
probability $P_{N_c}$ contains terms which increase in time,
following  the transition from $N_c-1$ and $N_c+1$ states to  the
$N_c$ state, as well as terms which decrease  since the state
$N_c$ can make transitions to $N_c+1$ and $N_c-1$ (see
Fig.~\ref{flow1}).
%%%%%%%%%%%%%%%%%%%%%%%%%%%%f66666666666666
%%%%%%%%%%%%%%%%%%%%%%%%%%%%%f66666666666666
%\vskip -0.5cm
\begin{figure}[htb]
 {\hskip 1.5cm
\includegraphics[width=18.9pc, height=4.pc]{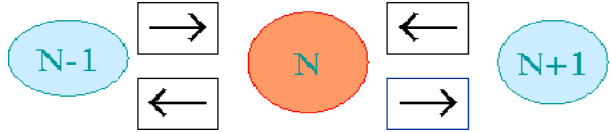}}\\
\vskip -0.6cm { \caption{A schematic view of the master equation
for the probability $P_N(\tau )$ due to  $a b  \leftrightarrow
c\bar c$ and the inverse process. } \label{flow1}}
\end{figure}
%%%%%%%%%%%%%%%%%%%%%%%%%%%

The  rate equation is  determined by the magnitude of the
transition probability per unit time due to the  production  $G/V$
and the absorption $L/V$    of $c\bar c$ pairs through $ab
\leftrightarrow    c\bar c$ process. The gain $(G= <\sigma_{a b\to
c\bar c}v_{ ab}>) $  and the loss  $(L= <\sigma_{ c\bar c\to
ab}v_{ c\bar c}> )$ terms represent the momentum average of
particle production and absorption cross sections.

The transition probability per unit time from  $N_c+1\to N_c$  is
given by the product of the probability $L/V$ that the single
reaction $c\bar c \to ab$ takes place multiplied  by the number of
possible reactions which is formally, $(N_c+1)(N_{\bar c}+1)$. In
the case when the  charge carried by particles $c$ and $\bar c$ is
exactly and locally conserved, that is if $(N_c+N_{\bar c}=0$),
this factor is just $(N_c+1)^2$. Similarly, the transition
probability from $N_c\to N_c+1$ is described by $G\langle
N_{a}\rangle\langle N_{b}\rangle /V$, where one assumes that
 particles $a$ and $b$ are not correlated and their multiplicity
 is governed by the thermal averages. One also assumes that
 the multiplicity of $a$ and $b$ is not affected by the $ab \to c\bar
 c$ process.
%\vskip -1cm
 \noindent The master equation for the time evolution
of the probability $P_{N_c}(\tau )$ can be written\cite{r20} in
the following form:
%A general rate equation for the production of particle pair $b_1 b_2$
%can be written in the form of iterative equations as the following:
\ber \frac{dP_{N_c}}{d\tau}&=&{G\over V} \langle
N_{a}\rangle\langle N_{b}\rangle P_{N_c-1} +\frac {L}{V} (N_c+1)^2
P_{N_c+1}
 \nonumber \\
&-& {G\over V}\langle N_{a}\rangle\langle  N_{b}\rangle
P_{N_c}~~~- \frac {L}{V} N_c^2 P_{N_c} . \label{eq1} \label{test}
\label{eq5} \eer
%%%%%%%%%%%%%%%
The first two terms in Eq.~(\ref{test}) describe the increase of
$P_{N_c}(\tau )$
 due to the transition from $N_c-1$ and $N_c+1$ to the $N_c$ state.
The last two terms, on the other hand, represent the decrease of
the probability function due to the transition from $N_c$ to the
$N_c+1$ and $N_c-1$ states, respectively.

%%%%%%%%%%%%%%%%%%%%%%%%%%%%%%%%%

 For a thermal
particle momentum distribution and under the Boltzmann
approximation the thermal averaged  cross sections are
obrained\cite{r36,r34} from

 \be
%<\sigma_{D\bar D\to \psi h}v_{D\bar D}> =
<\sigma_{a b\to c\bar c}v_{ ab}> = {\beta\over 8} {
{\int_{t_0}^\infty dt
%\sigma_{D\bar D\to \psi h}(t)
 \sigma_{ab\to c\bar c}(t)
 [t^2-(m_{ab}^+)^2]
 [t^2-(m_{ab}^-)^2]
 K_1(\beta t)}\over
 {m_a^2m_{b}^2K_2(\beta m_a)K_2(\beta m_{b})}
},\label{cs} \label{eq6}
 \ee
where  $K_1$, $K_2$ are modified Bessel functions of the second
kind, $m_{ab}^+ = m_a+m_{b}$ and $m_{ab}^- = m_a-m_{b}$, $t =
\sqrt {s}$ is the center-of-mass energy, $\beta$ the inverse
temperature, $v_{ab}=((k_ak_b)^2-m_a^2m_b^2)/E_aE_b$ is the
relative velocity of incoming particles and the integration limit
is taken to be $t_0=max [(m_a+m_{b}),(m_c+m_{\bar c})]$.

%%%%%%%%%%%%%%%%%%%%%%%%%%%

The rate equation for probabilities (\ref{eq1}) provides the basis
to calculate the time evolution of the momentum averages of
particle multiplicities and their arbitrary moments. Indeed,
multiplying the above equation by $N_c$ and summing over $N_c$,
one obtains the general kinetic equation for the time evolution of
the average number $\langle N_c\rangle = \sum_{N_c=0}^\infty
N_cP_{N_c}(\tau )$ of particles $c$ in a system. This equation
reads:

 \be
\frac{d\langle N_c\rangle }{d\tau}={G\over V} \langle
N_{a}\rangle\langle N_{b}\rangle - \frac{L}{V} \langle N_c^2
\rangle . \label{normal2} \label{eq7}
 \ee

The above equation cannot be  solved analytically as it connects
particle multiplicity $\langle N_c\rangle$ with its second moment
$\langle N_c^2\rangle$. However,  solutions can be obtained in two
limiting situations: { i)} for an abundant production of $c$
particles, that is when $\langle N_c\rangle\gg1$ or { ii)} in the
opposite limit of  rare particle production corresponding to
$\langle N_c\rangle\ll 1$. Indeed, since
 \be
{\langle N_c^2\rangle }=\langle N_c \rangle ^2+\langle \delta
N_c^2 \rangle, \label{5} \label{eq8}
 \ee
where $\langle \delta N_c^2 \rangle$ represents the  fluctuations
of the number of particles  $c$, one can make the following
approximations:

\noindent { i)} for  $\langle N_c\rangle \gg1$ one has a $\langle
N_c^2 \rangle \approx \langle N_c\rangle ^2, $ and Eq.~(\ref{eq7})
obviously reduces to the well known\cite{r34} form:
%Eq.(\ref{normal1}), i.e.,
 \be
\frac{d\langle N_c\rangle }{d\tau}\approx {G\over V} \langle
N_{a}\rangle \langle N_{b}\rangle - \frac{L}{V} \langle N_c
\rangle ^2. \label{normal1} \label{eq9}
 \ee

\noindent { ii)} however, for the rare  production,  particles $c$
and $\bar c$ are strongly correlated and thus,  for   $\langle
N_c\rangle\ll 1$ one takes $\langle N_c^2 \rangle \approx \langle
N_c\rangle$, consequently Eq.~(\ref{eq7}) takes the form:
%Eq.(\ref{normal1}), i.e.,
 \be
\frac{d\langle N_c\rangle }{d\tau}\approx {G\over V} \langle
N_{a}\rangle \langle N_{b}\rangle - \frac{L}{V} \langle N_c
\rangle, \label{normal3} \label{eq10}
 \ee
where the absorption term depends only linearly, instead of
quadratically, on the particle multiplicity.

From the above  it is thus clear that, depending on the thermal
conditions in  the system (that is its  volume and temperature),
% which amounts to getting different
%
we are getting different results for the equilibrium solution and
the time evolution of the number of produced particles $c$.
% in
%cases {\bf (i)} and {\bf (ii)}.
 This is very transparent when
solving the rate equations (\ref{normal1}) and (\ref{normal3}).

In the limit when $\langle N_c \rangle \gg 1$, the standard
Eq.~(\ref{normal1}) is valid and has the well known
solution\cite{r20,r34}: \ber \langle N_c(\tau) \rangle ^{}
=\langle N_c\rangle_{\rm eq}^{\rm } \tanh \left ( \tau/\tau_0^{\rm
} \right ), \label{eq11} \eer  where the equilibrium value
$\langle N_c\rangle_{\rm eq}^{}$ of the number of particles $c$
 and the relaxation time constant $\tau_0^{}$ are
given by: \ber \langle N_c\rangle_{\rm eq}^{\rm }= \sqrt
\epsilon~~, ~~ \tau_0^{\rm } = \frac {V}{L \sqrt \epsilon},
\label{eqgc} \label{eq12} \eer respectively, with    $\epsilon = G
\langle N_{a}\rangle\langle N_{b}\rangle/L$.

In the particular case when the particle momentum distribution is
thermal, the ratio of the  gain  ($G$) to the  loss  ($L$) terms
can be obtained \cite{r20} from Eq.~(\ref{cs}) as
%
% \ber
%\frac{G}{L}= \frac {\langle N_c\rangle_{\rm eq}\langle N_{\bar
%c}\rangle_{\rm eq}} {\langle N_a\rangle_{\rm eq}\langle N_{
%b}\rangle_{\rm eq}},\eer
%
%
 \ber
\frac{G}{L}= \frac { d_{c}m_c^2  K_2 (m_c/T) d_{\bar c}m_{\bar
c}^2 K_2 (m_{\bar c}/T)  }
 {  d_{a}m_a^2  K_2 (m_a/T) d_{ b}m_{
b}^2 K_2 (m_{ b}/T)    }, \label{eq13} \eer
where we have employed the detailed balance relation between the
cross sections for  production $\sigma_{ab}$ and for absorbtion
$\sigma_{c\bar c}$  for $ab
 \leftrightarrow c\bar c$ processes
%%%%%%%%%%%%%%%%%%%%%%%

 \be
 \sigma_{ab\to c\bar c}(t)=
{{d_{a}d_b}\over {d_{c}d_{\bar c}}}
 {{[t^2-(m_{c\bar c}^+)^2]
 [t^2-(m_{c\bar c}^-)^2]}\over
{[t^2-(m_{ab}^+)^2]
 [t^2-(m_{ab}^-)^2]}}\sigma_{ c\bar c\to ab}(t)
\label{csr} \label{eq14}
 \ee
%%%%%%%%%%%%%%%%%%%%%%%%%
with $d_i$ being the spin-isospin degeneracy factor and $m^\pm_{ij
}$ as in Eq.~(\ref{cs}).

In  Boltzmann approximation, the equilibrium average number of
particles $c$ in Eq.~(\ref{eqgc}) reads:

 \ber \langle N_c\rangle_{\rm eq}^{\rm }= { {d_{c}}\over
{2\pi^2}} VTm_c^2  K_2 (m_c/T).
 \label{15}
\label{eq15} \eer
 This is a well known result for the average number of
particles in the    Grand Canonical (GC) ensemble with respect to
the U(1) internal symmetry of the Hamiltonian. The chemical
potential, which is usually present in the GC ensemble, vanishes
in this case, because of the requirement of  charge neutrality of
the system. Thus, the solution of Eq.~(\ref{normal1})
 results in the expected value for the equilibrium limit in the GC
formalism where a   charge is conserved  on the average.

 In the
opposite limit, where $\langle N_c \rangle \ll 1$, the time
evolution of a particle abundance is described by
Eq.~(\ref{normal3}), that has the following solution: \ber \langle
N_c (\tau)\rangle ^{\rm C} = \langle N_c\rangle_{\rm eq}^{\rm C}
\left ( 1- e^{-\tau/\tau_0^{\rm C}} \right ),
 \label{16}
\label{eq16} \eer with  the equilibrium value and relaxation time
given by \ber \langle N_c\rangle_{\rm eq}^{\rm C}= \epsilon,~
\tau_0^{\rm C} = \frac {V}{L}. \label{eqc} \label{eq17}
 \eer

The above result, as will be shown in the next section,   is the
asymptotic limit of the particle multiplicity obtained in  the
canonical (C) formulation of the conservation laws\cite{r20,r20a}.
Here the charge related with the U(1) symmetry is exactly and
locally conserved, contrary to the GC formulation where this
conservation is only valid on the average.

%ensemble leads to suppression of particle multiplicity. It also clear
%that the volume dependence are different in the (GC) and canonical
%limit.
Comparing Eq.~(\ref{eqgc}) with Eq.~(\ref{eqc}), we first find
that, for  $\langle N_c \rangle \ll 1$, the equilibrium value  is
by far smaller than what is obtained in   the grand canonical
limit, i.e.

\ber \langle N_c\rangle_{\rm eq}^{\rm C}={\langle N_c\rangle_{\rm
eq}^{\rm }}^2 \ll {\langle N_c\rangle_{\rm eq}^{\rm }}.
\label{eq18}
 \eer
% holds since in the considered limit
%$<N_c><<1$.

 Secondly, we
can  conclude  that the relaxation time for a canonical system is
 shorter than the grand canonical value, i.e.
\ber \tau_0^{\rm C}=\tau_0^{\rm } \langle N_c\rangle_{\rm eq}^{\rm
} \ll \tau_0^{}, \label{eq19} \eer
 since in the limit $\bf (ii)$  the equilibrium
value  $\langle N_c\rangle_{\rm eq}^{\rm } \ll 1$.

 We note that the $ (\bf i)$ and
the  $(\bf ii)$  limits are essentially determined by the size of
$\langle \delta N_c^2 \rangle$, the  fluctuations of the number of
particles  $c$. The grand canonical results correspond to small
fluctuations, i.e. $\langle \delta N_c^2 \rangle/\langle N_c
\rangle^2\leq 1$, while large fluctuations  $\langle \delta N_c^2
\rangle/\langle N_c \rangle^2$ $ \geq  1$ require a canonical
description.

The  volume dependence of particle density obviously differs in
the C and in the GC limit. The particle density in the GC limit is
$V$-independent whereas in the canonical approach  it  can even
scale  linearly with $V$.
%\subsection{Master equation - general solution}
%%%%%%%%%%%%%%%%%%%%%%%%%&&&&&&&
%In the previous section, we formulated a general master equation
%(1) for the probability to find the number of U(1) charged
%particles being produced by the  binary $ab\to c\bar c$ process.
%The approximate solutions of Eq.~(1) were   considered in two
%limiting cases. For a large number of produced particles carrying
%U(1) charge, the time evolution and the equilibrium limit of
%Eq.~(1) correspond to the GC result.

The difference between the C and asymptotic GC result already seen
on the level of the rate equations (\ref{eq7},\ref{eq9}), is even
more transparent when comparing master equations for {\it
probabilities}. In the following we formulate  this equation for
the GC description of quantum number conservation.

%\subsubsection{Equilibrium multiplicity and relaxation time  in the asymptotic limits}
In  case of  abundantly produced particles $c$ and $\bar c$
through the $ab\to c\bar c$ process we do not need to worry about
strong particle correlations due to   charge  conservation. This
also means that, instead of imposing  charge neutrality conditions
through $N_c-N_{\bar c}=0$,  one assumes conservation on the
average, that is $\langle N_c\rangle -\langle N_{\bar c}\rangle
=0$. In this case the master equation (\ref{eq5}) can be
simplified.

In the derivation of Eq.~(\ref{eq5}) the absorption terms
proportional to $L$ were obtained by constraining the  charge
conservation to be local and  exact. For the conservation on the
average, the transition probability from $N_c$ to the $(N_c-1)$
state is no longer proportional to $(L/V) N_c^2$ but rather to
$(L/V) N_c\langle N_{\bar c}\rangle$, since the exact conservation
condition $N_c=N_{\bar c}$ is no longer  valid and the number of
$\bar c$ particles can only be    determined by its average value.
In the GC limit, the  master equation for the time evolution of
the probability $P_{N_c}(\tau )$ takes the following form:

%A general rate equation for the production of particle pair $b_1 b_2$
%can be written in the form of iterative equations as the following:
\ber \frac{dP_{N_c}}{d\tau}&=&{G\over V} \langle
N_{a}\rangle\langle N_{b}\rangle P_{N_c-1} +\frac {L}{V}
(N_c+1)\langle N_{\bar c}\rangle P_{N_c+1}
 \nonumber \\
&-& {G\over V}\langle N_{a}\rangle \langle N_{b}\rangle
P_{N_c}~~~- \frac {L}{V} N_c\langle N_{\bar c}\rangle P_{N_c} .
\label{eqq2} \label{eq20} \eer

Multiplying the above equation by $N_c$, summing over $N_c$ and
using the condition that $\langle N_c\rangle =\langle N_{\bar
c}\rangle $, one recovers Eq.~(\ref{normal1}), the rate equation
for $\langle N_c\rangle $ in the GC ensemble. The above equation
is thus indeed the general master equation for the probability
function in the GC limit. Comparing this equation with the more
general Eq.~(\ref{eq5}), one can see that the main difference is
contained in the absorption terms that  are linear in particle
number instead of being quadratic .

 Eq. (\ref{eqq2}) can be solved exactly. Indeed, introducing the
generating function $g(x,\tau)$ for $P_{N_c}$,

\be g(x, {\tau }) = \sum_{N_c=0}^{\infty} x^{N_c} P_{N_c}(\tau),
\label{eq21} \ee the  iterative equation (\ref{eqq2}) for the
probability can be converted into a differential equation for the
generating function:
 \be
  \frac{\partial g(x,\tau)}{\partial \tau}={{L}\over V}\sqrt\epsilon (1-x)[g^\prime -{\sqrt\epsilon}
  g],
\label{eq22}
 \ee
with the general solution\cite{r20}:
 \be
  g(x,\tau)=g_0(1-xe^{-\tilde\tau})\exp [{\sqrt{\epsilon}(1-x)(e^{-\tilde\tau}-1)}], \label{solu}
\label{eq23}
  \ee
%under  the initial conditions $g(x,0)=g_0(1-x)$ with
where    $g' =
\partial g / \partial x$,       $\tilde\tau=(L\sqrt\epsilon /V)\tau $ and
$\sqrt\epsilon =\langle N_c\rangle_{eq}$ given by Eq.~(9).
% The normalization $g(x=1,\tilde\tau)=\sum
%P_{N_c}=1$ also implies $g_0(0)=1$.

One can readily find out an equilibrium solution to the above
equation. Taking the limit $\tau=\infty$ in the Eq.~(\ref{eq23})
leads to
 \be
 g_{\rm eq}(x)=\exp [{-\sqrt{\epsilon}(1-x)}], \label{eq}
\label{eq24}
 \ee
with the corresponding equilibrium multiplicity distribution:
 \be
  P_{N_c,{\rm
  eq}}=\frac{(\sqrt{\epsilon})^{N_c}}{{N_c}!}e^{-\sqrt{\epsilon}}.
\label{eq25}
 \ee
This  is the expected Poisson distribution with  average
multiplicity $\sqrt{\epsilon}$.

\subsubsection{The equilibrium solution of the  general rate equation}

The master equation (\ref{eq20}), that  describes  the  evolution
of the probability function in the GC limit,  could be solved
analytically. The general equation (\ref{eq5}), however, because
of the quadratic dependence of the absorption terms, requires  a
numerical solution. Nevertheless, the equilibrium result for the
particle multiplicity can be given.

Converting Eq.~({\ref{eq5}})   for $P_{N_c}$ into a partial
differential equation for the  generating function
 \be
g(x,\tau) = \sum_{N_c=0}^\infty x^{N_c} P_{N_c} (\tau). \label{23}
\label{eq26}
 \ee
%Multiplying Eq.(\ref{generaln}) by $x^n$ and summing over $N_c$,
one  finds\cite{r20}
 \ber \frac {\partial g(x,\tau)}{\partial
\tau} = \frac {L}{V} (1-x) \left (x g''+ g'- \epsilon g \right ).
\label{eq27} \eer
%where .  The conservation of the total probability is justify
%as the $g(x=1,\tau)$ does not change with time.
%%%%%%%%%%
The equilibrium solution  $g_{\rm eq}(x)$ thus obeys the following
equation: \ber xg_{\rm eq}'' + g_{\rm eq}' - \epsilon g_{\rm eq} =
0. \label{equil} \label{eq28} \eer By a substitution of variables
($x = y^2\epsilon /4$), this equation is   reduced to the Bessel
equation, with the following solution:

\ber g_{\rm eq}(x) = \frac {1}{I_0 ( 2\sqrt \epsilon)} I_0 (
2\sqrt {\epsilon x} ), \label{eq29} \eer where the normalization
is fixed by $g(1) = \sum P_{N_c} = 1$.
% We
%have used an additional requirement
% that $g(0)=P_0 \leq 1$ to remove the
%$K_0(2\sqrt {\epsilon x})$ function as the solution of the above
%equation.

The equilibrium value for the probability function $P_{N_c}$ is
now written from      Eqs.~({\ref{eq26}--\ref{eq29}}) as:

\ber P_{N_c,\rm eq}=\frac {\epsilon^{N_c}}{I_0 ( 2\sqrt \epsilon )
(N_c!)^2}. \label{eq30} \eer
%%%%%%

We note that the equilibrium distribution of the particle
multiplicity is not  Poissonian. This fact was indicated first in
equilibrium studies  in Ref.~(\refcite{r39}). In our case  this is
a direct consequence of the quadratic dependence on the
multiplicity in the loss terms of the master equation (\ref{eq5}).
The Poisson distribution is obtained from Eq.~(\ref{eq30}) if
$\sqrt\epsilon \gg 1$, that is for large particle multiplicity
where  the C ensemble coincides with the GC asymptotic
approximation. In Fig.~(\ref{pois}) we compare the  Poisson
distribution from Eq.~(\ref{eq25}) with the distribution  from
Eq.~(\ref{eq30}) for two values of $\sqrt\epsilon$.

%%%%%%%%%%%%%%%%%%%%%%%%%%%%%%f222222222222222222222222222222222222222222%%%%%%%%%%%%%%%%%%%%
\begin{figure}[htb]
%\begin{minipage}[t]{80mm}
%\framebox[79mm]{\rule[-26mm]{0mm}{52mm}}
\vskip -0.4cm {\hskip -1.1cm
%\includegraphics[width=28.5pc, height=15.9pc]{figure2.eps}}\\
%hskip 2cm\includegraphics[width=15.5pc, height=21.9pc]{FKMUS.PS}}\\
\hskip 1.8cm\includegraphics[width=22.5pc, height=19.9pc,angle=180]{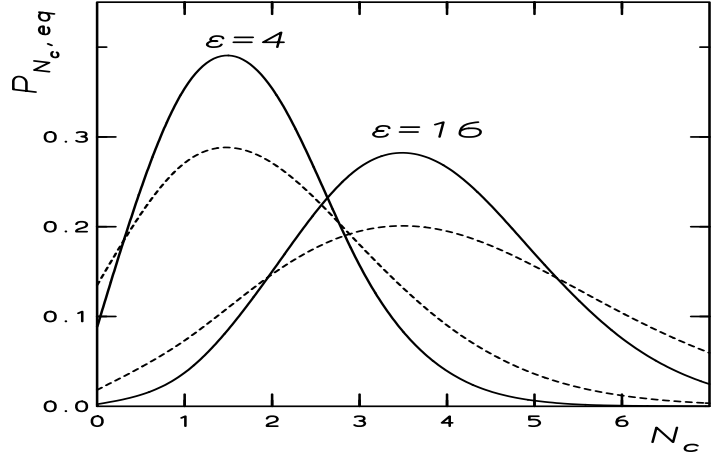}}\\
\vskip -0.7cm { \caption{ \label{pois} The probability function
from Eqs.~(\protect\ref{eq25},\protect\ref{eq30})  for two values
of $\epsilon =4$ and 16. The full lines represent Poisson
distribution.
 }}
%\end{minipage}
\end{figure}
%%%%%%%%%%%%%%%%%%%%%%%%%%%%%%1%%%%%%%%%%%%%%%%%%%%

The result for the  equilibrium average number of particles  $c $
 can be obtained  as:
%%%%%%%
  \ber \langle N_c \rangle_{\rm eq}= g'(1)= \sqrt
\epsilon \frac {I_1 ( 2\sqrt \epsilon)}{I_0 ( 2\sqrt \epsilon)}.
\label{neq} \label{eq31} \eer

 The above expression   will be shown in the next section
 to coincide with the one
expected  for the particle multiplicity in the canonical ensemble
with respect to  U(1)  charge conservation\cite{r37,r37a}. The
rate equation formulated in Eq.~(\ref{eq5}) is valid for arbitrary
values of $\langle N_c\rangle $ {\it and} obviously reproduces
(see Eqs.~(\ref{eq91} - \ref{eq94})) the standard grand canonical
result for a large $\langle N_c\rangle$. Thus, within the approach
developed above one can study  the chemical equilibration of
charged particles following Eq.~(\ref{eq5}), independent of
thermal conditions inside the system.

\subsubsection{The master equation in the presence of the net  charge.}

So far, in constructing the  evolution equation for probabilities,
we have assumed that  there is no  net  charge  in the system
under consideration. For the application of the statistical
approach to particle production in heavy ion and  hadron--hadron
collisions, the above assumption has to be extended to the more
general case of non--vanishing
  initial values of conserved charges.
  %baryon number and
%electric charge are small and non-vanishing.
%only justified when the initial state is charge neutral and when
%considering particle yields in a full phase--space. However,
%because of experimental limitations, one often deals with data in
%restricted kinematical windows where does not
%need to vanish. In addition, in peripheral heavy ion and in
%hadron--hadron collisions
%Thus, to take into account the overall  charge these relevant
%cases a generalization of the master equation (\ref{eq5}) is
%required.
In the following we construct the evolution equation for
$P_{N_c}^S(t)$ in a thermal medium assuming that  its net charge
$S$ is non--vanishing.

The presence of a non--zero net  charge requires modification of
the absorption terms in Eq.~({\ref{eq5}). The transition
probability per unit time from the $N_c$ to the $N_c-1$ state was
proportional to $(L/V)N_cN_{\bar c}$. Admitting an overall net
charge $S\neq 0$ the exact   charge conservation implies that
$N_c-N_{\bar c}=S$. The transition probability from $N_c$ to
$N_c-1$ due to pair annihilation is thus $(L/V)N_c(N_{ c}-S)$.
Following the same procedure as in Eq.~(\ref{eq5}) one can
formulate the following master equation for the probability
$P_{N_c}^S(t)$ to find $N_c$ particles $c$ in a thermal medium
with a  net charge $S$:

\ber \frac{dP_{N_c}^S}{d\tau}&=&{G\over V} \langle
N_{a}\rangle\langle N_{b}\rangle P_{N_c-1}^S +\frac {L}{V}
(N_c+1)(N_c+1-S) P_{N_c+1}^S
 \nonumber \\
&-& {G\over V}\langle N_{a}\rangle\langle  N_{b}\rangle
P_{N_c}^S~~~- \frac {L}{V} N_c(N_c-S) P_{N_c}^S , \label{eqS}
\label{eq32} \eer which  obviously reduces to Eq.~(\ref{eq5}) for
$S=0$.

To get the equilibrium solution for the probability and
multiplicity, we again convert the above equation to the
differential form for the generating function $ g^S(x,\tau) =
\sum_{N_c=0}^\infty x^{N_c} P_{N_c}^S (\tau)$:

\ber \frac {\partial g^S(x,\tau)}{\partial \tau} = \frac {L}{V}
(1-x) \left (x g_S''+ g_S'(1- S)-\epsilon g_S \right ).
\label{eq33} \eer
%%%%%%%%%%%%%
In equilibrium, ${\partial g^S(x,\tau)}{\partial \tau}=0$ and the
solution for $g^S_{\rm eq}$ can be found as follows:

\ber g^S_{\rm eq}(x) = \frac {x^{S/2}}{I_S ( 2\sqrt \epsilon)} I_S
( 2\sqrt {\epsilon x} ), \label{27} \label{eq34} \eer
%%%%%%%%%%%%
where   the normalization is fixed by $g(1) = \sum P_n = 1$.

The master equation for the probability to find $N_{\bar c}$
antiparticles $\bar c$, its corresponding differential form and
the equilibrium solution for the generating function can be
obtained by replacing $S$ with $ -S$ in
Eqs.~(\ref{eq32}--\ref{eq34})

%The corresponding results for the equilibrium value of the
%probability function $P_{N_c}^S$ can be now express  as:

%\ber P_{N_c,\rm eq}=\frac {\epsilon^n}{I_S ( 2\sqrt \epsilon )
%(n!)^2}.\eer

 The result for the  equilibrium average number of
particles  $\langle N_c\rangle_{\rm eq}$ and  antiparticles
$\langle N_{\bar c}\rangle_{\rm eq}$ is obtained from the
generating function using the relation:   $ \langle N_c
\rangle_{\rm eq}= g'(1)$. The final expressions read:
\ber \langle N_c \rangle_{\rm eq}=  \sqrt \epsilon \frac {I_{S-1}
( 2\sqrt \epsilon)}{I_S ( 2\sqrt \epsilon)}~~,~~ \langle N_{\bar
c} \rangle_{\rm eq}=  \sqrt \epsilon \frac {I_{S+1} ( 2\sqrt
\epsilon)}{I_S ( 2\sqrt \epsilon)}.
 \label{neq1}
\label{eq35}
 \eer
The charge conservation is explicitly seen by  taking the
difference of these equations that results in the net value of the
 charge $S$.

The  thermal average values of the particle number given through
Eq.~(\ref{neq1}) will be later derived from the equilibrium
partition function by using the projection method\cite{r37a,r38}.

\subsubsection{The kinetic equation for different particle species}

The rate equations  discussed until now were derived assuming that
there is only one kind of particle $c$ and its antiparticle $\bar
c$ that carry  conserved charge. To study equilibration of
particles in a strongly interacting environment one also needs to
include processes that involve different  species. In low energy
heavy ion collisions e.g.  the $K^+$ mesons are not only produced
in pairs together with $K^-$ but also with the strange hyperon
$\Lambda$ or $\Sigma^0$ due to the $\pi N\to \Lambda K^+$ process.
The contribution of $K^0$ and $\bar {K^0}$ has to be also included
as these particles are produced with similar  strength as charged
kaons.
%In addition to  $K^+$ carrying positive strangeness is
%also carrying by $K^0$ mesons.
To  account for this situation one  generalizes the rate equations
described in the last sections.

%%%%%%%%%%%%%%%%%%%%%%%%%%%%%f66666666666666
%\vskip -0.5cm
%\begin{figure}[htb]
% {\hskip .5cm
% \includegraphics[width=7.5pc, height=30.9pc,angle=270]{fig1n.eps}}\\
%%\includegraphics[width=20.9pc,height=6.pc]{RFIG27.PS}}\\
%\vskip -0.1cm { \caption{\label{flow2} A schematic view of the
%master equation for the probability $P_{N_{K^+},N_{\Lambda}}(\tau
%)$ due to  $m\bar m\to K^+K^- $ and $mN\to K^+\Lambda$ and  the
%inverse processes. }}
%\protect
%\end{figure}
%%%%%%%%%%%%%%%%%%%%%%%%%%%

Consider
$P_{N_K,N_{\Lambda}}(\tau )$ as the probability to find
$N_{K}$
and
$N_\Lambda$ number of $K^-$ mesons and $\Lambda$ baryons.
Including the production and absorption processes such  as: $m\bar
m\to K^+K^-$ and $mN\to K^+\Lambda$  this probability will
obviously change in time. Here $m(N)$ denotes a meson (nucleon).
%In Fig.~\ref{flow2} we  show a schematic view of
%the flow of the transition probability between different states.

%%%%%flow2
\noindent
 Following  a similar procedure  as  was explained
in Fig.~(\ref{flow1}) the master equation for the time evolution
of the probability $P_{N_K,N_{\Lambda}}(\tau )$ can be written
as:\cite{r20a,r22}

\ber \frac{P_{N_K,N_{\Lambda}} }{d\tau} &=&{G_m\over V} \langle
N_{m}\rangle\langle N_{\bar m}\rangle P_{N_K-1,N_\Lambda} +
\frac{L_m}{V}  (N_K+1)(N_K+1+N_\Lambda)
 \nonumber \\
&&\times P_{N_K+1,N_\Lambda} \nonumber \\
&-& {G_m\over V}\langle N_{m}\rangle\langle N_{\bar m}\rangle
P_{N_K,N_\Lambda}~~~- \frac {L_m}{V} N_K(N_K+N_\Lambda)
P_{N_K,N_\Lambda}\nonumber \\
&+&{G_N\over V} \langle N_{m}\rangle\langle N_{ N}\rangle
P_{N_K,N_\Lambda-1} + \frac{L_N}{V} (N_\Lambda+1)(N_\Lambda+1+N_K)
 \nonumber \\
&&\times P_{N_K,N_\Lambda+1} \\ &-& {G_N\over V}\langle
N_{m}\rangle\langle N_{N}\rangle P_{N_K,N_\Lambda}~~~- \frac
{L_N}{V} N_\Lambda(N_\Lambda+N_K) P_{N_K,N_\Lambda}\nonumber .
\label{eqf2} \label{eq36} \eer
%%%%%%%%%%%%%%%
with $G_m$ and $L_m$ being the production and absorption terms for
the $m\bar m \rightleftharpoons K^+K^-$ reaction  and  $G_N$ and
$L_N $ denote equivalent terms for the $mN \rightleftharpoons
K^+\Lambda$ process.

 The equilibrium solution for the probability function
$P_{N_KN_\Lambda}$
 can be found as follows\cite{r22}:
  \be
P_{N_K,N_\Lambda} = \frac{\epsilon_{tot}^{N_K+N_\Lambda}}{I_0(2
\sqrt{\epsilon_{tot}}) ((N_K+N_\Lambda)!)^2}
\frac{(N_K+N_\Lambda)! \epsilon_m^{N_K} \epsilon_N^{N_\Lambda}
}{\epsilon_{tot}^{N_K+N_\Lambda} N_K! N_\Lambda!}
\label{new_equil} \label{eq37} \ee
with $\epsilon_{tot} = \epsilon_m+\epsilon_N$ and
$\epsilon_{m(N)}= G_{m(N)} \langle N_{m_1(m)}\rangle \langle
N_{m_2(N)}\rangle/L_{m(N)}$.

 The equilibrium probability distribution is thus, according
to Eq. (\ref{eq37}),
 the product of
the distribution of the number of pairs  $(N_K+N_\Lambda)$ and a
binomial distribution that determines the relative weight of the
individual particles, in our case the $K^-$ and $\Lambda$.

The probability $P_{i,j}$ is obviously normalized such that
% \be
$\sum_{i,j}P_{i,j}=1$.
%
%\label{eq35} \ee
 The equilibrium value for the multiplicity $ \langle N_i\rangle$
with $i=K^-$ or $i=\Lambda$ can be obtained as:

% \be
%
\ber \langle N_{K^-} \rangle_{\rm eq}= {{\epsilon_m}\over  \sqrt
\epsilon_{tot}} \frac {I_{1} ( 2\sqrt {\epsilon_{tot}})}{I_0 (
2\sqrt {\epsilon_{tot}})}~~,~~
 \langle N_\Lambda \rangle_{\rm eq}= {{\epsilon_N}\over  \sqrt
\epsilon_{tot}} \frac {I_{1} ( 2\sqrt {\epsilon_{tot}})}{I_0 (
2\sqrt {\epsilon_{tot}})}
 \label{diff}
\label{eq38}
 \eer
%
%\label{eq37} \ee
with $\epsilon_m, \epsilon_N$ and $\epsilon_{tot}$ defined as
above.

The average value of $K^+$ can  be obtained applying strangeness
conservation  leading  to:

 \be
  \langle N_{K^+}\rangle_{eq}=
\langle N_{K^-}\rangle_{eq} + \langle N_{\Lambda}\rangle_{eq} .
\label{eq39}
 \ee

% The master equations derived here can be used to study
% the time evolution and equilibration of particles being produced
%in a thermal medium
% This equations  account for U(1) charge conservation
%for both strongly correlated and uncorrelated particles.

 The results presented here can be extended\cite{r22} to an even  more general case
where there is an arbitrary number of  different particle species
carrying  the quantum numbers related with  U(1) symmetry of the
Hamiltonian.

\subsection{The canonical description of an internal symmetry - projection method }
%\subsection{Partition function in the canonical ensemble }

 Using the above  kinetic  analysis of charged particle
 production probabilities  we have demonstrated
 that   equilibrium distributions  does not necessarily
coincide with the GC value. It is thus natural to ask what is the
corresponding partition function that can reproduce the kinetic
results obtained in Eqs.~(\ref{eq31},\ref{eq35},\ref{eq38}). The
main step in deriving these equations was an assumption of an {\it
exact} conservation of quantum numbers in the kinetic master
equations (\ref{eq5},\ref{eq32},\ref{eq36}). Thus,  one should
account for this important constraint  in constructing the
partition function.

The exact treatment of quantum numbers in statistical mechanics
has been well established\cite{r37,r37a} for some time now. It is
in general obtained\cite{r38,r38f} by projecting the partition
function onto the desired values of the conserved charge by using
group theoretical methods. In this section we develop these
methods and show how one gets the partition function that accounts
for exact conservation of quantum numbers. The  derivation will be
not only restricted to the charge conservation related  with an
Abelian U(1) internal symmetries and their direct products, but it
will
 include also symmetries  that are imposed by any semi-simple compact
Lie group.

The usual way of treating the problem of quantum number
conservation in  statistical physics is by introducing the grand
canonical partition function, as in Eq.~(\ref{eq3}). For only one
conserved  charge, e.g.  strangeness $S$,
\be
Z(\mu_S,T)={\rm Tr}[e^{-\beta (\hat{H}-\mu_{S}  \hat{ S})}]
\label{eq40} \ee
The chemical potential $\mu_S$ is then fixed by the condition that
the average value of strangeness of a thermodynamical system  is
conserved and has  the required value $\langle S\rangle$ such
that:
\be \langle S\rangle=T { {\partial\ln Z(\mu_S,T)}\over {\partial
\mu_S}} \label{eq41} \ee

This method, as shown in the previous sections, is only adequate
if the number of particles carrying strangeness is very large and
their fluctuations  can be neglected.

In order to derive a partition function that is free from the
above requirements let us first reorganize Eq.~(\ref{eq40}).
Denoting the states under the trace as $\vline s\rangle$ such that
$\hat H\vline s\rangle=E_s\vline s\rangle$ and $\hat S\vline
s\rangle=s\vline s\rangle$ one writes

\be Z(\mu_S,T)=\sum_{s=-\infty}^{s=+\infty} e^{-\beta E_s}
e^{s\beta\mu_S}= \sum_{s=-\infty}^{s=+\infty} Z_S \lambda_S^s
\label{eq42}
 \ee
where we have  introduced the fugacity $\lambda_S=e^{\beta\mu_S}$
and where \be
 Z_S={\rm Tr_S}[e^{-\beta \hat H}]
\label{eq43}
 \ee
  is just the  partition function  that is restricted to a specific
  total value $S$ of the conserved charge.
  This is  the  {\it canonical }
partition function with respect to  strangeness conservation.
 Thus, $Z_S$ is a coefficient in the Laurent series in the
fugacity. Our goal is to calculate $Z_S$. This is an easy task:
starting from  Eq.~(\ref{eq42}) we apply the  Cauchy formula and
take an inverse transformation to obtain

\be Z_S(T,V)={{1}\over {2\pi i}}\oint {{d\lambda_S}\over
{\lambda_S^{s+1}}} Z(\lambda_S,T,V) \label{eq44} \ee
Choosing the integration path as the unit circle and
parameterizing it as $\lambda_S=\exp{(i\phi)}$ we can convert the
contour integral into the angular one as
\be Z_S(T,V)=\int_{-\pi}^{+\pi} {{d\phi}\over {2\pi}} \tilde
Z(\phi ,T,V) \label{eq45}
 \ee
where the generating function $\tilde Z(\phi,T,V)= Z(\lambda_S
=e^{i\phi },T,V)$ is  obtained from the grand canonical partition
function by a  Wick rotation of  the chemical potential $\mu_S\to
i\phi$. This generating function is the same for all canonical
partition functions with an arbitrary but fixed value of the
conserved charge.  Eq.~(\ref{eq45}) is the projection formula onto
the canonical partition function that accounts for the  exact
conservation of an Abelian charge. This is  a projection procedure
as $Z_S$ is obtained from
\be
Z_S(T,V)={\rm Tr_S}[e^{-\beta \hat H}] = {\rm Tr}[e^{-\beta \hat
H}P_S] \label{eq46} \ee where $P_S=P_S^2$ is the projection
operator on the states with the  exact value of $S$. For an
Abelian symmetry, $P_S$ is the $\delta$--function
$P_S=\delta_{\hat S,S}$. Introducing the Fourier decomposition of
delta into Eq.~(\ref{eq46}) one can reproduce the projected result
(\ref{eq45}).

The conservation of additive quantum numbers like baryon number,
strangeness, electric charge or charm is related to the invariance
of the Hamiltonian under the  U(1) Lie group. In many applications
it is important to generalize the projection method to symmetries
that are related with a non-Abelian Lie group $G$. An example is
the special unitary group SU(N) that plays an essential role in
the theory of strong interactions. Generalization of the
projection method would require to specify the projection operator
or generating function. Consequently, the partition function
obtained with the specific eigenvalues of the Casimir operators
that fixes the multiplet of the irreducible representation of the
symmetry group $G$ could be determined.

To find the generating function for the  canonical partition
function with respect to the symmetry group $G$, let us introduce
the quantity $\tilde Z(g)$ via

 \be
 \tilde Z(g)={\rm Tr }[U(g)e^{-\beta\hat H}] \label{gen}.
\label{eq47}
 \ee
This expression is  a  function on the group $G$ with U(g) being
the unitary representation of the group with  $g\subset G$. The
quantity U(g) can be decomposed into irreducible representations
$U_\alpha(g)$
 \be
 U(g)=\sum_\alpha^\bigoplus U_\alpha(g)
\label{eq48}
 \ee
where  $\alpha$ is labelling these representations. From
Eq.~(\ref{eq47}) and (\ref{eq48}) one has
 \ber
 \tilde Z(g)&=&\sum_\alpha{\rm Tr_\alpha }[U_\alpha (g)e^{-\beta\hat H}]
 \nonumber  \\
&=&\sum_\alpha\sum_{\nu_\alpha,\xi_\alpha} \langle
\nu_\alpha,\xi_\alpha\mid U_\alpha (g)e^{-\beta\hat H}\mid
\nu_\alpha,\xi_\alpha\rangle
 \label{gen2}
\label{eq49}
 \eer
where $\nu_\alpha$ labels the states  within the representation
$\alpha$ and  $\xi_\alpha$ are degeneracy parameters.

Introducing the unit operator $1=\mid \rangle\langle \mid$ into
the above equation the expression factorizes
 \ber
 \tilde Z(g)
&=&\sum_\alpha\sum_{\nu_\alpha,\xi_\alpha} \langle
\nu_\alpha,\xi_\alpha\mid U_\alpha (g)\mid
\nu_\alpha,\xi_\alpha\rangle  \langle \nu_\alpha,\xi_\alpha\mid
e^{-\beta\hat H}\mid \nu_\alpha,\xi_\alpha\rangle
\nonumber  \\
 &=&\sum_\alpha\sum_{\nu_\alpha,\xi_\alpha} \langle
\nu_\alpha\mid U_\alpha (g)\mid \nu_\alpha\rangle \langle
\xi_\alpha\mid e^{-\beta\hat H}\mid \xi_\alpha\rangle,
 \label{gen4}
\label{eq50}
 \eer
where we have  used that, due to the exact  symmetry, the only
non--vanishing matrix elements of $e^{-\beta \hat H}$  are those
diagonal in $\nu_\alpha$. The matrix elements of $U_\alpha(g)$ are
only non-zero if they are diagonal in $\xi_\alpha$. Finally, the
matrix elements of the Hamiltonian are independent  of the states
within representation (since due to symmetry they are dynamically
equivalent)  and those of $U(g)$ of degeneracy factors (since
$U(g)$ does not distinguish dynamically different states that
transform under the same representation).

 The last two sums in Eq.~(\ref{eq50}) can be further
simplified   as
 \be
\sum_{\nu_\alpha} \langle \nu_\alpha\mid U_\alpha (g)\mid
\nu_\alpha\rangle ={\rm Tr_{\alpha}}[U_\alpha(g)]=\chi_\alpha (g).
 \label{gen5}
\label{eq51}
 \ee
The quantity $\chi_\alpha$  is by definition the character of the
irreducible $U_\alpha(g) $ representation and
 \be
\sum_{\xi_\alpha}  \langle \xi_\alpha\mid e^{-\beta\hat H}\mid
\xi_\alpha\rangle ={ {1}\over {d(\alpha)}} {\rm Tr_\alpha}
e^{-\beta\hat H}={ {1}\over {d(\alpha)}}Z_\alpha(T,V),
 \label{gen6}
\label{eq52}
 \ee
where $Z_\alpha$ is  the {\it canonical partition function} with
respect to the $G$ symmetry of the Hamiltonian and $d(\alpha)$ is
the dimension of the representation $\alpha$. Calculating
$Z_\alpha$ one considers  under the trace only those states that
transform with respect to a given irreducible representation of
the symmetry group.

We have thus connected,  through Eq.~(\ref{eq50}) and
(\ref{eq51}--\ref{eq52}),  the canonical partition function with
the generating functional on the group

 \be
 \tilde Z(g)=\sum_\alpha {{\chi_\alpha(g)}\over {d(\alpha)}}
Z_\alpha(T,V)
 %\label{las}
\label{eq53}
 \ee
The canonical partition function is the coefficient in the cluster
decomposition of the generating function with respect to the
characters of the representations.

\noindent The character functions satisfy the orthogonality
relation
\be
  {1\over {d(\alpha})}\int
  d\mu(g)\chi_\alpha^*(g)\chi_\gamma(g)=\delta_{\alpha,\gamma}
  \label{or}
\label{eq54}
 \ee
where $d\mu(g)$ is an invariant Haar measure on the group.

The orthogonality relation for characters allows to find the
coefficients, the canonical partition function, in  this cluster
decomposition. From Eq.~(\ref{eq53}) and (\ref{eq54}) one gets
 \be
Z_\alpha(T,V)=d(\alpha)\int d\mu(g) {\chi_\alpha^*(g)} \tilde Z(g)
  \label{maing}
\label{eq55}
 \ee
This result is a generalization of Eq.~(\ref{eq45}) to an
arbitrary symmetry group that is a compact Lie group. The formula
holds  for any dynamical system described by the Hamiltonian $H$.

 To find the
canonical partition function we have to determine  first the
generating function $\tilde Z(g)$ defined on   the symmetry group
$G$. If the symmetry group is of rank $r$, then the character of
any irreducible representation   are the functions of $r$
variables $\{\gamma_1,_{....},\gamma_r\}$. Denoting as $J_k$ the
commuting generators  of $G$  with $k=1,...,r$ the character
function
 \be
\chi_\alpha(\gamma_1,..,\gamma_r)=\sum_{\nu_\alpha} \langle
\nu_\alpha\mid e^{i\sum_{i=1}^{i=r}\gamma_iJ_i}\mid \nu_\alpha
\rangle
  \label{las1}
\label{eq56}
 \ee
is obtained. Here, $\nu_\alpha$ labels  the state within the
representation $\alpha$. With the above  form of the characters we
can write Eq.~(\ref{eq53}) as
 \be
\tilde Z(\gamma_1,..,\gamma_r)={\rm Tr}[ e^{-\beta\hat H+
i\sum_{i=1}^{i=r}\gamma_iJ_i}]
  \label{la1}
\label{eq57}
 \ee
Through the Wick rotation $\gamma_i=-i\beta\mu_i$ the generating
function $\tilde Z$ is just the GC partition function with respect
to the conservation laws given by all commuting generators of the
symmetry group $G$.

The  equations (\ref{eq55}) and (\ref{eq57}) are the basis that
permits to obtain the canonical partition function for systems
restricted to any symmetry. The simplicity of the projection
formula (\ref{eq55}) is   that the operators that appear in the
generating function are additive they  are generators of the
maximal Abelian subgroup of $G$. Thus, the problem of extracting
the canonical partition function with respect to an arbitrary
semi-simple compact Lie group $G$ is reduced to the projection
onto a maximal Abelian subgroup of $G$.

The calculation of the generating function from the
Eq.~(\ref{eq57}) can be  done applying    standard perturbative
diagrammatic methods or a mean field approach. However, if
interactions can be omitted or effectively  described by a
modification of the particle dispersion relations by implementing
an effective mass, then the trace in  Eq.~(\ref{eq57})  can be
worked out\cite{r38} exactly, leading to

 \be
\tilde Z(\vec \gamma)=\exp [{\sum_\alpha {{\chi_\alpha(\vec
\gamma)}\over {d(\alpha)}}Z^1_\alpha}]
%  \label{la3}
\label{eq58}
 \ee
where $\vec \gamma =(\gamma_1,_..,\gamma_r)$ and $Z^1_\alpha=\int
(gVdp/2\pi^2)p^2\exp {(-\sqrt {p^2+m_\alpha^2}/T)}$ is just the
thermal particle phase--space in  Boltzmann approximation
belonging to a given irreducible multiplet of a symmetry group
$G$.  The sum is taken over all particle representations that are
constituents of the thermodynamical system.

\subsubsection{Canonical models with a non-Abelian symmetry }

To illustrate how the projection method described above works, we
discuss a statistical  model that accounts for the canonical
conservation of non-Abelian charges related with the
$SU_c$(N)$\times U_B(1)$ symmetry with $N=3$ and $B$ being the
baryon number and $c$ denoting the global gauge colour symmetry.

Let us consider a thermal fireball that is composed of quarks and
gluons at temperature T and  volume V . We  describe the canonical
partition function that is projected on the global color singlet
and exact value of the baryon number. The interactions between
quarks and gluons  are implemented effectively, resulting in
dynamical particle masses that are temperature dependent, e.g.
through $m_{q,(g)}\sim g T$. Since the interactions are only
trivially modifying the dispersion relations one can  still use
the free particle momentum phase-space. Thus, under this
assumption, Eq.~(\ref{eq58}) provides the correct description of
the generating function. The sum in the exponents in (\ref{eq58})
gets  contributions from quarks, antiquarks  and gluons that
transform under the fundamental (0,1), their conjugate  (1,0) and
adjoint (1,1) representation of the $SU_c(N)\times U_B(1)$
symmetry group. Thus,
 \be
\ln \tilde Z(T,V,\vec\gamma,\gamma_B) = {{\chi_Q}\over {d_Q}}Z_Q^1
 + {{\bar\chi_Q}\over {d_Q}}Z_{\bar Q}^1 + {{\chi_G}\over
{d_G}}Z_G^1
  \label{las4}
\label{eq59}
 \ee
where $\vec\gamma =(\gamma_1,..,\gamma_{N-1})$ are the parameters
of the  $SU_c(N)$ and $\gamma_B$ of the   $U_B(1)$ symmetry group.

Through an explicit calculation of  one-particle partition
functions for massive quarks and gluons the corresponding
generating function  is obtained as
 \ber
\ln\tilde Z_Q(T,V,\vec\gamma,\gamma_B)&=&\frac {g_Q}{d_Q}\frac
{m_Q^2VT}{2\pi^2}\sum_{n=0}^\infty  \frac {(-1)^{n+1}}{{n^2}}
K_2(m_Q/T)
\nonumber\\
&[&e^{i\gamma_Bn/T}\chi_Q(n\vec \gamma)+
e^{-i\gamma_Bn/T}\chi_Q^*(n\vec \gamma) ]\label{las6} \label{eq60}
 \eer
where the two  terms in the bracket represent the contribution of
quarks and antiquarks, respectively. The corresponding result for
massive gluons reads
 \ber
\ln\tilde Z_G(T,V,\vec\gamma,\gamma_B)=\frac {g_G}{d_G}\frac
{m_G^2VT}{2\pi^2}\sum_{n=0}^\infty         \frac {1}{{n^2}}
K_2(m_G/T)
%\nonumber\\
[\chi_G(n\vec \gamma)+ \chi_Q^*(n\vec \gamma) ]~~~~\label{las7}
\label{eq61}
 \eer
where $g_G,g_Q$ and $d_Q=N,d_G=N^2-1$, are respectively, the quark
and gluon degeneracy factors and dimensions of the
representations.

Now we can apply this generating function in  the  projection
formula (\ref{eq55})     to get  the canonical partition function.
Of particular interest  is  the color singlet partition function
that represents global colour neutrality (phenomenological
confinement) of a quark-gluon plasma droplet. The conjugate
character for the $SU_c(N)$ singlet representation is particulary
simple, $\chi^{(0,0)}=1$. The baryon number  be treated grand
canonically  requiring  a substitution $\gamma_B=-i\mu_B/T$ in
Eq.~(\ref{eq60}). To find $\tilde Z$  one still  needs an explicit
form of the fundamental and adjoint characters and the Haar
measure on the $SU_c(N)$ group. Here we quote their structure for
the $SU_c(3)$ group. The real $L_R$  and the imaginary $L_I$ parts
of the character in the fundamental (quark) representation are
 \ber
&&L_R=\cos\gamma_1+\cos\gamma_2+\cos(\gamma_1+\gamma_2) \nonumber
\\
&&L_I=\sin\gamma_1+\sin\gamma_2-\sin(\gamma_1+\gamma_2) \nonumber.
\label{las8} %\label{eq61}
 \eer
For the adjoint (gluon) representation
 \ber
\chi_G=2[\cos(\gamma_1-\gamma_2)+\cos(2\gamma_1+\gamma_2)+\cos(2\gamma_2+\gamma_1)+1].
 \label{eq62}
 \eer
 The invariant Haar measure on the $SU_c(3)$ internal symmetry
 group
 \ber
d\mu(\gamma_1,\gamma_2)=\frac {8}{3\pi^2} \sin^2(\frac {
\gamma_1-\gamma_2}{2})
 \sin^2(\frac {
2\gamma_1+\gamma_2}{2}) \sin^2(\frac { 2\gamma_2+\gamma_1}{2}).~~
 \label{eq63}
 \eer

From Eqs.~(\ref{eq60})-(\ref{eq63}) and  (\ref{eq55})     we write
the final result for the $SU_c(3)$ color singlet partition
function that for non-vanishing  baryon chemical potential $\mu_B$
reads
 \ber
Z^0(\mu_B,T,V)=\int d\mu(\gamma_1,\gamma_2) &&\exp \{c_1\chi_G+
\nonumber \\ && c_2[L_R\cosh(\beta\mu_B)+iL_I\sinh(\beta)] \}
  \label{las10}
\label{eq64}
 \eer
where the constants $c_1$ and $c_2$   can be extracted from
Eqs.~(\ref{eq60}--\ref{eq61}).

The above partition function shows a complex structure of the
integrand. However, due to its  symmetry it is straightforward to
show that the partition function is real. The thermodynamical
properties of this color singlet canonical partition function and
other thermodynamical observables   can be
studied\cite{r64,r64e,r64n} by a numerical analysis.

In   finite temperature gauge theory the zero component of the
gauge field $A_0$ takes on the role of the Lagrange multiplier
guaranteeing   that all states satisfy  Gauss law. In  Euclidean
space one can choose a gauge in such a way that
$A_0^\nu(x,\tau)\lambda_\nu$ is a constant in  space-time.
%so that
%
% \be
%
%A_0^{ab}=g^{-1}\alpha^{ab}\delta_{ab}.
%  \label{las}
%\label{eq65}
% \ee
%
In such a gauge the Wilson loop defined as
 \be
L(x)={1\over N} {\rm Tr}P\exp[ig\int_0^\beta A_0(x,\tau)d\tau]
\label{eq66}
 \ee
represents the character of the fundamental representation of the
$SU_c(N)$ group\cite{r64}.

The effective potential of the   $SU(N)$ spin model for the Wilson
loop in the above gauge coincides\cite{r64,r64p} essentially with
the generating function given in Eq.~(\ref{eq64}). In addition,
this generating function could be also related to the strong
coupling effective free energy of the lattice gauge theory with a
finite chemical potential\cite{r64}. Thus, the effective model
formulated above connects the colored quasi-particle degrees of
freedom with the Wilson loop.
% In view of recent findings that the
%Wilson loops are an important degrees of freedom near
%deconfinement\cite{r66} and quasi-particles  above the critical
%region\cite{r67q}, the above model and the projection method could
%be of a general interest. In the following we shall concentrate,
%however, on the application of this method to describe particle
%production in heavy ion collisions.  For this purpose  of
%particular interest is the canonical approach related with  the
%abelian U(1) symmetries and their direct products.

% \be
%
%  \label{las}
% \ee
%
% (\ref{?})
%\subsubsection{Color singlet partition function and LGT
%thermodynamics}

\subsubsection{The canonical partition function for Abelian charges}

In this section we show how the  projection method described above
leads to a description of particle yields under the constraints
imposed by the Abelian  $U_B(1)$ symmetry. In this case the
formalism is particularly transparent due  to a simple structure
of the symmetry group.

The U(1) group is of rank one, thus the characters of the
representations, numbered by the eigenvalues of the conserved
charge B, depend only on one parameter $\phi$. They are of the
exponential type:
\be
 \chi^B_{U_B(1)}=e^{iB\phi}.
\label{character} \label{eq67}
 \ee
For the conservation  of a few Abelian charges inside the system
like stran\-ge\-ness (S),   baryon number (B) or  electric charge
(Q) and charm (C) one needs to account for the products of the
U(1) symmetries: $U_S(1)\times U_B(1)\times U_Q(1)\times U_C(1)$.
In this case the characters are numbered by the values of all
conserved charges and they  are expressed as the products of the
corresponding characters of U(1) groups. For  simultaneous
 conservation of baryon number and strangeness the characters
read:
\be
 \chi^{S,B}_{U_S(1)\times U_B(1)}=
\chi^S_{U_S(1)}  \cdot \chi^B_{U_B(1)}=
 e^{i(S\psi +B\phi}).
\label{character2} \label{eq68}
 \ee

The invariant measure   on the $U_S(1)\times U_B(1)$ group is just
the product of the differentials $
d\mu(\phi_S,\phi_B)=(d\phi_S/2\pi)\cdot (d\phi_B/2\pi)$.

In nucleus-nucleus collisions  the absolute values of the baryon
number, electric charge and strangeness are fixed by the initial
conditions. Modelling the particle production using  statistical
thermodynamics, in general, requires a canonical formulation of
all these quantum numbers. We restrict our discussion only to the
case when at most  two conserved charges can be simultaneously
canonical (e.g. the strangeness and baryon number) and all others
are treated using the GC formulation. The corresponding canonical
partition functions can be obtained from
Eqs.~(\ref{eq55},\ref{eq58}) as:

 \be
Z_S={1\over {2\pi }}\int_0^{2\pi}d\phi e^{-iS\phi } {\tilde
Z}(T,V,\phi ) \label{e:zq} \label{eq69}
 \ee
and
 \be
Z_{B,S}={1\over {4\pi^2 }}\int_0^{2\pi}d\phi e^{-iQ\phi }
\int_0^{2\pi}d\psi e^{-iS\psi } {\tilde Z}(T,V,\phi,\psi
)\label{c2} \label{eq70}
 \ee
\noindent where $\tilde Z$ is obtained from the  grand canonical
(GC) partition function replacing the fugacity parameter
$\lambda_B$, $\lambda_S$  by the factors $e^{i\phi }$ and
$e^{i\psi }$ respectively,
 \be
{\tilde Z}(T,V,\phi )=
       Z^{GC}(T,V,\lambda_B \to e^{i\phi }, \lambda_S \to e^{i\psi })
\label{pc2} \label{eq71}
 \ee
The particular form of the generating function $\tilde Z$ in the
above equation is model dependent. In  applications of the above
statistical partition function to the description of particle
production in heavy ion and hadron-hadron collisions   we
calculate $\tilde Z$ in the hadron resonance gas model. In our
analysis we neglect interactions between a hadron and resonances
as well as any medium effects on particle properties. In general,
however, already in the low-density limit, the modifications of
the resonance width or particle dispersion relation
%, in this particular for $\Delta$ and
%$\pi$ \cite{nor,rus},
    could   be of importance\cite{gery,hatsuda,r32,r67}.
For the sake of simplicity, we use a  classical    statistics,
i.e.~we assume a  temperature and density regime such  that all
particles can be treated  using  Boltzmann statistics.

Within the approximations described above and neglecting  the
contributions of multi-strange baryons, the generating function in
Eq.~(\ref{e:zq}),
 has the
following form
 \be
\tilde Z(T,V,\mu_Q,\mu_B,\phi )=\exp (N_{s=0}+N_{s=1}e^{i\phi
}+N_{s=-1} e^{-i\phi }) \label{e:zs0} \label{eq72}
 \ee
\noindent where $N_{s=0,\pm 1}$ is defined as the sum over all
particles and resonances
 having
strangeness $0,\pm 1$,
 \be
N_{s=0,\pm 1}=\sum_k Z_k^1 %\label{e:N}
 \label{eq73}
 \ee
\noindent and  $Z^1_k$ is the one-particle partition function
defined as
 \be
Z_k^1= {{Vg_k}\over {2\pi^2}} \, m_k^2 \, T \, K_2(m_k/T)  \, \exp
(B_k\mu_B+Q_k\mu_Q) \label{e:zk1} \label{eq74}
 \ee
\noindent with the mass $m_k$, spin-isospin degeneracy factor
$Q_k$, particle baryon number $B_k$ and electric charge $Q_k$. The
volume of the system is $V$ and the chemical potentials related to
the charge and  baryon number  are determined by $\mu_Q$ and
$\mu_B$, respectively.

With the particular form  of the generating function (\ref{e:zs0})
the canonical partition function $Z_S$ is obtained from
Eqs.~(\ref{eq69}--\ref{eq71}) as
%%%%%%%%%%%%%%%
%
 \be
Z_S=   Z_{0}  {1\over {2\pi }}\int_0^{2\pi}d\phi e^{-iS\phi }
e^{S_{1}e^{i\phi }+S_{-1} e^{-i\phi }},
 \label{z1}
\label{eq75},
 \ee

\noindent where $Z_0=\exp {(N_{S=0})}$ is the partition function
of all particles having  zero strangeness and where we introduce
$S_{\pm 1}= N_{s=\pm 1}$ with $N_{s=\pm 1}$ defined as in
Eq.~(\ref{eq73}).

 To calculate the canonical partition function (\ref{eq75}) one can expand
each term in the power series and then perform the $\phi$
integration\cite{r68}. Rewriting  the above equation as

 \be
Z_S=  Z_{0}  {1\over {2\pi }}\int_0^{2\pi}d\phi e^{-iS\phi }
e^{\sqrt {S_1S_{-1}}(\sqrt {{S_1}\over S_{-1}}e^{i\phi }+ \sqrt
{{S_{-1}}\over S_{1}}e^{-i\phi })},
 \label{z1i}
\label{eq76}
 \ee
and using the following relation for the modified Bessel functions
$I_S(x)$,
\be e^{{x}\over 2}(t+{1\over
t})=\sum_{-\infty}^{+\infty}t^SI_S(x), \label{relation}
\label{eq77}
 \ee
one gets after  the $\phi$-integration  the  canonical partition
function for a gas with  the net  strangeness $S$:
 \be
Z_{S}(T,V,\mu_B,\mu_Q)= Z_{0}(T,V,\mu_B,\mu_Q)  ({{S_1}\over
S_{-1}})^{S/2} I_S(x) \label{eq78}
 \ee
\noindent where  the argument of the Bessel function
 \be
x= 2\sqrt {S_1S_{-1}}. \label{e:x} \label{eq79}
 \ee
\noindent

The calculation of the particle density $n_k$ of species $k$ in
the canonical formulation is straightforward. It amounts to the
replacement
 \be
Z_k^1 \mapsto \lambda_k \, Z_k^1 \label{eq80}
 \ee
of the corresponding one-particle partition function  in equation
(\ref{eq72}) and taking the derivative of the canonical partition
function  (\ref{eq75}) with respect to the particle fugacity
$\lambda_k$

 \be
n_k^C= \lambda_k \left.{{\partial}\over
 {\partial\lambda_k}}\ln Z_S(\lambda_k)\right|_{\lambda_k=1}
\label{eq81}
 \ee

As an example, we quote the canonical result for the density of
kaons $K^+$ and anti-kaons $K^-$  in  an environment with a net
overall strangeness $S$,
 \be
n_{K^+}^C={{Z_{K^+}^1}\over V} {{S_{-1}}\over {\sqrt {S_1S_{-1}}}}
{{I_{S-1}(x)}\over {I_S(x)}} ~~~~ n_{K^-}^C={{Z_{K^-}^1}\over V}
{{S_1}\over {\sqrt {S_1S_{-1}}}} {{I_{S+1}(x)}\over {I_S(x)}}
 , \label{e:nkc1}
\label{eq82}
 \ee
where $x=\sqrt {S_1S_{-1}}$ and $Z^1$ are as in (\ref{eq74}) and
(\ref{eq75}).

For the particular case when  $S_1=S_{-1}$ the above equation
coincide with (\ref{eq35}). Thus, the master equation (\ref{eq32})
represents  the rate  for the time evolution  of the probabilities
for which  the equilibrium limit corresponds to the canonical
ensemble.

The partition function (\ref{eq76})  and the corresponding results
for particle densities (\ref{eq82}) were derived neglecting the
contribution of multistrange baryons to the generating functional
(\ref{eq72}). Multistrange baryons are, however, an  important
characteristics of the collision fireball created in heavy ion
collisions. Thus, the canonical formalism described above should
be extended  to   account for these  particles.
 Under the constraints of the global  strangeness neutrality
condition $S=0$ and including hadrons with stran\-ge\-n\-ess
content $s=\pm 1, \pm 2, \pm 3$ the canonical partition function
in Eq.~(\ref{eq75}) is replaced\cite{r21,r68} by

 \be
Z^C_{S=0}={1\over {2\pi}}
       \int_{-\pi}^{\pi}
    d\phi~ \exp{\left(\sum_{n=- 3}^3S_ne^{in\phi}\right)},
\label{eq83}
 \ee
where  $S_n= \sum_k Z_k^1$  and the sum is over all particles and
resonances that carry   strangeness $n$ with $Z_k^1$ defined as in
Eq.~(\ref{eq74}).

The  integral representation of the partition function in
Eq.~(\ref{eq83}) is not convenient for a numerical analysis as the
integrant  is a strongly oscillating function. The partition
function, however, after  $\phi$ integration, can be obtained in a
form that is free from oscillating terms. Indeed, rewriting
Eq.~(\ref{eq83}) to

 \be
Z^C_{S=0}={1\over {2\pi}}e^{S_0}
       \int_{-\pi}^{\pi}
                   d\phi~ \prod_{n=1}^3
\exp{\left[{{x_s}\over 2} \left(a_ne^{in\phi}+
a_n^{-1}e^{-in\phi}\right)\right]},  \label{eq84}
 \ee
and using the relation  (\ref{eq77}) one finds, after
integration\cite{r42}

 \be
Z^C_{S=0}=e^{S_0}
\sum_{n=-\infty}^{\infty}\sum_{p=-\infty}^{\infty} a_{3}^{p}
a_{2}^{n} a_{1}^{{-2n-3p}} I_n(x_2) I_p(x_3) I_{-2n-3p}(x_1)
 ,
\label{eq85}
 \ee
where
\be a_i= \sqrt{{S_i}/{S_{-i}}}~~,~~ x_i = 2\sqrt{S_iS_{-i}}
\label{eq86} \ee
and $I_n$ are the modified Bessel functions.

The expression for the  particle density, $n_i$, can be obtained
from  Eq.~(\ref{eq81})  and  Eq.~(\ref{eq83}). For  a particle $i$
having  strangeness $s$
 \be
n_{i}={{Z^1_{i}}\over {Z_{S=0}^C}}
\sum_{n=-\infty}^{\infty}\sum_{p=-\infty}^{\infty} a_{3}^{p}
a_{2}^{n}
 a_{1}^{{-2n-3p- s}} I_n(x_2) I_p(x_3) I_{-2n-3p- s}(x_1)
 .
\label{eq87}
 \ee

In the limit of   $x_2\to 0$ and $x_3\to 0$ it is sufficient to
take only   terms with  $n=0$ and $p=0$ in Eq.~(\ref{eq85}) and
(\ref{eq87})\cite{r21}. In this case   the density of particle
$n_s^C$ and antiparticle $n_{\bar s}^C$ with strangeness content
$s$ and $\bar s= -s$ respectively, reads

 \be
n_{s}^C\simeq {{Z_{s}^1}\over V} \langle {{S_{1}}\over {
{S_{-1}}}}\rangle^{s/2} {{I_{s}(x)}\over {I_0(x)}} ~~,~~ n_{\bar
s}^C\simeq {{Z_{\bar s}^1}\over V} ({{S_{-1}}\over {{S_{1}}}})^{
{\bar s}/2} {{I_{\bar s}(x)}\over {I_0(x)}}
 , %\label{e:nkc1}
\label{eq88}
 \ee
with $x=\sqrt {S_1S_{-1}}$ and $Z_{s}^1$ as in (\ref{eq74}).

The above equation is  an approximation and can be only used  for
a qualitative discussion.   The quantitative  description of
multistrange particle   production requires the  exact result
given in  Eq.~(\ref{eq85}) and (\ref{eq87}).

\subsubsection{The equivalence of the canonical formalism in  the grand canonical limit}

Discussing  the   strangeness kinetics in Section 4.1 we have
already indicated that the canonical description of the
conservation laws is valid  over the whole parameter range. The
grand canonical formulation, on the other hand,   is the
asymptotic realization of the exact canonical approach. This can
be indeed verified  when directly  comparing particle densities
obtained in the C and GC ensemble. Consider first a thermal system
that contains only strangeness 1 particles and their
antiparticles. In such an environment  the GC result for the
strangeness $s=\pm 1$ hadrons  is obtained from (\ref{eq5}) as
\be n_{s=\pm 1}^{GC}={{Z_{s=\pm 1}^1}\over V}\lambda_s^{\pm 1}.
\label{eq89}
 \ee
with the fugacity $\lambda_s= \exp{( {{{\mu_S}/ T}})}.$

 Comparing the
above GC and C result of Eq.~(\ref{eq82}) with $S=0$ one sees that
 \be
n_{s=\pm 1}^{C}=n_{s=\pm 1}^{GC}\left(\tilde\lambda_s\right).
\label{e:nkc2} \label{eq90}
 \ee
where the {\it effective} fugacity parameter
 \be
\tilde\lambda_s={{S_{\mp 1 }}\over {\sqrt {S_1S_{-1}}}}
{{I_1(x)}\over {I_0(x)}}.
 \label{eq900}
 \ee

 In the limit of large $x\to \infty$
 the canonical and the grand canonical formulations are
equivalent. In the opposite limit, however,  the differences
between these two descriptions  are large. This can be seen in the
most transparent way, when directly comparing the  two limiting
situations of  the large and small $x$ in the Eq.~(\ref{eq82}).
For  $x\to \infty $
 \be
\lim_{x\to \infty } {{I_{1}(x)}\over {I_0(x)}}\to 1
\label{e:i1i01} \label{eq91}
 \ee
\noindent and the ratio $S_{-1}/\sqrt {S_1S_{-1}}$ corresponds
exactly to the fugacity $\lambda_S$ in the GC formulation
(\ref{eq89}). Indeed, a strangeness neutrality condition in the GC
ensemble requires that $\langle S\rangle =0$, thus   through
Eqs.~(\ref{eq71}--\ref{eq72}) one has:

\be
\lambda_s S_1- {{\lambda_s^{-1}}{S_{-1}} } =0, \label{eq92}
 \ee
that is  $\lambda_s=S_{-1}/\sqrt {S_1S_{-1}}$.

Thus,  {\it  neglecting  multistrange baryons} in the generating
functional (\ref{eq83}) one gets
\be
 n_{s=\pm 1}^{C}=
n_{s=\pm 1}^{GC}{{I_{1}(x)}\over {I_0(x)}}.
\label{eq93} \ee
Comparing Eq.~ (\ref{eq88}) with the GC result (\ref{eq89}) for
the density $n_s$ of multistrange particles one  finds that
 \be
 n_{s}^{C}\simeq
n_{s}^{GC}{{I_{s}(x)}\over {I_0(x)}}.
\label{eq94} \ee
 However, one needs to remember that the above relation is only
 valid if a thermal phase space of all multistrange hadrons is
negligibly  small. This assumption is, however, questionable
particularly when   approaching the thermodynamical limit.% Thus
%Eq.~(\ref{eq94}) is only a crude approximation.

%%%%%%%%%%%%%%%%%%%%%%%%%%%%%%1%%%%%%%%%%%%%%%%%%%%
\begin{figure}[htb]
\begin{minipage}[t]{124mm}
\vskip -1.5cm\hskip 0.7cm
\includegraphics[width=22.5pc,height=22.2pc,angle=180]{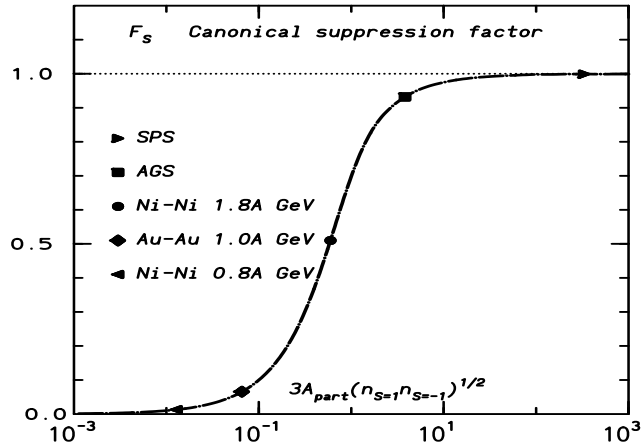}\\
\end{minipage}
%\begin{minipage}[t]{125mm}
{\vskip -1.5 true cm \caption{\label{ff7} The  canonical
strangeness suppression factor  (see  text for the explanation).
The SPS and the AGS values are shown for Pb--Pb and Au--Au
collisions, respectively. } }
%\end{minipage}
\end{figure}

%%%%%%%%%%%%%%%%%%%%%%%%%%%%%%%%%%%%%%%

From Eq.~(\ref{eq93}) one concludes that the relevant parameter
$F_S$ that describes deviations
 of particle multiplicities from their grand canonical value reads
 \be
F_S=  {{I_1(x)}\over {I_0(x)}}. \label{eq95}
 \ee
\noindent
The largest differences appear in the limit of a small $x$ where
 \be
\lim_{x\to 0 } {{I_{1}(x)}\over {I_0(x)}}\to {x/2} \label{e:i1i02}
\label{eq96}
 \ee

\noindent
This limit reproduces the solution  of our kinetic equation
(\ref{eq17})  for large particle number fluctuations.

 The argument
of the Bessel functions in Eq.~(\ref{eq96}) describes the size of
the  thermal  phase-space that is available for strange particles.
For a system free of multistrange hadrons  the argument $x$ can be
also identified as being proportional to the total number of
strange particle-antiparticle pairs in the GC limit.

The canonical suppression factor $F_S(x)$ is quantified in
Fig.~(\ref{ff7}).
% To relate the initial volume of the system
%with the number of participants we use the approximate relation
%$V\sim 1.9\pi A_{part}$. The corresponding values of $x$ at SIS,
%AGS and SPS energies are calculated with the     baryochemical
%potential and temperature extracted from the measured particle
%multiplicity ratios.
Typical values of $x$ expected for the SIS, AGS  and SPS  energies
for central collisions of  different nucleus are also indicated in
this figure. Fig.~(\ref{ff7}) shows the importance of the
canonical suppression of particle phase-space at SIS energies
where it  can even exceed an order of magnitude. In central heavy
ion collisions at the AGS and particularly at higher energies
(SPS, RHIC, LHC),   the canonical  suppression is  seen in
Fig.~(\ref{ff7}) to be  negligible. Thus, here the GC formalism is
adequate. In general, the canonical statistical interpretation of
the particle production in {\it central} heavy ion collisions is
important if the
  CMS collisions  energy per nucleon pair becomes less than about  $(2-4)$ GeV.
%This is mainly because in this energy range the  event mean values
%of strangeness multiplicities   are still too small to neglect
%their fluctuations, through, have never been measured.
%Consequently, the large-argument expansion of the Bessel functions
%in equation (\ref{e:i1i01}) is not justified.
However, as we show later,  the canonical suppression effect can
also  be important at high energy for  {\it non-central } heavy
ion collisions or for the description of heavy quark production.

At the end of this section we also formulate  thermodynamics of
the canonical ensemble with strangeness and baryon number being
exactly conserved. The corresponding partition function was
already presented  in   Eq.~(\ref{eq70}). Neglecting the
contribution from multistrange baryons and the
particle-antiparticle charge asymmetry, the generating functional
${\tilde Z}(T,V,\phi,\psi )$ in Eq.~(\ref{eq71}) reads
\ber
\ln {\tilde {Z}_S(T,V,\mu_Q=0,\phi,\psi
)}&=&N_{s=0,b=0}+2N_{s=1,b=0}\cos {\psi}\\\nonumber  &+&
2N_{s=0,b=1}\cos {\phi } +2N_{s=1,b=1}\cos (\phi-\psi )
\label{gsb}\label{eq103} \eer
 \noindent where $N_{s,b}$ is defined
as the sum over all particles and resonances
 having the
strangeness $s$ and baryon number $b$,
 \be
N_{s,b}=\sum_k Z_k^1 %\label{e:N}
 \label{eq104}
  \ee
\noindent and % $Z^1_k$ is the particle  thermal phase--space
$ Z_k^1= {{Vg_k}\over {2\pi^2}} \, m_k^2 \, T \, K_2(m_k/T) \ $ is
the thermal phase--space  available for a  particle that carries
strangeness $s$ and baryon number  $b$.

The above generating functional can be  applied  in
Eq.~(\ref{eq70}) to get  the canonical partition function of the
fireball with net  value of  strangeness $S$ and baryon number $B$

\ber Z_{B,S}(T,V)&=&{{Z_0(T,V)}\over {4\pi^2 }}\int_0^{2\pi}d\phi
e^{-iB\phi } \exp [z_N\cos\phi ]
\\ \nonumber &&
 \int_0^{2\pi}d\psi e^{-iS\psi } \exp
[z_K\cos\psi +z_Y\cos (\phi -\psi)] d\phi d\psi \label{cbsi}
\label{eq105}
 \eer
\noindent
 where $Z_0=\exp (N_{s=0,b=0})$, $z_K=N_{s=1,b=0}$,
$z_N=N_{s=0,b=1}$ and $z_Y=N_{s=1,b=1}$. This notation indicates
the type of particles that carry corresponding quantum numbers:
 charge neutral hadrons,  strange mesons, non-strange and
strange baryons. \noindent In the exponent of the $\psi$ integral
we write
\ber z_K\cos\psi +z_Y\cos (\phi-\psi ) = z(\phi )\cos
(\psi-\alpha(\phi) ) \label{eq106}
  \eer
where $\alpha(-\phi)=-\alpha(\phi)$ and
\ber && z(\phi )=(z_K^2+2z_Kz_Y\cos\phi +z_Y^2)^{1/2}\nonumber
\\  &&
 e^{i\alpha(\phi)}={{z_K}\over {z(\phi)}}+
{{z_Y}\over {z(\phi)}}e^{i\phi}. \label{107}
 \eer
 Since the $\psi$ integral goes over the whole period, we may
shift the integration by $\alpha$ and perform the $\psi$ integral
exactly to yield
\ber Z_{B,S}(T,V)={{Z_0(T,V)}\over {\pi }}&\int_0^{\pi}& \cos (
B\phi+S\alpha(\phi))  \exp [2z_N\cos\phi ]
\\ \nonumber &&
I_S(2z(\phi)) d\phi  %\label{cbsf}.
\label{108} \eer \noindent However, the $\phi$  integration cannot
be solved analytically.

Starting from the above partition function one can find  the  mean
multiplicity of particle species $i$. To get it one  simply (i)
separates for these  species the particle and anti-particle term
in Eq.~(\ref{eq103}), (ii) multiplies the relevant one by
$\lambda$, (iii) differentiates with respect to $\lambda$
following  Eq.~(\ref{eq81}) and  puts $\lambda=1$ afterwards. The
result for the particle $i$ with the strangeness $sS_i$ and baryon
number $B_i$ reads\cite{r37a}

\be \langle N_i\rangle_{B_i,S_i} = Z_i^1(m_i,T,V)
{{Z_{B-B_i,S-S_i}(T,V)}\over {Z_{B,S}(T,V)}} \label{eq109}
 \ee

In view of further applications of these results  in heavy ion
collisions we  restrict  the  discussion only to non-strange
systems, that is these   with    overall strangeness   $S=0$.

The  results of Eq.~(\ref{eq109}) should coincide with the GC
value in the limit of large $B$ and $V$, however, with a fixed
baryon density $B/V$. This can be shown\cite{r37a} explicitly
using a Chebyshev approximation of the corresponding integrals.

\section{The canonical statistical model and its applications}

The results discussed in the last section indicate that  the major
difference between the C and GC treatment of the conservation laws
appears through a strong suppression of thermal particle
phase--space in the canonical
approach\cite{r21,r42,r20,r20a,r37,r37a,r38,r58} as well as in an
explicit  volume dependence of particle densities. Deviations from
the asymptotic GC limit are thus expected to be large for low
temperature\footnote{ The temperature T should be low relative to
the lowest particle mass that carries the conserved charge.}
and/or  small volume. In a thermal fireball created in heavy ion
collisions these parameters are related to the CMS collision
energy and the number of participating (wounded) nucleons,
respectively. It is thus clear that the canonical formulation of
quantum number conservation should be of importance in low energy
central and high energy peripheral heavy ion collisions as well as
in hadron--hadron collisions. In the following section the
applications of the canonical statistical model in the above
collision scenarios and for different conserved  quantum numbers
will be discussed. A special case is the production of hadrons
containing charm quarks. This will be dealt with in Section 5.3.
%%%%%%%%%%%%%%%%%%%%%%%%%%%%%%f3333333%%%%%%%%%%%%%%%%%%%%
%\vskip -0.8cm
\begin{figure}[h]
\vskip -.2 true cm
\begin{minipage}[t]{120mm}
\hskip 1.1cm \includegraphics[width=25.5pc, height=17.2pc]{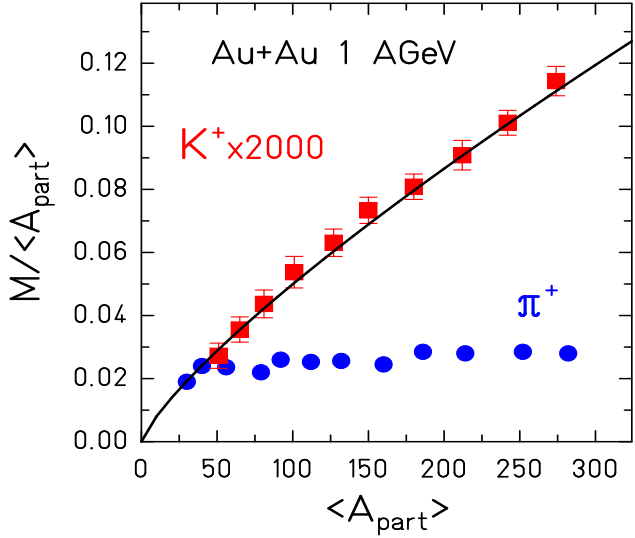}\\
%\caption{ }
\end{minipage}
\begin{minipage}[t]{114.mm}
{\vskip -0.1 true cm
 \caption{\label{ff8} The yield of  kaons and pions  measured in
Au--Au collisions at beam kinetic energy of 1 A GeV from
Ref.~(\protect\refcite{r57}) versus centrality given  by the
number of participants $A_{part}$. The line represents the
$A_{part}^\alpha$ fit\protect\cite{r57} to experimental data}. }
\end{minipage}
\end{figure}
%%%%%%%%%%%%%%%%%%%%%%%%%%%%%1%%%%%%%%%%%%%%%%%%%%
\subsection{Central heavy ion collisions at SIS energies}

The number of  strange particles produced in  heavy ion collisions
depends on the  energy and { centrality} of the collision. In low
energy  A--A collisions in  the  SIS/GSI energy range from 1 to 2
A$\cdot$GeV, the average number of strange particles produced in
an event is of the order of $10^{-3}$. Thus, following the kinetic
analysis presented in Section 4.1, a statistical description would
require   the canonical  treatment of strangeness conservation.
However, the conservation of baryon number and isospin can be
treated grand canonically. Consequently, one   expects a different
centrality dependence of strange and non-strange particle yields.
Fig.~(\ref{ff8}) shows experimental data on $K^+$ and $\pi^+$
yields divided by the number of participants $A_{\rm part}$ as a
function of $A_{\rm part}$ measured\cite{r57}  in Au--Au
collisions at beam kinetic energy of 1  A$\cdot$GeV. The data
indeed exhibit the behavior expected in the canonical statistical
model: a strong increase of the $K^+$ yield per participant and an
almost constant $\pi^+$ yield per participant with centrality.
%  corresponds to an asymptotic
%%HO asymptotic
%regime of canonical ensemble with respect to strangeness
%conservation.

In the canonical model the particle  densities depend on four
parameters:
 the  chemical potentials, $\mu_Q$ and $\mu_B$,
related with the   GC description of the electric charge and
baryon number conservation, the temperature $T$ and the  volume
parameter appearing through the   canonical treatment of the
strangeness conservation. Constraints on these variables arise
from the isospin
 asymmetry  measured  by the baryon number divided by twice the
 charge,  $B/2Q$. For an isospin symmetric system this ratio is
 simply 1, for Ni+Ni it is 1.04 while for Au+Au this ratio is 1.25.
When considering particle multiplicity ratios we are thus left
with three independent parameters. The volume parameter $V$ that
is responsible for the canonical suppression, the freeze--out
temperature $T$  and freeze--out baryon-chemical  potential
$\mu_B$ of the  fireball.
% could
%be in general consider as a free parameter in the canonical model.
%%%%%%%%%%%%%%%%%%%%%%%%%%%%%%%%%%%%%%%%%
%%%%%%%%%%%%%%%%%%%%%%%%%%%%%%1%%%%%%%%%%%%%%%%%%%%
\begin{figure}[htb]
%\vskip .cm
\begin{minipage}[t]{55mm}
%\framebox[79mm]{\rule[-26mm]{0mm}{52mm}}
%\hskip -1.0cm
\includegraphics[width=13.pc, height=13.3pc]{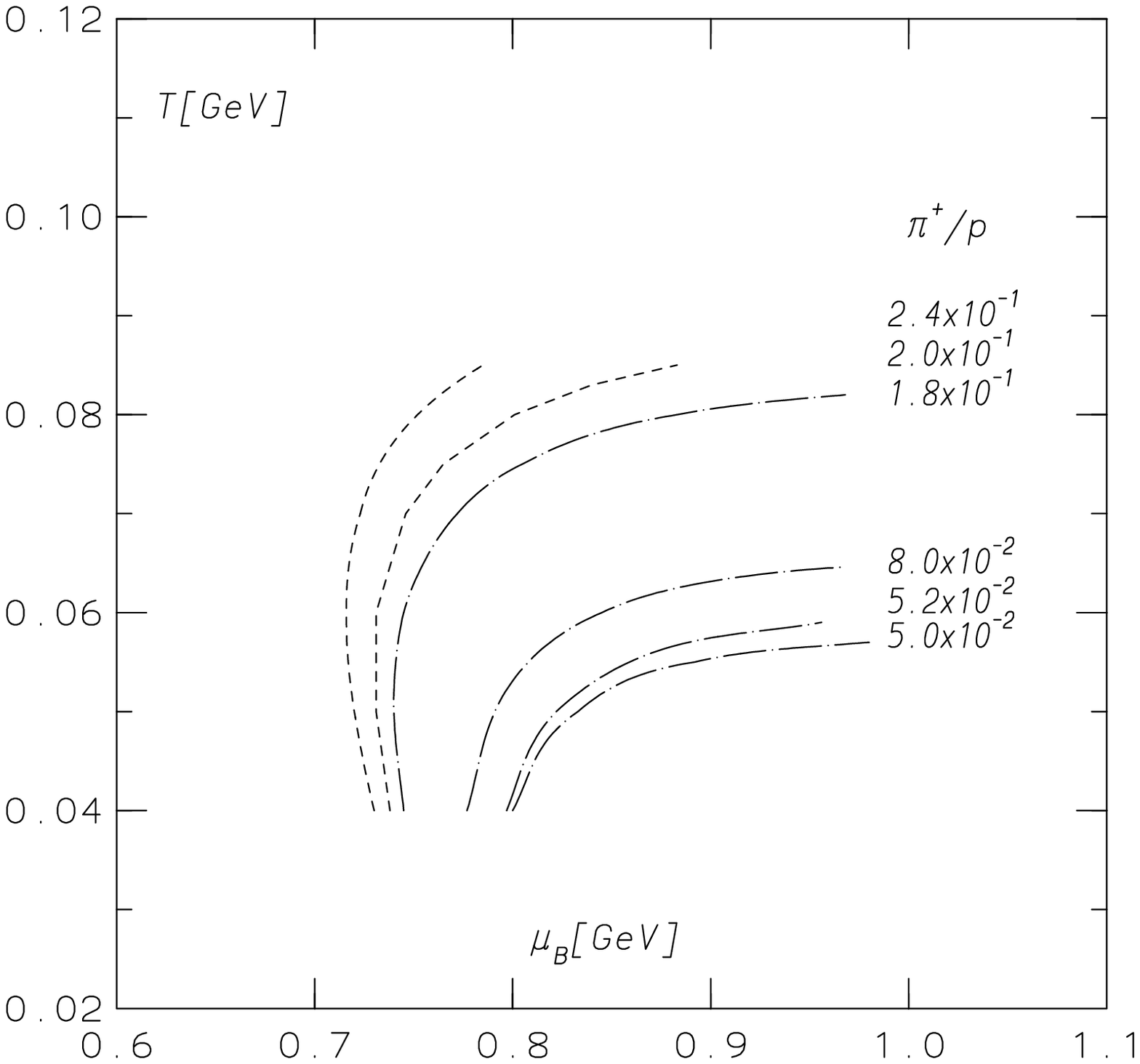}\\
\vskip -.5 true cm \caption{ \label{ff9} Lines of a constant
$\pi^+/{\rm proton}$ ratio in the $T$--$\mu_B$ plane
obtained\protect\cite{r31} in the statistical model. }
\end{minipage}
\hspace{\fill}
\begin{minipage}[t]{54mm}
\includegraphics[width=13.pc, height=13.3pc]{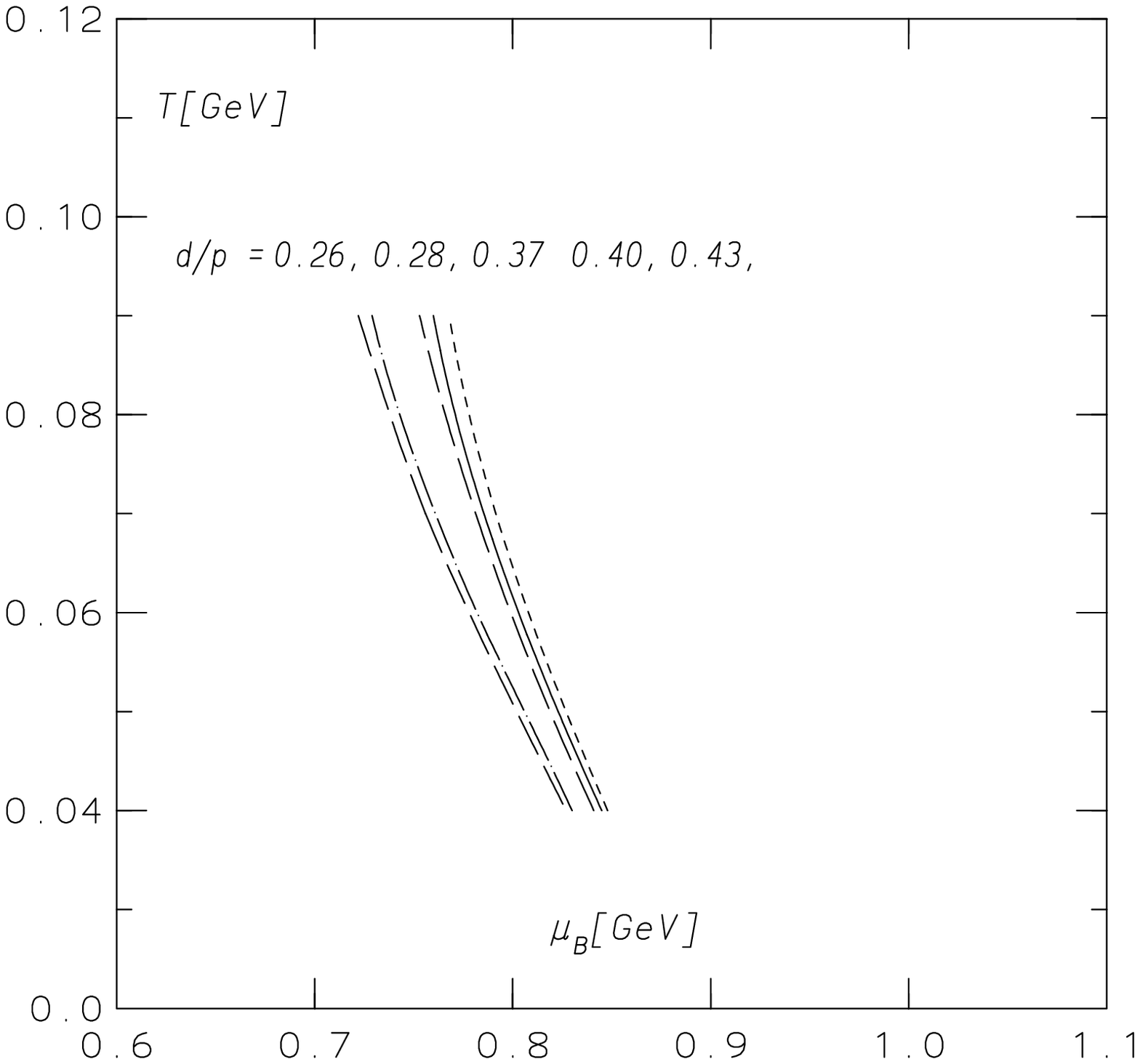}\\
\vskip -0.5 true cm \caption{ \label{ff10} Lines of constant
deutron/proton ratio in the $T$--$\mu_B$ plane
obtained\protect\cite{r31} in the statistical model. \hfill}
\end{minipage}
\end{figure}
%%%%%%%%%%%%%%%%%%%%%%%%%%%%%%1%%%%%%%%%%%%%%%%%%%%
%%%%%%%%%%%%%%%%%%%%%%%%%%%%%%%%%%

 Figs.~(\ref{ff9} - \ref{ff12})   show
the location and sensitivity  of the freeze--out  parameters for
different particle ratios
%
%$T_f$ and of the freeze-out $\mu_B^f$
%
% Fig.~.. shows the results
%Fig.~..  shows  the sensitivity of  thermal parameters in the
in the $(T-\mu_B)$ plane
 when
varying these ratios in the range obtained at SIS energy.
% Discussing these trends requires
%different treatment of strange and non-strange particles.
%
 The deuteron to proton  $d/p$ and the
$\pi^+/p$ ratios  provide  a good determination of the range
 of  thermal parameters. The $\pi^+/p$ curve
in the $(T-\mu_B)$ plane
 shows   temperature saturation
for a large $\mu_B$  that establishes the upper limit of the
freeze--out temperature $T$. On the other hand,  the $d/p$ ratio
fixes   the range of the freeze--out value for $\mu_B$ as
 it shows a steep dependence
on  the temperature.\footnote{The  deuteron, is the composite
object as it is the proton-neutron bound state. It is most likely
produced by nucleon coalescence at kinetic freeze--out. Thus, one
could question if deuteron yield can be used to fix chemical
freeze--out parameters (see Section 3.04). At SIS energies,
however, chemical and thermal freeze--out coincide and the
deuteron multiplicity seems to follow a statistical order with the
same thermal parameters as its constituents.} The $K^+/K^-$ ratio
in Fig.~(\ref{ff11}) exhibits a similar behavior as $d/p$
% i.e.
%shows a  very strong dependence on the temperature but a rather
%weak dependence on the baryon chemical potential, $\mu_B$. The
%$K^+/K^-$ ratio
and is also independent of the volume parameter as is evident from
Eq.~(\ref{eq35}) when requiring that $S=0$.

%%%%%%%%%%%%%%%%%%%%%%%%%%%%%%%%%%%%%%%%
%%%%%%%%%%%%%%%%%%%%%%%%%%%%%%1%%%%%%%%%%%%%%%%%%%%
\begin{figure}[htb]
%\vskip .cm
\begin{minipage}[t]{55mm}
%\framebox[79mm]{\rule[-26mm]{0mm}{52mm}}
%\hskip -1.0cm
\includegraphics[width=13.pc, height=13.3pc]{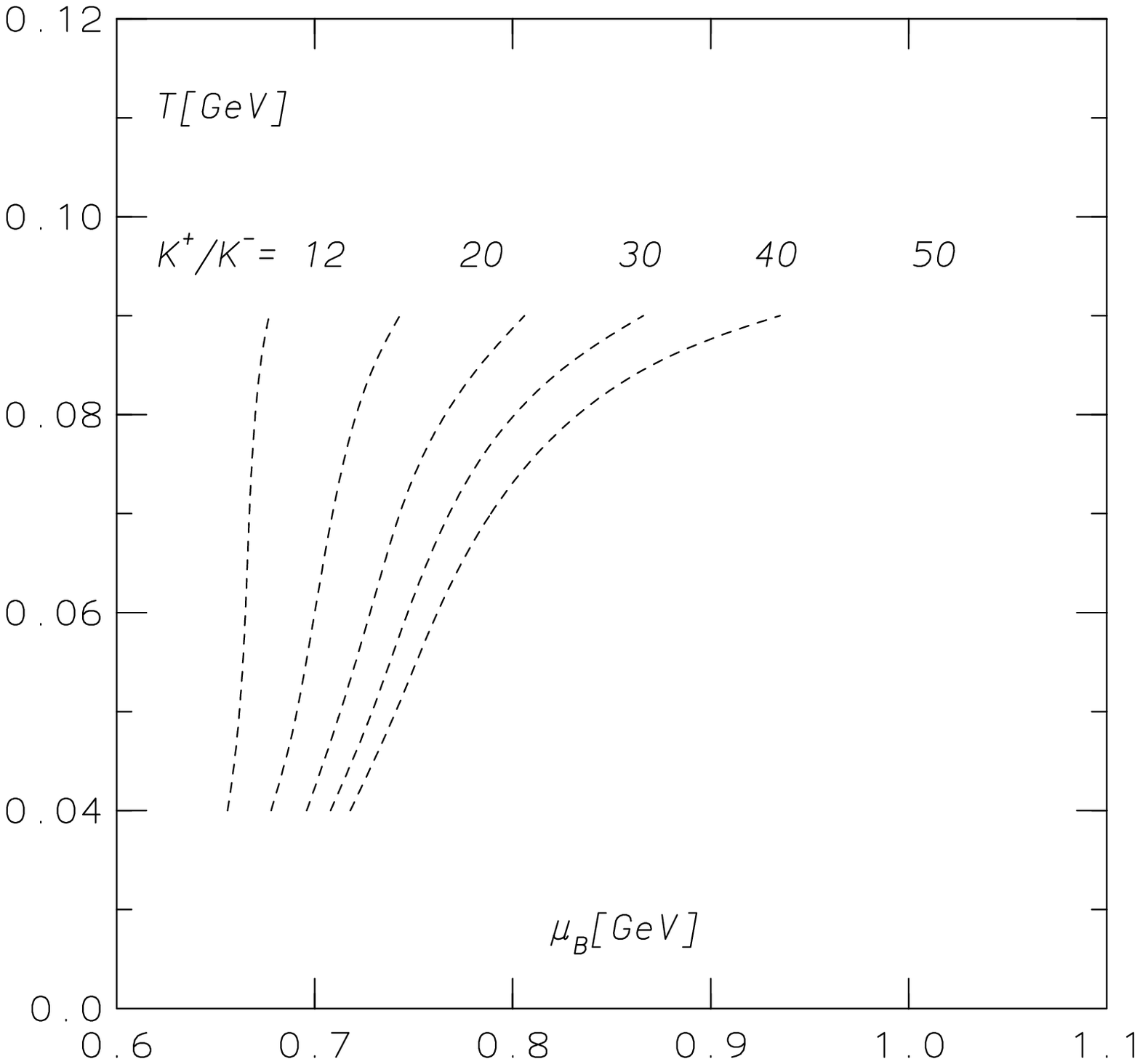}\\
\vskip -.5 true cm \caption{\label{ff11} Lines of a constant
$K^+/K^-$ ratio in the $T$--$\mu_B$ plane
obtained\protect\cite{r31} in the statistical model. }
\end{minipage}
\hspace{\fill}
\begin{minipage}[t]{54mm}
\includegraphics[width=13.pc, height=13.3pc]{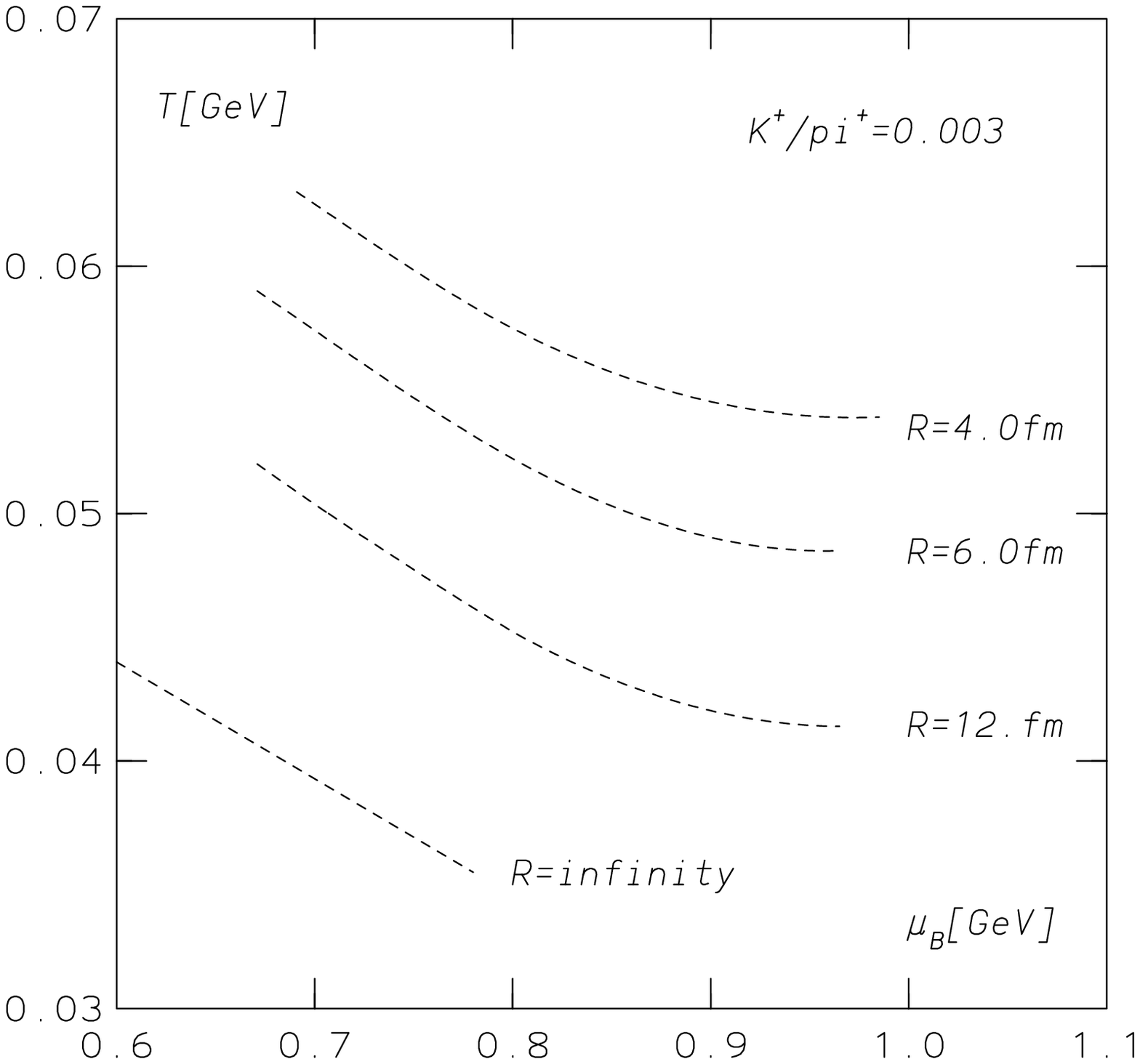}\\
\vskip -0.5 true cm \caption{\label{ff12}
 The freeze--out
parameters in the $T$--$\mu_B$ plane calculated\protect\cite{r31}
for a fixed value of the ratio $K^+/\pi^+=0.003$  and  for
different correlation volumes $V=4\pi R^3/3$ with $R=4,6,12$ fm
and $R=\infty$.}
\end{minipage}
\end{figure}
%%%%%%%%%%%%%%%%%%%%%%%%%%%%%%1%%%%%%%%%%%%%%%%%%%%
%%%%%%%%%%%%%%%%%%%%%%%%%%%%%%%%%%

% In the  SIS-energy
%range the  canonical  thermal model predicts  a different
%dependence of the strange and non-strange particle yields on the
%volume parameter that  is also on  the number of participating
%nucleons.
The variation of thermal parameters with the system size is shown
in Fig.~(\ref{ff12})
% Following ouThe ratio of
% strange to non-strange particle  should exhibit
% a strong variation of thermal parameters with the system size.
% Fig.~4 shows the $K^+/\pi^+$ ratio for different volumes. As
using as an example the   $K^+/\pi^+$ ratio. For   a   given
system size the $K^+/\pi^+$  ratio  clearly determines a lower
limit of the freeze--out temperature  as it saturates for a large
$\mu_B$.  Changing the volume parameter  $V=4/3\pi R^3$ implies a
substantial modification of the line in the $(T-\mu_B)$ plane
calculated for a  fixed value of the $K^+/\pi^+$ yields. Thus, in
the canonical model, the strange to non-strange particle ratio
requires an additional consideration of the range of correlation
of strange particles  that is quantified by   the  volume
parameter $V$. This
  parameter  is assumed to be
       related  to  the number of nucleons participating
in A--A collisions. From a  detailed analysis of  experimental
data from SIS up to AGS energies it was  shown that $V$ can be
identified\cite{r31} as the initial overlap volume of the system
created
 in A--A collisions.
 Thus, it is
obtained from the atomic number of the colliding nuclei and from
the impact parameter by simple
 geometric arguments. In heavy ion collisions at SIS
 energies a good description of  Ni--Ni and Au--Au
 data was obtained\cite{r31}$^,$\footnote{ This volume
parameter $V$ can be in general $\sqrt s$ dependent. One way to
include this dependence would be to replace the spherical
symmetric $V$ by cylinder with its longitudinal size being Lorentz
contracted.}
 when choosing
$V\simeq V_0A_{part}/2$ with $V_0\simeq 7$fm$^3$ i.e. of the same
order as  the volume of the nucleon.

%%%%%%%%%%%%%%%%%%%%%%%%%%%%%%%%%%%%%%%%%%%%%%%%%%%%%%%%%%%%%%%%%%%%%%
%%%%%%%%%%%%%%%%%%%%%%%%%%%%%%%%%%%%%%%%%%%%%%%%%%%%%%%%%%%%%%%%%%%%%%%%

%%%%%%%%%%%%%%%%%%%%%%%%%%%%%%%%%%%%%%%%
%%%%%%%%%%%%%%%%%%%%%%%%%%%%%%1%%%%%%%%%%%%%%%%%%%%
\begin{figure}[htb]
\vskip -0.7cm
\begin{minipage}[t]{55mm}
\vskip -6.2cm\includegraphics[width=13.pc, height=14.9pc,angle=180]{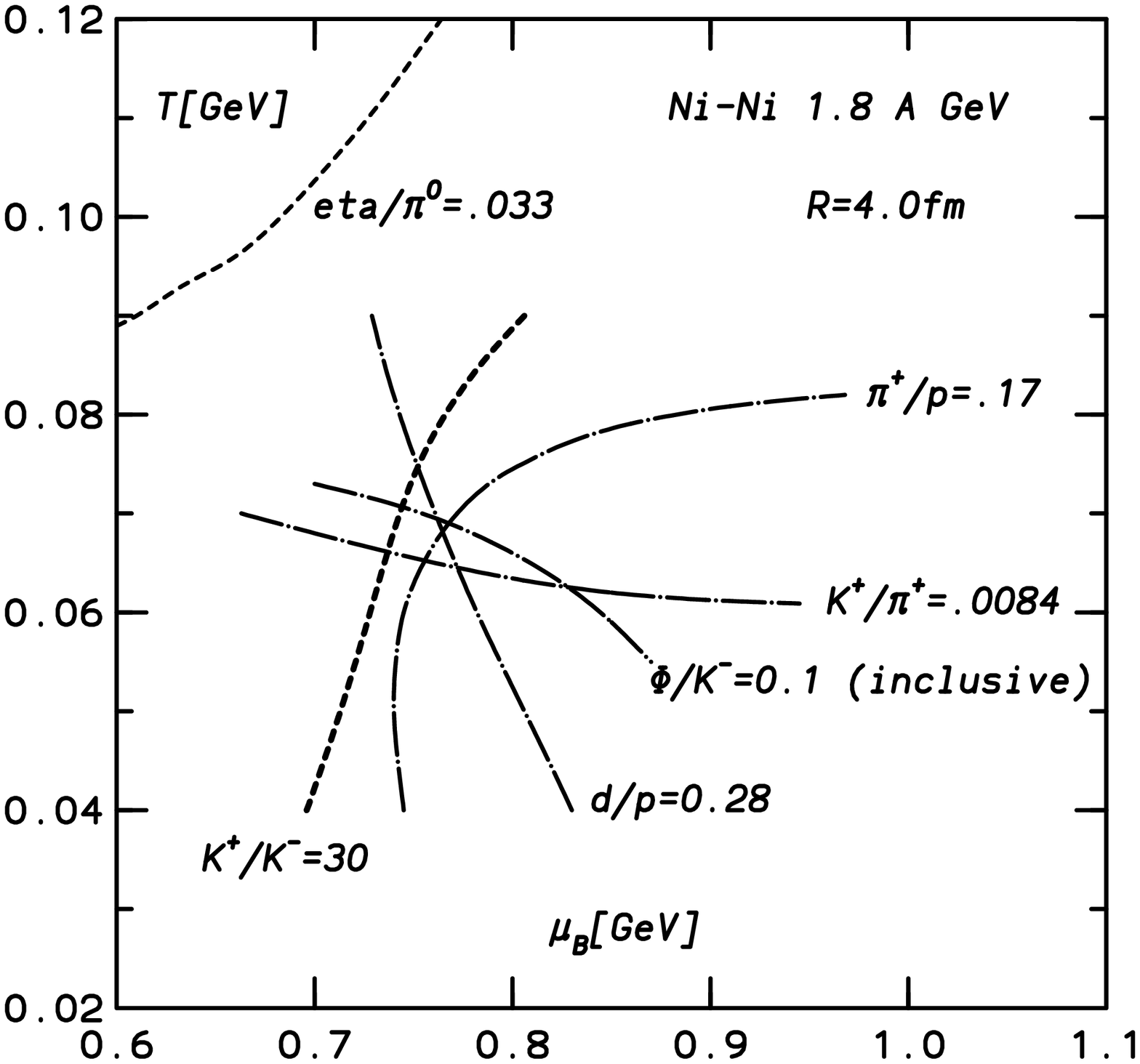}\\
\vskip -.5 true cm \caption{\label{ff13} The lines in the
$T$--$\mu_B$ plane calculated\protect\cite{r31} in the statistical
model for different particle ratios obtained in central Ni--Ni
collisions at 1.8 A$\cdot$GeV . }
\end{minipage}
\hspace{\fill}
\begin{minipage}[t]{54mm}%\vskip 0.5cm
\includegraphics[width=13.pc, height=14.6pc]{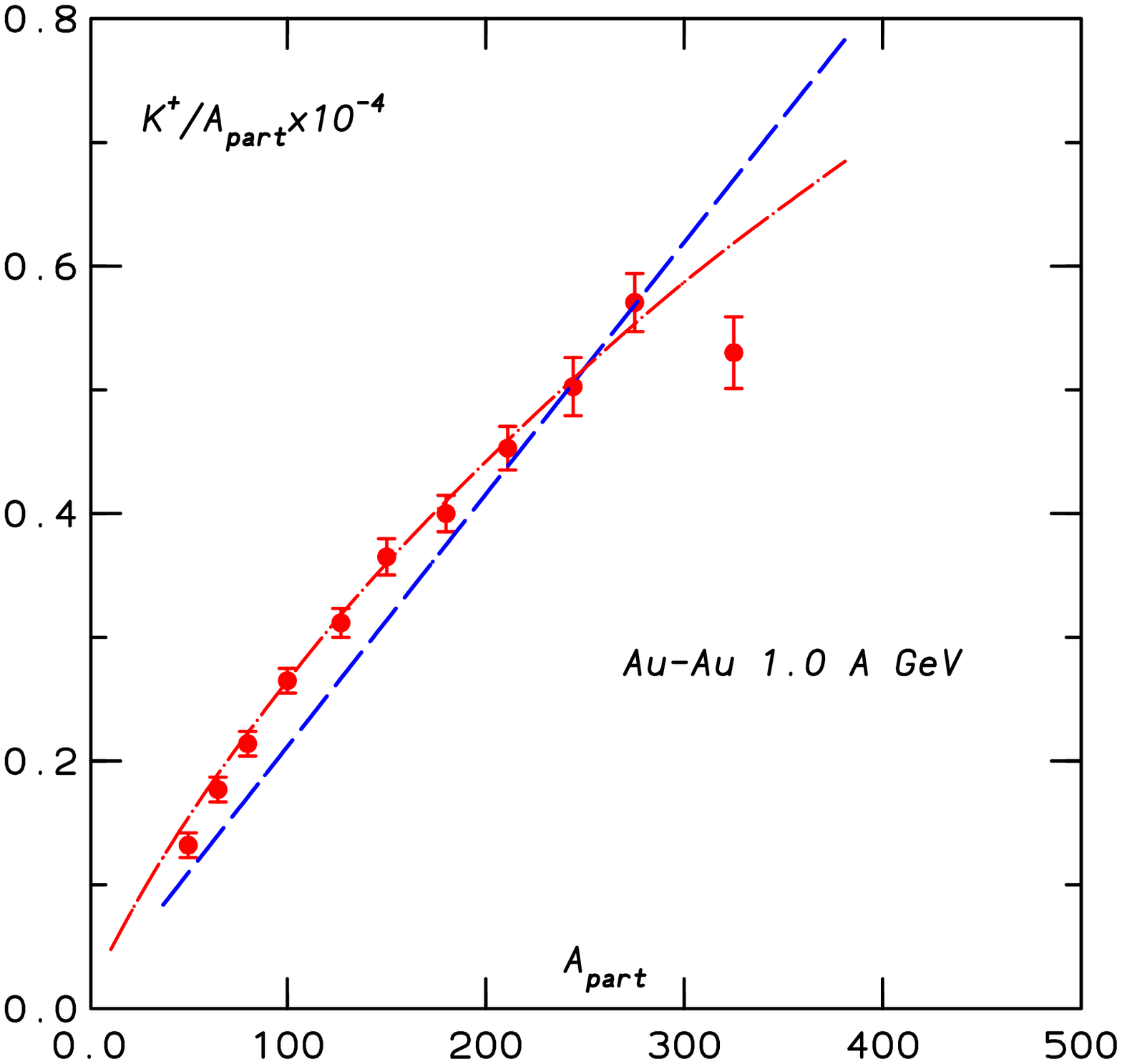}\\
\hskip 0.3cm \vskip -0.85 true cm \caption{\label{ff14} A
comparison of the statistical model results\protect\cite{r31} for
the $K^+/A_{part}$ ratio with the data from Fig.~(\ref{ff8}). The
dashed and dashed--dotted lines represent predictions of the
statistical model without and with a small $A_{part}$ dependence
of $\mu_B$ and $T$. For more details see text. \hfill}
\end{minipage}
\end{figure}
%%%%%%%%%%%%%%%%%%%%%%%%%%%%%%1%%%%%%%%%%%%%%%%%%%%
%%%%%%%%%%%%%%%%%%%%%%%%%%%%%%%%%%
The comparison of the thermal model with experimental data from
AGS, SPS up to RHIC energy was discussed in Section  3 and it was
shown that there are common freeze--out parameters  which describe
simultaneously  all measured particle multiplicity ratios. In
order to illustrate that this is also the case in low energy heavy
ion collisions we show in Fig.~(\ref{ff13})
%   in Fig.~2 right
%clearly indicate that both the magnitude of the yield and the
%strong, almost quadratic, dependence of the kaon yield on the
%number of participants is well reproduced by the canonical model.
the lines in the $(T-\mu_B)$ plane corresponding to different
particle multiplicity ratios measured\cite{Barth} in  Ni-Ni
collisions at 1.8 $A\cdot$GeV. The experimental errors are for
simplicity not shown in the figure. All lines, except  the one for
$\eta /\pi^0$, have a common crossing point  around $T\sim 70$ MeV
and $\mu_B\sim 760$ MeV. A value of $R\sim 4$ fm is needed to
describe the measured $K^+/\pi^+$ ratio with the freeze--out
parameters
 extracted from $\pi^+/p$ and $d/p$ ratios.
This radius  is compatible with that  expected for a central
Ni--Ni collision and  was found\cite{r31} to be the same  in the
whole energy range from 0.8 up to 1.8 A$\cdot$GeV.

The corresponding results for a thermal description of   Au-Au
collisions at two different incident kinetic energies 1.0  and
1.5 $A\cdot$GeV can be found in Ref.~(\refcite{r31}). As     for
Ni--Ni data, the particle ratios,
    $\pi^+/p$, $K^+/\pi^+$, $\pi^+/\pi^-$, $K^+/K^-$ and $d/p$,
     with exception of  $\eta/\pi^0$, could be described with  the same value of
     the
 freeze--out parameters.
The temperature $T\sim 53$ MeV and $\mu_B\sim 822$ MeV were found
in Au--Au collisions at 1.0 A$\cdot$GeV.\footnote{With these
thermal parameters the total density of particles   at chemical
freezeout corresponds to $n_B\sim 0.05/$fm$^3$. Due to rather
large value of the  baryochemical potential and leading
contribution of baryons to total particle density this value  also
corresponds to the total baryon density of the system.} Thus, the
freeze--out temperature is obviously decreasing whereas the baryon
chemical potential is an increasing function of the collision
energy. A small variation of freeze--out parameters with A in
central A--A collisions for the same collisions energy was also
extracted from the data\cite{r31}.

The observed scaling of the volume parameter that determines the
canonical suppression of the strange particle  phase--space with
the number of participants  was also found to be valid  in Au--Au
collisions. Here,  in the most  central collisions  a radius of
$\sim 6.2$ fm is required to reproduce the measured $K^+/A_{part}$
and $K^-/A_{part}$ yields. The  larger radius value obtained for
Au compared to Ni data is compatible with the larger size of Au
and corresponds to $A_{part}\sim 330$ in the most central Au--Au
collisions.

The importance of  strangeness suppression due to  canonical
treatment of the conservation laws is particularly  transparent in
the comparison of the thermal model with the  Au-Au  data at 1
$A\cdot$GeV . Using the {\it grand canonical} formulation of the
strangeness conservation
 one would get a value  of  $K^+/\pi^+\sim 0.04$
 that
overestimates the data by more than an order of magnitude. This
shows that  the  thermal particle phase--space at SIS energies is
far from the grand canonical limit and that the exact and local
treatment of   strangeness conservation  is of crucial importance.
%%%%%%%%%%%%%%%%%%%%%%%%%%%%%%%%%%%%%%%%%%%%%%%%%%%%%%%%%%%%%%%%%%

 The multiplicity of $K^+$ per participant
($K^+/A_{part}$) was indicated in Fig.~(\ref{ff8})  to  increase
strongly with centrality while the corresponding ratio for pions
$\pi^+/A_{part}$, is constant  with $A_{part}$\cite{Barth}.
Consequently, the pion yield is proportional to the number of
participants  while the multiplicity of $K^+$ scales with
$A_{part}$ as $K^+\sim A_{part}^\alpha$ with $\alpha \sim 1.8$.
The canonical treatment of the strangeness conservation predicts
the yield of strange particles to increase quadratically with the
number of participants, see e.g. Eq.~(\ref{eq18}). In
Fig.~(\ref{ff14}) the experimental data from Fig.~(\ref{ff8}) are
compared with the thermal model. The parameters were chosen  to
reproduce the $\pi /p$ and $d/p$  ratios. The dashed-line in
Fig.~(\ref{ff14}) describes the results of a thermal model under
the assumption that both $T$ and $\mu_B$ are independent of
$A_{part}$. One sees that already under this approximation the
 agreement of the model and the experimental data is very
satisfactory as the model describes the magnitude and the
centrality dependence of these data.
 Some differences between     the model and the data seen in
 Fig.~(\ref{ff14})
 can be accounted    for when including a small
variation of the freeze--out parameters with $A_{part}$. A smooth
and almost linear increase of the temperature with centrality by a
few MeV  and a corresponding decrease of $\mu_B$ from very
peripheral to central collisions is sufficient to get a very good
description of the data. In Fig.~(\ref{ff14}) the dashed--dotted
line shows thermal model results that include the  variation of
thermal parameters with centrality. Such a small   dependence  of
the freeze--out temperature on impact parameter comes at first
glance  as a surprise since the experimental result on the
apparent inverse slope parameter $T_{app}$ of particle yields as a
function of $p_t$  in Au--Au collisions shows a strong dependence
on $A_{part}$\cite{MUE97,HON98}. This difference, however, can be
accounted for by including the concept of centrality dependent
transverse collective flow of the collision fireball. A detailed
analysis of particle spectra in Au--Au collisions at 1 A$\cdot$GeV
has shown that keeping the chemical freeze--out temperature at
$53$ MeV and including collective transverse flow  reproduces the
transverse momentum distributions of pions, protons and kaons as
well as their centrality dependence\cite{r31}. This result
indicates that chemical and thermal freeze--out coincide at  SIS
energy.

 The canonical thermal model provides a consistent description
of  the  experimental  data in the GSI/SIS energy range.  The
abundances of  $K^+, K^-$, $p ,d,$ $\pi^+ $ and $\pi^-$ hadrons
(with the notable  exception of $\eta$ and possibly $\phi$)
\footnote{ The observed deviations of the model from   the
measured yield of $\eta$ mesons require\cite{tabs} further
studies. It is conceivable that the canonical model described in
Section 4 does not account correctly for hidden strange particle
production. The recent result of     $\phi /K^-= 0.44\pm 0.16\pm
0.22$ ratio obtained\cite{hern}  in Ni--Ni collisions at 1.8
A$\cdot$GeV could also be larger then  the  statistical model
value of $\phi /K^-\simeq 0.1$. However, within large experimental
uncertainties the model prediction is still not excluded. } seem
to come from a common hot source and with well defined
temperature, $T\approx 50,54,70$ MeV, and baryon chemical
potential $\mu_B\approx 825,805,750$ MeV for central Ni-Ni
collisions at 0.8, 1.0, 1.8 $A\cdot$GeV and correspondingly
$T\approx 53$ MeV and $\mu_B\approx 822$ MeV for central Au-Au
collisions at 1.0 $A\cdot$GeV. These temperatures are lower than
the ones observed in the particle spectra  but here a common
explanation is possible in  terms of hydrodynamic flow. The flow
differentiates between  particles of different mass since they
acquire the same boost in velocity but very different boosts in
momentum.

The common freeze--out condition for almost all particle species
is  very strong evidence  for
 chemical equilibrium in low energy heavy ion collisions.
A satisfactory agreement of the model and the data could be only
obtained when including a canonical description of the strangeness
conservation. The relevant parameter that quantifies the canonical
suppression of the particle phase--space was found to be the
initial volume of the collision fireball that scales with the
number of projectile participants. This volume parameter describes
the range of strange particle correlations and  is smaller than
the radius of the fireball at chemical freeze--out. In central
Au--Au collisions at 1 A$\cdot$GeV the correlation radius was
found to be 6.2 fm roughly  corresponding  to the size of Au
whereas the radius required to reproduce measured particle yields
is almost two times larger\cite{r31}. The appearance  of two
different space--like scales in the canonical description of
particle production can be possibly understood from the kinetics
of strangeness production. Introducing the locality of strangeness
conservation in the kinetic equation (\ref{eq5}) implies that the
volume parameter in the loss   terms can be different (smaller)
than in the gain terms. Consequently, strange particle yields
depend on two volume parameters: (i) the volume of the fireball
which  is also an overall normalization factor that determines the
total strange and non-strange particle yields originating from the
collisions fireball and  (ii) the volume that parameterizes the
space-like correlations of strange particles that is required to
satisfy  exactly  the strangeness conservation. The second
parameter is also related with the initial number of nucleons
participating in the  collision.

The apparent chemical equilibration of particle yields measured at
SIS energies and the kinetic theory developed in Section 4 has
recently inspired a more complete dynamical study of the problem
in terms of microscopic  transport
models\cite{urqmd}$^-$\cite{r48}. Recently the relativistic
transport model was applied\cite{hart01,r40} to describe the
chemical equilibration of kaon and anti-kaon  at energies that are
below the N--N threshold. The results of these microscopic studies
indicate   that $K^+$ can possibly  appear\footnote{However, this
equilibration can  be obtained if the  $K^+$ mass is substantially
changed in the hot and dense medium.} in   chemical equilibrium
during the lifetime of the collision fireball. The $K^-$, on the
other hand, approaches chemical equilibrium even at a earlier
times However, it may eventually fall out of equilibrium at a
later time due to the large annihilation cross sections in nuclear
matter\cite{r40}. Thus, the results of transport models do not
exclude chemical equilibration in low energy heavy ion collisions.
The level of equilibration in these models is strongly related to
the magnitude of  production and absorption cross section of kaons
inside the nuclear medium\cite{g11}. Although significant
progress has been made in the theoretical description and
understanding of in-medium kaon cross sections, the results are
still far from  complete. In particular, recently it was suggested
that the coupling of kaons with the p-wave $\Sigma (1385)$
resonance can substantially increase the $K^-$ production cross
section that could influence\cite{pcas} the approach of kaons
towards an equilibrium.

\subsection{Particle production in high energy p--p  collisions }
%%%%%%%%%%%%%%%%%%%%%%%%%%%%%f66666666666666

The  success of the statistical approach in  the description of
particle production in heavy ion collisions discussed in Sections
3 and  4 prompts  the question if a statistical order of the
secondaries is  also observed   in  elementary collisions such as
high energy p--p interactions, to which we restrict our
discussion. The results of Section 4 make it clear that one should
in this case apply a model that accounts for the canonical
conservation  of the quantum numbers. This is particularly the
case for strangeness since  even in very high energy p--p
collisions the number of produced strange particles per event is
of the order of unity. Thus, large, event by event, strange
particle multiplicity fluctuations prevent the applicability of
the GC approach. Whether or not the GC treatment of the baryon
number and electric charge is adequate, would require a detailed
study of the relative error between the canonical and the grand
canonical results.

 The application of the  statistical model to the elementary
hadron--hadron reactions was first proposed by Rolf
Hagedorn\cite{r37} in order to describe the exponential shape of
the $m_t$-spectra of produced particles in p--p collisions.
Hagedorn also pointed out  phenomenologically the importance of
the canonical treatment of the conservation laws for rarely
produced particles. The first  application of the canonical model
to strangeness  production  in p--p collisions was  done by Edward
Shuryak\cite{r37} in the context of ISR data. Recently a complete
analysis of hadron yields in p--p as well as in $\bar {\rm p}$--p,
$e^+e^-$,  $\pi$--p and in ${\rm K}$--p collisions at several
center-of-mass energies has been done in
Refs.~(\refcite{r17,becalast,r41}). This detailed analysis has
shown that particle abundances in elementary collisions  can be
also described by a statistical ensemble with   maximized entropy.
In fact, measured yields are consistent with the  model assuming
the existence of equilibrated fireballs at a temperature $T
\approx $160-180 MeV.
%%%%%%%%%%%%%%%%%%%%%%%%%%%%%f66666666666666
%\vskip -0.8cm
\begin{figure}[htb]
 {\vskip -0.5cm
{\hskip -.6 cm
%\center\includegraphics[width=27.5pc, height=15.5pc,angle=0]{rfig25.ps}}}\\
\center\includegraphics[width=27.5pc, height=18.5pc,angle=0]{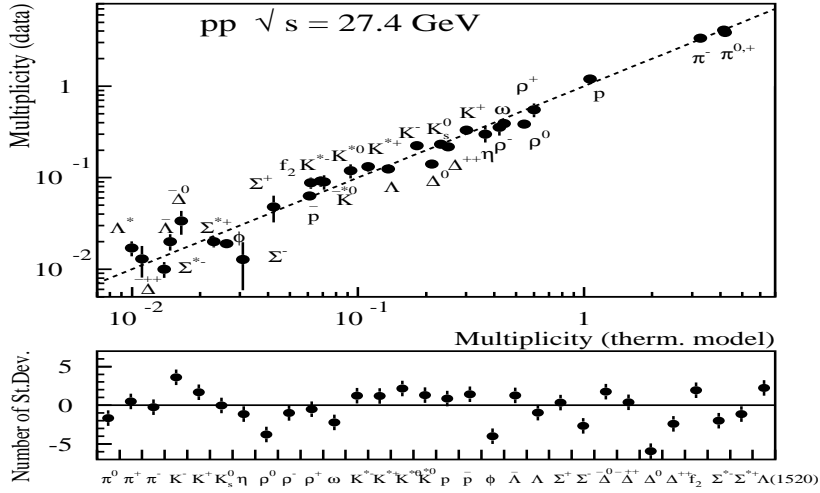}}}\\
{\vskip -0.8cm \caption{\label{ff15} A
comparison\protect\cite{r41} of the p--p multiplicity data at
$\sqrt s=27.4$ with the  statistical model that accounts for the
exact conservation of baryon number, electric charge, strangeness
and includes the strangeness undersaturation factor
$\gamma_s\simeq 0.51$. }}
\end{figure}
%%%%%%%%%%%%%%%%%%%%%%%%%%%

The most general partition function $Z_{B,Q,S}(V,T)$ that is
applied to test the chemical composition of the secondaries in
elementary collisions should account for the canonical
conservation of baryon number $B$, strangeness $S$, and electric
charge $Q$. It can be constructed  applying   the projection
method (see Section 4.2) for the $U_B(1)\times U_S(1) \times
U_Q(1)$ symmetry. Following Eq.~(\ref{eq70}) one has

 \ber
Z_{B,Q,S}(V,T)= &&\int_0^{2\pi}{{d\phi_B}\over
{2\pi}}e^{-iB\phi_B} \int_0^{2\pi}{{d\phi_Q}\over
{2\pi}}e^{-iQ\phi_Q}
 \nonumber \\ && \int_0^{2\pi}{{d\phi_S}\over
{2\pi}}e^{-iS\phi_S} e^{\ln \tilde
Z(T,V,\phi_B,\phi_Q,\phi_S)},\label{eq110a}
 \eer
where   $B,Q$,  and $S$ are  the initial values of the quantum
numbers in  hadron--hadron collisions which are    in    p--p
scattering B=Q=2 and S=0.

The generating function $\tilde Z$ in Eq.~(\ref{eq110a}) can be
obtained   from the grand canonical partition function of a hadron
resonance gas (\ref{eq3}) by a Wick rotation of  all appropriate
chemical potentials
 \ber
&& \tilde Z(T,V,\phi_B,\phi_Q,\phi_S)= \nonumber \\
 &&
Z^{GC}(T,V,\mu_B\to -i\beta\phi_B, \mu_Q\to -i\beta\phi_Q,
\mu_S\to -i\beta\phi_S).\label{eq111a}
 \eer

 Different
particle multiplicities and their ratios  are obtained from the
partition function (\ref{eq110a}) following the procedure that was
described in Section 4.2.3.

Fig.~(\ref{ff15}) shows an example of the comparison of the
canonical model prediction (\ref{eq110a}) with the  experimental
data for different particle yields obtained\cite{r41} in p--p
collisions at $\sqrt s= 27.4$--$ 27.6$ GeV. The agreement of the
model with most of the experimental  data is reasonable.
%%%%%%%%%%%%%%%%%%%%%%%%%%%%%%1%%%%%%%%%%%%%%%%%%%%
\begin{figure}[htb]
\vskip -1.0cm
\begin{minipage}[t]{55mm}
%\framebox[79mm]{\rule[-26mm]{0mm}{52mm}}
\vskip 0.3cm
\includegraphics[width=13.pc, height=15.3pc,angle=180]{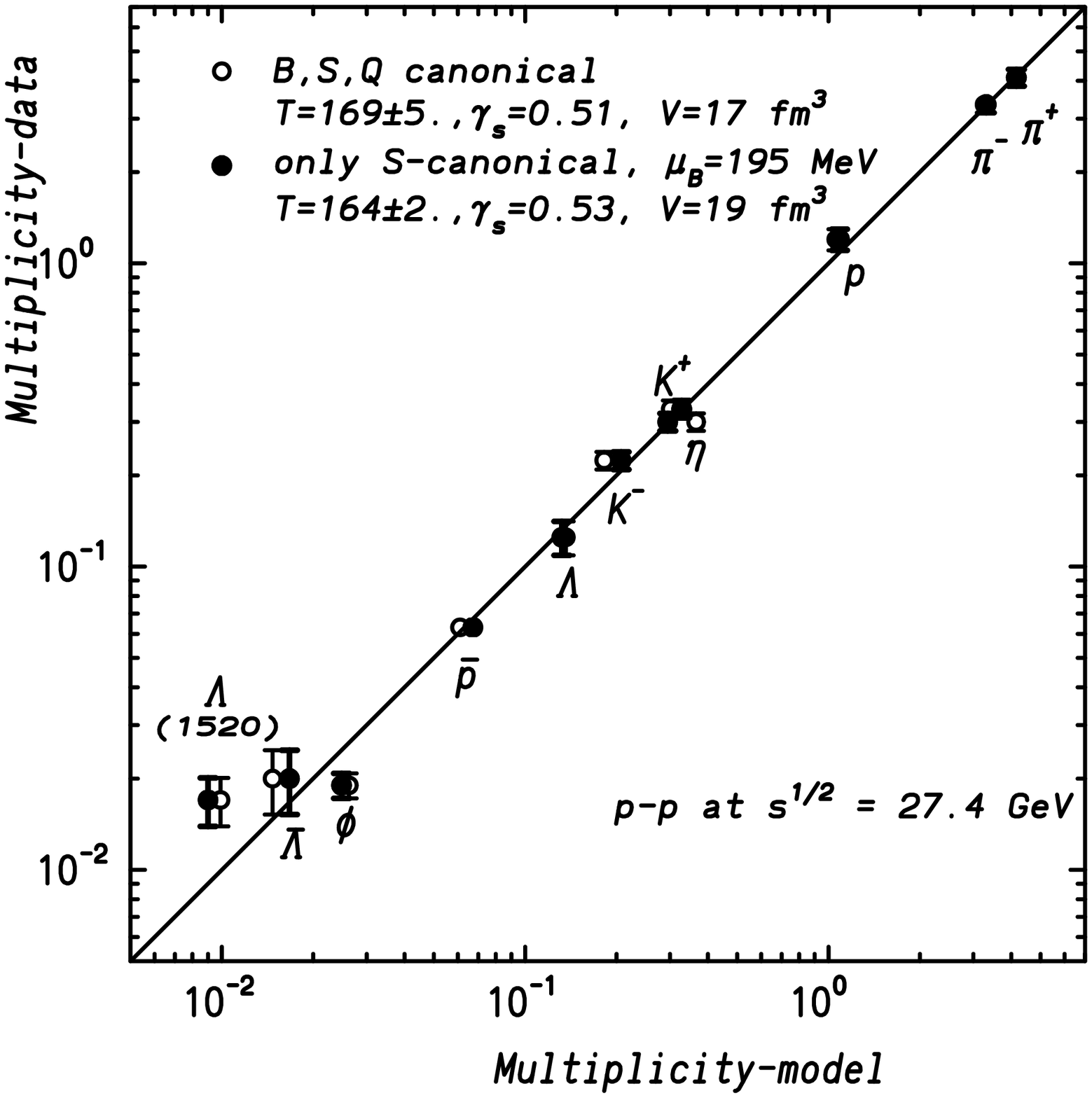}\\
\vskip -.6 true cm \caption{\label{ff16} A comparison of the p--p
multiplicity obtained  at  $\sqrt s=27.4$ with the  canonical
models. The open symbols represent the result of the model that
accounts for $S,B$ and $Q$ being exactly
conserved\protect\cite{r41}. The filled  symbols are the
statistical  model results with only $S$ being exact whereas $B$
and $Q$ are treated grand canonically. }
\end{minipage}
\hspace{\fill}
\begin{minipage}[t]{55mm}\vskip 0.3cm
\includegraphics[width=13.pc, height=15.3pc,angle=180]{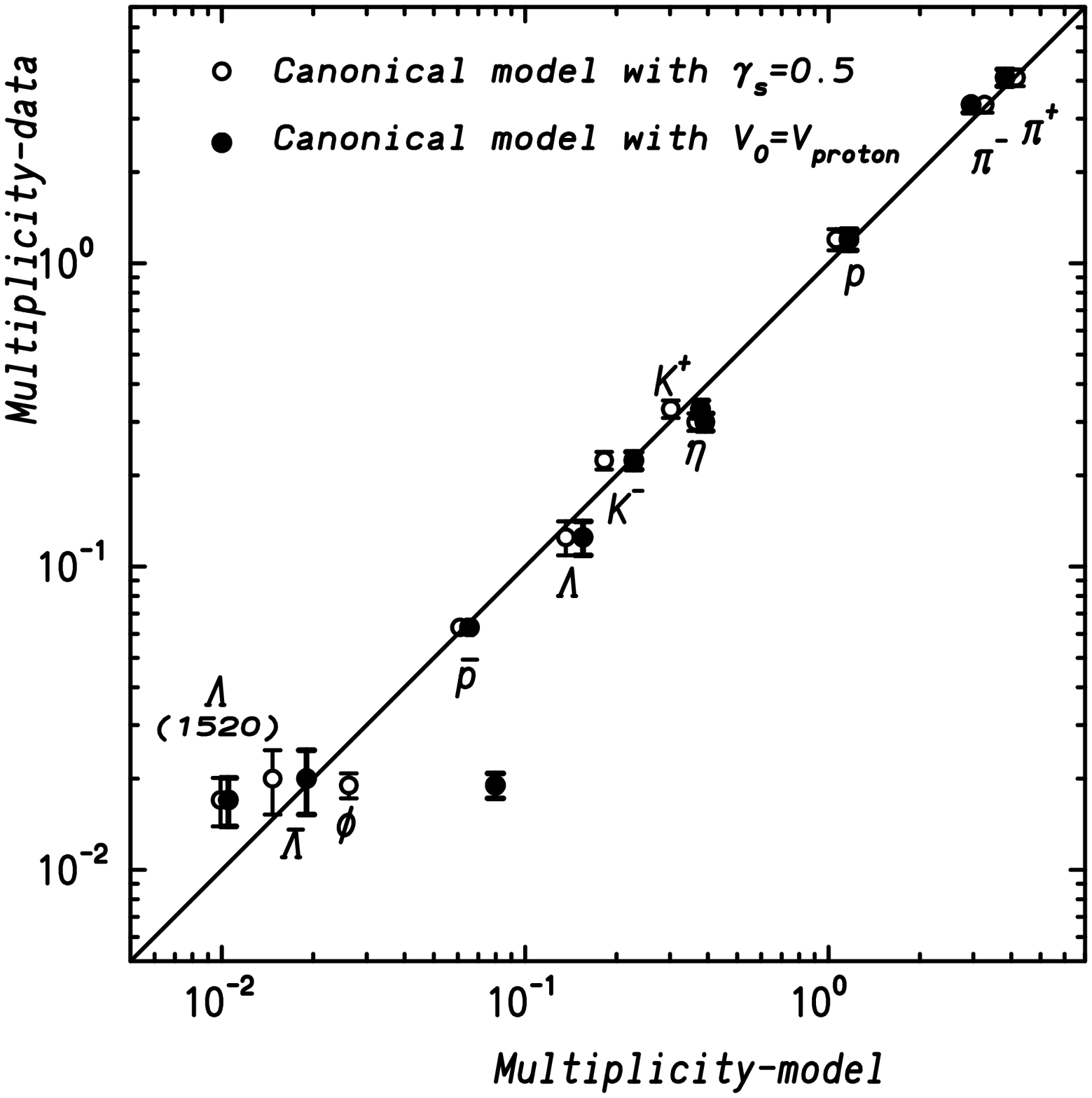}\\
\hskip 0.3cm \vskip -0.6 true cm  \caption{\label{ff17} As in
Figure (\protect\ref{ff16}) but the filled circle are obtained in
the canonical model that accounts for the strangeness
undersaturation to be controlled by the correlation volume,
$V_p=4\pi R_p^3/3$ with $R_p\simeq 1.1$ fm instead of  $\gamma_s$.
\hfill}
\end{minipage}
\end{figure}
%%%%%%%%%%%%%%%%%%%%%%%%%%%%%%1%%%%%%%%%%%%%%%%%%%%
%%%%%%%%%%%%%%%%%%%%%%%%%%%%%%%%%%
%%%%%%%%%%%%%%%%%%%%%%%%%%%%%f66666666666666
 However,
 the yields of the  resonances like $\Delta^0,\rho^0$ and $\phi$
can differ by   a few standard deviations from the data leading to
rather large reduce $\chi^2$ values of $\simeq 5$.

The agreement of the model
 with the strange particle  yields     could only be,
however, achieved  in Ref.~(\refcite{r41}) when
introducing\cite{r37,r41}  an additional parameter $\gamma_s$ into
the canonical partition function (\ref{eq110a}). This parameter
suppresses a thermal phase--space of particles composed of $n_s$
strange or antistrange quarks by a factor $(\gamma_s)^{n_s}$. In
p--p and $\bar {\rm p}$--p collisions and in a very broad energy
range from $\sqrt s\sim 20$ up to $\sqrt s\sim 900$ GeV the same
value of $\gamma_s\simeq 0.51$ was needed\cite{r41} to reproduce
the measured strange particle yields.

In high energy p--p collisions the canonical effects due to the
baryon number and isospin conservation are  expected to be small.
Fig.~(\ref{ff16}) shows the comparison of p--p data with the
canonical model that accounts  for only strangeness being
conserved exactly. The baryon number and electric charge
conservation are  treated in the GC ensemble, thus are  controlled
by  chemical potentials. A satisfactory description of
experimental data with $\mu_B\sim 195$ MeV, $\mu_Q\sim 30$ MeV,
$T\sim 165$ MeV and $\gamma_s\sim 0.53$ seen in Fig.~(\ref{ff16})
shows that canonical effects related with charge and baryon number
conservation are indeed small\cite{gazdzickil}.
 This is, however,  not the
case in $\bar {\rm p} $--p and $e^+e^-$ collisions since  the
initial values of $B=Q=0$ are there obviously too small to use a
GC approximation.

The experimental data shown in Fig.~(\ref{ff15}) can be also
described in terms of the canonical model that was successfully
applied in low energy heavy ion collisions (see Section  5.1).
There, instead of the strangeness undersaturation factor
$\gamma_s$,    space-like correlations of strange particles were
introduced. Consequently, there were  two volume parameters   that
determined particle yields: (i) the volume of the fireball and,
(ii) the correlation volume. Fig.~(\ref{ff17}) shows that choosing
the correlation volume $V_p=4\pi R_p^3/3$ and $R_p\sim 1.1-1.2$ fm
the p--p data are well reproduced     with the {\it exception of}
$\phi$. The $\phi$-meson is not canonically suppressed because it
only carries zero strangeness.
 Taking the
correlation volume instead of $\gamma_s$   gives a deviation of
the measured $\phi$ abundance from a thermal fit of $12\sigma$,
thus increasing even more the already large deviation of $4\sigma$
seen in Fig.~(\ref{ff15}). The canonical model that introduces a
$\gamma_s\sim 0.5$  factor\cite{r41} is obviously better
reproducing
 the yields of hidden strange particles obtained in
elementary collisions.

 The
value of $\gamma_s$ extracted from   p--p data is smaller than
required\cite{r14} in central Pb--Pb at 160 A$\cdot$GeV  where for
fully integrated yields   $\gamma_s = 0.75 \pm 0.05$ was
fitted\footnote{ We have to point out, however, that midrapidity
data in A--A collisions at SPS are consistent\cite{r13} with the
value of  $\gamma_s=1$. See the more detailed discussion in
Section 2.2. }. This result shows that the undersaturation of
strangeness in p--p and A--A collisions differs by almost $50\%$.
This difference alone already indicates\cite{uh} that strangeness
in A--A is enhanced relative to p--p collisions.\footnote{Looked
at from a different angle we conclude that in central
nucleus--nucleus collisions at AGS energy and higher strangeness
production reaches values consistent with complete chemical
equilibrium. At lower energies, and in particular in elementary
particle collisions, strangeness is strongly undersaturated. This
will be discussed further in Section 6.}

The temperature parameter extracted from particle yields  in high
energy elementary collisions is at first glance surprisingly
compatible with the chemical freeze--out temperature extracted
from heavy ion data\cite{r14,r41}. At SPS and RHIC energies, $T$
 can be considered as a  measure of
thermal excitations of a non--perturbative QCD vacuum due to the
particle scattering. Thus, $T$  should be mostly correlated with
the collision energy and not with the system size. The charge
chemical potential, however, due to the isospin asymmetry,
differers substantially in p--p and A--A collisions. In Pb--Pb
collisions at the SPS the $\mu_Q\sim -7$ MeV\cite{r12} whereas in
p--p at $\sqrt s \simeq 27$ the $\mu_Q\sim 35$ MeV  is required.
%\footnote{ One
%needs to note, however, that some differences of chemical
%freezeout temperature in p--p and A--A collisions could appear due
%to a collective transverse motion in the former. In A--A
%collisions, however, this motion is mostly developed after the
%system freed chemically, that is why these differences could be
%small.}

%%%%%%%%%%%%%%%%%%%%%%%%%%%%f10101010\end{figure}
%%%%%%%%%%%%%%%%%%%%%%%%%%%%%f66666666666666
%\vskip -0.8cm
%\begin{figure}[htb]
% {\vskip -0.5cm
%{\hskip 1. cm
%\includegraphics[width=22.5pc, height=19.5pc,angle=0]{RFIG15.PS}}}\\
%{\vskip -2.cm \caption{ p--A results}} \protect\label{freezeout}
%\end{figure}
%%%%%%%%%%%%%%%%%%%%%%%%%%%

\subsubsection{ Statistical hadronization and string dynamics in  p--p
collisions}
%%%%%%%%%%%%%%%%%%%%%%%%%%%%%%f222222222222222222222222222222222222222222%%%%%%%%%%%%%%%%%%%%

The apparent  agreement of the  canonical statistical model with
 experimental data on particle production in elementary
collisions  leads to the interpretation that hadronization in
particle collisions is a statistical process. This result is
difficult to reconcile with the popular picture that hadron
production in hadron--hadron collisions is due to the decay of
color flux tubes\cite{colflux}, a model that  has explained many
dynamical features of these collisions. In the following we
address the question, how one can  possibly distinguish the string
hadronization via the break up of a color flux tube from the
statistical hadronization.
% model
%description of p--p experimental data on particle yields and their
%ratios.
We argue following Ref.~(\refcite{markus})
 that the
$\overline{\Omega}/\Omega = \Omega^+/\Omega^-$ ratio in elementary
proton--proton collisions is a sensitive probe to
 differentiate possibly  these two  scenarios.

Color flux tubes, called strings, connect two SU(3) color charges
[ $3$ ] and [ $\overline 3$ ] with a linear confining potential.
If the excitation energy of the string is high enough it is
allowed to decay via the Schwinger mechanism\cite{schwinger}, i.e.
the rate of newly produced quarks is given by:
 \be
 \frac {{\rm d}N_{\kappa}}{{\rm d}p_{\perp}} \sim {\rm
\exp}\left[-\pi m_{\perp}^2/\kappa \right], \label{eq110b}
 \ee
where $\kappa$ is the string tension and $m_{\perp} =
\sqrt{p_{\perp}^2 +m^2}$ is the transverse mass of the produced
quark with mass $m$.

However, specific string models may differ in their philosophy and
the types of strings that are created:
\begin{itemize}
\item
In UrQMD\cite{urqmd} the projectile and target protons become
excited objects due to the momentum transfer in the interaction.
The resulting strings, with at most two strings being formed,  are
of the diquark--quark type.
\item
In NeXuS\cite{nexusmodel}, the p--p interaction is described in
terms of pomeron exchanges or ladder diagrams. Both hard and soft
interactions take place in parallel. Energy is equally shared
between all cut pomerons and the remnants. The endpoints of the
cut pomerons (i.e. the endpoints of the strings) may be valence
quarks, sea quarks or antiquarks.
\item
In PYTHIA\cite{pythiamodel}, a  scheme similar to that  in UrQMD
is employed. However, hard interactions may create additional
strings from scattered gluons and sea quarks. Most strings are
also of a  diquark--quark form.
\end{itemize}

Fig.~(\ref{pythia} -left)
 depicts the antibaryon to baryon ratio at
midrapidity in proton--proton interactions at 160 GeV. The results
of the   calculations by NeXuS, UrQMD and PYTHIA are included in
this figure\cite{markus}. In all these models, the $\overline B
/B$ ratio increases strongly with the strangeness content of the
baryon. For strangeness $|s|=3$ the ratio significantly exceeds
the unity. In UrQMD and PYTHIA the hadronization of the
diquark--quark strings leads directly to the overpopulation of
$\overline \Omega$. In NeXuS, however, the imbalance of quarks and
anti quarks in the initial state leads to the formation of $q_{\rm
val}-\overline s_{\rm sea}$ strings (the $s_{\rm val}-\overline
q_{\rm sea}$ string is not possible). These strings  result then
in the overpopulation of $\overline \Omega^,$s.

%............................................

In order to   understand  the large  $\overline{\Omega}/\Omega$
values predicted by string models we include   in
Fig.~(\ref{string}), the color flux tube break-up mechanism.
\par\noindent
Fig.~(\ref{string}) shows the fragmentation of the color field
into quark-antiquark pairs, which then coalesce into hadrons.
While in large strings $\Omega^,$s and $\overline \Omega^,$s are
produced in equal abundance (a), low-mass strings in UrQMD
suppress $\Omega$ production at the string ends (b), while in
NeXuS $\overline \Omega$$^,$s are enhanced (c). Thus, the
microscopic method of hadronization leads to a strong imbalance in
the $\overline{\Omega}/\Omega$ ratio in low-mass strings.

%............................................
\begin{figure}[h]
\vskip 0mm \vspace{0cm}
\centerline{\psfig{figure=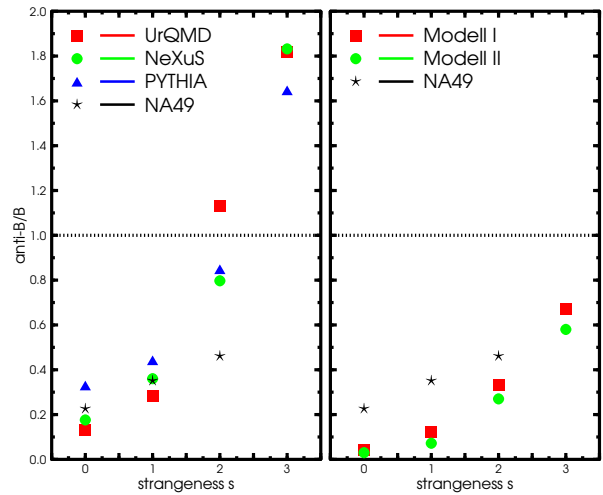,width=3.2in}}
\caption{\label{pythia} The left hand figure: the anti baryon to
baryon ratio at $|y-y_{\rm cm}|<1$ in p--p interactions at 160 GeV
as given by PYTHIA, NeXuS and UrQMD. The right hand figure: the
anti baryon to baryon ratio for the same reaction as given by the
statistical models. Stars depict preliminary NA49 data for the
$\overline B /B$ ratio at midrapidity. } \vspace{-0.4cm}
\end{figure}
%................................................

In Fig.~(\ref{pythia}) the string model results are compared with
the predictions of two statistical models (SM)\cite{colflux} and
the preliminary experimental data obtained\cite{na49pp} by the
NA49 Collaboration. The main difference between these models is
the implementation of baryon number and electric charge
conservation and the way an additional strangeness suppression is
introduced. In model ({\bf i}) the calculation\cite{becalast} is a
full canonical one with fixed baryon number, strangeness and
electric charge identical to those of the initial state.  An extra
strangeness suppression is introduced to reproduce the
experimental multiplicities. This is done by considering the
number of newly produced $\langle s \overline s \rangle$ pairs as
an additional charge to be found in the final hadrons. The
$s\overline s$ pairs fluctuate according to a Poisson distribution
and their mean number is considered as a free parameter to be
fitted\cite{becalast}. The parameters used for the prediction of
the $\Omega^+/\Omega^-$ ratio ($T$, the global volume $V$ sum of
single cluster volumes and $\langle s \overline s \rangle $) have
been obtained by a fit to preliminary NA49 p--p data\cite{na49pp}
yielding $T=183.7 \pm 6.7$ MeV, $VT^3 = 6.49 \pm 1.33$ and
$\langle s \overline s \rangle = 0.405 \pm 0.026$ with a
$\chi^2/{\rm dof}= 11.7/9$. It must be pointed out that the
$\Omega^+/\Omega^-$ ratio is actually independent of the $\langle
s \overline s \rangle$ parameter and only depends on $T$ and $V$.
%............................................

\begin{figure}[h]
\vskip 0mm \vspace{0cm}
\centerline{\psfig{figure=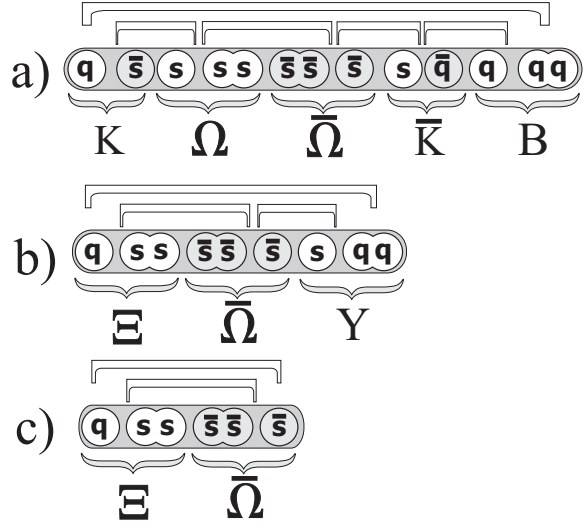,width=3in}} \vskip 2mm
\caption{\label{string} Fragmentation of a color field into quarks
and hadrons. While in large strings $\Omega^,$s and $\overline
\Omega^,$s are produced in an equal abundance (a), small
di$^,$quark strings suppress $\Omega^,$s at the string ends (b),
sea-$\overline s$ quarks enhance $\overline \Omega^,$s (c).}
\vspace{-0.4cm}
\end{figure}
%................................................

In model ({\bf ii})  the conservation of the baryon number and
electric charge is approximated by using the GC ensemble.
% by at
%most 20 30$\%$ \cite{marek}.
The strangeness conservation  is, however, implemented on the
canonical level following the procedure that  accounts for strong
correlations of produced strange particles.  In p--p collisions
the strangeness is assumed not to be distributed in the whole
volume of the fireball but to be locally strongly correlated. A
correlation volume parameter $V_0=4\pi R_0^3/3$ is introduced,
where $R_0\sim 1$ fm is a typical scale of QCD interactions.
% The
%previous analysis of WA97 pA data yields: $R_0\sim 1.12$~fm
%corresponding to $V_0\simeq 6$ fm$^3$. Note that, however, hidden
%strange particles are not canonically suppressed in this approach.
%The analysis of experimental data in AA collisions has shown that
%$T$ and $\mu_B$ are almost entirely determined by the collision
%energy and only depend weakly on the number of  participants
%\cite{cleymans}. The $4\pi$ results of NA49 on $\bar p/\pi$ and
%$\pi /A_{part}$ ratios in p-p and Pb-Pb collisions coincide within
%20 30$\%$. In terms of the SM this can be understood if $T$ and
%$\mu_B$ in p-p and Pb-Pb collisions have similar values. We take
The temperature  $T\simeq 158$ MeV and $\mu_B\simeq 238$ MeV were
taken  as obtained\cite{r14} from the SM analysis of a  full
phase-space Pb--Pb data of NA49 Collaboration. The volume of the
fireball $V\sim 17$ fm$^3$ and the charge chemical potential in
p--p was then found to reproduce the average charge and baryon
number in the initial state.

The predictions of the statistical models are shown in
Fig.~(\ref{pythia} -right).  In these approaches the $\overline B
/B$ ratio is seen to exhibit a significantly weaker increase with
the strangeness content of the baryon than that expected in the
string fragmentation models. For comparison, both figures include
preliminary data on the $\overline B /B$ ratios
obtained\cite{na49pp} at midrapidity by the NA49 Collaboration.
\noindent Note that the predictions of the statistical models in
Fig.~(\ref{pythia}) refer to full phase-space particle yields
whilst measurements of $\overline B /B$ ratios in p--p collisions
have been performed at midrapidity, where they are expected to be
the largest. Therefore, sizeable deviations of the model results
from the data seen in Fig.~(\ref{pythia})  are to be expected.
However, admitting the applicability of SM for midrapidity one
reproduces\cite{crn} the experimental data quite well.

In macroscopic string models the $\overline{\Omega}/\Omega$ ratio
depends in a strongly non-linear fashion on the mass (energy) of
the fragmenting string.  All these models predict a strong
enhancement of $\overline {\Omega}$ production at low energies,
while for large string masses the ratio approaches the value  of
$\overline{\Omega}/\Omega = 1$ (which should be reached in the
limit of an infinitely long color flux tube).

Statistical models, on the other hand, are not able to yield a
ratio of $\overline{\Omega}/\Omega > 1$. This can be easily
understood in the GC formalism, where the $\overline{B}/B$ ratio
is very sensitive to the baryon chemical potential $\mu_B$. For
finite baryon densities and including 100$\%$ feeding from
resonances, the $\overline{B}/B$ ratio will always be $< 1$ and
only in the limit of $\mu_B = 0$ may $\overline{\Omega}/\Omega =
1$ be approached. These features survive in the canonical
framework, where the GC fugacities are replaced by the ratios of
partition functions\cite{r37,r37a,r38}.

From the above discussion and from Fig.~(\ref{pythia}) it is thus
clear that  within the fragmenting color flux tube models the
$\overline \Omega/\Omega$ ratio is significantly above the unity.
This is in strong contrast to statistical model results, that
always imply that $\overline B / B $ ratios are below or equal to
a  unity in proton--proton reactions. Since this observable is
accessible to NA49 measurements at the SPS it can provide an
excellent test to distinguish the statistical model hadronization
scenario from that of a microscopic color-flux tube dynamics.

We have to point out, however, that the classical string models
considered above do not account for the so called string junction
mechanism that allows for diffusion of baryon number towards
midrapidity. This mechanism, recently included\cite{r1007} in Dual
Parton Model, was shown to be  very important for (multi)strange
baryon and antibaryon production. Thus, it would be of importance
to study the energy dependence of $\bar B/B$ ratio in p--p
collisions in terms of the model that includes this baryon number
transport.
%............................................

%%%%%%%%%%%%%%%%%%%%%%%%%%%%%%%%%%%%%%%%%%
%%%%%%%%%%%%%%%%%%%%%%%%%%%%%%%%%%%%%%%%%%
%%%%%%%%%%%%%%%%%%%%%%%%%%%%%%%%%%%%%%%%%%
%%%%%%%%%%%%%%%%%%%%%%%%%%%%%%%%%%%%%%%%%%
%%%%%%%%%%%%%%%%%%%%%%%%%%%%%%%%%%%%%%%%%%
\subsection{Heavy quark production } Charm quarks are heavy ($m_c
\gg T_c$) thus,  thermal production of charm quarks and charmed
hadrons is strongly suppressed in ultra-relativistic heavy ion
collisions\cite{krpbm}. The situation has been recently
discussed\cite{r611,r612} with the conclusion that, compared to
direct hard production, thermal production of charm quarks can be
neglected at SPS energies and is small even at LHC energy.
However, these investigations led to a new  scenario for the
production of hadrons which contain charm quarks in which
production of heavy quarks through hard collisions is combined
with a statistical procedure to produce open and hidden charm
hadrons at hadronization. This  idea of statistical hadronization
of charm quarks\cite{r611,r612} has sparked an intense activity in
this field\cite{gor,gra}. Initial interest focussed on the
available SPS data on $J/\psi$ production and their interpretation
in terms of a conventional statistical model\cite{goro}. As we
show below, these data can be well described, but only assuming a
charm cross section which is enhanced compared to predictions
within the framework of perturbative QCD.
%both concerning absolute cross sections as well as centrality dependence.
However, the largest differences between results from the
statistical coalescence scenario (or a similar\cite{the} model)
and more conventional models are expected at collider energies.
For example, in the Satz-Matsui approach\cite{satz}, one would
expect very strong suppression compared to direct production of
$J/\psi$ mesons (up to a factor\cite{vogt99} of 20) for central
Au-Au collision at RHIC energy. In the present approach this
suppression is overcome by statistical recombination of $J/\psi$
mesons from the same or different $c \bar c$ pairs, so that much
larger yields are expected. We therefore focus in this section on
predictions\footnote{This section is based on work by the authors
and A. Andronic and reported in
Ref.~(\refcite{andronic,andronic1}).} for open and hidden charm
mesons at RHIC and LHC energy, with emphasis on the centrality
dependence of rapidity densities.

\subsubsection{Statistical Recombination Model}

%The assumptions of the model are (see ref.~\cite{pbm1,pbm2} for more details):
In this model it is assumed that all charm quarks are produced in
primary hard collisions and (thermally) equilibrate in the
quark-gluon plasma (QGP); in particular, no $J/\psi$ is preformed
in QGP (complete screening) and there is no thermal production of
charm quarks. For a description of the hadronization of the c and
$\bar {\rm c}$ quarks, i.e. for the determination of the relative
yields of charmonia, and charmed mesons and baryons, we employ the
statistical model, with parameters as determined by the analysis
of all other hadron yields\cite{r80}. The picture we have in mind
is that all hadrons form within a narrow time range at or close to
the phase boundary. All charmed hadrons (open and hidden) are
formed at freeze-out (at SPS and beyond, freeze-out is at the
phase boundary\cite{r80}) according to the statistical laws.

Another interesting point concerns the $\psi^{'}$/(J/$\psi)$
ratio. As is well known, this ratio is, in hadron-proton and
p-nucleus collisions, close to 12 \%, independent\cite{na50_3} of
collision system, energy, transverse momentum etc.. In the thermal
model, the ratio is 3.7 \%, including feeding of the J/$\psi$ from
heavier charmonium states. A temperature of about 280 MeV would be
necessary to explain the ratio found in pp and p-nucleus
collisions in a thermal approach. Clearly, J/$\psi$ and $\psi^{'}$
production in pp and p-nucleus collisions  are manifestly
non-thermal. This was previously realized in
Refs.~(\refcite{gerschel,shuryak}). Similar considerations apply
for the $\chi$ states. In fact, feeding from $\chi_1$ to J/$\psi$
is less than 3 \% if the production ratios are thermal.

The experimental situation concerning the evolution with
participant number of the $\psi^{'}$/(J/$\psi)$ ratio in
nucleus-nucleus collisions, multiplied with the respective
branching into muon pairs,  is presented in Fig.~(\ref{fig:psi'}).
The data are from the NA38/50
collaboration\cite{na38_1,na50_4,na50_5,na38_2}. With increasing
N$_{part}$ the $\psi^{'}$/(J/$\psi)$ ratio drops first rapidly
(away from the value in pp collisions) but seems to saturate for
high N$_{part}$ values at a level very close to the thermal model
prediction, both for S+U and Pb--Pb collisions

\begin{figure}[thb]
\vspace{-1cm} \epsfxsize=10cm
\begin{center}
\hspace*{0in} \epsffile{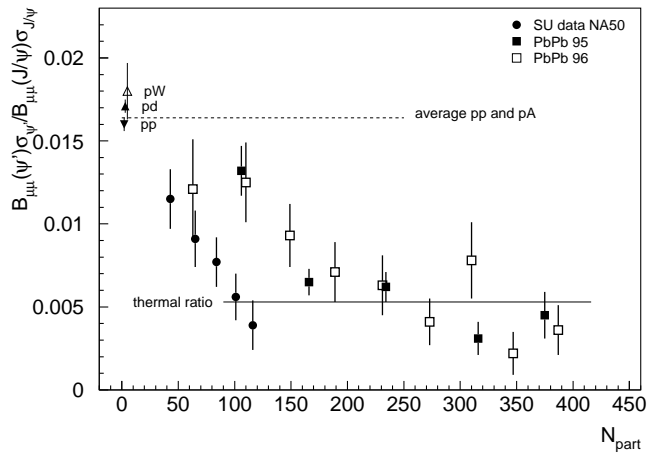}
%\vspace{-2ex}
\end{center}
\vspace{-1.5cm} \caption{\label{fig:psi'} Comparison of the
dependence of the measured $\psi^{'}$/(J/$\psi)$ ratio on the
number of participating nucleons with the prediction of the
thermal model. The data of  NA38 and NA50 Collaborations are from
Refs.~(\protect\refcite{na38_1,na50_4,na50_5,na38_2}). See text
and Refs.~(\protect\refcite{r611,r612}) for more details. }
\end{figure}

Taking this into account we note that predictions of the model
should only be trusted from about N$_{part}> 150$ on, where also
the $\psi'/(J/\psi)$ ratio is close to the thermal value for
Pb--Pb data. In our approach, ratios for all higher charmonia
states including the $\chi_c$ should approach the thermal value
from N$_{part} > 150$ on, implying that for those N$_{part}$
values feeding to J/$\psi$ should be small. In this picture, there
should thus not be different ``thresholds'' for the disappearance
of different charmonia.

The total number of open charm hadrons expected in a purely
thermal approach , $N_{oc}^{th}$, is then readjusted  to the
number of directly produced $c\bar{c}$ pairs, $N_{c\bar{c}}^{dir}$
as (neglecting charmonia): $N_{c\bar{c}}^{dir}=\frac{1}{2}g_c
N_{oc}^{th} {I_1(g_cN_{oc}^{th})}/{I_0(g_cN_{oc}^{th})}$, from
which the charm enhancement factor $g_c$ is extracted. Here, $I_n$
are modified Bessel functions. Note that we use here the canonical
approach as, depending on beam energy, the number of charm quark
pairs maybe smaller than 1. The grand-canonical limit will likely
only be reached at LHC energy. For a detailed study of the
transition from the canonical to the grand-canonical regime see
Section 3.2 and Ref.~(\refcite{r612}).  The yield of a given
species $X$ is then determined by
$N_X=g_cN_X^{th}{I_1(g_cN_{oc}^{th})}/{I_0(g_cN_{oc}^{th})}$ for
open charm mesons and hyperons and $N_X=g_c^2N_X^{th}$ for
charmonia (see Refs.~(\refcite{r611,r612}) for more details).

The inputs for the above procedure are: i) the total charged
particles yields (or rapidity densities), which are taken from
experiments at SPS\cite{na49aa,na50a} and RHIC\cite{phobos} and
extrapolated for LHC;
%NA50: $\ud N_{ch}/\ud \eta$=428 (PLB 530 (2002) 33)
%LHC... ``terra incognita'' \cite{xxx}.
%$ \ud N_{ch}/\ud y=n_{ch}^{therm}\cdot V N_{oc}^{therm}=n_{oc}^{therm}\cdot V$
and ii) $N_{c\bar{c}}^{dir}$, which is taken from next-to-leading
order (NLO) perturbative QCD (pQCD) calculations for pp\cite{vogt}
(the yield from MRST HO parton distributions was used here) and
scaled to AA via the nuclear
overlap function. % for a given centrality.
%$N_{c\bar{c}}^{dir}(N_{part})=
%\sigma({pp}\rightarrow {c\bar{c}})\cdot T_{AB}(N_{part})$
A constant temperature of 170 MeV and the baryonic chemical
potential $\mu_b$ according to the parameterization
$\mu_b$(MeV)=$1270/(1+\sqrt{s_{NN}}/4.3$ have been used for the
calculations\cite{r42}.

\subsubsection{Results}

We first compare predictions\cite{andronic,andronic1} of the model
to 4$\pi$-integrated $J/\psi$ data\cite{na50} at the SPS from NA50
Collaboration replotted as outlined in Ref.~(\refcite{r612}).  In
Fig.~(\ref{aa:fig1}) we present the model results  for two values
of $N_{c\bar{c}}^{dir}$: from NLO calculations\cite{vogt}
(dashed--line) and scaled up by a factor of 2.8 (continuous line).
The dashed--dotted line in Fig.~(\ref{aa:fig1}) is obtained with
the NLO cross section for charm production scaled--up by a factor
1.6, which is the ratio of the open charm cross section
estimated\cite{na504} by NA50 for p--p collisions at 450 GeV/c and
the NLO values from Ref.~(\refcite{andronic1}). For this case the
$N_{part}$ scaling is not the overlap function, but is taken
according to measured\cite{na504} dimuon enhancement as a function
of $N_{part}$.

 The results  of  Fig.~(\ref{aa:fig1}) indicate that
the observed centrality dependence of $J/\psi$ for
$100<N_{part}<$350 is well reproduced by the statistical model
using the NLO cross sections for charm production scaled by the
nuclear overlap function. However, to explain the overall
magnitude of the data a $N_{c\bar{c}}^{dir}$ increase by a factor
of 2.8 compared to NLO calculations is needed. The drop of the
$J/\psi$ yield per participant observed in the data for
$N_{part}>$350 ( see Fig.~(\ref{aa:fig1})) is currently
understood\cite{bla} in terms of energy density fluctuations for a
given overlap geometry.

 We mention in this context that the
observed enhancement of the dimuon  yield at intermediate masses
has been interpreted\cite{na50b} by NA50 as possible indication
for an anomalous increase in the charm cross section. We note,
however, that other plausible explanations\cite{rapp,vesa2} exist
of the observed enhancement in terms of thermal radiation.

%\vspace{-4mm}
\begin{figure}[htb]
\begin{tabular}{cc}
\begin{minipage}{.6\textwidth}
%exe plot#jpsi-npart-sps
\centering\includegraphics[width=1.\textwidth,height=.95\textwidth]{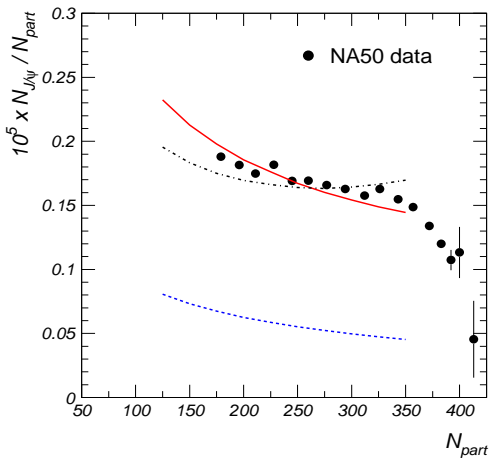}
\end{minipage}
& \begin{minipage}{.35\textwidth} \vspace{-4mm}
\caption{\label{aa:fig1}The centrality dependence of $J/\psi$
production at SPS. Model predictions are compared to
4$\pi$-integrated NA50 data\protect\cite{na50,gos}.
%, replotted as outlined in \cite{pbm2}.
Two curves for the model correspond to values of
$N_{c\bar{c}}^{dir}\sim 0.137$ from NLO
calculations\protect\cite{vogt} (dashed line)
 and scaled up by a factor of 2.8
(continuous line). The dashed--dotted curve is obtained when
considering the possible NA50 $N_{part}$--dependent  charm
enhancement\protect\cite{na50b} over their
extracted\protect\cite{na504} p--p cross section (see text). }
%taking $N_{ch}=1530$ (NA49) leads to
\end{minipage}
\end{tabular}
\end{figure}

We turn now to discuss model  predictions for collider energies.
For comparison we include in this study also results at SPS
energy.  The input parameters for these calculations for central
collisions ($N_{part}$=350) are presented in Table~\ref{aa:tab1}.
Notice that from now on we focus on rapidity densities, which are
the relevant observables at the colliders. The results  are
compiled in Table~\ref{aa:tab2} for a selection of hadrons with
open and hidden charm. All predicted yields increase strongly with
energy, reflecting the increasing charm cross section and the
concomitant importance of statistical recombination. Also ratios
of open charm hadrons evolve with increasing energy, reflecting
the corresponding decrease in charm chemical potential.

%%%%%%%%%%%%%%%%%%%%%%%%%%%%%%%%%%%%%%%%%%%%%%%%%%%%%%%%%%%%%%
%%%%%%%%%%%%%%%%%%%%%%%%%%%%%%1%%%%%%%%%%%%%%%%%%%%
\begin{table}
\tbl{ \label{aa:tab1} Input parameters for model calculations at
SPS, RHIC and
LHC.} {\tabcolsep34.8pt\begin{tabular*} %{l l l l l l}
{\typewidth}{@{}lllll@{}}\hline

$\sqrt{s_{NN}}$ (GeV) &  17.3  &  200 &  5500 ~  & \\ \hline %

$T$ (MeV)        & 170   & 170   & 170  & \\ %

$\mu_b$ (MeV)    & 253   & 27    & 1     & \\ \hline %

$\ud N_{ch}/\ud y$  &  430   &  730    &  2000  &  \\

$V_{\Delta y=1}$ (fm$^{3}$) &  861   &  1663   &  4564  & \\ \hline %%

$\ud N_{c\bar{c}}^{dir}/\ud y$  & 0.064 & 1.92  & 16.8  &  \\ %

$g_c$                           & 1.86  & 8.33  & 23.2  & \\ \hline %% %

\end{tabular*}}
%\begin{tabnote}[Note]
%\end{tabnote}
\begin{tabnote}[Source]
A. Andronic et al. from Ref.~(\refcite{andronic}).
\end{tabnote}
\end{table}
%%%%%%%%%%%%%%%%%%%%%%%%%%%%%%%%%%%%%%%%%%%%%%%%%%%%%%%%%%%%%%
%%%%%%%%%%%%%%%%%%%%%%%%%%%%%%%%%%%%%%%%%%%%%%%%%%%%%%%%%%%%%%

%%%%%%%%%%%%%%%%%%%%%%%%%%%%%%%%%%%%%%%%%%%%%%%%%%%%%%%%%%%%%%
%%%%%%%%%%%%%%%%%%%%%%%%%%%%%%1%%%%%%%%%%%%%%%%%%%%
\begin{table}
\tbl{  \label{aa:tab2} Results of model calculations at SPS, RHIC
and LHC for $N_{part}$=350.
} {\tabcolsep28.8pt\begin{tabular*} %{l l l l l l}
{\typewidth}{@{}lllll@{}}\hline

~$\sqrt{s_{NN}}$ (GeV) &  17.3  &  200 &  5500& \\ \hline %

~$\ud N_{\mathrm{D}^+}/\ud y$                  & 0.010 & 0.404& 3.56 & \\ %

~$\ud N_{\mathrm{D}^-}/\ud y$                  & 0.016 & 0.420 & 3.53 & \\ \hline %

~$\ud N_{\mathrm{D}^0}/\ud y$                  & 0.022 & 0.89 & 7.8 &  \\

~$\ud N_{\Lambda_c}/\ud y$                     & 0.014 & 0.153 & 1.16 & \\ \hline %%

~$\ud N_{J/\psi}/\ud y$  &  2.55$\cdot$10$^{-4}$ & 0.011 & 0.226&  \\ %

~$\ud N_{\psi'}/\ud y$ & 0.95$\cdot$10$^{-5}$ & 3.97$\cdot$10$^{-4}$ & 8.46$\cdot$10$^{-3}$  & \\ \hline %% %

\end{tabular*}}
%\begin{tabnote}[Note]
%\end{tabnote}
\begin{tabnote}[Source]
A. Andronic et al. from Ref.~(\refcite{andronic}).
\end{tabnote}
\end{table}

%%%%%%%%%%%%%%%%%%%%

%\vspace{-4mm}
\begin{figure}[htb]
\begin{tabular}{cc}
\begin{minipage}{.6\textwidth}
%exe plot#jpsi-npart-sps
\centering\includegraphics[width=1.\textwidth,height=1.15\textwidth]{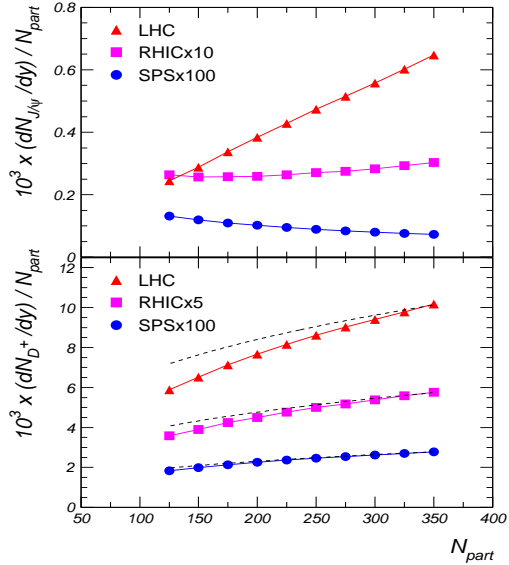}
\end{minipage}
& \begin{minipage}{.35\textwidth} \vspace{-5mm}
\caption{\label{aa:fig2} Centrality dependence of $J/\psi$ (upper
figure)  and of  D$^+$  (lower figure) rapidity density at SPS,
RHIC and LHC energies.}
%taking $N_{ch}=1530$ (NA49) leads to
\end{minipage}
\end{tabular}
\end{figure}
%%%%%%%%%%%%%%%%%
Predictions for the centrality dependence of $J/\psi$ production
are presented in Fig.~(\ref{aa:fig2}). In addition to the dramatic
change in magnitude (note the scale-up by factors of 10 and 100
for RHIC and SPS energy, respectively) the results exhibit a
striking change in centrality dependence, reflecting the
transition from a canonical to a grand-canonical regime (see
Ref.~(\refcite{r612}) for more details). The preliminary PHENIX
results on $J/\psi$ production at RHIC\cite{nagle} agree, within
the still large error bars, with our predictions. A stringent test
of the present model can only be made when high statistics
$J/\psi$ data are available. Another important issue in this
respect is the accuracy of the charm cross section, for which so
far only indirect measurements are available\cite{aver}. In any
case, very large suppression factors as predicted, e.g., by
Ref.~(\refcite{vogt99}) seem not supported by the data. In
Fig.~(\ref{aa:fig2}) we present the predicted centrality
dependence of charged D$^+$-meson production for the three
energies. The expected approximate scaling of the ratio
D$^+$/$N_{part}$ like $N_{part}^{1/3}$ (dashed lines in lower
Fig.~(\ref{aa:fig2})) is only roughly fulfilled due to departures
of the nuclear overlap function from the simple $N_{part}^{4/3}$
dependence.

\subsubsection{Charmonium Production from  Secondary  Collisions at
    LHC Energy}

     Another possibility to produce charmonium states is
due to reactions among D mesons in the hadronic and mixed phase of
the collision. This has been investigated in
Refs.~(\refcite{krpbm,ko}). As demonstrated there, this mechanism
does not lead to appreciable charmonium production at SPS and RHIC
energies. However, the large number of $c\bar c$ pairs and
consequently $D$,$\bar D$ mesons produced in Pb-Pb collisions at
LHC energy can lead to an additional production of charmonium
bound states due to reactions such as: $D\bar {D^*} +D^*\bar D +
D^*\bar {D^*}\to J/\psi +\pi$ and $D^*\bar {D^*} +D\bar D \to
J/\psi +\rho$. These processes were studied within a kinetic model
taking into account the space-time evolution of a longitudinally
and transversely expanding medium. The results\cite{krpbm}
demonstrate that secondary charmonium production appears almost
entirely during the mixed phase and is very sensitive to the
charmonium dissociation cross section with co-moving hadrons.
Within the most likely scenario for the dissociation cross section
of the $J/\psi$ mesons their regeneration in the hadronic medium
will be negligible, even at LHC energy.  Secondary production of
$\psi^,$ mesons however, due to their large cross section above
the threshold, can substantially exceed the primary yield.

\subsubsection{Conclusions on Heavy Quark Production} We have
demonstrated that the statistical coalescence approach yields a
good description of the measured centrality dependence of $J/\psi$
production at SPS energy, albeit with a charm cross section
increased by a factor of 2.8 compared to current NLO calculation.
Rapidity densities for open and hidden charm mesons are predicted
to increase strongly with energy, with striking changes in
centrality dependence. First RHIC data on $J/\psi$ production
support the current predictions, although the errors are too large
to make firm conclusions. For LHC energies we predict cross
sections for charm production in central Pb-Pb collisions
significantly exceeding the values predicted by scaling results
for N-N collisions with the nuclear thickness function. The
statistical coalescence implies travel of charm quarks over
significant distances in QGP. If the model predictions will
describe consistently precision data then J/$\psi$ enhancement
(rather than suppression)  would be a clear signal for the
presence of a deconfined phase.

Regeneration of charmonia in the mixed and hadronic phase has also
been studied.  For J/$\psi$ mesons this will likely only be a
small effect, even at LHC energy. However, secondary production of
$\psi^,$ mesons may be significant at LHC energy.

%%%%%%%%%%%%%%%%%%%%%%%%%%%%
%%%%%%%%%%%%%%%%%%%%%%%%%%%%
%%%%%%%%%%%%%%%%%%%%%%%%%%%%
%%%%%%%%%%%%%%%%%%%%%%%%%%%%
%%%%%%%%%%%%%%%%%%%%%%%%%%%%
%%%%%%%%%%%%%%%%%%%%%%%%%%%%
%%%%%%%%%%%%%%%%%%%%%%%%%%%%

\section{Unified conditions of  particle freeze--out in heavy ion collisions}

 A detailed analysis of
experimental data in heavy ion collisions from SIS through AGS,
SPS up to RHIC energy discussed  in Section 3 and 5  makes it
clear that the canonical or grand canonical statistical model
reproduces most of the measured hadron yields.
%%%%%%%%%%%%%%%%%%%%%%%%%%%%%f66666666666666
%\vskip -0.8cm
\begin{figure}[htb]
 {\vskip -0.5cm
{\hskip 0.4 cm
\includegraphics[width=24.5pc, height=22.5pc,angle=-180]{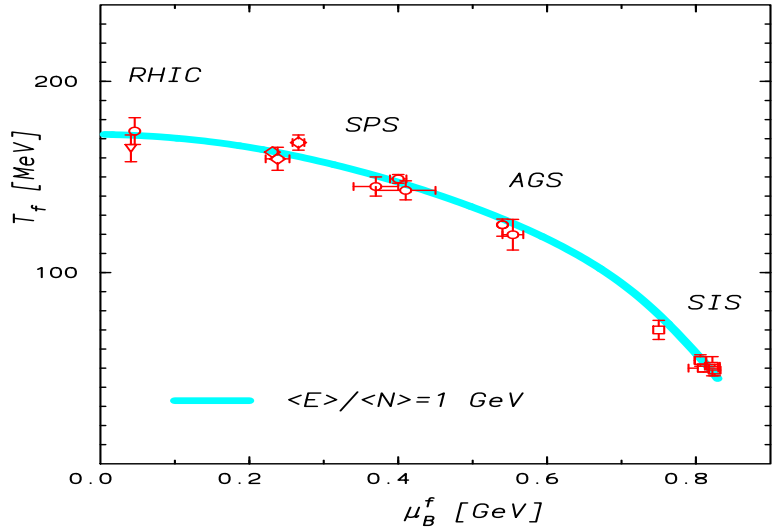}}}\\
{\vskip -1.cm \caption{\label{ff20} A compilation of chemical
freeze--out parameters appropriate for  A--A collisions at
different  energies:  SIS results are from
Ref.~(\protect\refcite{r31}), AGS  from
Ref.~(\protect\refcite{r14}),  SPS at 40 A$\cdot$GeV from
Refs.~(\protect\refcite{r15,r16,r17,r80}), SPS at 160 A$\cdot$GeV
from Refs.~(\protect\refcite{r12,r13,r14}), and RHIC from
Refs.~(\protect\refcite{r10,r53,r54}). The full line represents
the phenomenological condition of a chemical freeze--out at the
fixed energy/particle $\simeq 1.0$ GeV.\protect\cite{r9} }}
\end{figure}
%%%%%%%%%%%%%%%%%%%%%%%%%%%

  Figure~(\ref{ff20}) is a compilation of chemical freeze--out
parameters that are required to reproduce the measured particle
yields in central A--A collisions at SIS, AGS, SPS and RHIC
energy. The GSI/SIS results have the lowest freeze--out
temperature and the highest baryon chemical potential. As the beam
energy increases a clear shift towards higher $T$ and lower
$\mu_B$ occurs. There is a common feature to all these points,
namely that the average energy $\langle E\rangle$ per average
number of hadrons  $\langle N\rangle$   is approximately 1 GeV.
{\it A chemical freeze--out} in A--A collisions is thus
reached\cite{r9} {\it when the energy per particle $\langle
E\rangle$/$\langle N\rangle$ drops below 1 GeV} at all collision
energies.

In  cold nuclear matter the $\langle E\rangle$/$\langle N\rangle$
is approximately determined by the nucleon mass. For thermally
excited nuclear matter, in the non--relativistic approximation
 \be
 {{\langle E\rangle}\over {\langle N\rangle}}\simeq \langle m\rangle
+{3\over 2}T.\label{eq110}
 \ee
with $ \langle m\rangle$ being the thermal average mass in the
collisions fireball.  This result makes it clear why at SIS  the
energy/particle at chemical freeze--out is of the order of 1 GeV
since $T\simeq 53$ MeV. At SPS and RHIC energy the leading
particles in the final state (at  thermal freeze--out) are pions.
However, at chemical freeze--out most of the pions are still
hidden in the mesonic and baryonic resonances. Thus, here the
average thermal mass
%$\langle m \rangle$
is  larger than the pion mass and corresponds approximately  to
%$\langle m
%\rangle\sim m_\rho$
the $\rho$--meson  mass. Consequently, since $\langle m\rangle
>>T$,  Eq.~(\ref{eq110}) can be still used to justify
approximately that $\langle E\rangle/\langle N\rangle\simeq 1$ GeV
at  the SPS. Actually, this argument  holds, to a large extent, in
the whole energy range from SIS up to RHIC energy.

The physical origin of  the phenomenological freeze--out condition
of  fixed energy/particle would require  further dynamical
justification and interpretation.  Recently, this question  has
been investigated in central  Pb--Pb collisions at the SPS  in
terms of the Ultra--relativistic Quantum Molecular Dynamics model
(UrQMD)\cite{r72,r72n}. %UrQMD  is a microscopic transport approach
%that simulates multiple interactions of ingoing and newly produced
%particles, the excitation and fragmentation of color strings and
%the formation and decay of particles and resonances.
 A detailed
study has shown that there is a clear correlation between the
chemical break--up in terms of inelastic scattering rates and the
rapid decrease in energy per particle. If $\langle
E\rangle/\langle N\rangle$ approaches the value of  1 GeV the
inelastic scattering rates drop substantially and  further
evolution is due to elastic and pseudo-elastic collisions that
preserved the chemical composition of the collision fireball.
Following  the above UrQMD results and  the previous
suggestions\cite{r73} one could consider the phenomenological
chemical freeze--out of $\langle E\rangle/\langle N\rangle\simeq
1$ GeV as the condition of inelasticity in heavy ion collisions.

 Unified freeze--out conditions were also considered\cite{r74s} in the
context of hydrodynamical models  for particle production and
evolution in heavy ion collisions. There, it  was suggested,  that
the condition for chemical freeze--out,
 $\langle
E\rangle/\langle N\rangle\simeq 1$ GeV, selects the softest point
of the equation of state, namely the point where the ratio of the
thermodynamical pressure $P$ to the  energy density $\epsilon$ has
a minimum.  The  considerations were  essentially based on the
  proposed\cite{r74}
mixed phase model that  seems to be  consistent with the available
QCD lattice data. The quantity $P/\varepsilon$ is closely related
to the square of the velocity of sound  and characterizes the
expansion speed \cite{r75} of the reaction zone. Thus,  the system
lives for the longest time around the softest point that allows to
reach the chemical equilibrium  of its constituents. The above
interpretation, however, crucially depends on the type of the
equation of state used in the model.
%%%%%%%%%%%%%%%%%%%%%%%%%%%%%f66666666666666
%\vskip -0.8cm
\begin{figure}[htb]
 {\vskip -0.5cm
{\hskip 0.6 cm
\includegraphics[width=22.5pc, height=21.5pc,angle=-180]{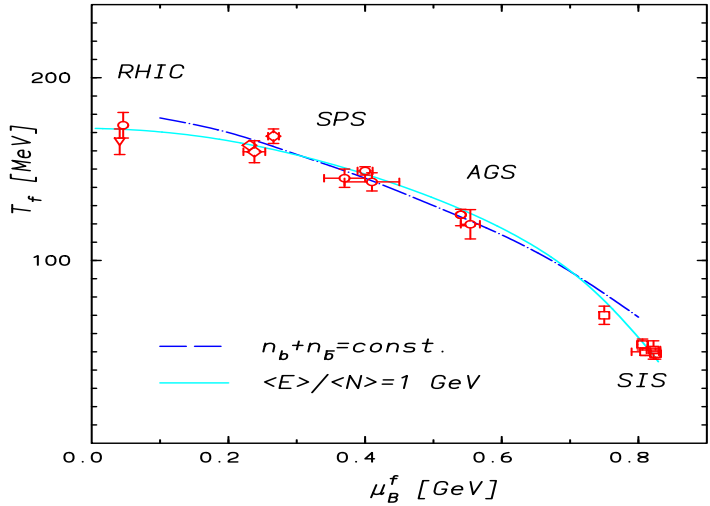}}}\\
{\vskip -1.cm \caption{\label{ff200} The broken line describes the
chemical freeze--out conditions of fixed total density of baryons
 plus antibaryons, $n_b+n_{\bar b}=0.12/$fm$^3$ from Ref.~(\protect\refcite{r31}).
  The full line represents the condition of
the fixed energy/particle $\simeq 1.0$  GeV from
Fig.~(\protect\ref{ff20}). The freeze--out points are as in
Fig.~(\protect\ref{ff20}). }}
\end{figure}

Chemical freeze--out  in heavy ion collisions can also be
determined\cite{ceres2,r85} by the condition of  fixed density of
the total number of baryons plus  antibaryons.  As it is seen in
Fig.~(\ref{ff200}), within statistical uncertainties on the
freeze--out parameters the above condition provides a good
description of  experimental data from the top AGS up to RHIC
energy. However, in the energy range from SIS to AGS it slightly
overestimates the freeze--out temperature for a given chemical
potential. Consequently, e.g. the yield of the strange/non-strange
particle ratios obtained at SIS turns out to be too large. The
freeze--out conditions determined by the extensive thermodynamical
observables are  in addition  very sensitive to the size and the
model that describes repulsive interactions between hadronic
constituents.

The condition of fixed
 $\langle
E\rangle/\langle N\rangle\simeq 1$ GeV, is very insensitive to
repulsive interactions. Independently on how the repulsive
interactions are implemented, that is through a mean field
potential\cite{r77}, an effective hard core\cite{r60} or a
thermodynamically consistent implementation \cite{r12,r79}, the
freeze--out line in        Fig.~(\ref{ff20})        is hardly
modified. However, the energy per particle is being sensitive to
the composition of the collision fireball. Considering heavy
fragments like e.g. the He or  Li as being  the constituents of a
thermal fireball would change the line shown in Fig.~(\ref{ff20}).
In general these fragments would make the  line steeper below the
AGS energy. This is an open question whether  at the SIS energy,
the multiplicity and the spectra of such composite objects are of
thermal origin and can be reproduced with the same parameters as
all other hadrons. For higher energies beyond AGS, however, the
above is not excluded as discussed in Section 3.04 and in Ref.~
(\refcite{r80}).

For the phenomenological determination of freeze--out parameters
for different collision energies we use in the following the
requirement $\langle E\rangle/\langle N\rangle\simeq 1$ GeV.

\subsection{Chemical freeze--out and the  QCD phase boundary}
The chemical freeze--out temperature,  found  from a thermal
analysis\cite{r12,r14,r10} of experimental data in Pb--Pb
collisions at the SPS and in Au--Au collisions at RHIC energy is
remarkably consistent, within errors, with the critical
temperature $T_c\simeq 173\pm 8$ MeV obtained\cite{r23} from
lattice Monte-Carlo simulations of QCD at a vanishing net baryon
density. Thus, the observed hadrons seem to be originating from a
deconfined medium and the chemical composition of the system is
most likely to be established during hadronization of the
quark-gluon plasma\cite{r3,r4,uh}. The observed coincidence of
chemical and critical conditions in the QCD medium at the SPS and
RHIC energy open the question  if this property is also valid in
heavy ion collisions at lower collision energies where the
statistical order of the secondaries is phenomenologically well
established.
%%%%%%%%%%%%%%%%%%%%%%%%%%%%%f66666666666666
%\vskip -0.8cm
\begin{figure}[htb]
 {\vskip -0.1cm
{\hskip 0.6 cm
\includegraphics[width=22.5pc, height=17.5pc]{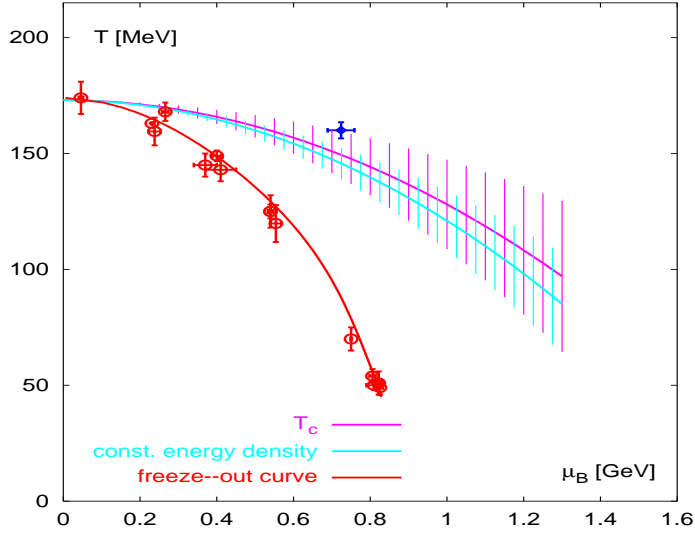}}}\\
{\vskip -0.2cm \caption{\label{ff21} A Comparison of the chemical
freeze--out curve from Fig.~(\ref{ff20}) with the phase
boundary line. The upper thin line represents the LGT results
obtained in Ref.~(\protect\refcite{r81}) and the lower thin line
describes the conditions of constant energy density that was fixed
at $\mu =0$. The upper  point with crossed error bars denotes the
end-point of the crossover
transition from Ref.~(\protect\refcite{r82}). }}
\end{figure}
%%%%%%%%%%%%%%%%%%%%%%%%%%%

Recently, first attempts have been made to extent lattice
calculations into the region of finite $\mu_B$. This provided an
estimate\cite{r81,r82} of the location of the phase boundary at
finite baryon density. The generic problem of the Monte-Carlo
simulation of QCD with the finite chemical potential, related with
a complex structure of the fermionic determinant, was partly
overcome. The reweighting method, in which the physical
observables at finite $\mu_B$ are computed by simulating the
theory at vanishing $\mu_B$\cite{r82} was successfully applied and
first results on the phase boundary were obtained\cite{r82}. The
region of applicability of this approach and uncertainties on the
results due to a small lattice size and large strange quark mass
are still, however, not well established. Another efficient
method, at least  for low baryon density, is based on the Taylor
expansion in $\mu_B$ of any physical observable\cite{r81}. The
coefficients of the series are calculated at vanishing $\mu_B$,
and thus could be obtained using a standard Monte-Carlo method.
This procedure was recently applied to get a series expansion of
the critical temperature in terms of the critical
$\mu_B$\cite{r81}.

Fig.~(\ref{ff21}) shows the results on the position of the phase
boundary that were obtained using the methods indicated above
together with the  freeze--out curve from Fig.~(\ref{ff20}). The
upper  thin-line represents an extrapolation of the leading,
$(\mu_B)_c^2$ order, term in the Taylor expansion of $T_c$ to a
larger values of the chemical potential\cite{r81}. It is
interesting to note that, within statistical uncertainties, the
energy density along this line is almost constant and corresponds
to $\epsilon_c\sim 0.6$GeV/fm$^3$ (thin lower line in
Fig.~(\ref{ff21})), that is the same value as found on the lattice
at $\mu_B=0$. It is thus, conceivable that the critical
$(\mu_B^c-T_c)$ surface  is determined by the condition of  fixed
energy density. This can be also argued phenomenologically. The
transition from a confined to deconfined phase could appear if the
particle (like in percolation models) or energy density is so
large that  hadrons start to overlap. It should  not be important
if this density is achieved by heating or compressing the nuclear
matter. Thus, since the percolation type argument\cite{r83} is
well describing critical conditions at $\mu_B=0$ it could be also
valid at finite $\mu_B$.

Fig.~(\ref{ff21}) shows that  the chemical freeze--out points at
SPS and RHIC energy are indeed lying on the phase boundary. The
results of SPS at 40 A$\cdot$GeV and top AGS are already below the
boundary line. However, it is not excluded that also at these
lower energies the collision fireball in the initial state appears
in the  deconfined phase. The initial energy density expected at
AGS is of the order of 1 GeV/fm$^3$ (see Section 1.1) thus,  it is
larger than the critical energy density along the  boundary line
in Fig.~(\ref{ff21}).

The canonical suppression effects for strangeness production, were
shown to be negligible  already at the top AGS energies. Here,
strangeness was uncorrelated and well described by the GC
approach. It is quite possible that the asymptotic GC formulation
and the maximal thermal phase space for strangeness is  achieved
if in the initial state  the system was  created in a deconfined,
QGP phase. The abundant production of strangeness in the
QGP\cite{r34} together with a long range correlations during a
non-perturbative  hadronization results in strangeness population
that maximizes the entropy in the GC limit. In this context the
energy range between SIS and 40 A$\cdot$GeV is of particular
interest and it is expected to  be covered by the
planned\cite{r84} for the future  heavy ion experiments at GSI.

\section{Particle yields and their energy dependence}

The  hadronic composition in the final state obtained in heavy ion
collisions is determined solely by an energy per hadron to be
approximately 1 GeV per hadron in the rest frame of the system
under consideration. This phenomenological freeze--out condition
provides the relation between the temperature and the chemical
potential at all collision energies.

%%%%%%%%%%%%%%%%%%%%%%%%%%%%%%1%%%%%%%%%%%%%%%%%%%%
\begin{figure}[htb]
%\vskip .cm
\begin{minipage}[t]{114mm}
%\framebox[79mm]{\rule[-26mm]{0mm}{52mm}}
%\hskip -1.0cm
\vskip -0.8cm \center\includegraphics[width=22.pc, height=19.6pc,angle=180]{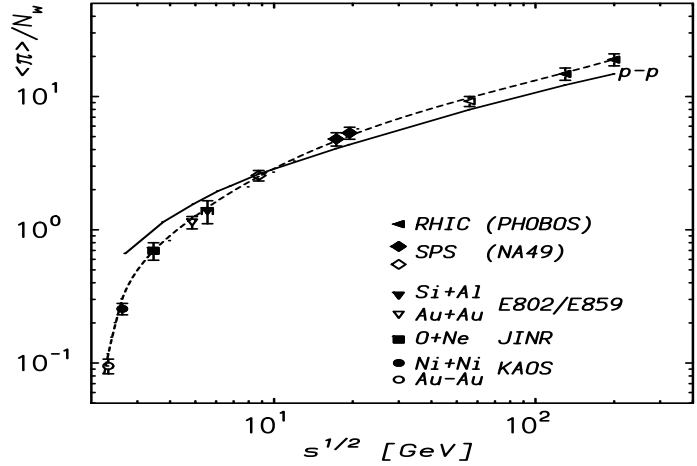}\\
\vskip -0.8 true cm \caption{\label{ff22} The total number of
pions per wounded nucleon  ($\langle \pi\rangle /N_{\rm w}$)
versus the center-of-mass energy. The data at lower energies in
A--A  as well as in p--p collisions are from
Refs.~(\protect\refcite{gr,na49aa}). The RHIC results are from
Ref.~(\protect\refcite{rhic}). The short-dashed and dashed lines
are a fit to the data. }
\end{minipage}
\end{figure}
%%%%%%%%%%%%%%%%%%%%%%%%%%%%%%1%%%%%%%%%%%%%%%%%%%%
%%%%%%%%%%%%%%%%%%%%%%%%%%%%%%%%%%%%%

 The above  relation together with
only one measured particle ratio, e.g. the ratio of
pion/participant\footnote{ The mean number of pion multiplicity is
defined as:  $\langle\pi\rangle\equiv
1.5(\langle\pi^+\rangle+\langle\pi^-\rangle)$ whereas the number
of participant is calculated as the number of wounded nucleons}
 as shown in Fig.~(\ref{ff22})
 establishes\cite{r21,r42}
the energy dependence of the two thermal parameters $T$ and
$\mu_B$.
%%, the temperature and baryon chemical potential.
Consequently, predictions of particle excitation functions can be
given in terms of the canonical statistical model. An alternative
approach would be to interpolate and/or parameterize the energy
dependence of the $\mu_B$ and then using the unified freeze--out
condition of $\langle E\rangle/\langle N\rangle\simeq 1$ to get
the energy dependence of $T$. The energy dependence of the
chemical potential was shown\cite{r42} to be well parameterized as
\begin{equation}
\mu_B(s) \simeq {a~{\mathrm{}}\over(1+ \sqrt{s}/b)}
\end{equation}
where  $a\simeq 1.27$ GeV and $b\simeq 4.3$ GeV. The result of
this parameterization is   shown by the full line in
Fig.~(\ref{ff23}) together with the  energy dependence of the
freeze--out temperature.

In the statistical approach,  the knowledge of $T(\sqrt s)$ and
$\mu_B(\sqrt s)$   determines the energy dependence of different
observables. Of particular interest are  the ratios of strange to
non-strange particle multiplicities as well as the relative
strangeness content of the system that is  measured by the
Wr\`oblewski factor\cite{r51}.
%The thermal model calculations for
%Au--Au and Pb--Pb collisions of different observables and their
%energy dependence were performed using a canonical correlation
%volume of  $R\sim$7 fm \cite{..}. The radius could, in general,
%vary with energy, however, for simplicity this was not taken into
%account.

\begin{figure}
\hspace*{1cm}
\includegraphics[width=21.5pc, height=15.5pc]{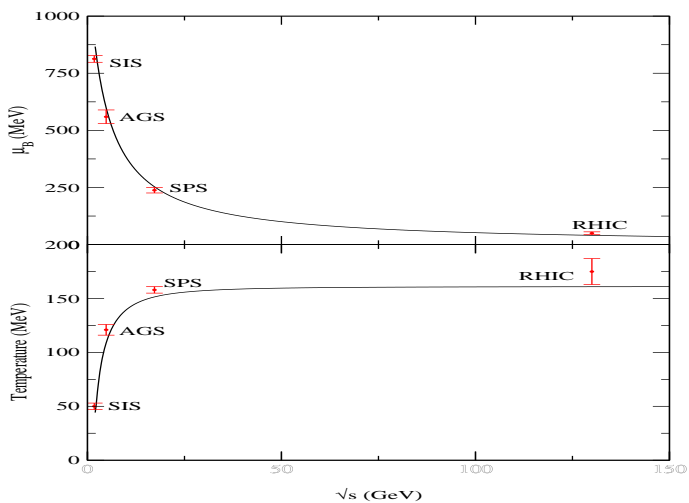}
\caption{ \label{ff23} { Behavior of the freeze--out baryon
chemical potential $\mu_B$ (upper curve) and the temperature $T$
(lower curve) as a function of energy from
Ref.~(\protect\refcite{r42}). The temperature $T$ as a function of
beam energy is determined from the  unified freeze--out conditions
of fixed energy/particle.  }}
\end{figure}
We turn our attention first to a study of the energy dependence of
the Wr\`oblewski ratio defined as
\begin{equation}
\lambda_s \equiv {2\bigl<s\bar{s}\bigr>\over \bigl<u\bar{u}\bigr>
+ \bigl<d\bar{d}\bigr>},
\end{equation}
%%%%%%%%%%%%%%%%%%%%%%%%%%%%%%%%%%%%%%%%
%%%%%%%%%%%%%%%%%%%%%%%%%%%%%%1%%%%%%%%%%%%%%%%%%%%
\begin{figure}[htb]
%\vskip .cm
\begin{minipage}[t]{55mm}
%\framebox[79mm]{\rule[-26mm]{0mm}{52mm}}
%\hskip -1.0cm
\vskip 2.2cm
\includegraphics[width=13.pc, height=12.6pc]{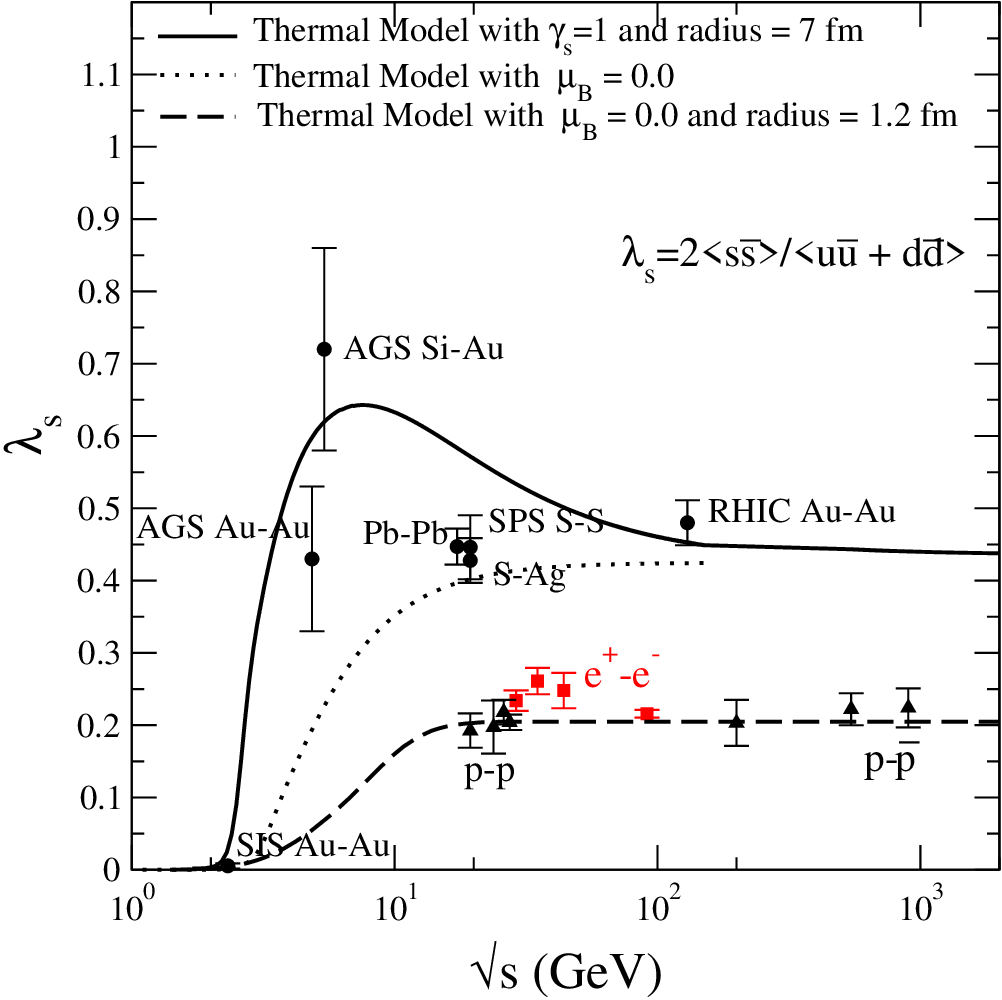}\\
\vskip -0.5 true cm \caption{\label{ff24}  The Wr\`oblewski ratio
$\lambda_s$ as a function of $\sqrt s$. For the description of the
lines see text. The points are the statistical model results
calculated with thermal parameters obtained from the fit to
measured particle yields.
%The thick solid line has been calculated using the freeze-out
%values of the temperature and the baryon chemical potential. The
%dashed line has been calculated using canonical corrections,
% keeping $\mu_B$=0 and taking the energy dependence of the temperature
%as determined previously. The radius has been taken as 1.2 fm. The
%dotted line has been calculated using $\mu_B=0$.
}\label{fig38}
\end{minipage}
\hspace{\fill}
\begin{minipage}[t]{54mm}\vskip 2.2cm
\hskip -0.1cm \includegraphics[width=13.pc, height=12.3pc]{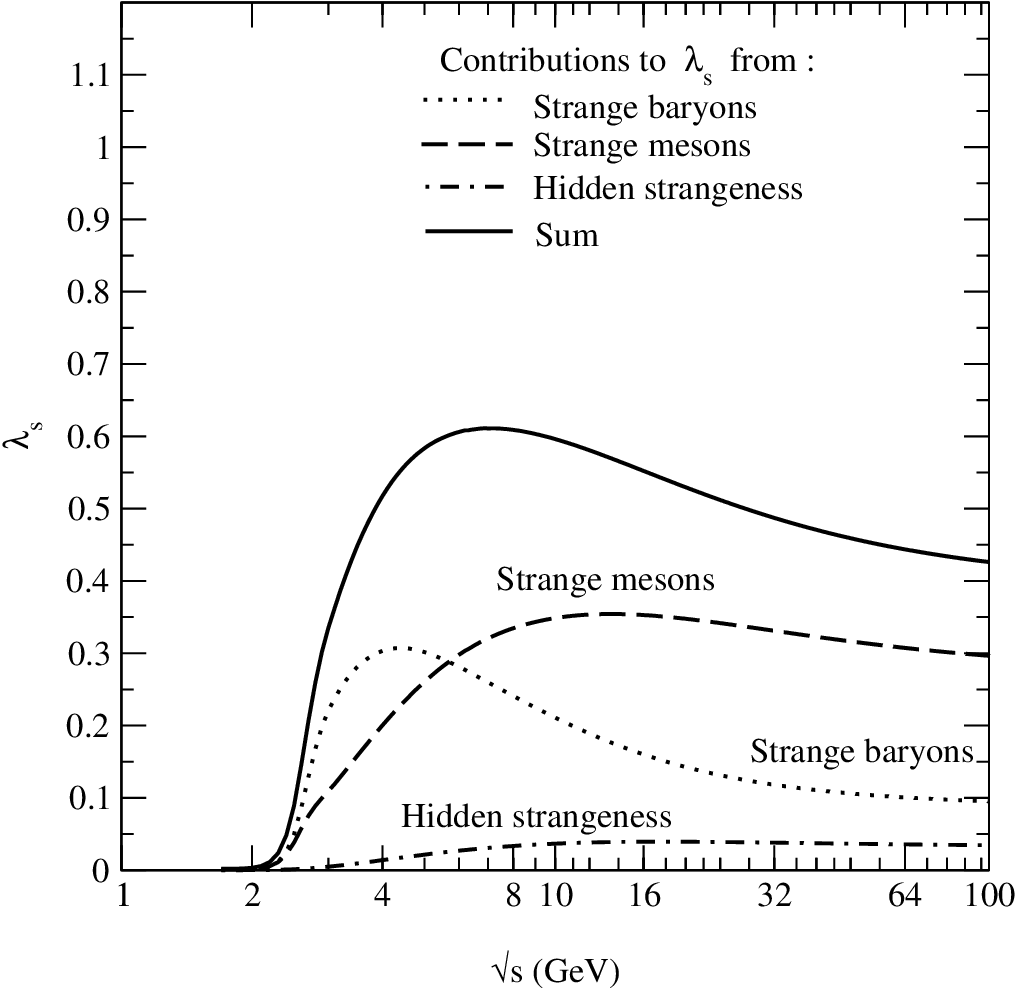}\\
\vskip -0.4 true cm \caption{\label{ff25}      Contributions to
the Wr\`oblewski factor from strange baryons, strange mesons and
hidden strange particles.  Full line is a sum of all these
contributions. \hfill}
\end{minipage}
\end{figure}
%%%%%%%%%%%%%%%%%%%%%%%%%%%%%%1%%%%%%%%%%%%%%%%%%%%
%%%%%%%%%%%%%%%%%%%%%%%%%%%%%%%%%%%%%
where the  quantities in angular brackets refer to the number of
newly formed quark-antiquark pairs, i.e., it excludes all the
quarks that are present in the target and projectile.

 The quark content used in this ratio is
determined at the moment
 of {\it {chemical freeze--out}}, i.e.
from  hadrons and especially, hadronic resonances, before they
decay.
%Thus, this factor counts the newly formed quark-antiquark
%pairs before e.g. the $\Delta$'s, the
%$\rho$'s and the $K^*$'s decay into nucleon plus pion, pions, kaon plus pion
%respectively.
This ratio is thus not  easily measurable unless one can
reconstruct all resonances from the final-state particles.

 The results are shown in Fig.~(\ref{ff24}) as a function of
center of mass energy  $\sqrt{s}$. The values calculated from the
experimental data at chemical freeze--out  in central A--A
collisions have been taken from reference
(\refcite{r14}).\footnote{ Here the statistical model was fitted
with an extra parameter $\gamma_s$ to account for possible
chemical undersaturation of strangeness. At the SPS,
$\gamma_s\simeq 0.75$ was required to get the best agreement with
4$\pi$ data. See the discussion in chapter  2 concerning this
issue.} All values of $\lambda_s$ were extracted from fully
integrated data besides RHIC where the STAR collaboration results
%%%%%%%%%%%%%%%%%%%%%%%%%%%%f888888888888888888888888
\begin{figure}[htb]
%\vskip 1.0 true cm
\vskip -0.5cm
\begin{minipage}[t]{60mm}
%\framebox[79mm]{\rule[-26mm]{0mm}{52mm}}
{%\vskip -8.3cm
\includegraphics[width=13.pc,height=16.3pc,angle=180]{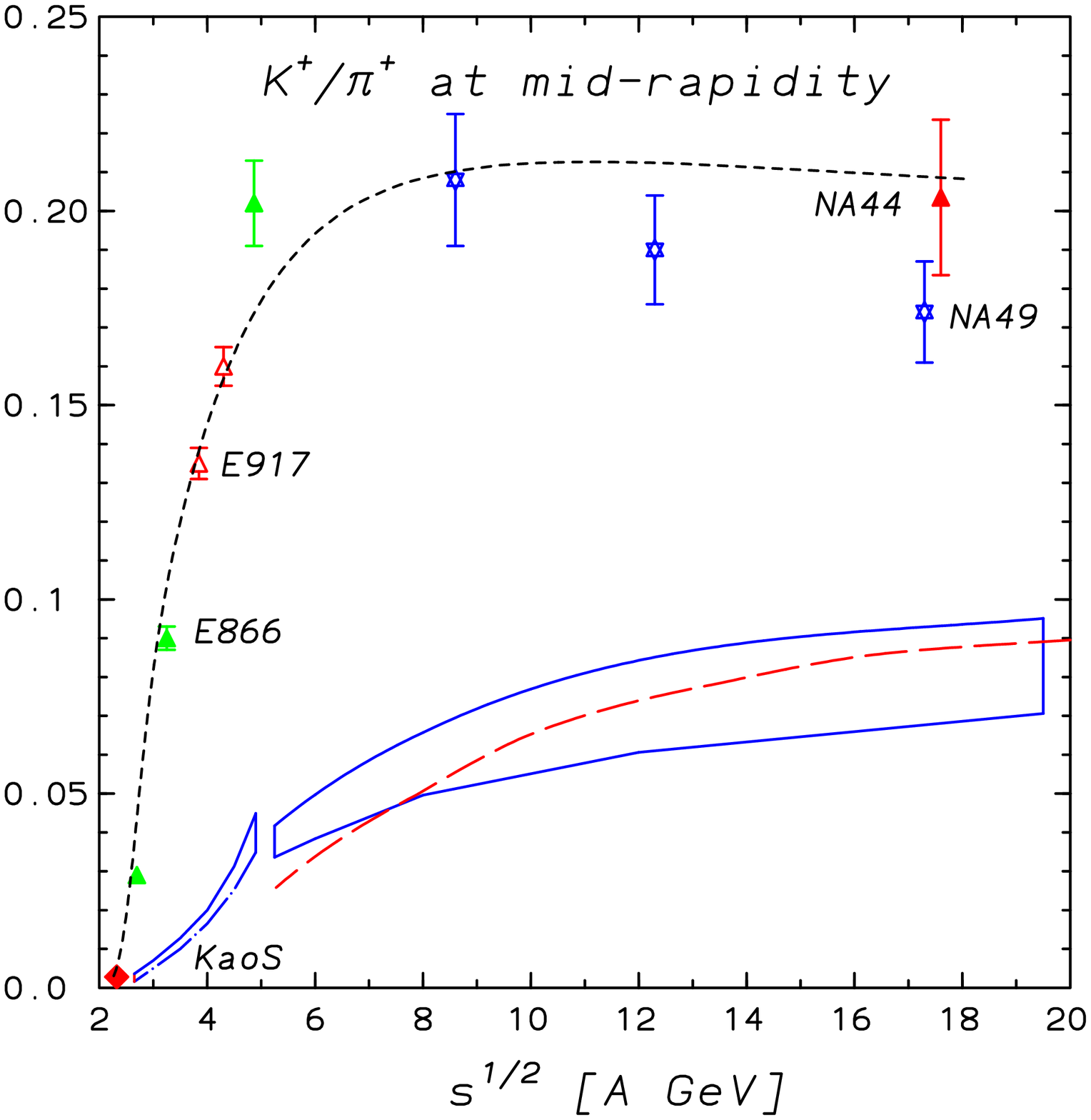}}\\
\end{minipage}
\hspace{\fill}
\begin{minipage}[t]{60mm}
\hskip -0.3cm \includegraphics[width=13.pc,height=16.3pc,angle=180]{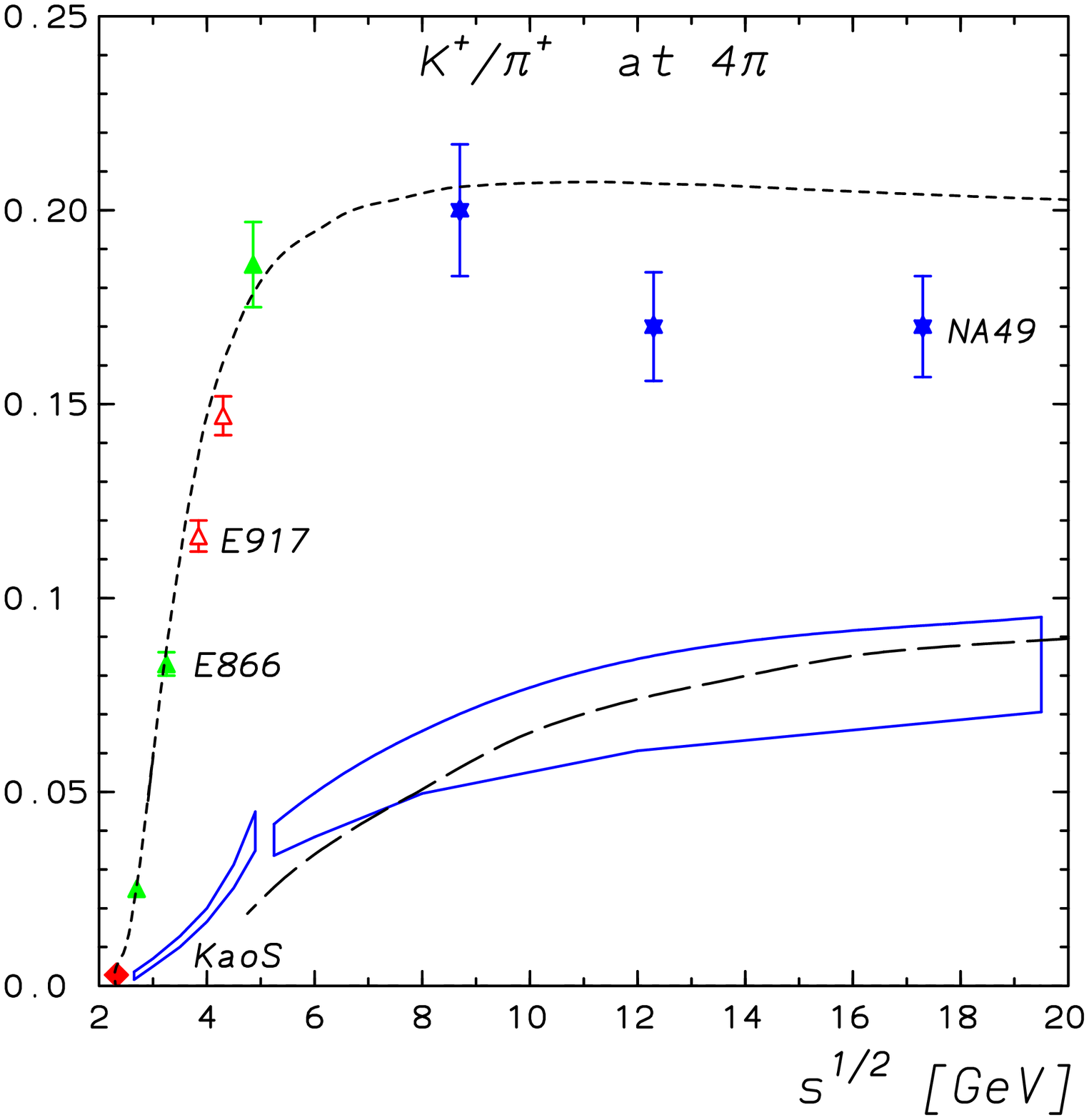}\\
\end{minipage}
\hskip -0.9cm  \begin{minipage}[t]{113.5mm}
 {\vskip -0.7 true cm \caption{\label{ff26}
 The ratio of kaon to pion measured in heavy ion collision at different
 collisions energies. The left-hand figure describes  midrapidity
 data, whereas the right hand figure represents the ratio of fully
 integrated yields. Data at SIS, AGS, SPS and RHIC energy are taken from
 Refs.~(\protect\refcite{na49aa,r85,r49}). The short-dashed line  describes the statistical model
 predictions along the unified
  freeze--out curve. The right hand figure also shows the parameterization
 of the
 p--p data (full-lines) from Ref.~(\protect\refcite{ogilvie}) and the canonical model results (dashed-line).
 }}
\end{minipage}
\end{figure}
%%%%%%%%%%%%%%%%%%%%%%%%%%%%%%1%%%%%%%%%%%%%%%%%%%%
on particle ratios measured\cite{r49} at mid-pseudorapidity were
used. The solid line in Fig.~(\ref{ff24}) describes the
statistical model calculations\cite{r9} in  complete equilibrium
along the unified freeze--out curve and with the energy dependent
thermal parameters shown in Fig.~(\ref{ff23}). From
Fig.~(\ref{ff24}) one sees that around 30 $A\cdot$GeV lab energy
the relative strangeness content in heavy ion collisions reaches a
clear and well pronounced maximum. The Wr\`oblewski factor
decreases towards higher incident energies and reaches a limiting
value of about 0.43.

The appearance of the maximum can be related to the specific
dependence of $\mu_B$ on
the beam energy. %To verify the above
In Fig.~(\ref{ff24}) we also show   $\lambda_s$ calculated under
the assumption that only the temperature  varies with collision
energy but the baryon chemical potential is kept fixed at zero
(dotted line). In this case the Wr\`oblewski factor is indeed seen
to be a smooth function of energy. The assumption of vanishing net
baryon density is close to the prevailing situation in e.g.
p--$\bar{\rm{p}}$ and e$^+$--e$^-$ collisions. In
Fig.~(\ref{ff24}) the results for $\lambda_s$ extracted from the
data in p--p, $\bar {\rm p}$--p and e$^+$---e$^-$ are also
included\cite{r41}. The dashed line represents the results
obtained with  $\mu_B= 0$ and a canonical radius of 1.2 fm. There
are two important differences in the behavior of $\lambda_s$ in
elementary compared to heavy ion collisions. Firstly, the
strangeness content is smaller by a factor of two. This is mainly
because in the elementary collisions particle multiplicities
follow the values given by the canonical ensemble with radius
1.1-1.2 fm whereas in A-A collisions the grand canonical ensemble
can be used, thus strangeness is uncorrelated and distributed in
the whole fireball. Secondly, there is no evidence, at the moment,
of a significant maximum in the behavior of $\lambda_s$ in
elementary collisions.

The importance of finite net baryon density on the behavior of
$\lambda_s$ is demonstrated in Fig.~(\ref{ff25}) showing
 separately the  contributions to $\left<s\bar{s}\right>$
coming from strange baryons, strange mesons and from hidden
strangeness, i.e., from hadrons  like $\phi$ and $\eta$.
 As can be seen in Fig.~(\ref{ff25}),
the origin of the maximum in the Wr\`oblewski
 ratio can be traced  to the contribution
of strange baryons  that is strongly enhanced in the energy range
up to 30 A$\cdot$GeV.

The appearance of the maximum in the strangeness content of the
collisions fireball can be also justified on the level of
different ratios that includes strange particles. The
measured\cite{r85,r49} $K^+/\pi^+$ ratio (see also
Fig.~(\ref{ff26})) is a very abruptly increasing function of the
collision energy between SIS up to the top AGS energy. At higher
energies it reaches  a broad maximum between 20 A$\cdot$GeV - 40
A$\cdot$GeV and gradually decreases up to RHIC energy. In
microscopic transport models\cite{r47} the increase of the kaon
yield with collision energy is qualitatively expected as being due
to a change in the production mechanism from associated to direct
kaon emission. However, hadronic cascade transport models do not,
until now, provide quantitative explanation of the experimental
data in the whole energy range. This is evident in
Fig.~(\ref{ff27}) where the comparisons of RQMD, URQMD and BUU
microscopic transport
%%%%%%%%%%%%%%%%%%%%%%%%%%%%%%%%%%%%
%%%%%%%%%%%%%%%%%%%%%%%%%%%%f888888888888888888888888
\begin{figure}[htb]
%\vskip 1.0 true cm
\vskip -0.3cm
\begin{minipage}[t]{60mm}
%\framebox[79mm]{\rule[-26mm]{0mm}{52mm}}
{%\vskip -8.3cm
%\hskip -0.6cm \includegraphics[width=16.pc, height=14.3pc]{gaz.prn}}\\
\hskip -0.0cm \includegraphics[width=14.5pc, height=13.3pc]{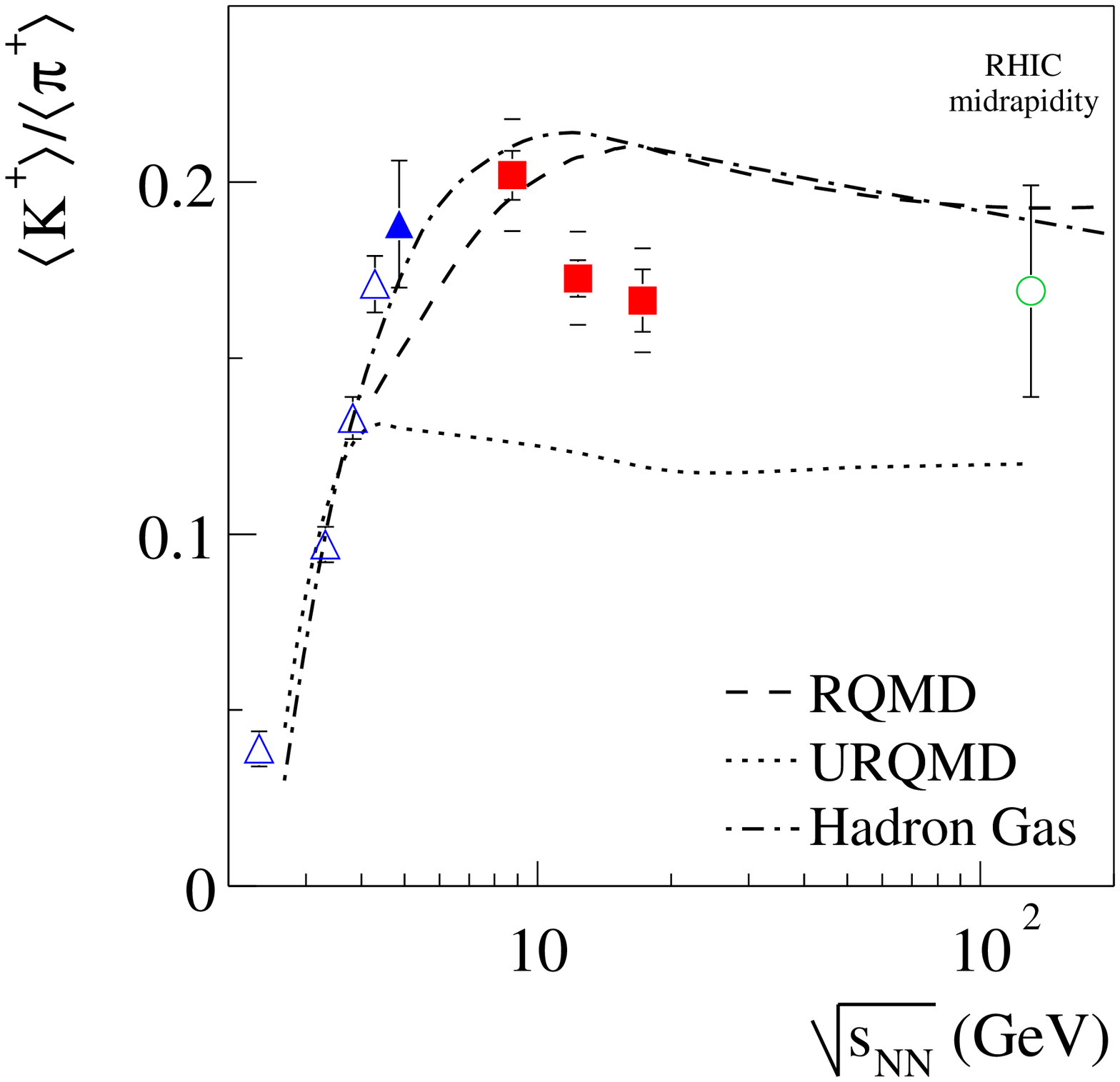}}\\
\end{minipage}
\hspace{\fill}
\begin{minipage}[t]{60mm}
\hskip -0.3cm \includegraphics[width=13.1pc, height=14.3pc]{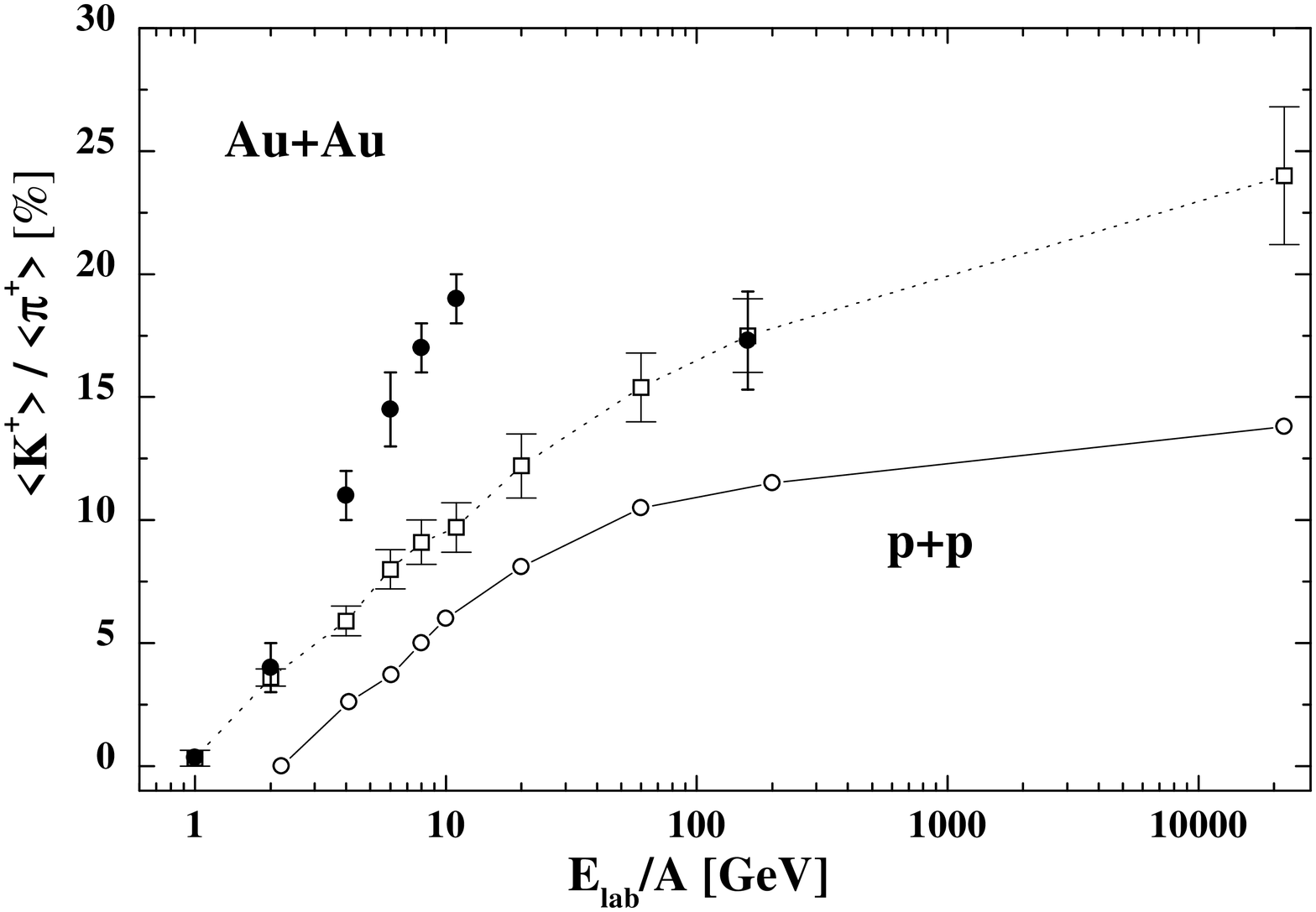}\\
\end{minipage}
 \begin{minipage}[t]{113.5mm}
 {\vskip -0.5 true cm \caption{\label{ff27}
The ratio of kaon to pion measured in heavy ion collision at
different collisions energies in comparison with microscopic
transport models. The left-hand figure represents the
RQMD\protect\cite{r86}, UrQMD\protect\cite{r87,urqmd} and the
statistical model predictions. The right hand figure shows the BUU
results\protect\cite{r47}. For the description of data see
Refs.~(\protect\refcite{r59,r48}).
 }}
\end{minipage}
\end{figure}
models with experimental data is presented. The RQMD provides a
good description of the high energy data. The URQMD works quite
well at the low energy, however, underestimates the yields  at the
SPS and RHIC energy. The BUU on the other hand is well suitable
for SIS energy range, however, shows different energy dependence
than that obtained in  experiments. The statistical model in the
canonical formulation\cite{r59},on the other hand, provides a good
description of the $K/\pi$ {\it midrapidity} ratio  from SIS up to
AGS as seen in Fig.~(\ref{ff26}). The abrupt increase from SIS to
AGS and the broad maximum of this ratio are   consequences of the
specific dependence of thermal parameters on the collision energy
and the canonical strangeness suppression at SIS. A drop in the
$K^+/\pi^+$ ratio for 4$\pi$ yields  reported\cite{r85,r49}  by
the NA49 Collaboration at 158 A$\cdot$GeV (see Fig.~(\ref{ff27})
is, however, not reproduced by the statistical model without
further modifications, e.g. by introducing an additional parameter
$\gamma_s\sim 0.75$\cite{r14} that accounts for additional
suppression of strangeness. We also note that an abrupt drop in
the $K^+/\pi^+$ ratio is predicted\cite{r88} in a special model
with particular conditions on the early stage of the collisions.
This model, however, neglects completely the production of strange
particles from the hadronization of gluons.

%%%%%%%%%%%%%%%%%%%%%%%%%%%%%%1%%%%%%%%%%%%%%%%%%%%
\begin{figure}[htb]
\begin{minipage}[t]{115mm}
%\framebox[79mm]{\rule[-26mm]{0mm}{52mm}}
\vskip -4.1cm\hskip .5cm
\center\includegraphics[width=29.5pc, height=29.2pc,angle=180]{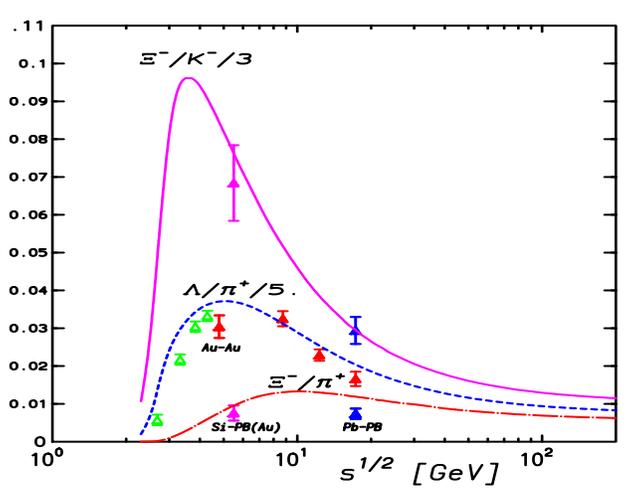}\\
\end{minipage}
\hspace{\fill}
\begin{minipage}[t]{115mm}
{\vskip -2.5 true cm \caption{\label{ff28} Particle ratios in A--A
collisions versus the center-of-mass energy. Data at the SPS are
fully integrated NA49 results. The corresponding ratio at the top
AGS was obtained from E810 results on $\Xi^-$
measured\protect\cite{r55} in Si--Pb collisions in the rapidity
interval $1.4<y<2.9$, normalized to the full phase--space values
of $\pi^+$ and $K^-$ yield obtained\protect\cite{r56}in Si--Au
collisions by E802 Collaboration. The lines represent statistical
model results\protect\cite{r42} along the unified freeze--out
curve.
 }}
\end{minipage}\vskip -0.5cm

\end{figure}
%%%%%%%%%%%%%%%%%%%%%%%%%%%%%%%%%%%%%%%
 The appearance of the maximum in the relative strange/non-strange
particle multiplicity ratios already seen in $K^+/\pi^+$ is even
more pronounced\cite{r42} for strange baryon/meson ratios.
Fig.~(\ref{ff28}) shows the energy dependence of $\Lambda /\pi^+$
and $\Xi^- /\pi^+$. There is a very clearly pronounced maximum,
especially in the $\Lambda/\pi^+$ ratio. This maximum is related
with a rather strong decrease of the chemical potential coupled
with only a moderate increase in the associated temperature as the
 energy increases\cite{r42}. The relative enhancement of
$\Lambda$ is stronger than that of $\Xi^-$. There is also a shift
of the maximum to higher energies for particles with  increasing
number of  strange quarks.  This is because an  enhanced
strangeness content of the baryon suppresses the dependence of the
corresponding ratio on $\mu_B$. This is also seen for   $\Xi^-
/K^-$ ratio that shows a substantially narrower maximum since the
strangeness dilution effect is compensated by the strangeness
content of the $K^-$. The actual experimental data
 for both $\Lambda /\pi^+$ and $\Xi^- /\pi^+$ ratios shown in
Fig.~(\ref{ff28}) are following the predictions of the statistical
model. However, as in the case of kaons, midrapidity results  are
better reproduced by the model than $4\pi$ data.

%%%%%%%%%%%%%%%%%%%%%%%%%%%%f888888888888888888888888
\begin{figure}[htb]
%\vskip 1.0 true cm
\vskip -0.1cm
\begin{minipage}[t]{60mm}
%\framebox[79mm]{\rule[-26mm]{0mm}{52mm}}
{%\vskip -8.3cm
\hskip -0.2cm \includegraphics[width=17.5pc, height=14.3pc]{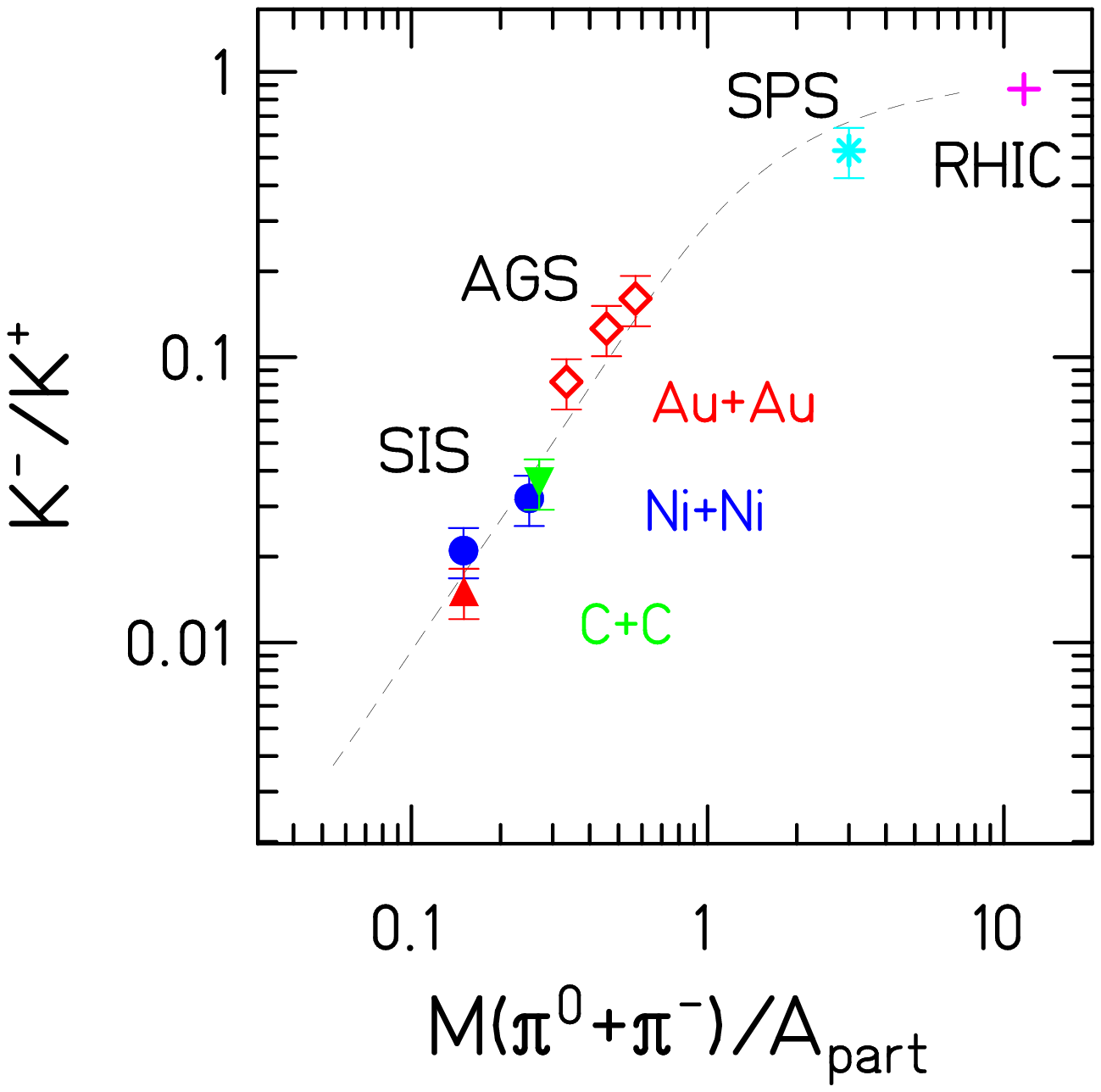}}\\
\end{minipage}
\hspace{\fill}
\begin{minipage}[t]{60mm}
\hskip -0.99cm \includegraphics[width=17.5pc, height=14.3pc]{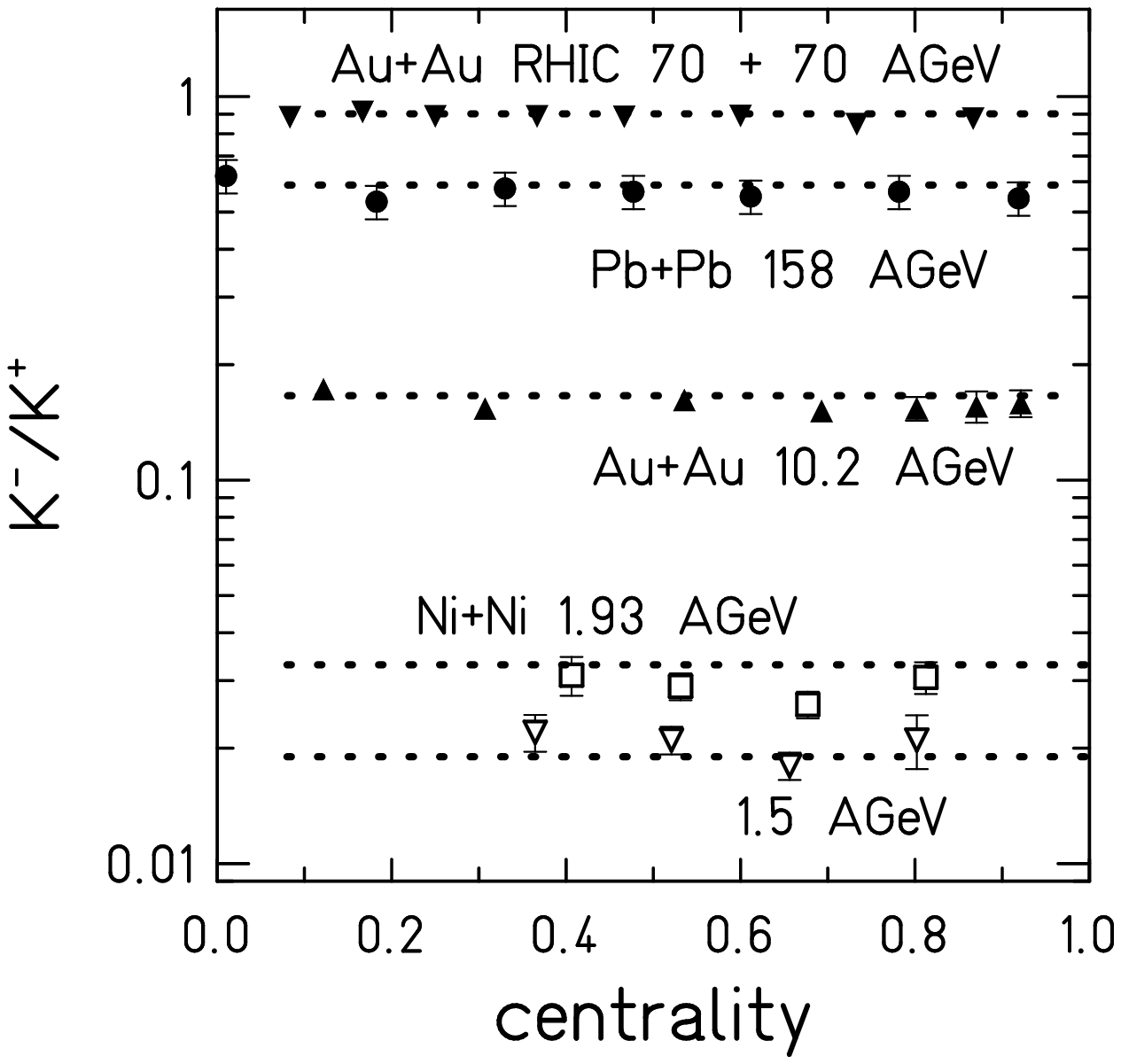}\\
\end{minipage}
 \begin{minipage}[t]{113.5mm}
 {\vskip -0.0 true cm \caption{\label{ff29} The left-hand figure,
 shows the  ratio of $K^+/K^-$
as a function of $(\pi^+-\pi^0)/A_{part}$. Points are the
experimental results, a  line is the statistical model result
along the unified  freeze-out curve. The right-hand figure shows
  $K^-/K^+$ ratio that appears to be
constant as a function of centrality from SIS up to RHIC energy.
Data are from the STAR, NA49, E866, and KaoS Collaborations. The
broken lines are statistical model results.
 }}
\end{minipage}
\end{figure}

%%%%%%%%%%%%%%%%%%%%%%%%%%%%%%1%%%%%%%%%%%%%%%%%%%%

The statistical model predicts  that if at least two different
ratios of non strange particles are  constant with centrality then
also strange particle/antiparticle ratio should be centrality
independent. Dynamically this is a rather surprising result as
strange particles and their antiparticles are generally produced
and absorbed in the surrounding nuclear medium in a different way.
This is particularly the case at lower energies (up to AGS) where
e.g. $K^+$ and $K^-$ are predominantly produced due to $ \pi N\to
\Lambda K^+$ and $\pi\Lambda\to K^-N$ processes. In addition
$K^+$,  mesons feel  a repulsive potential whereas $K^-$ mesons
are   attracted.
% nuclear potential with different absorption cross
%sections.
 Thus, the prediction of the thermal model that the
$K^+/K^-$ ratio is centrality independent was dynamically
unexpected. Figure (\ref{ff29}) represents the energy and
centrality dependence of the $K^+/K^-$ ratio from SIS to RHIC
energy. The statistical model predictions are seen in
Fig.~(\ref{ff29} -right)  to agree remarkably well with the data.
The results of Fig.~(\ref{ff29} -right) could be considered as the
evidence of an apparent chemical equilibrium population of kaons
in the final state. This behavior  of data is also seen on a
different level.  The chemical equilibration of the associated
production of $K^+$ with a hyperon and strangeness exchange
production of $K^-$, indicated above, should result in a linear
dependence of the $K^+/K^-$ ratio on $\pi /A_{part}$\cite{r89}.
Fig.~(\ref{ff29} - left)  shows that, indeed, in the energy range
from SIS up to AGS, and almost independently from the colliding
system, the above prediction is valid. It is also clear from this
figure that between AGS and SPS the production mechanism of
strange mesons is changing.

The results for the $K^+$ and $K^-$ excitation function show an
interesting behavior when  expressed as a function of  available
energy   $\sqrt {s} - \sqrt{s_{th}}$ (see Fig.~(\ref{ff30}
-left)\cite{r31}. The threshold energy $ \sqrt{s_{th}}$
corresponds to the production threshold in N--N collisions. For
$K^+$ mesons  $\sqrt {s_{th}}$ = 2.548 GeV, whereas for $K^-$ the
corresponding value is $\sqrt{s_{th}}$ = 2.87 GeV. It turned out
that in this representation the measured yields of $K^+$ and
$K^-$, close to threshold, in heavy ion collisions are about equal
while they differ in p--p collisions by a factor of 10 --
100\cite{r90}. Fig.~(\ref{ff30} -left) shows the experimental data
together with the canonical statistical model results\cite{r31}.
In the range where $\sqrt{s} - \sqrt{s_{th}}$ is less than zero,
the excitation functions for $K^+$ and $K^-$, obtained in the
model, cross each other,  leading to the observed equality of
$K^+$ and $K^-$ at SIS energies. The yields differ at AGS energies
by a factor of five. The difference in the rise of the two
excitation functions can be understood in the statistical model as
being due to the different dependence of $K^+$ and $K^-$ yields on
$\mu_B$. The $K^+$ yield is strongly $\mu_B$ dependent through
associated production with $\Lambda$ whereas $K^-$ yield is not
directly effected by $\mu_B$. Consequently, the excitation
functions, i.e.~the variation with $T$, exhibit a different rise
for kaons and  anti-kaons.

%%%%%%%%%%%%%%%%%%%%%%%%%%%%f888888888888888888888888
\begin{figure}[htb]
%\vskip 1.0 true cm
\vskip -0.5cm
\begin{minipage}[t]{60mm}
%\framebox[79mm]{\rule[-26mm]{0mm}{52mm}}
{%\vskip -8.3cm
%\hskip -0.6cm \includegraphics[width=16.5pc, height=14.3pc]{pal1.prn}}\\
\hskip -0.1cm{\includegraphics[width=17.5pc, height=15.3pc]{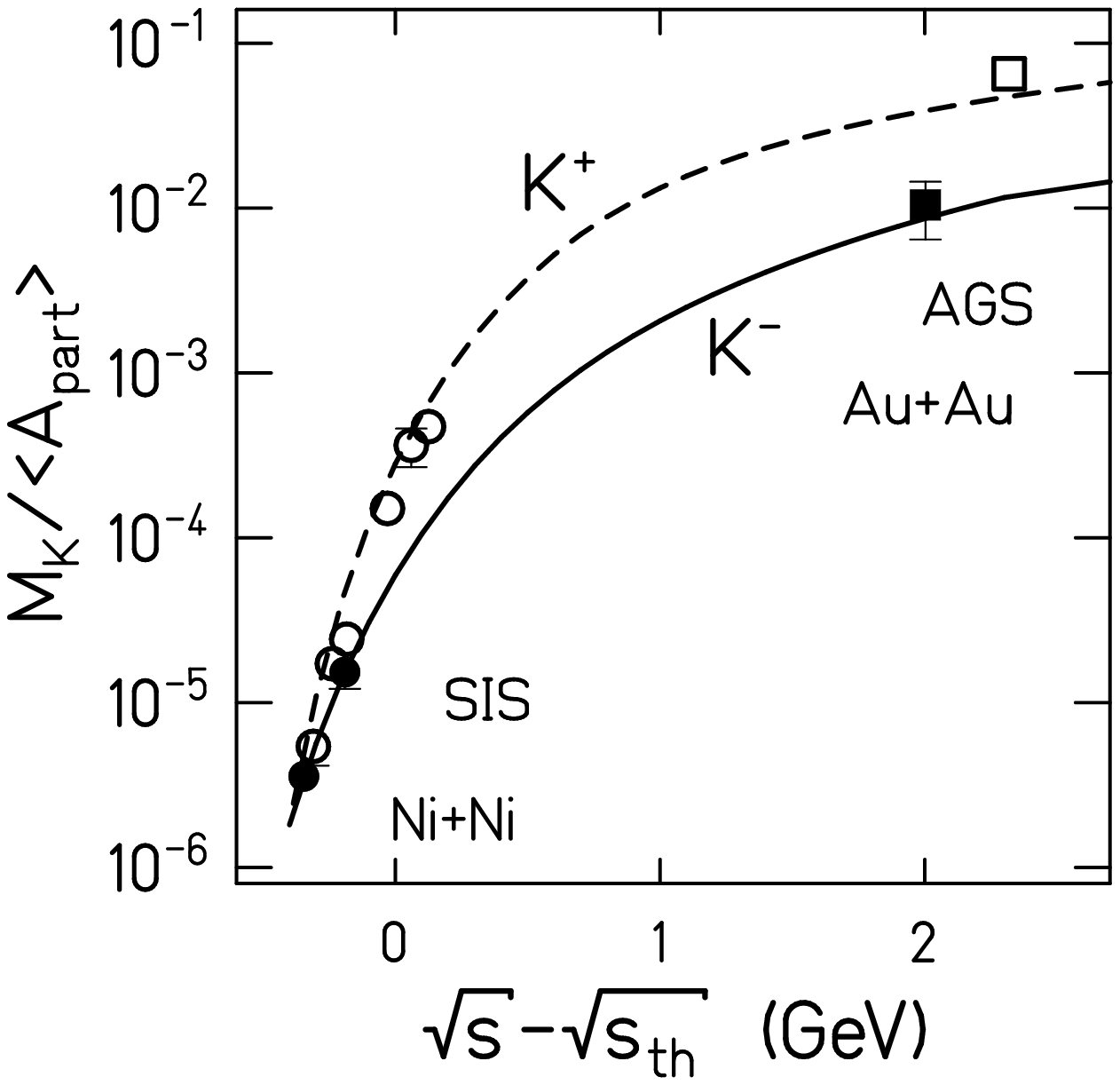}}}\\
\end{minipage}
\hspace{\fill}
\begin{minipage}[t]{60mm}
%\hskip -0.6cm
 \vskip -6.2cm\includegraphics[width=11.5pc, height=15.3pc,angle=180]{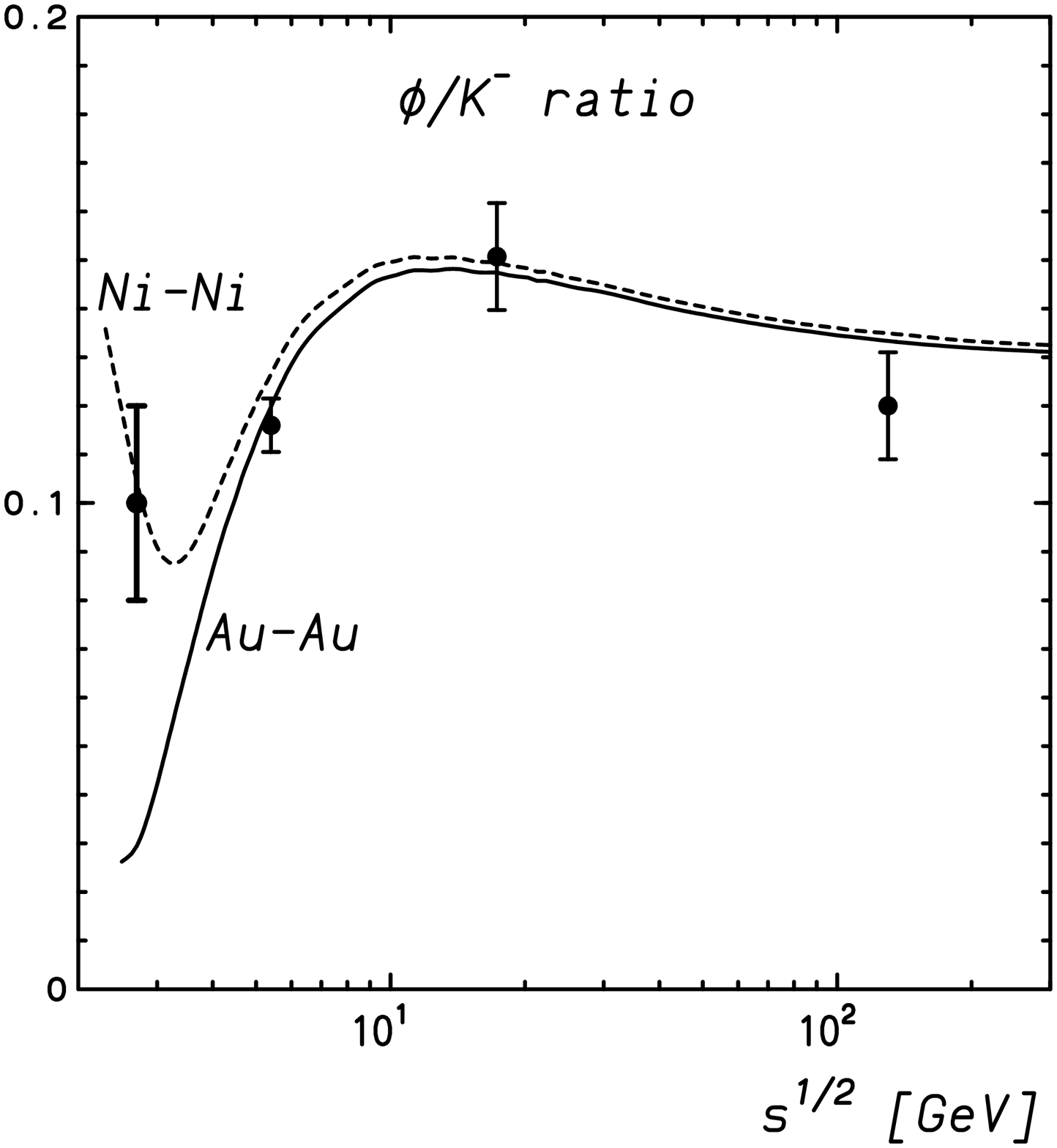}\\
\end{minipage}
 \begin{minipage}[t]{113.5mm}
 {\vskip -0.4true cm \caption{\label{ff30}
 The left--hand figure: calculated $K^+/A_{part}$ and
$K^-/A_{part}$ ratios of yield/participant in the statistical
model as a function of $\sqrt{s} - \sqrt{s_{th}}$ for Ni+Ni
collisions. The points are the results for  Ni+Ni collisions at
SIS\protect\cite{Barth,Marc}  and for Au+Au collisions at 10.2
$A\cdot$GeV (AGS)\protect\cite{Ahle} energies. The right-hand
figure: the yield of $\phi/K^-$ ratio in Ni--Ni and Au--Au
collisions calculated\protect\cite{r98} in the statistical model
along the unified freeze--out curve. The data points are from
Ref.~(\protect\refcite{r99})} \label{KP_KM_sthr_therm}
 }
\end{minipage}
\end{figure}

In the statistical model all particle species are considered to be
on shell. Thus, the statistical model reproduces  the kaon yields
and their   excitation functions with current particle masses. A
transport calculation\cite{r40,r47,r48}, on the other hand,
required in--medium modifications (a reduction) of the $K^-$ mass
(as expected for kaon in the nuclear medium) in order to describe
the measured yields.
 These differences are, however not necessarily in
contradiction, as the transport model describes a time evolution
of particle production, whereas the statistical models are only
valid for particles that are measured in the final state, where
the on-shell conditions are to be expected.\footnote{
Nevertheless, in transport models the kaons and anti-kaons are
produced in different time during the collision, thus also at
different temperatures. In the statistical model there is a common
temperature for kaons and their antiparticles.} It is conceivable
that the apparent chemical equilibration observed in the data at
SIS energies could be a direct consequence of in--medium effects.
One such possibility is an increase of the in--medium production
cross section of strange particles.

In the  context of  particle production in heavy ion collisions a
particular role has been attributed to the vector meson
resonances. The measurements of these particles could possibly
provide an information  on the chiral symmetry
restoration\cite{gery,hatsuda,medium,medium1,medium2} and in
medium effects due to the collision broadening\cite{r67} of their
decay widths. The production of $\phi$ meson is of particular
interest due to its $s\bar s$ content that could make it  a very
sensitive probe of strangeness production in the early stage. The
yield of $\phi$ mesons at the SPS was found to be very different
when it was reconstructed\cite{r100,r94}  from the $K^+K^-$ or
$\mu^+\mu^-$ decay channels. Possible scenarios of this difference
were extensively discussed\cite{r95} in the literature. In view of
the above problems it is interesting to check if the production of
$\phi$ mesons  follows the statistical order. Figure (\ref{ff30}
-right) shows experimental values for the ratio of $\phi/K^-$
obtained at SIS\cite{r99,her}, AGS\cite{r96}, SPS\cite{r100} and
RHIC energies\cite{r97}. These results are compared\cite{r98} with
the canonical model for Ni--Ni and Au--Au collisions. The
 $\phi/K^-$
ratio is seen in Fig.~(\ref{ff30} -right) to exhibit a broad
maximum as a function of collision energy that appears  around  40
A$\cdot$GeV. Thus, the maximum strangeness content indicated in
Fig.~(\ref{ff28}) is also visible  here. It is surprising that the
yield of $\phi /K^- $ within 40$\%$ is the same in a very broad
energy range from AGS up to RHIC. The Ni--Ni results seems to
follow this systematics. However, we have to point out that the
most recent, still preliminary results\cite{r99} of the FOPI
Collaboration indicate a much larger value of the ratio than it is
shown in Fig.~(\ref{ff30} -right). Also at the SPS the
results\cite{r94} of the NA50 Collaboration, extrapolated with the
measured slope to lower $p_t$, would give the value that is by a
factor of $3-5$ larger then that obtained\cite{r100} by NA49
Collaboration. The above uncertainties notwithstanding, the
statistical model follows the trend of  data and reproduces their
magnitude remarkably well. It is interesting to note the
differences in the canonical model on the energy dependence of the
$\phi/K^-$  for Ni--Ni and Au--Au collisions. The Ni--Ni line in
Fig.~(\ref{ff30} -right) shows a sharp minimum at $3< \sqrt s <4$
that appears due to a strong decrease of the $K^-$ yield through
the canonical suppression. In the canonical model, applied
here\footnote{The results were obtained in the canonical model
without strangeness suppression factor.} the $\phi$ meson yield is
unconstraint. This would  not be the case if the strangeness
undersaturation factor, instead of the correlation volume, would
be implemented.

Another interesting future of the $\phi$ yield is its centrality
dependence. The ratio of $\phi/K^-$ measured at  AGS\cite{r96},
SPS\cite{r100} and RHIC\cite{r97} energies  is centrality
independent whereas both $\phi$ and $K^-$ are showing a strong
centrality dependence that increases with decreasing beam energy.
Thus, the $\phi$ meson shows a similar centrality dependence as
kaons.

 The centrality dependence of the $K/\pi$ ratio at the
SPS was recently described\cite{r101,r101n}, in the context of a
statistical model, as being due to centrality dependence of
strangeness undersaturation parameter $\lambda_s$. This is quite a
conceivable scenario. The  $K^+/K^-$ as well as $\bar p/p$ ratios
are only very weakly centrality dependent, thus  the temperature
and baryon chemical potential, should be rather constant with
$A_{part}$. The change of the  $K/\pi\sim \lambda_s$ with
$A_{part}$  could be thus attributed to $\lambda_s$. The same
arguments could be also applied to the AGS and RHIC results.
However, the $\phi/K^-$ is to be proportional to  $\lambda_s$,
consequently the $\phi/K^-$ ratio should show similar  $A_{part}$
dependence as $K/\pi$ ratio. The above, however, is not confirmed
experimentally. This discrepancy calls   into question the concept
of a $\gamma_s$ factor as the relevant parameter characterizing
strangeness production in heavy ion collisions (see also the
discussion in Section 2).
% The firm up this conclusion,
%would require further study, particularly on centrality dependence
%of the $\phi/K^-$ ratio at midrapidity and in the full phase
%space.
%%%%%%%%%%%%%%%%%%%%%%%%%%%%%%%%%%%%%%%%%%%%%%%%%
%%%%%%%%%%%%%%%%%%%%%%%%%%%%%%%%%%%%%%%%%%%%%%%%%%%%%%%%%%%%%%%%%%%%%%%%%%%%%%%%%%%%%%%%%%%%%%%%%%%%%%%%
%%%%%%%%%%%%%%%%%%%%%%%%%%%%%%%%%%%%%%%%%%%%%%%%%%%%%%%%%%%%%%%%%%%%%%%%%%%%%%%%%%%%%%%%%%%%%%%%%%%%%%%%%%

\section{Lifting of the strangeness suppression in heavy ion collisions}

An enhanced production of strange particles compared to the
suppressed strangeness yield observed in collisions between
elementary particles  was long suggested\cite{r34,rn13,r50} as a
possible signal of the QGP formation in heavy ion
collisions.\footnote{ Strangeness enhancement in heavy ion
collisions was recently discussed and interpreted  at the parton
level in Ref.~(\refcite{hwa}) without requirements of the  QGP
formation in the initial state. } In the QGP the production and
equilibration of strangeness is very efficient due to  a large
gluon density and a low energy threshold for dominant QCD
processes of  $s\bar s$ production\cite{r34,r36}. In hadronic
systems the higher threshold for strangeness production was argued
in Ref.~(\refcite{r34}) to make the strangeness yield considerably
smaller and the equilibration time much longer.\footnote{
Recently,
 it was
argued in Refs.~(\refcite{r48,r35}), that multi-mesonic reactions
could accelerate the equilibration time of strange antibaryons
especially when the hadronic system is hot and very dense. This
argument does not apply, however, for strange baryon, where also
strong enhancements are seen.}

Based on such arguments predictions have been developed for
experimental signatures of deconfinement. Key predictions
 are\cite{r34,r33}:

i) {\it the disappearance of the strangeness suppression} observed
in collisions among elementary particles leading to e.g. an
enhancement of multistrange baryons and anti-baryons in central
A--A collisions, with respect to proton induced reactions.

ii) {\it chemical equilibration of secondaries}: the appearance of
the QGP being close to a chemical equilibrium and subsequent phase
transition should   in general  drive hadronic constituents
produced from hadronizing QGP  towards  chemical equilibrium.

Heavy ion experiments at the CERN SPS  reported\cite{r43,r44}
actually a global increase of  stran\-ge\-ness production    from
p-p, p-A to A-A collisions. This effect is seen e.g. in
Fig.~(\ref{ff26} -right) that shows the enhancement of the
$K^+/\pi^+$ ratio  in Pb--Pb relative to p+p collisions. There is
indeed  an increase by a factor of about two in strangeness
content when going from p+p to heavy ion collisions. This is also
seen on the level of the Wr\`oblewski factor in Fig.~(\ref{ff24}).
%%%%%%%%%%%%%%%%%%%%%%%%%%%%%%f44444444444
%\vskip 0.2cm
\begin{figure}[htb]
%\vskip 0.2cm
 {\hskip .2cm
\includegraphics[width=26.5pc, height=12.9pc]{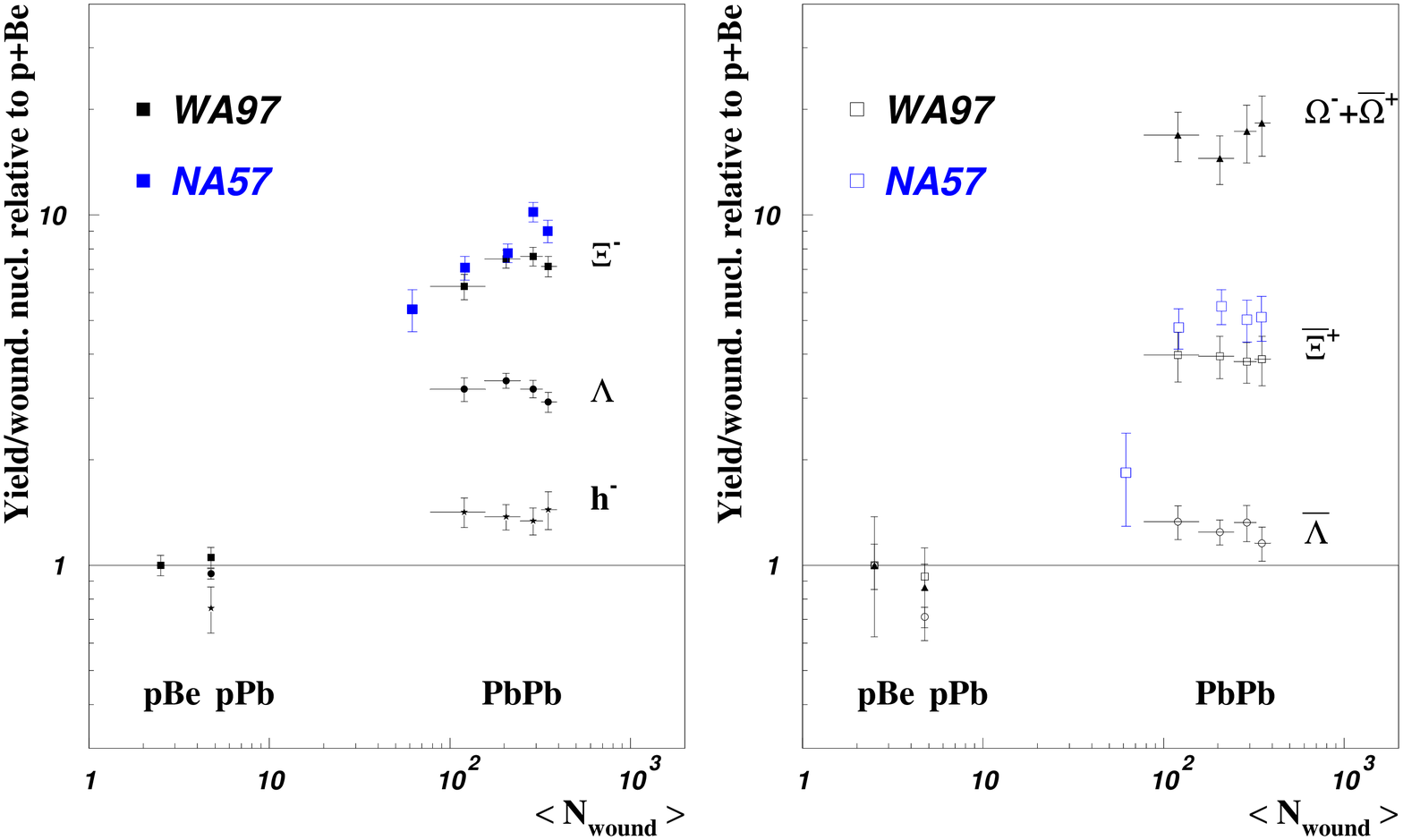}}\\
{ \caption {\label{ff31} Particle yields per participant in Pb--Pb
relative to p--Be and p--Pb collisions centrality dependence. The
data are from WA97\protect\cite{r43} and NA57\protect\cite{r44}
Collaborations.}}
\end{figure}
%%%%%%%%%%%%%%%%%%%%%%%%%%%%%%%%%%%

A large strangeness content of the QGP plasma should, following
(ii), be reflected in a very specific hierarchy of multistrange
baryons\cite{r34}: implying an enhancement of  $\Omega > \Xi
>\Lambda$. Fig.~(\ref{ff31})   shows the yield/participant in
Pb--Pb relative to p--Be and  p--Pb collisions
measured\cite{r43,r44} by the WA97 and NA57 Collaborations. Indeed
the enhancement pattern of the antihyperon yields is seen to
increase with strangeness content of the particle and in WA97 data
there is a saturation of this enhancement for $N_{wound} > 100$.
The recent results  of the NA57 collaboration are showing in
addition an abrupt  change of the anti-cascade enhancement for a
lower centrality. Similar behavior was previously seen on the
level of $K^+$ yield measured\cite{r104} at $\theta_{lab}=0^ø$ by
the NA52 experiment in Pb-Pb collisions. These results are very
interesting as they might be interpreted as an indication of the
onset of a new dynamics. However, a more detailed experimental
study and theoretical understanding are still required here. Until
now there is no quantitative understanding of this exceptional
threshold behavior of $\bar\Xi$.

A number of different mechanisms were considered   to describe the
magnitude of the enhancement and centrality dependence of
(multi)strange baryons
measured\cite{r21,at,r101,soff}$^-$\cite{lin} by the WA97
Collaboration. Studies using microscopic transport models make it
clear that data shown in Fig.~(\ref{ff31}) can not be explained by
pure final state hadronic interactions. The combination of the
former with additional pre-hadronic mechanisms    like sting
formation and their subsequent hadronization,   baryon junction
mechanism, color ropes or a color flux tubes overlap improves the
agreement with the measured enhancement pattern and the magnitude
for the most central collisions. However, the detailed centrality
dependence is still not well reproduced within the microscopic
models.

The results of Section 2 have shown that the statistical model
provides a satisfactory description of strange and multistrange
particle yields in A--A collisions. The midrapidity data were
argued in Section. 1 to be reproduced by the statistical model in
a full equilibrium. The fully integrated  results, on the other
hand can be successfully reproduced when including a strangeness
undersaturation parameter $\gamma_s\simeq 0.75$ that accounts, at
the SPS, for a 25$\%$ deviation  from  equilibrium.

Strangeness production in p--p collisions was  shown in Section 5
to be consistent with the canonical statistical model. The
abundance of single-strange  particles could be described by this
model by including,  as in heavy ion collisions, the strangeness
undersaturation factor $\gamma_s\simeq 0.51$ or a correlation
volume $V\sim V_p$ that accounts for the locality of strangeness
conservation. Consequently, the enhancement from p--p to A--A
collisions of strangeness 1 particles could be well described in
terms of the statistical model as the transition from the
canonical to the asymptotic GC limit\cite{r21}.
%%%%%%%%%%%%%%%%%%%%%%%%%%%f888888888888888888888888
\begin{figure}[htb]
%\vskip 1.0 true cm
\vskip -0.8cm
\begin{minipage}[t]{60mm}
{%\vskip -8.3cm
{\hskip -.4cm \includegraphics[width=20.1pc, height=19.3pc]{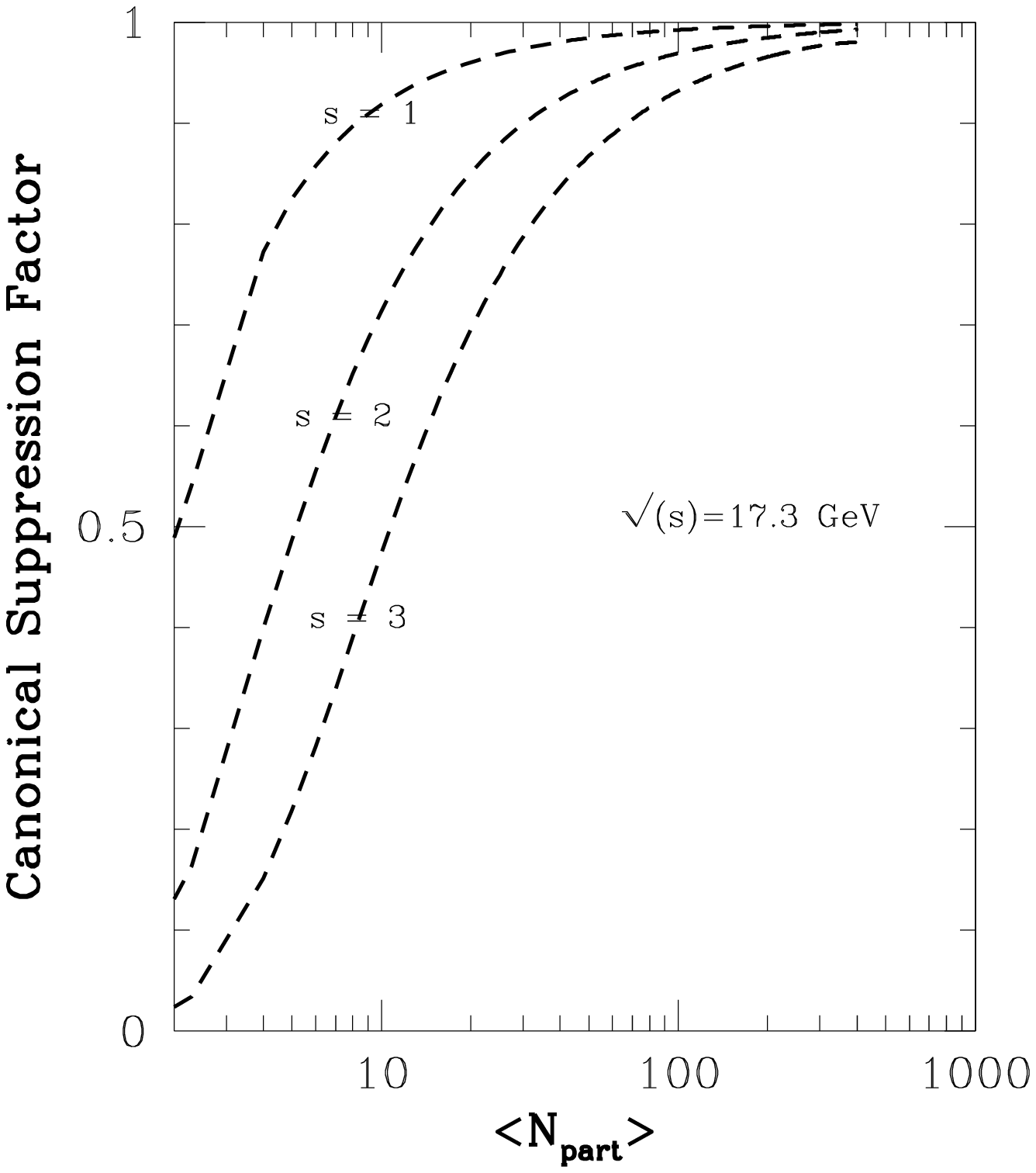}}}\\
\end{minipage}
\hspace{\fill}
\begin{minipage}[t]{65mm}
{\vskip -7.55cm
\includegraphics[width=12.2pc,height=19.5pc,angle=180]{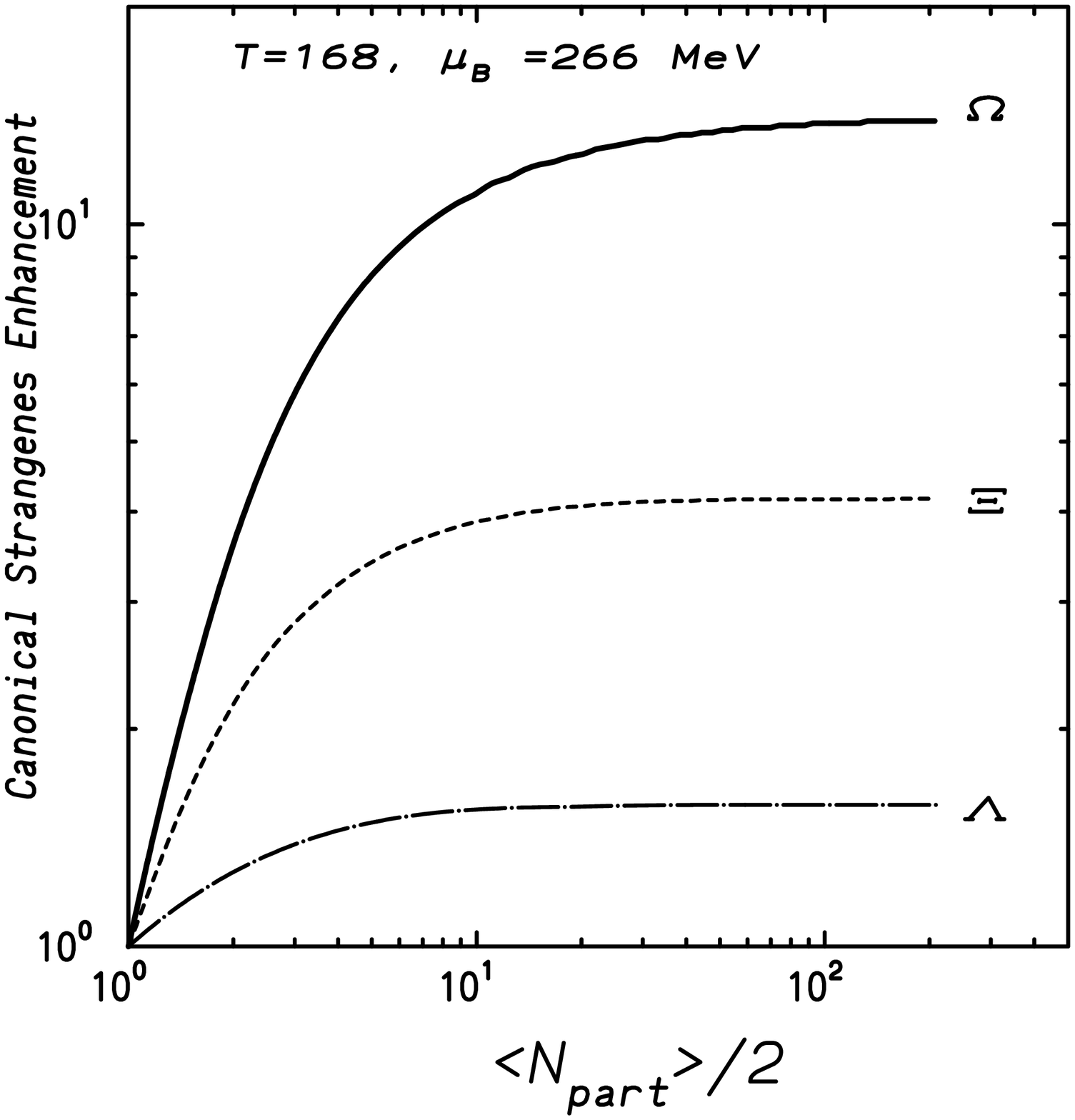}}
\end{minipage}
 \begin{minipage}[t]{113.5mm}
{\vskip -0.8 true cm \caption{ \label{ff32} The left hand figure:
the canonical suppression factor $F_s\sim I_s(Vx)/I_0(Vx)$  as a
function of $V\sim \langle N_{part}\rangle$ with the argument $x$
calculated  for a hadron resonance gas at $T=168$ MeV and
$\mu_B=266$ MeV. The right hand figure: statistical model results
on centrality dependence of the relative enhancement of $4\pi$
particle yields/participant in central Pb--Pb to p--p collisions
expected   at $\sqrt{s}\simeq 17$ GeV.}}
\end{minipage}
\end{figure}

%%%%%%%%%%%%%%%%%%%%%%%%%%%%%%%%%%%%%

%%%%%%%%%%%%%%%%%%%%%%%%%%%%%%%%%%%%%%%%%%%%%%%%%%%%%%%%%%%%%%%%%%%%%%

%%
One of the consequences of the canonical model is a very
particular volume dependence of multistrange particle densities.
In heavy ion collisions this volumes is usually assumed to  scale
with the number of participants.  From Eq.~(\ref{eq88}) it is
clear that the canonical suppression should increase with
strangeness content of the particles. Indeed the approximate
strangeness suppression factor $F_s\sim I_s(Vx)/I_0(Vx)$ for a
fixed $Vx$ is a decreasing function of $s$. This is particularly
evident in the limit of $(Vx)<<1$ where $I_s(Vx)/I_0(Vx)\sim
(Vx)^s$. The suppression factor is quantified in Fig.~(\ref{ff32}
-left) that indicates the expected suppression pattern.

 Fig.~(\ref{qm10} -left)  shows  predictions\cite{r21} of the canonical model for the
multiplicity/participant of $\Omega,~ \Xi ,$ and $\Lambda$
relative to their value in p--p collisions. Thermal parameters,
$T=168$ MeV and $\mu_B=266$ MeV, that are
appro\-pria\-te\cite{r12} for a description of  central Pb--Pb
collisions at   160 A$\cdot$GeV, were used here and assumed to be
centrality--independent.\footnote{ For a detailed discussion of
the possible centrality dependence of thermal parameters see e.g.
Ref.~(\refcite{r101}).} The canonical volume parameter was taken
to be proportional to the number of projectile participants,
$V\simeq V_0N_{\rm part}/2$ where $V_0=4\pi R_p^3/3$ with
$R_p=1.15\pm 0.5$. The volume $V_0$ is then, depending on the
particular choice of $R_p$, of the order of the  volume of a
nucleon. Figure (\ref{ff32} -right) indicates that the statistical
model in the canonical ensemble reproduces the basic features of
 WA97 data: the enhancement pattern and enhancement saturation
for large $A_{\rm part}$. The appearance of the saturation of the
enhancement indicates that the grand canonical limit was reached.
It is also clear from Fig.~(\ref{qm10}-left) that this saturation
is shifted towards a larger centrality with
 increasing strangeness content of the particle.
%%%%%%%%%%%%%%%%%%%%%%%%%%%%%%%%%%%%%%%%%%%%%%%%%%%%%%%
\begin{figure}[htb]
\vskip -1.5cm
\begin{minipage}[t]{49mm}
\includegraphics[width=18.5pc, height=19.7pc]{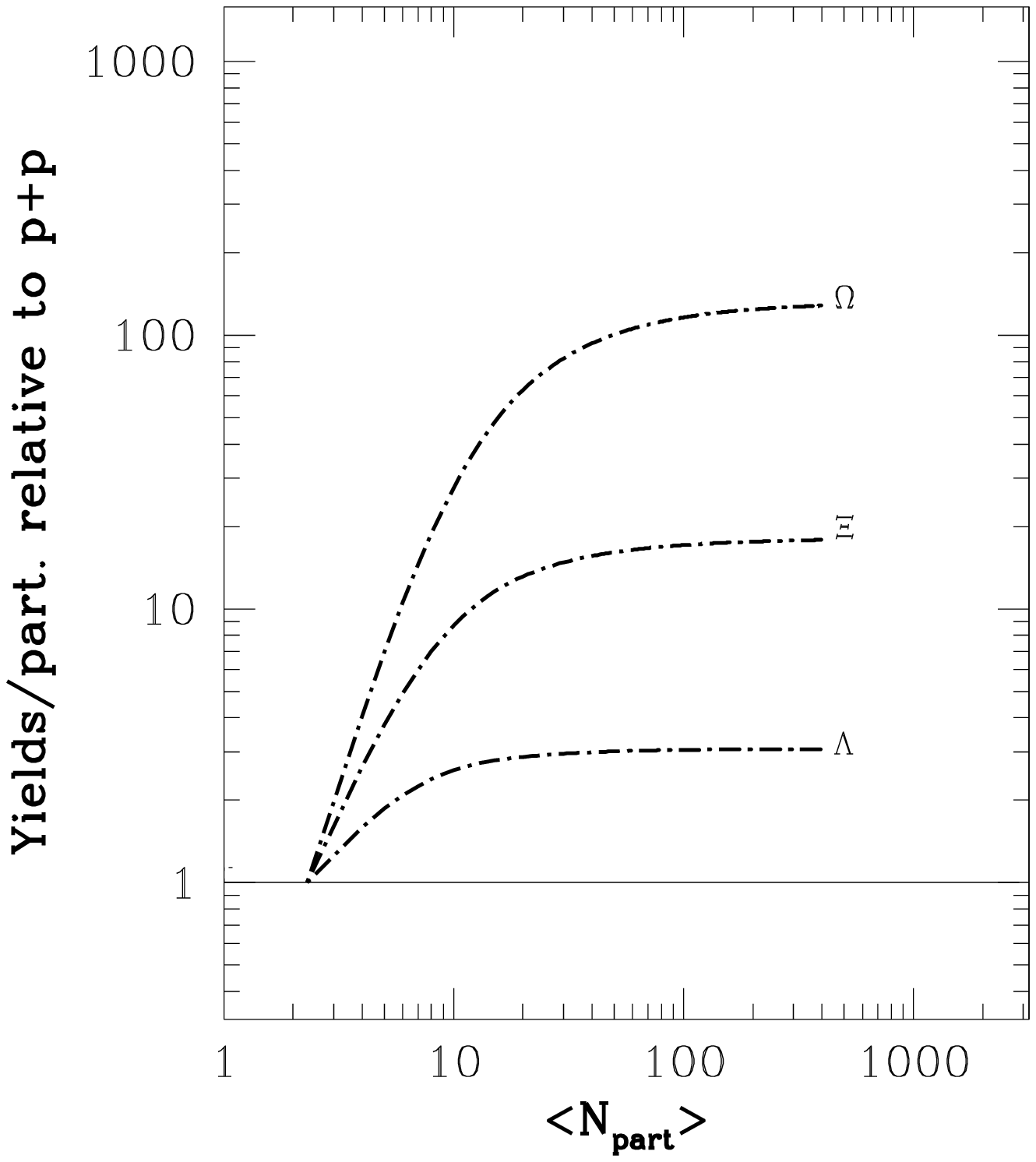}
\end{minipage}
\hskip 0.4cm
\begin{minipage}[t]{65mm}
{%\vskip 0.5cm
\hskip 0.6cm  \includegraphics[width=12.pc, height=18.3pc]{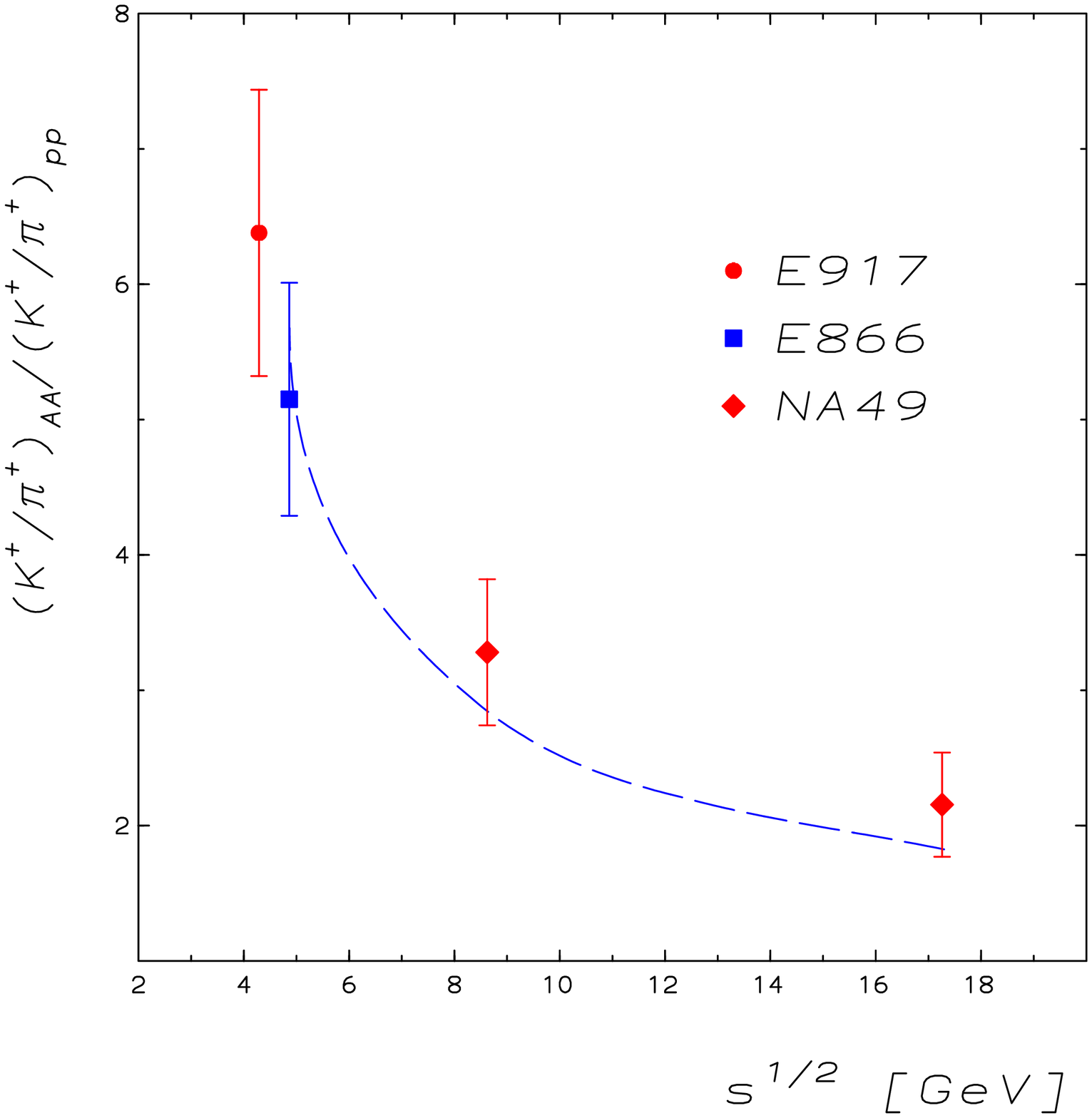}}\\
\end{minipage}
\vskip -0.8cm \caption{ The left hand figure: the statistical
model results on the centrality dependence of the relative
enhancement of $\Lambda$, $\Xi$ and $\Omega$ yields/participant in
central Pb--Pb to p--p collisions at $\sqrt{s}=8.73$ GeV. The
right hand figure: $K^+/\pi^+$ ratio  in A+A relative to p+p
collisions. For the compilation of data, see
Refs.~(\protect\refcite{ogilvie,r49}). The dashed line represents
the statistical model results.\label{qm10}
 \hfill}
\end{figure}

The essential prediction\cite{r15} of the canonical statistical
model is that the strangeness enhancement from p--p to A--A
collisions should increase with  decreasing collision energy. This
is a direct consequence of e.g. Eq.~(\ref{eq88}) where the
canonical suppression factor is seen to be a   decreasing function
of temperature and  thus also    collision energy.
Fig.~(\ref{qm10} -right)  shows the compilation of  data on the
$K^+/\pi^+$ ratio in A--A relative to p--p collisions from
Ref.~(\refcite{ogilvie,r49}). This double ratio could be referred
to as a strangeness enhancement factor. The enhancement is seen in
 data to be the largest at the smallest beam energy and is
decreasing towards higher energy. The line is a smooth
interpolation between the canonical  model results for $\sqrt s
=17.3,12.3,8.73, 5.56$ GeV, calculated  with the thermal
parameters $T$ and $\mu_B$ that were extracted from
Fig.~(\ref{ff20}). The canonical volume parameter  was taken the
same as used in Fig.~({\ref{ff12}}- right). The enhancement seen
in Fig.~( \ref{qm10} -right) is due to the suppression of the
$K^+/\pi^+$ ratio in p--p collisions with decreasing energy and
not due to a dilution of this ratio by excess pions in the A--A
system. The $K^+/\pi^+$ ratio is known experimentally not to vary
within $30\%$  in the energy range from $\sqrt s\sim 5$ GeV at AGS
up to $\sqrt s=130$ GeV at RHIC\cite{r85,r49,ogilvie}.
 Recent data of NA49\cite{r85} and CERES\cite{r16,at} Collaborations  on
 $\Lambda$
yields exhibit a similar energy dependence of the enhancement
factor as is seen in Fig.~(\ref{qm10}- right)  for kaons.

The canonical model (see  Eq.~(\ref{eq88})) also predicts that the
multistrange baryon enhancement from p--p to A--A should be larger
at lower  collision energy.
  Fig.~(\ref{qm10}- left) shows the
canonical model results for the strangeness enhancement and its
centrality dependence for Au--Au collisions at 40 A$\cdot$GeV. The
qualitative behavior of the enhancement is like that  at the SPS.
However, the {\it strength} of the enhancement is seen to be
substantially larger. For $\Omega$ it can be as large as  a factor
of 100. The enhancement of $\Xi$ in A--A relative to p--p can  be
 deduced from  data. Indeed, combining the Si--Pb results
for $\Xi^-$ production obtained\cite{r55} by  E810 Collaboration
and the Si--Au  results for pion or $K^-$ yields
obtained\cite{r56} by E802 Collaboration in  collisions at the top
AGS energy, one can estimate that $\Xi^-/\pi^+\sim 0.0076$. Within
errors this agrees with the value of $\Xi^-/\pi^+\sim 0.0074$
obtained\cite{na49pp} by NA49 in Pb--Pb at $\sqrt s =17.3$ GeV, as
seen in Fig.~(\ref{ff28}). In p--p collisions the $\Xi^-/\pi^+$
ratio is obviously a strongly decreasing function of beam energy.
From the above one would therefore expect that the relative
enhancement of $\Xi^-$ from p--p to A--A collisions should be
larger at AGS than at SPS energy. This is seen in Fig.~(\ref{qm10}
-left). In addition ratios containing only newly produced
particles, such as e.g. $\Xi^-/K^-$, are also larger at AGS than
at SPS energy, see Fig.~(\ref{ff28}). The above heuristic argument
should be, however, verified by more complete experimental data.

 Recently, the production of $\Xi^-$ was  study\cite{koxi} in the
relativistic transport model in the energy range from 2-11
A$\cdot$GeV . The authors discussed the equilibration and the
influence of a phase transition on $\Xi$ yields in A--A
collisions.  There, it was argued that, for beam energies above 4
A$\cdot $GeV, there  should be  a sharp increase of the $\Xi^-$
yield if there is a deconfinement transition in the collisions
fireball. This is an interesting conjecture. However,  with the
presently available experimental data this prediction cannot be
tested. The canonical statistical model, predicts continuous
increase of $\Xi$ yield with collisions energy as it is seen in
Fig.~(\ref{ff28}).

 The prediction of the statistical model on
the energy dependence of strange particle yields   is also in
contrast with the UrQMD results\cite{bleicher}. There, the
production of strangeness is very sensitive to the initial
conditions. In UrQMD the early stage multiple scattering may imply
an increase of the colour electric field strength due to an
overlap of produced strings\cite{bleicher}. Consequently,
according to the Schwinger mechanism, this should increase the
production of (multi)strange baryons. Under similar kinematical
conditions as at the SPS, the UrQMD model predicts\cite{bleicher}
at RHIC an increase in relative strength of $\Omega$ yields from
p--p to A--A by a factor of 5. A recent analysis of multistrange
baryon yields in Au--Au collisions at RHIC energy in the context
of Dual Parton Model, leads\cite{r1007} to a smaller  increase of
the enhancement at RHIC energies.

The interpretation that the  strangeness enhancement is explained
by canonical effects  is not necessarily in contradiction with the
heuristic pre\-di\-ct\-ions\cite{r34} that strangeness enhancement
and its pattern are due to a
 quark--gluon plasma formation in  A--A collisions.
This is particularly the case if one connects the asymptotic grand
canonical description of strangeness production in A--A collisions
with the formation of a  QGP in the initial state.

\section{Conclusions and outlook}

This article has focussed almost exclusively on the use of the
statistical approach to understand yields of different particle
species that have been measured in heavy ion collisions. We have
discussed  a statistical description of the conservation laws and
described their kinetic implementations. We have argued, both on
the qualitative and quantitative level, that exact conservation of
quantum numbers is of crucial importance when applying the methods
of statistical physics in the context of heavy ion and particle
collisions. We have presented the systematics of particle
production in heavy ion collisions from SIS up to LHC energy  and
discussed particular properties of strangeness production. One of
the most intriguing results that comes from these investigations
is the observation that particles seem to be produced according to
the principle of the maximal entropy. In a very broad energy
range, from $\sqrt s\sim 2.5 $  up to 200 GeV per nucleon pair,
hadronic yields and their ratios observed in heavy ion collisions
resemble a chemical equilibrium population along a unified
freeze--out curve determined by the conditions  of  fixed
energy/hadron $\simeq 1$GeV or complementary above SIS energy by
fixed total density of baryons. Strangeness production follows
this systematics from low to very high energy. However, there are
some characteristic features of the system at chemical freeze--out
in high energy central A+A collisions regarding strangeness
production  that are not present in  low energy heavy ion and
collisions among elementary particles. In nucleus--nucleus
collisions, strangeness is un-correlated and redistributed in the
macroscopic volume of a collision  fireball and is conserved on
the average. In hadron--hadron collisions the thermal phase space
available for strange particles is strongly suppressed since, with
only few particles produced per event, strangeness is strongly
correlated  in a volume that
 approximately  coincides with the size of a nucleon, i.e. a distance over which color is confined.  Thus, following
 the statistical kinetics, strangeness   has to be
conserved exactly and locally.
% The exact conservation of
%strangeness  requires that strange particles have to appear
%together with strange antiparticles to exactly and locally
%neutralize strangeness.
 The associated production and locality of
strangeness conservation is, in the context of the statistical
model, the  origin of the suppression of the thermal phase space
for produced particles. Within this context the strangeness
suppression observed in collisions among elementary particles
finds its natural explanation. We also note that the suppression
increases with the strangeness content of the particle as well as,
for A--A collisions, with decreasing collision energy. A further
consequence of the transition from the canonical to the
grand-canonical regime is that strangeness production should be
enhanced in  A--A collisions compared to p--p collisions.
% and
%this enhancement is  larger at a low, rather  than at high
%center-off-mass energy.
%The asymptotic saturation of strangeness
%seen in high energy central nucleus--nucleus collisions could be
%possibly related with the appearance of the quark gluon plasma in
%the initial state.

At  SPS and RHIC energies  the freeze--out points approach the
calculated QCD phase boundary. This fact lends
 strong support to the interpretation
 that the matter
produced in nuclear collisions at SPS and RHIC energies was first
thermalized in the deconfined  quark-gluon plasma  phase and
subsequently expanded through the phase boundary into a thermal
gas of mostly elastic and quasi--elastic  interacting hadrons. The
above  connection between the QCD phase boundary and the observed
chemical freeze--out points is sometimes called into question\cite{volker}
 since hadron production in $e^+e^-$
 and p--p or p--$\bar{\rm{p}}$ can also be described in thermal
 models
 yielding a (apparently universal) temperature
T$_e \approx 170$ MeV. While the fact that T$_e$ is close to T
values determined for heavy ion collisions at top SPS and RHIC
energies might indeed reflect the fundamental hadronization scale
of QCD, we have already noted above  that there is an essential
difference between thermal descriptions of central heavy ion
collisions and elementary particle reactions.  Strange particle
densities and their ratios can be, for heavy ion collisions at
full AGS energy and higher, well described in the grand-canonical
ensemble.  In contrast, for a description of particle production
in elementary
 collisions, local quantum number conservation, that is
 the canonical description is   need. Consequently,
 in central nucleus-nucleus collisions at
ultra-relativistic energies, strangeness percolates freely over
volumes of thousands of fm$^3$! whereas in elementary processes is
approximately restricted to the size of nucleon.   At top SPS and
RHIC energies it is natural to conclude that in nucleus--nucleus
collisions  the percolation has its origin in the quark-gluon
phase, lending further strong support to the interpretation that
the ``coincidence'' between experimentally determined chemical
freeze--out points and the calculated phase boundary implies that
a deconfined phase was produced in such collisions.

In the context of statistical physics the fact that the measured
particle yields coincides with a  thermal multiplicities
calculated with a given statistical operator is a necessary and
sufficient conditions to maintained thermalization  of the
collisions fireball. In the sense of Gibb$^´$s interpretation of
thermodynamics this implies that the parameters $T$ and $\mu_B$
reflect the thermal properties of the fireball.
%Obviously, one would like to have a further indications that
%extracted parameters $T$ and $\mu_B$ are indeed a thermal
%quantities and not only a Lagrange multipliers that guarantee the
%energy and charge quantum number conservations.\footnote{For a
%recent discussion of this point see e.g. Ref.~(\refcite{volker}).}
Furthermore in heavy ion collisions there are experimental
observables indicating the appearance of thermodynamical pressure
and correlations that are expected in a thermalized medium. The
build--up  of  pressure and collectivity is seen in heavy ion
collisions on the level of particle transverse momentum
distributions and elliptic flow parameters. An increase of average
particle transverse momentum with particle mass, seen from SIS up
to RHIC energy, is a typical property of a transversally expanding
thermal medium. The appearance of strong elliptic flow is an
indication of thermodynamical pressure that develops in the early
stage in the collision. Finally, the measured particle number
fluctuations are consistent with thermal expectations. Taken
together, these observations lend  strong support to the
thermodynamic interpretation of $T$ and $\mu_B$ with the
concomitant QGP interpretation.

% An interesting question is what leads
%to the percolation at SIS energies, where the freeze-out points
%are far away from where we think the phase boundary is. We
%speculate that, in this case, the percolation takes place in the
%high density hadronic phase indicated by the hatched area in Fig.
%~\ref{fig:phase}.

\section*{Acknowledgments}
\addcontentsline{toc}{section}{Acknowledgements}

K.R acknowledges stimulating discussions with  R. Baier, S. Bass,
M. Bleicher, J. Cleymans, B. Friman,  F. Karsch, V. Koch, W.
N\"orenberg, H. Oeschler, H. Satz, R. Stock,
 A. Tounsi, Nu Xu  and the support of the Alexander von Humboldt
Foundation.


\begin{thebibliography}{000}

\bibitem{r1} For a  review see eg. H. Satz,  Rep. Prog. Phys. {
63} (2000) 1511; S.A. Bass, M. Gyulassy, H. St\"ocker, and W.
Greiner, J. Phys. G25 R1 (1999). E.V. Shuryak, Phys. Rep. 115
(1984) 151; M.G. Alford, hep-ph/0209287, H. Satz, hep-ph/0209181.


\bibitem{r1j}
J. Cleymans, R.V. Gavai and E. Suhonen, Phys. Rep. 130 (1986) 217.
\bibitem{r1a} B. Alessandro, et al.,
CERN-ALICE-INTERNAL-NOTE-2002-025.

\bibitem{gery} G.E. Brown and M. Rho,
Phys. Rep. 363 (2002) 85.

\bibitem{gery1} W. Cassing and H. Bratkovskaya,
Phys. Rep. 308 (1999) 65.


\bibitem{r3} P. Braun-Munzinger, and J. Stachel, Nucl. Phys.
A606 (1996) 320; Nucl. Phys. A683 (1998) 3;  J. Stachel, Nucl.
Phys. A654 (1999) 119c.

\bibitem{r4}  R. Stock, { Phys. Lett.} B{456} (1999) 277.

\bibitem{r5}
R. Baier, A.H. Mueller, D. Schiff, and D.T. Son, Phys. Lett. B {
502} (2001) 51; Nucl. Phys. A698 (2002) 217.

\bibitem{uh}
 U. Heinz, Nucl. Phys.  A685 (2001) 414 and
 A661 (1999) 349.
\bibitem{r28} K. J. Eskola, K. Kajantie, P. Ruuskanen and K.
Tuominen, Nucl. Phys. B570 (2000) 379.


%%%%%%%%%%%%%%%%%%%%%%%%%%%%%



\bibitem{alam} J. Alam, et al., hep-th/010802; Ann. Phys. 286 (2001)
159.

\bibitem{dinesh} D. K. Srivastava, Eur. Phys. J. 10 (1999) 487.



\bibitem{rapp} R. Rapp, Phys. Rev.
 C63 (2001) 054907.





\bibitem{vesa}
V. Ruuskanen, Acta. Phys. Polon. B18 (1987) 551; R. Rapp, and E.
Shuryak, Phys. Lett. B{473}{2000}{13}; I. Krasnikova, Ch. Gale,
and D.K. Srivastava, Phys. Rev. C65 (2002) 064903, hep-ph/0112139;
Ch. Gale, Nucl.Phys. A698 (2002) 143c, hep-ph/0104235.

\bibitem{vesa1}
R. Rapp and J. Wambach, Phys. Rev. C55 (1998) 916; Nucl. Phys.
A638 (1998) 171c; Eur. Phys. J. A6 (1999) 415.

\bibitem{vesa2}
K. Gallmeister, B. K\"ampfer, and O.P. Pavlenko, Phys. Lett. B473
(2000) 20.

\bibitem{red} F. Karsch, K. Redlich, and L. Turko,
Z. Phys. C60  (1993) 519.

\bibitem{rednn} K. Kajanie, J. Kapusta, L.D. McLerran, and A. Mekjian,
Phys. Rev.  D34 (1986) 2746 .



\bibitem{hatsuda} T. Hatsuda, Nucl. Phys. A698 (2002) 243; K.
Yokokawa, et al., Phys. Rev. C66 (2002) 022201; D. Jido, et al.,
Phys. Rev. D63 (2001) 011901.





\bibitem{r2}
D. Teaney, J. Lauret, and E. Shuryak, Phys. Rev. Lett.  86 (2001)
4783.



\bibitem{wideman} U.A. Wiedemann and U. Heinz, Phys. Rep. 319
(1999) 145, and references therein.

\bibitem{tomasik} B. Tomasik, U.A. Wiedemann and U. Heinz,
Nucl. Phys. A663 (2000) 753; R. Nix, Phys. Rev. C58 (1998) 2303.
\bibitem{xus}
Nu Xu, nucl-ex/0211012.
\bibitem{medium} V. Metag, Nucl. Phys. A690 (2001) 140.

\bibitem{medium1} W. Weinhold, B.L. Friman and W. N\"orenberg,
Acta. Phys. Polon. B27 (1996) 3249; Phys. Lett. B433 (1998) 236;
M. Hermann, B.L. Friman and W. Weinhold, Nucl. Phys. A545 (1992)
2676.

\bibitem{medium2} G. Boyd, et al., Nucl. Phys. Proc. Suppl. 42
(1995) 469; Phys. Lett. B349 (1995) 170; E. Laermann, et al.,
Nucl. Phys. Proc. Suppl. 34 (1994) 292.

\bibitem{medium2} M.F.M. Lutz, G. Wolf and B. Friman,
 Nucl. Phys. A706 (2002) 431; Nucl. Phys. A661
(1999) 526.

\bibitem{satz} T. Matsui and H. Satz, Phys. Lett. {\rm B178} (1986)
416.



\bibitem{knol}
H. W. Barz, H. Schulz, B.L. Friman and J. Knoll, Phys. Lett. B254
(1991) 315, Nucl. Phys. A545 (1992) 259.
\bibitem{biro}
T.S.  Biro, hep-ph/0005067, hep-ph/0205049; T.S. Biro, Phys. Lett. B487
(2000) 133.
\bibitem{r34} P.
Koch, B. M\"uller, and J.~Rafelski, { Phys. Rep.} { 142} (1986)
167; J. Rafelski, J. Letessier, and A. Tounsi, { Acta Phys.
Polon.} B27 (1996) 1037; J. Rafelski Phys. Lett. B262 (1991) 333.
\bibitem{r36} T. Matsui, B. Svetitsky and L.D.
McLerran, Phys. Rev.  D34  (1986) 783.


\bibitem{r35}
R. Rapp and E. Shuryak, Phys. Rev. Lett. 86 (2001) 2980; C.
Greiner and S. Leupold, J. Phys. G27 (2001) L95.

\bibitem{r9} J.  Cleymans, and K. Redlich,
Phys. Rev. Lett. { 81} (1998)  5284.

\bibitem{r10} P. Braun-Munzinger, D. Magestro, K. Redlich, and
J. Stachel, { Phys. Lett.} B{518} (2001) 41  { and references
therein}.
\bibitem{r10n}  D. Magestro,
 J. Phys. G28 (2002) 1745.

\bibitem{r11}  P. Braun-Munzinger, J.
Stachel, J.P. Wessels and N. Xu, Phys. Lett. B344 (1995)  43 and
 B365 (1996) 1.




\bibitem{r12}P.  Braun-Munzinger, I. Heppe and J.
Stachel, Phys. Lett. { B465} (1999) 15.

\bibitem{r13} S.V. Akkelin, P. Braun-Munzinger and Yu. M.
Sinyukov, Nucl. Phys. A710 (2002) 439.

\bibitem{r14} F. Becattini, J. Cleymans, A. Keranen, E.
Suhonen and K. Redlich,  Phys. Rev. C64 (2001) 024901.

\bibitem{anti}  A. Keranen,  and F. Becattini, J. Phys. G28 (2002)
2041; Phys. Rev. C65 (2002) 044091.

\bibitem{rn13} J. Rafelski and J. Letessier, nucl-th/0209080.

\bibitem{r50} M. Gazdzicki,  et al., Z. Phys. C65 (1995) 215,
Acta Phys. Polon.  B30 (1999) 2705.

\bibitem{r15} A. Tounsi and K. Redlich,  Eur. Phys. J. C24
(2002) 529;  J. Phys. G28 (2002) 2095.

\bibitem{r17} F. Becattini, hep-ph/0202071.

\bibitem{r18} J. Sollfrank, et al., Phys. Rev. C55 (1997) 392.
\bibitem{r19} P. Huovinen, et al., Phys. Lett. B503 (2001) 58.


\bibitem{r21}
J. S. Hamieh, K. Redlich and A. Tounsi, Phys. Lett. { B486} (2000)
61; J. Phys.  G27  (2001) 413.

\bibitem{r31} J.  Cleymans and K. Redlich,
Phys. Rev.  C60 (1999) 054908; J. Cleymans, H. Oeschler and K.
Redlich, Phys. Rev.  C59 (1999) 1663; Phys. Lett. { B485} (2001)
27.

\bibitem{r31n} J. Cleymans, D. Elliott, A. Keranen and E. Suhonen,

 Phys. Rev.  C57 (1998) 3319.

\bibitem{r32} D. Zschiesche et al., Nucl. Phys. A681 (2001)
34.
\bibitem{r33}
 J. Letessier, and J. Rafelski,
{ Int. J.  Mod. Phys.} E9 (2000) 107.


\bibitem{r42} P. Braun-Munzinger, J. Cleymans, H. Oeschler and K.
Redlich,  Nucl. Phys. A697 (2002) 902.
\bibitem{r53} N. Xu, M. Kaneta, Nucl. Phys. A698 (2002)306,  nucl-ex/0104021;
M. Kaneta and N. Xu, J. Phys. G27 (2001) 589; Nucl. Phys. A698
(2002) 306.


\bibitem{r54} W. Broniowski and W. Florkowski, Phys. Rev.
Lett. 87 (2001) 272302.

\bibitem{becalast}
F. Becattini and  G. Passaleva, Eur. Phys. J. C23 (2002) 551, hep-ph/0110312.

\bibitem{r59} K. Redlich, Nucl. Phys. A698 (2002) 94.

\bibitem{r60} J. Cleymans and H. Satz, Z. Phys. C57 (1993) 135.


\bibitem{magestro3} D. Magestro, P. Braun-Munzinger, and J. Stachel,
 {\it in preparation}.


\bibitem{diplomarbeit} I. Heppe, Diploma thesis, Heidelberg(1998).

\bibitem{exclv2} G. Yen,
M.I. Gorenstein, W. Greiner, S.N. Yang, Phys. Rev. C  56, (1997)
2210.

\bibitem{bohrmott} see, e.g., A. Bohr and B. Mottelson, Nucl. Structure
(Benjamin, New York 1969), Vol. 1, p. 266.

\bibitem{brahms} D. Ouerdane et al., BRAHMS Coll., QM2002
contribution, Nucl. Phys. A (in print), nucl-ex/0212001.
\bibitem{ullrich} T. Ullrich, QM2002 contribution, nucl-ex/0211004,
Nucl. Phys. A (in print).

\bibitem{cleymans} J. Cleymans in 3rd International Conference on
Physics and Astrophysics of Quark Gluon Plasma (ICPAQGP 97),
Jaipur, India, 17-21 Mar 1997.

\bibitem{appels1} H. Appelsh\"auser et al., NA49 Coll., Phys.Rev.Lett.  82
(1999) 2471.

\bibitem{biro1} T.S. Biro, P. Levai, J. Zimanyi,  J.Phys. G28
(2002) 1561, hep-ph/0112137; J. Zimanyi, et al., in S.A. Bass, et
al.,  Nucl. Phys. A661 (1999) 205c.

\bibitem{rafelski}J. Rafelski, J. Letessier, Qm2002 contribution,
nucl-th/0209084, Nucl. Phys. A(in print); J. Rafelski, et al., in
S.A. Bass, et al.,  Nucl. Phys. A661 (1999) 205c.

\bibitem{fachini} P. Fachini et al.,  STAR Coll., QM2002
contribution,  nucl-ex/0211001, Nucl. Phys. A (in print).

\bibitem{ceres1} D. Adamova et al., CERES coll., Nucl.Phys.  A714
 (2003) 124, nucl-ex/0207005.

\bibitem{ceres2} D. Adamova et al., CERES coll., Phys.Rev.Lett.
  90 (2003) 022301, nucl-ex/0207008.

\bibitem{leuwen} M. van Leeuwen et al., NA49 coll., QM2002
contribution, Nucl. Phys. A  (in print),  nucl-ex/0208014.


\bibitem{agsplb} J. Barrette et al., E814 Coll., Phys. Lett. B333
(1994) 33.
\bibitem{nu} N. Xu et al., E814 Coll., Nucl. Phys. A566 (1994) 585c.


\bibitem{r611} P. Braun-Munzinger and J. Stachel, Phys. Lett.
B490 (2000)196 ; Nucl. Phys. A690 (2001) 119c.
\bibitem{r612} P. Braun-Munzinger and J. Stachel,  Nucl. Phys. A690 (2001) 119c.


\bibitem{r23} F. Karsch, E. Laermann, and A. Peikert,  Nucl. Phys. B605
(2001)579.

\bibitem{r26}
J.D. Bjorken, Phys. Rev. D27 (1983) 140.
\bibitem{fkk} F. Karsch, K. Redlich, and A. Tawfik, hep-ph/0303108.

\bibitem{fkk1} P. Braun-Munzinger, J. Stachel, and C. Wetterich, {\it in preparation}.
\bibitem{r24} J.P. Blaizot, E. Iancu, and A. Rebhan,
Phys. Rev. D63 (2001) 065003.

\bibitem{r79} E. Suhonen and S. Sohlo, J. Phys. G13 (1987) 1487;
 D. H. Rischke, M. I. Gorenstein, H. St\"ocker and W. Greiner, Z.
Phys. C51 (1991) 485.
\bibitem{r66} A. Peshier, B. K\"ampfer, and G. Soff, Phys. Rev. C61
(2000) 045203; Phys. Rev. D66 (2002) 094003;
 B. K\"ampfer, and O.P. Pavlenko, Phys. Lett. B391
(1997) 185.

\bibitem{r25} R. Pisarski, Phys. Rev. D62 (2000) 111501.

\bibitem{r65} A. Dumitru and R.D. Pisarski, Phys. Lett. B525 (2002)
95; A. Dumitru and R.D. Pisarski, Phys. Rev. D66 (2002) 096003, hep-ph/0204223.



\bibitem{400}
M. Agarwal, et al., WA98 Coll., Eur. Phys. J. C18 (2001) 651.

\bibitem{r27} L. V. Gribov, et al., Phys. Rep. 100 (1983) 1;
A.H. Mueller and J. Qiu, Nucl. Phys. B268 (1986) 427.


\bibitem{r29} L.D. McLerran, and R. Venugopalan,
Phys. Rev.  D49 (1994) 2233; Phys. Rev. D49 (1994) 3352; Phys.
Rev. D50 (1994) 2225.


\bibitem{r30} D. Kharzeev and M. Nardi, Phys. Lett. B507
(2001) 121.

\bibitem{hot} E.V.  Shuryak, and L. Xiong, Phys. Rev.  Lett. 70
(1993) 2241; J. Kapusta, L. McLerran, and D.K. Srivastava, Phys.
Lett. B283 (1992) 145.
\bibitem{particle}
Particle Properties


\bibitem{lhci}
D.~K. Srivastava, M.~G. Mustafa, and B. M\"uller, Phys. Rev. C56
(1977) 1064; R.J. Fries, B. M\"uller and D.~K. Srivastava,
nucl-th/0208001.



\bibitem{pc} K. Geiger and D. K. Srivastava, Nucl. Phys A661
(1999) 592, nucl-th/9808042.

\bibitem{huang}
K. Huang, Statistical Mechanics, 2nd Edition, Wiley, New York,
1987, sect. 3.4.

\bibitem{col1} T. A. Armstrong, et al.,  E864 Coll., Phys. Rev. C61
(2000) 064908.

\bibitem{col2} T. A. Armstrong, et al., E864 Coll., Phys. Rev. Lett. 85
(2000) 2685.

\bibitem{col3} P. Braun-Munzinger, and J. Stachel, J. Phys. G:
Nucl. Part. Phys. 21 (1995) L17.

\bibitem{col4} A.J. Baltz et al., Phys. Let. B325 (1994) 7.


\bibitem{r20}  C.M. Ko, V. Koch, Z. Lin, K.
Redlich, M. Stephanov and X.N. Wang, Phys. Rev. Lett. 86 (2001)
5438.
\bibitem{r20a} S. Jeon,  V. Koch,  K. Redlich and X.N. Wang, Nuc. Phys.
A697 (2002) 546; K. Redlich, V. Koch and A. Tounsi,  Nucl. Phys.
A702 (2002) 326;  K. Redlich, J. Cleymans, H. Oeschler, and A.
Tounsi, Acta Phys. Polon. B33 (2002) 1609.






\bibitem{r22} Z. Lin, and C.M. Ko, Phys. Rev. C64 (2001)
041901.

\bibitem{r6}
H.-Th. Elze, M. Gyulassy, and D. Vasak, Nucl. Phys. B{276} (1986)
706.

\bibitem{r7}
J.P. Blaizot, and  E. Iancu, Nucl. Phys. B{  570} (2000) 326.

\bibitem{r8}
Q. Wang, K. Redlich, H. St\"ocker, and W. Greiner,  Phys. Rev.
Lett. 88 (2002) 132303; Nucl. Phys. A714 (2003) 293;  J. Phys. G28
(2002) 2115.

\bibitem{r37}
R. Hagedorn, Thermodynamics of strong interactions, CERN Report
71-12 (1971); E.V. Shuryak, Phys. Lett. { B42} (1972) 357;
Rafelski and M. Danos, Phys. Lett. B97 (1980) 279; R. Hagedorn and
K. Redlich, Z. Phys. { C27} (1985)  541.

\bibitem{r37a}
R. Hagedorn and K. Redlich, Z. Phys. { C27} (1985)  541.

\bibitem{r38} K. Redlich and L. Turko,
Z. Phys.  { B97} (1980) 279; L. Turko, Phys. Lett. B104 (1981)
153; L. Turko and J. Rafelski, Eur. Phys. J. C18 (2001) 587.
\bibitem{r38f} K. Redlich, F. Karsch, and A. Tounsi,
hep-ph/0302245.


\bibitem{r39} J. Cleymans and P. Koch, Z. Phys. C{52} (1991)
137.




\bibitem{r64}
D. Miller and K. Redlich, Phys. Rev. D37 (1988) 3716; Phys. Rev.
D35 (1987) 2524; H. Th. Elze, D. Miller and K. Redlich, Phys. Rev.
 D35 (1987) 748.

\bibitem{r64e}
H. Th. Elze, W. Greiner and J. Rafleski, Phys. Lett. B124 (1983)
515; Z. Phys. C24 (1984) 361; Ch. Derreth, W. Greiner, H. Th. Elze
and J. Rafelski, Phys. Rev. C31 (1985) 1360; H. Th. Elze and W.
Greiner, Phys. Rev. A33 (1986) 1879.
\bibitem{r64n}
D.H. Rischke, M.I. Gorenstein, A. Schafer, H. St\"ocker and W.
Greiner, Phys. Lett. B278 (1992) 19; D.H. Rischke, J. Schaffner,
M.I. Gorenstein, A. Schaefer, H. St\"ocker and W. Greiner, Z.
Phys. C56 (1992) 325.
\bibitem{r64p}
A. Gocksch, R.D. Pisarski, Nucl. Phys. B402 (1993) 657.

\bibitem{r67q}
J. Engels, et al., Z. Phys. C42 (1989) 341; A. Peshier, et al.,
Phys. Rev. C61 (2000) 045203 and references therein.





\bibitem{r67} R. Rapp and J. Wambach, Adv. Nucl. Phys. 25 (2000) 1.



\bibitem{r68} J. Cleymans, K. Redlich and E. Suhonen, Z. Phys. C58
 (1993) 347.

%\bibitem{r69} J. Rafelski and J. Letessier, J. Phys. G28 (2002)
%1819.


\bibitem{r58} K. Zalewski, Acta. Phys. Polon. B62 (1965) 933.



\bibitem{r57} A. Wagner et al., KaoS Coll., Phys. Lett. B420 (1998)
20, C. M\"untz et al., KaoS Coll., Z. Phys.  C357 39.




\bibitem{Barth} R. Barth et al., KaoS  Coll.,
Phys. Rev. Lett. 78 (1997) 4007; F. Laue %, C. Sturm
et al., KaoS  Coll., Phys. Rev. Lett. 82 (1999) 1640.


\bibitem{Marc} M. Menzel, Ph.D.Thesis, Universit\"at Marburg, 2000.
\bibitem{Ahle}
L.~Ahle et al., E-802 Coll., Phys. Rev. C58 (1998) 3523;





\bibitem{tabs} R. Averbeck, R. Holzmann, V. Metag and R.S. Simon,
nucl-ex/0012007; A.R. Wolf et al.,Phys. Rev. Lett. 80 (1998) 5281;
By TAPS Coll., Z. Phys. A359 (1997) 65; Phys. Rev. C56 (1997)
2920.


\bibitem{HON98} B. Hong et al., FOPI Coll.,
Phys. Rev. C57 1998 244.

\bibitem{MUE97} C. M\"untz et al., KaoS Coll.,
Z. Phys. C357 1997 399; J. Phys. G28 (2002) 1895; Euro. Phys. J.
A9 (2000) 397.

\bibitem{urqmd}
M.~Bleicher { et al.}, J.\ Phys.\  { G25}  (1999) 1859; S.~A.~Bass
{ et al.}, Prog.\ Part.\ Nucl.\ Phys.\  { 41} (1998) 225.
\bibitem{nexusmodel}
H.J. Drescher { et al.}, Phys. Rep. { 350} (2001) 93.
\bibitem{pythiamodel}
H.-U. Bengtsson and T. Sj\"ostrand, { Comput. Phys. Commun.} { 46}
 (1987) 43.

\bibitem{r48}
C. Greiner, hep-th/0209021; J. Phys. G28 (2002) 1631.

\bibitem{hart01}
C.~Hartnack, H.~Oeschler, J.~Aichelin, Phys. Rev. Lett. (in print),
nucl-th/0109016.
\bibitem{r40} S. Pal, C.M. Ko, and Z. Lin,  Phys. Rev. C64 (2001)
042201.
\bibitem{g11} G.E.  Brown, M. Rho, and Ch. Song, Nucl. Phys. A698
(2002) 483; Nucl. Phys. A690 (2001) 184.

\bibitem{pcas} W. Cassing and M. Lutz, privat communication.


\bibitem{r41}
F. Becattini,  Z. Phys. C69 (1996) 485; F. Becattini and U. Heinz,
Z. Phys. C76 (1997)  269.

\bibitem{gazdzickil} M. Gazdzicki, and M. Gorenstein, Phys. Lett.
B483 (2000) 60.



\bibitem{colflux}
A.~Casher, H.~Neuberger and S.~Nussinov,
%``Chromoelectric Flux Tube Model Of Particle Production,''
Phys. Rev.   D20 (1979) 179.
%%CITATION = PHRVA,D20,179;%%

\bibitem{markus} M. Bleicher, et al., Phys. Rev. Lett. 88 (2002)
202501; S.A. Bass, et al., nucl-th/0204049



\bibitem{schwinger}
%\cite{Schwinger:1951nm}

%\bibitem{Schwinger:1951nm}
J.~S.~Schwinger, Phys. Rev.  { 82} (1951) 664.


 \bibitem{na49pp}
J. B\"achler et al., NA49 Coll.,  Nucl. Phys. { A661} (1999) 45;
R.A. Barton et al., NA49 Coll., J. Phys. { G27} (2001) 367; V.
Afanasev et al., NA49 Coll., Phys. Lett. { B491} (2000) 59.


\bibitem{r1007} A. Capella, C. A. Salgado, and D. Sousa,
nucl-th/0205014.

\bibitem{capellan} A. Capella and C. A. Salgado, Phys. Rev. C60
(1999) 054906.
\bibitem{crn} J. Cleymans  and K. Redlich, {\it in preparation}.
%%%%%%%%%%%%%%%%%%%
\bibitem{krpbm} P. Braun-Munzinger, and K. Redlich, Eur. Phys. J.
 C16 (2000) 519; Nucl. Phys. A661 (1999) 546.

%\bibitem{pbm1} P. Braun-Munzinger and J. Stachel, Phys. Lett. {\bf B490} (2000) 196. %nucl-th/0007059
%\bibitem{pbm2} P. Braun-Munzinger and J. Stachel, Nucl. Phys. {\bf A690} (2001) 119c.
\bibitem{gor} M. Gorenstein et al., Phys. Lett. {\rm B509} (2001)
277; K.A. Bugaev, M. Gazdzicki and M. Gorenstein, Phys. Lett. B544
(2002) 127; Phys. Rev. Lett. 88 (2000) 132301; Phys. Lett. B544
(2002) 127.
\bibitem{gra} L. Grandchamp and R. Rapp, Phys. Lett.
{\rm B523} (2001) 60; Nucl.Phys. A709 (2002) 415, hep-ph/0205305.
\bibitem{goro}
 M. Gorenstein, and M. Gazdzicki,  Phys. Rev. Lett. 83 (1999)
 4003.
\bibitem{the} R.L. Thews, M. Schroedter and J. Rafelski, Phys. Rev.  C63 (2001) 054905.


\bibitem{vogt99} R. Vogt (p. 250c) in S.A. Bass et al.,
Nucl. Phys. {\rm A661} (1999) 205c.
\bibitem{andronic} A. Andronic, P. Braun-Munzinger, K. Redlich,
J. Stachel, QM2002 contribution, Nucl. Phys. A (in print), nucl-th/0209035.
%\bibitem{pbm3} P. Braun-Munzinger and J. Stachel, J. Phys. {\rm G28}
%(2002) 1971.

\bibitem{andronic1} A. Andronic, P. Braun-Munzinger, K. Redlich,
J. Stachel, nucl-th/0303036.
\bibitem{bla} J.-P. Blaizot, P.M. Dinh, J.-Y. Ollitrault,
Phys. Rev. Lett. {85} {2000} {4012} [nucl-th/0007020].

\bibitem{shuryak}H. Sorge, E. Shuryak, Phys. Rev. Lett.  79 (1997) 2775.
\bibitem{na50_3} M.C. Abreu et al., NA50 Coll., Phys. Lett.
 B438 (1998) 35, and references therein.
\bibitem{na504} M.C. Abreu et al., NA50 Coll., Nucl. Phys. A698
(2002) 539c.
 B438 (1998) 35, and references therein.

\bibitem{gerschel} C. Gerschel, Acta Phys. Pol.  B30 (1999) 3585.

\bibitem{na38_1} M.C. Abreu et al., NA38 Coll., Phys. Lett.
 B449 (1999) 128.

\bibitem{na50_4} M. Gonin et al., NA50 Coll., Proc. 3rd
Conf. on Physics and Astrophysics of Quark-Gluon Plasma, Jaipur,
India, March 1999, B. C. Sinha, D.K Srivastava, Y.P. Viyogi,
editors, Narosa Publ. House 1998, p. 393.


\bibitem{na50_5} L. Ramello et al., NA50 Coll.,
Nucl. Phys.  A638 (1998) 261c.

\bibitem{na38_2}  M.C. Abreu et al., NA38 Coll., Phys. Lett.
B466 (1999) 408.

%\bibitem{na49} NA49 collaboration, Phys. Rev. C66  (2002) 054902,
%nucl-ex/0205002.
\bibitem{na50a} NA50 Coll., Phys. Lett.  B530 (2002) 33.
%\bibitem{pho} PHOBOS collaboration, Phys. Rev. Lett. {\rm 88} (2002) 022302.
\bibitem{vogt} R. Vogt, hep-ph/0203151; hep-ph/0111271.
%\bibitem{pbm4} P. Braun-Munzinger et al., Nucl. Phys. {\rm A697} (2002) 902.
\bibitem{na50} NA50 Coll., Phys. Lett.  B450 (1999) 456;
Phys. Lett.  B477 (2000) 28.
\bibitem{gos} J. Gosset et al., Eur. Phys. J.  C13 (2000) 63.
\bibitem{na50b} NA50 Coll., Nucl. Phys.  A698 (2002) 539c.
\bibitem{rappb} R. Rapp and E. Shuryak, Phys. Lett.  B473 (2000) 13.
%\bibitem{gall} K. Gallmeister, B. K\"ampfer and O.P. Pavlenko, Phys. Lett. {\rm B473} (2000) 20.
\bibitem{nagle} J. Nagle, PHENIX Coll., Nucl. Phys. A698 (2002)
599.
\bibitem{aver} R. Averbeck, PHENIX Coll., Nucl. Phys. A698 (2002)
39.

\bibitem{ko} C.M. Ko, X.N. Wang,  B. Zhang and X.F. Zhang,
Phys. Lett. B178 (1998) 237.



%%%%%%%%%%%%%%%%%%%
\bibitem{r16} W. Schmitz et al., CERES Coll., J. Phys. G28 (2002)
1861.
\bibitem{r72} M. Bleicher and J. Aichelin, Phys. Let. B530 (2002)
81.

\bibitem{r72n} see also: L. Bravina, et al., Nucl. Phys. A698 (2002) 383;
Phys. Rev. C66 (2002) 014906

\bibitem{r73} M. Gyulassy, privat communication.

\bibitem{r74s} V.D.~Toneev, J. Cleymans, E.G.~Nikonov, K. Redlich, and
A.A.~Shanenko, J. Phys. G27 (2001) 827; nucl-th/9904048.

\bibitem{r74} E.G.~Nikonov, A.A.~Shanenko and V.D.~Toneev,
Heavy Ion Physics  8 (1998) 89.


\bibitem{r75}  C.M.~Hung and  E.V.~Shuryak,
 Phys. Rev. Lett.  75 (1995) 4003; Phys. Rev. C57 (1998)
      1891.



%\bibitem{r76} By CERES Coll., nucl-ex/0207008.
\bibitem{r80} P. Braun-Munzinger and J. Stachel, J. Phys. G28
(2002) 1971.
\bibitem{r77}  K.A. Olive, Nucl. Phys. B190 [FS3] (1981) 483;
 K.A. Olive, Nucl. Phys. B198 (1982) 461;
 J.I. Kapusta and K.A. Olive, Nucl. Phys. A408 (1983) 478;
 A.L. Fetter and J. D. Walecka, {\em Quantum Theory of
Many-Particle Systems}, McGraw-Hill, 1971.






\bibitem{r81} S. Ejiri, et al., hep-lat/0209012; C. Schmidt, et
al., hep-lat/0209009.

\bibitem{r82} Z. Fodor and S.D. Katz, Phys. Lett. B534 (2002) 87.

\bibitem{r83} S. Fortunato, P. Petreczky, and H. Satz, Phys. Lett.
B502 (2001) 321; hep-lat/0207021.

\bibitem{r84} See e.g. P. Senger, J. Phys. G28 (2002) 1869, and
GSI Report, Nov. 2001.

\bibitem{gr} M. Gazdzicki, and D. Roehrich, Z. Phys. C66
(1995) 77.


\bibitem{rhic} The 4$\pi$ results at RHIC are obtained from
Ref.~(\refcite{brahms1}) and Ref.~(\refcite{phobos}) using the
extrapolation procedure described in Ref.~(\refcite{na49aa}).
% $\sqrt s=56$ GeV were estimated from
%PHOBOS mid-rapidity data from \cite{phobos}, scaled by the same
%factor as required in p--p collisions when going from mid-rapidity
%to full phase space. The results at $\sqrt s=130$ GeV are from
%\cite{brahms1}. The 4$\pi$ results at $\sqrt s=200$ GeV were
%estimated by multiplying the $\sqrt s=130$ GeV results by  the
%same factor (modulo an increase in rapidity interval) as recently
%measured by PHOBOS at mid--rapidity \cite{phobos}.

\bibitem{brahms1}
I.G. Bearden, et al., nucl-ex/0108016.

\bibitem{phobos}
B.B. Back, et al., PHOBOS Coll., Phys. Rev. Lett. 85, (2000) 3100;
Phys. Rev. Lett. 87, (2001) 102303; Phys. Rev. Lett. 88, (2002)
022302.

\bibitem{na49aa} By Na49 Coll., nucl-ex/0205002.

\bibitem{r51} K. Wr\`oblewski, Acta. Phys. Polon. B16 (1985) 379.



\bibitem{r85} By STAR Coll., J. Phys. G28  (2002) 1535; Nucl.
Phys. A698 (2002) 64.

\bibitem{r85c}
V. Afanasev et al., NA49 Coll., nucl-ex/0208014; 0209002.


\bibitem{at} K. Redlich and A. Tounsi, hep-ph/0105201;
hep-ph/0111159;  A. Tounsi, A. Mischke, and K. Redlich, QM2002
contribution, Nucl. Phys. A (in print), hep-ph/0209284.

\bibitem{r49}
Ch. Blume  et al., NA49 Coll., Nucl. Phys. A698 (2002) 104.


\bibitem{ogilvie} J.C. Dunlop,  et al., Phys. Rev. C61
(2000) 031901.



\bibitem{r47}
 W. Cassing, Nucl. Phys. A661 (1999) 468c.


\bibitem{r86} F. Wang, H. Liu, H. Sorge, N. Xu  and J. Yang, Phys.
Rev. C61 (2000) 064904.

\bibitem{r87} H. Weber, E.L. Bratkovskaya and H. St\"ocker,
nucl-th/0205030.



\bibitem{r88} M. Gazdzicki and M.I. Gorenstein, Acta. Phys. Polon.
B30 (2000) 965; M. Gazdzicki and M. Gorenstein, Phys. Lett. B483
(2000) 60.


\bibitem{r55} S. E. Eiseman et al., E810 Coll., Phys. Lett. B297
(1992) 44 and  B325 (1994) 322.

\bibitem{r56} T. Abbott et al., E802 Coll., Phys. Rev. C60 (1999)
044904; Y. Akiba et al., E802 Coll., Nucl. Phys. A590 (1995) 179c.

\bibitem{r89} H. Oeschler, et al.,  {\it in preparation}.


\bibitem{r90} F. Laue, C. Sturm et al.,
Phys. Rev. Lett.  82 (1999) 1640.

\bibitem{r96} B. Holzman, E917 Coll., Nucl. Phys. A698 (2002) 643.



\bibitem{r100} By NA49 Coll., Phys. Lett. B491 (2000) 59;
H. Bia\l kowska., and W. Retyk, J. Phys. G27 (2001) 397.

\bibitem{r97} By STAR Coll., Phys. Rev. C65 (2002) 041901;
nucl-ex/0206008.


\bibitem{r99} R. Kotte, FOPI Coll.,  Proceedings of Hirschegg
XXVIII International Conference, Hadrons in Dense Matter (2000).

\bibitem{r94} M.C. Abbreu et al., NA50 Coll., Nucl. Phys. A661 (1999) 534c.


\bibitem{r95} S. Pal, C.M. Ko, and Z. Lin, Nucl. Phys. A707 (2002)
525.

\bibitem{her} N. Herrmann, FOPI Coll., Nucl. Phys. A610 (1996) 49.

\bibitem{hern} by  FOPI Coll., ncl-ex/0209012.



\bibitem{r98} N. Xu and K. Redlich, {\it in preparation}.

\bibitem{r101} J. Cleymans, B. K\"ampfer, and S. Wheaton, QM2002
contribution, Nucl. Phys. A (in print), hep-ph/0208247; Contributed to
30th International Workshop on Gross Properties of Nuclei and Nuclear
Excitation: Hirschegg 2002: Ultrarelativistic Heavy Ion Collisions,
Hirschegg, Germany, 13-19 Jan 2002. hep-ph/0202134; S. Yacoob and
J. Cleymans, hep-ph/0208246.


\bibitem{r101n} J. Cleymans, J. Phys. G28 (2002) 1575.

\bibitem{hwa} R. Hwa, and C.B. Yang, Phys. Rev. C66 (2002) 064903.

\bibitem{r43} E. Andersen, et al., WA97 Coll.,
Phys. Lett. B449 (1999) 401.

\bibitem{r44}
 N. Carrer, NA57 Coll.,  Nucl. Phys. A698 (2002) 29.

\bibitem{r104} S. Kabana,  et al., NA52 Coll., J.
Phys. G27 (2001) 495.

\bibitem{soff} S. Soff, et al., J. Phys. G27 (2001) 449.

\bibitem{bravina}  L. Bravina,  J. Phys. G27 (2001) 449.

\bibitem{vance} S.E. Vance,  et al., Phys. Rev. Lett. 83
(1999) 1735; J. Phys. G27 (2001) 627.

\bibitem{koxi} S. Pal, C.M. Ko, J.M. Alexander, P. Chung, and R.A. Lacey,
nucl-th/0211020.
\bibitem{bleicher} M. Bleicher, W. Greiner, H. St\"ocker and N. Xu,
Phys. Rev. C62 (2000) 061901.





\bibitem{lin} Z. Lin,  et al., Phys. Rev. C64 (2001) 011902.

\bibitem{volker} V. Koch, hep-th/0210070.




%%%%%%%%%%%%%%%%%%%%%%%%%%%%%
%


\end{thebibliography}
\end{document}